\documentclass{book}
\usepackage[utf8]{inputenc}

\usepackage[bookmarks=true,
     pdfnewwindow=true,      
     colorlinks=true,    
     linkcolor=black,     
     citecolor=xlinkcolor,     
     filecolor=xlinkcolor,  
     urlcolor=xlinkcolor,      
     final=true,
 ]{hyperref}

\usepackage{wasysym}
\usepackage{siunitx}

\usepackage[letterpaper, portrait, margin=1in]{geometry}

\usepackage{setspace}
\usepackage{lipsum}
\usepackage{float}
\usepackage{graphicx}
\usepackage{lscape}
\usepackage[printonlyused]{acronym}

\makeatletter
\newcommand*{\org@overidelabel}{}
\let\org@overridelabel\@verridelabel
\@ifpackagelater{acronym}{2015/03/21}{
  \renewcommand*{\@verridelabel}[1]{%
    \@bsphack
    \protected@write\@auxout{}{\string\AC@undonewlabel{#1@cref}}%
    \org@overridelabel{#1}%
    \@esphack
  }%
}{
  \renewcommand*{\@verridelabel}[1]{%
    \@bsphack
    \protected@write\@auxout{}{\string\undonewlabel{#1@cref}}%
    \org@overridelabel{#1}%
    \@esphack
  }%
}
\makeatother

\acrodef{ac}{Acronym}

\usepackage{amssymb} 

\usepackage{xcolor}
\definecolor{xlinkcolor}{cmyk}{0,0,0,1} 
\PassOptionsToPackage{hyphens}{url}

\usepackage[percent]{overpic}
\usepackage{color}

\usepackage{url}

\usepackage{natbib}
\usepackage{enumitem}

\usepackage{makecell} 
\usepackage{longtable} 

\newcommand{\dissertationTitle}{CHASING TAILS: ACTIVE ASTEROID, CENTAUR, AND\\ QUASI-HILDA DISCOVERY WITH ASTROINFORMATICS AND CITIZEN SCIENCE}

\newcommand{\degr}{^\circ}
\newcommand{\arcsec}{'}

\newcommand{\allthumbsSAFARI}{15,600}
\newcommand{\uniquethumbsSAFARI}{11,703}

\newcommand{\fitscountSAFARI}{35,640}
\newcommand{\gifcountSAFARI}{3029}

\newcommand{\sublimate}{$\wr\hspace{-0.5mm}\wr\hspace{-0.5mm}\wr$}

\newcommand{\farcsec}{\hbox{$.\!\!^{\prime\prime}$}}
\newcommand{\farcmin}{\hbox{$.\!\!^{\prime}$}}
\newcommand{\fdegree}{\hbox{$.\!\!^{\circ}$}} 

\newcommand{\og}{2014~OG$_{392}$}
\newcommand{\ogColor}{$1.64\pm0.4$} 

\usepackage{sectsty} 
\chapternumberfont{\large} 
\chaptertitlefont{\large}
\chapterfont{\large}
\allsectionsfont{\large}
\subsectionfont{\large}
\subsubsectionfont{\normalsize}
\paragraphfont{\normalsize}

\usepackage{calc}
\newlength{\okinalen}
\newcommand{\omuamua}{\hbox to.666\okinalen{\hss`\hss}Oumuamua}

\newcounter{rownumber}
\newcommand{\rlabel}[1]{\refstepcounter{rownumber}\label{#1}}
\newcounter{tabfoot}

\newcommand{\fakechapter}[1]{%
  \par\refstepcounter{chapter}
  \addcontentsline{toc}{chapter}{\protect\numberline{\thechapter}#1}
}

\usepackage[figuresleft]{rotating} 
\usepackage{everypage}
\def\PageTopMargin{1in}
\def\PageLeftMargin{1in}
\newcommand\atxy[3]{%
 \AddThispageHook{\smash{\hspace*{\dimexpr-\PageLeftMargin-\hoffset+#1\relax}%
  \raisebox{\dimexpr\PageTopMargin+\voffset-#2\relax}{#3}}}}

\begin{document}

\begin{centering}

\dissertationTitle{}

\thispagestyle{empty} 

\vspace{10mm}

\textit{By} Colin Orion Chandler

\vspace{10mm}

A Dissertation

Submitted in Partial Fulfillment

of the Requirements for the Degree of 


Doctor of Philosophy

in Astronomy and Planetary Science

\vspace{10mm}

Northern Arizona University

\vspace{5mm}

August 2022

\vspace{8mm}

\textit{Approved:}

Chadwick Aaron Trujillo, Ph.D., Chair

Jonathan G. Fortney, Ph.D.

Michael Gowanlock, Ph.D.

Henry H. Hsieh, Ph.D.

Tyler D. Robinson, Ph.D.

David E. Trilling, Ph.D.

\end{centering}

\doublespacing

\fakechapter{Abstract}

\begin{center}
    \large
    ABSTRACT
    
    \large
    \dissertationTitle{}
    
    \large
    \vspace{2mm}
    COLIN ORION CHANDLER
    
\end{center}

The discovery of activity emanating from asteroid (4015)~Wilson-Harrington in 1950 \citep{harrisCometNotesComet1950} prompted astronomers to realize that comet-like activity, such as comae and tails, is not limited to comets. Fewer than 30 of these ``active asteroids'' have been discovered \citep{chandlerSAFARISearchingAsteroids2018} in the last 70 years, yet they promise to hold clues about fundamental physical and chemical processes at play in our solar system \citep{jewittActiveAsteroids2012,hsiehMainbeltCometsPanSTARRS12015}. Activity is attributed to sublimation for roughly half of these objects, highlighting asteroids as a ``volatile reservoir.'' In this context a volatile reservoir is any dynamical group of bodies in the solar system that is known to harbor volatiles. Understanding the past and present volatile distribution in the solar system has broad implications ranging from informing future space exploration programs to helping us understand how planetary systems form with volatiles prerequisite to life as we know it, especially water. Notably, the origin of Earth's water is essentially unknown, although it is now believed that asteroids account for at least some of the terrestrial volatile budget \citep{alexanderOriginInnerSolar2017}.

A second volatile reservoir came to light following the 1977 discovery of Centaur (2060)~Chiron \citep{kowalSlowMovingObjectKowal1977}. Centaurs, found between the orbits of Jupiter and Neptune, are thought to be icy objects originating from the Kuiper Belt, a circumstellar region between the orbit of Neptune (30~au) and about 50~au from the Sun \citep{jewittActiveCentaurs2009}. The Kuiper Belt is roughly 200 times more massive than the Asteroid Belt. Nevertheless, active Centaurs are also rare, with fewer than 20 discovered to date \citep{chandlerCometaryActivityDiscovered2020a}.

We set out to increase the number of known active objects in order to (1) enable the study of these active objects as populations, and (2) search for new volatile reservoirs. I proposed to the \acs{NSF} \ac{GRFP} to create a Citizen Science project designed to carry out an outreach program while searching through millions of images of known asteroids in order to find previously unknown active objects. My proposal was selected for funding, and on 31 August 2022 we successfully launched \textit{Active Asteroids} (\url{http://activeasteroids.net}), a \acs{NASA} Partner, and discoveries have been abundant ever since.

In this dissertation I present (1) \ac{HARVEST}, a pipeline that extracts images of known solar system objects for presentation to Citizen Scientists, (2) our proof-of-concept demonstrating \ac{DECam} images are well-suited for activity detection \citep{chandlerSAFARISearchingAsteroids2018}, (3) how we discovered a potential new recurrent activity mechanism \citep{chandlerSixYearsSustained2019}, (4) a Centaur activity discovery plus a novel technique for estimating which species are sublimating \citep{chandlerCometaryActivityDiscovered2020a}, (5) how our discovery of an additional activity epoch for an active asteroid enabled us to classify the object as a member of the \acf{MBC} \citep{chandler2483702005QN2021}, a rare ($<10$) active asteroid subset that orbits in the Asteroid Belt that is known for sublimation-driven activity, (6) a dynamical pathway that can explain the presence of some of the active asteroids, and (7) the Citizen Science project \textit{Active Asteroids}, including initial results.

\clearpage
\singlespacing
\chapter*{Copyright}
\section{Published Works}
The following copyright statements apply to each respective published manuscript.

\vspace{3mm}
\noindent\textbf{Chapter \ref{chap:SAFARI} -- Manuscript I} 
\\
\noindent \textit{Searching Asteroids for Activity Revealing Indicators (SAFARI)} \citep{chandlerSAFARISearchingAsteroids2018}:

\begin{quote}
    \textit{This is the Accepted Manuscript version of an article accepted for publication in Publications of the Astronomical Society of the Pacific.  IOP Publishing Ltd is not responsible for any errors or omissions in this version of the manuscript or any version derived from it. The Version of Record is available online at} \url{https://iopscience.iop.org/article/10.1088/1538-3873/aad03d}\textit{.}
\end{quote}

\noindent\textbf{Chapter \ref{chap:Gault} -- Manuscript II}
\\
\noindent \textit{Six Years of Sustained Activity from Active Asteroid (6478)~Gault} \citep{chandlerSixYearsSustained2019}

\begin{quote}
    \textit{This is the Accepted Manuscript version of an article accepted for publication in Astrophysical Journal Letters.  IOP Publishing Ltd is not responsible for any errors or omissions in this version of the manuscript or any version derived from it.  The Version of Record is available online at}  \url{https://iopscience.iop.org/article/10.3847/2041-8213/ab1aaa}\textit{.}
\end{quote}

\noindent\textbf{Chapter \ref{chap:2014OG392} -- Manuscript III}
\\
\noindent \textit{Cometary Activity Discovered on a Distant Centaur: A Nonaqueous Sublimation Mechanism} \citep{chandlerCometaryActivityDiscovered2020a}

\begin{quote}
    \textit{This is the Accepted Manuscript version of an article accepted for publication in Astrophysical Journal Letters.  IOP Publishing Ltd is not responsible for any errors or omissions in this version of the manuscript or any version derived from it.  The Version of Record is available online at }\url{https://iopscience.iop.org/article/10.3847/2041-8213/ab7dc6}\textit{.}
\end{quote}

\noindent\textbf{Chapter \ref{chap:2005QN173} -- Manuscript IV}
\\
\textit{Recurrent Activity from Active Asteroid (248370) 2005~QN173: A Main-belt Comet} \citep{chandlerRecurrentActivityActive2021a}
\begin{quote}
    \textit{This is the Accepted Manuscript version of an article accepted for publication in Astrophysical Journal Letters.  IOP Publishing Ltd is not responsible for any errors or omissions in this version of the manuscript or any version derived from it.  The Version of Record is available online at }\url{https://iopscience.iop.org/article/10.3847/2041-8213/ac365b}\textit{.}
\end{quote}

\section{Unpublished Works}
The remaining chapters will be submitted to the American Astronomical Society journals for publication as soon as possible: 
\\

\noindent\textbf{Chapter \ref{chap:282P} -- Manuscript V}
\\
\textit{Active Asteroid Origin Insights from Transition Object (323137) 2003 BM80} 




\doublespacing

\clearpage
\chapter*{Acknowledgements}

\textit{Note: Manuscript-specific acknowledgements are found at the end of each corresponding chapter.}

\section{In Memoriam}
\textit{In memoriam} Jean E. Buethe (Flagstaff, Arizona), Nadine G. Barlow (Northern Arizona University), Adam P. Showman (University of Arizona) and Kazuo Kinoshita (Japan), all of whom encouraged and influenced my work.

\section{Indigenous Land Acknowledgement} Virtually all of the work described in this dissertation took place on lands important to indigenous cultures, but which has overwhelmingly been stolen from them. Here is an incomplete list of communities impacted in places where I have done work for this dissertation. \textbf{You can help by supporting organizations such as the 	Southwest Indian Relief Council (}\url{http://www.nativepartnership.org}) https://www.yakama.com (\url{https://engage.collegefund.org}\textbf{).}

The \ac{NAU} Mountain Campus is in Flagstaff, Arizona at the base of the San Francisco Peaks. This area is home to many peoples, including the Havasupai Tribe\footnote{\url{https://theofficialhavasupaitribe.com}}, Hopi Tribe\footnote{\url{https://www.hopi-nsn.gov}}, Hualapai Tribe\footnote{\url{https://hualapai-nsn.gov}}, Kaibab Band of Paiutes\footnote{\url{https://kaibabpaiute-nsn.gov}}, the Navajo Nation\footnote{\url{https://www.navajo-nsn.gov}}, the San Juan Southern Paiute Tribe\footnote{\url{https://www.sanjuanpaiute-nsn.gov}}. 
I have traveled through areas of the Zuni Tribe\footnote{\url{https://www.ashiwi.org}} and the Yavapai-Prescott Indian Tribe\footnote{\url{https://www.ypit.com}}. 
I have observed at \ac{MGIO}, a mountain stolen from the San Carlos Apache\footnote{\url{http://www.sancarlosapache.com/home.htm}} and sacred to others, including the White Mountains Apache Tribe\footnote{\url{http://www.wmat.us/}}. 
I have traveled through and/or attended conferences and workshops on land in Arizona taken from the Tohono O'Odham Nation\footnote{\url{http://www.tonation-nsn.gov/}}, Gila River Indian Community\footnote{\url{https://www.gilariver.org/}} and undoubtedly many others. Agencies that have provided me with funding also benefit directly and indirectly from ceded lands, such as those of the Yakama Nation\footnote{\url{https://www.yakama.com}}.

\section{Individuals and Groups}
I thank my partner, Dr.\ Mark Jesus Mendoza Magbanua (University of California San Francisco), for his frequent feedback, critical insights, and constant encouragement. I thank my parents, Arthur and Jeanie Chandler, for their continued support that enabled my academic endeavors, and for having the foresight to give me ``Orion'' as a middle name. I also appreciate Corey and Atsuko, Charles and Van, and Julie and Praxis for their enthusiasm all along the way. Many thanks to my best friend, Robert Reich, for his continuous encouragement and compassion. Thanks also to Jerome, Rushmore and Ziggy Diaz for being great friends. 
Thanks to my former business partner and great friend Bill Bowker (Fog City Mac, LLC) for allowing me to move on and pursue my path in astronomy.

Many thanks to all of my committee members for guiding me through this entire process. A special thanks to Chad, David, and Ty, all of whom provided crucial guidance and support during my time at \ac{NAU}. A special thank you to Prof. Cristina Thomas of \ac{NAU} for including me in \ac{DART} work that helped me to become a better photometrist.

Many thanks to Dr. Annika L. Gustafsson of \ac{NAU}, Lowell Observatory, and \ac{SwRI}, for countless hours together in classes, teaching, sharing world-class telescopes, attending conferences, and collaborating on research. Much appreciation goes to Aaron Weintraub of \ac{NAU} for being the best of friends and a supportive colleague from start to finish of this work. Thanks to both for encouraging me to go ``all in'' on my \ac{NSF} \ac{GRFP}. I am eternally grateful for James D. Windsor of \ac{NAU} for great times, helping me balance my life again through cycling and, along with his wife Cheyann, for literally saving my life.

Many, many thanks to Will Oldroyd for collaborating on papers, proposals, observing, hosting workshops, attending conferences, cooking, praying, and catching mice. A very special thanks to Jay Kueny (Steward Observatory, University of Arizona) for his commitment to the Citizen Science project and all that you taught me along the way. Thanks also for stepping in to help observe at the last minute on more than one occasion and at more than one observatory. Thanks also to Will Burris of \ac{SDSU} for your help with \textit{Active Asteroids} classification analysis.

A special thank you to \textit{Active Asteroids} forum Moderator Elisabeth Baeten (Belgium) who has greatly enhanced the success of our Citizen Science. Thank you also Cliff Johnson (Zooniverse) and Marc Kuchner (NASA), both of whom provided invaluable insights into Citizen Science and encouraged this project to move forward.

Thank you to the \ac{LSST} community for welcoming me, especially Ranpall Gill (Rubin Observatory), Henry Hsieh (Planetary Science Institute), Meg Schwamb (Queen's University Belfast), Agata Rożek (University of Edinburgh), and Mario Jurić and Andrew Connolly of the \ac{DiRAC} Institute and University of Washington).

Thank you Nathan Smith and Lori Pigue for independently inspiring and fueling my interest in meteorites and minerals. Thank you Schuyler Borges (\ac{NAU}) for helping me to become a better member of the community and inspiring me to continue advocating for diversity, equity and inclusivity. Thanks Tony and Catherine (\acs{NAU}) for being great colleagues and for the fun times on the slopes and in the pool. Thank you Trevor Cotter (NAU, McGill University) for all the adventures and, especially, for introducing me to cross country skiing. Thanks to fellow \acs{NAU} students Erin Aadland, Haylee Archer, Dan Avner, Lauren Biddle, Helen Eifert, Anna Engle, Oriel Humes, Joel Johnson, Ari Koeppel, David Kelly, Sarah Lamm, Audrey Martin, Lauren McGraw, Robyn Meier, Alissa Roegge, Raaman Nair, Kathryn Neugent, Garrett Thompson, and Mike Zeilnhofer for being supportive colleagues and classmates. A special thanks to Christian Joey Tai Udovicic for listening to my qualifying examination talk many times.

Thank you to my friends from the \ac{SFSU} \ac{PAC} for your encouragement, especially Daniel Steckhahn of \ac{UCB}, Ryan Rickards-Vaught of \ac{UCSD}, Sarah Deveny of {TTU}, and Michelle Howard. Many thanks to \ac{SFSU} faculty for their inspiration, especially Joseph Baranco, Adrienne Cool, Jeff Greensite, Ron Marzke, Weining Man, Chris McCarthy, and Barbara Neuhauser. Caroline Alcantra also helped me with all things administrative at \ac{SFSU} and I am eternally grateful. Thank you Alan Fisk (\ac{SFSU}) and the \ac{DPRC} for your guidance and helping me succeed.

Thank you Phil Massey (Lowell Observatory) for teaching me to become a better astronomer. Thank you Will Grundy, Stephen Levine, Jenn Hanley, Michael West, Diedre Hunter, Jeff Hall, Nick Moskovitz, Audrey Thirouin, Joe Llama, Lisa Prato, Gerard van Belle, Amanda Bosch, and Dave Schleicher for welcoming me at Lowell and vastly improving my experience in astronomy.

Thank you to Guy Consolmagno, Rich Boyle, Paul Gabor, Gary Gray and Chris Johnson for all of your support at the Vatican Observatory. Many thanks to Jenny Power of \ac{MGIO} for all the support with \ac{LBT} observation planning and execution. Thanks to everyone at the \ac{LBT} for accommodating us on-site on multiple occasions, especially Rick Hansen and David Huerta. A special thanks to all of the telescope operators at the \acf{LDT} for all their help through the years. Thanks also to Matt Holman (Harvard) and Jerome Berthier (\ac{IMCCE}) for their support with the \ac{MPC} and SkyBot, respectively.


Thank you to \ac{NAU} Pres. Rita Cheng for supporting the Astronomy and Planetary Science Ph.D. program at \ac{NAU}, the \ac{NAU}--\ac{LDT} partnership, and the Presidential Fellowship Program, all of which were essential to this work. Thank you Maribeth Watwood for overseeing this program and many others.

Thank you Ed Anderson (\ac{NAU}), Lisa Chien (\ac{NAU}), John Kistler (\ac{NAU}), David Koerner (\ac{NAU}), and Mark Salvatore (\ac{NAU}) for helping me to become a better teacher. Thanks Anna Engle (\ac{NAU}) for collaborating with teaching during the pandemic. Thanks to my many undergraduate teaching assistants, especially Anna Ross-Browning (University of Iowa) and Lisa Matrecito (\ac{NAU}), for directly enabling me to be a better teacher.  Thanks to all of my astronomy and physics students at \ac{NAU} who all enriched my experience at \ac{NAU}. Thanks to department administrators Elizabeth Massey, Judene McLane, and Alix Ford for all their help throughout my time at \ac{NAU}. Thanks also to Lara Schmit (Merriam Powell Institute, \ac{NAU}) for being a great neighbor during the pandemic and for all your advice in navigating \ac{NAU} systems. 
The unparalleled support given by Monsoon high performance computing cluster administrator Christopher Coffey of \ac{NAU}, as well as the High Performance Computing Support team, truly facilitate the scientific process throughout all of my work.

Thank you to Barry Lutz (\ac{NAU}) for funding and supporting the \ac{BLT} which was part of my education as well as my teaching endeavors. Thank you Stephen Tegler (\ac{NAU}) for your insights into Centaur colors and observational techniques, and Mark Loeffler (\ac{NAU}) for encouraging me to question sources to their very beginnings and the value of astrochemistry. Thank you Christopher Edwards (\ac{NAU}) for helping me better understand spectroscopy and advocating for me and my colleagues. Thanks also Chris Mann (\ac{NAU}) for helping me better understand the optics behind the telescopes I rely upon. Thanks to Devon Burr and Josh Emery for all your encouragement and advocacy at \ac{NAU}.

The Trilling Research Group at \ac{NAU} has always been an invaluable resource for insights which substantially enhanced this work. A special thanks to members Annika Gustaffson (\ac{SwRI}), Andy J López Oquendo (\ac{NAU}), Ryder Sluss (\ac{NAU}), Joey Chatelain of \ac{LCO}, Samuel Navarro-Meza (University of Mexico, \ac{NAU}), Daniel Kramer (\ac{NAU}), Andrew McNeil (\ac{NAU}), Maggie McAdam of \ac{SOFIA}, Connor Auge (University of Hawaii), and Michael Mommert (\ac{NAU}).

The HabLab research group (Ty Robinson) at \ac{NAU} and \ac{UA} is full of science and support. A special thanks to members Arnaud Salvadore, Amber Young, Megan Gialluca (\ac{UW}), Malik Bossett, Chris Wolfe, Patrick Tribbett, and Shih-Yun Tang.

Thank you Stephen Kane (University of California Riverside) for encouraging me to pursue my academic dreams and laying the foundation for successful research. 
Thank you Erik Mamajek (NASA \ac{JPL})for invaluable advice that helps keep my focus on science goals.
Thank you Michael Way (NASA Goddard), Brandon Cruickshank (\ac{NAU}), Greg McGuffey, Kathy \& John Eastwood, all of whom inspire balance between outdoor activity and other pursuits.

Thank you to Hannah Nuñez and Dr. Matt Wise of \ac{NAU} for helping to keep me healthy during my time at \ac{NAU}. Thanks also Fr. Matt Lowry and the \ac{NAU} Catholic Jacks for providing a welcoming spiritual environment on campus.

\section{Citizen Scientists}
\label{ack:sec:citSci}
I wish to thank the following individuals for their efforts on the \textit{Active Asteroids} project. These individuals represent our most active classifiers, classified thumbnail images included in this work, or both. A special thanks goes to our top classifier, 
Michele T. Mazzucato, FRAS (Sesto Fiorentino, Italy). 

Thank you A. J. Raab (Seattle, USA), 
Alice Juzumas (São Paulo, Brazil), 
Angelina A. Reese (Sequim, USA), 
Arttu Sainio (Järvenpää, Finland), 
Bill Shaw (Fort William, Scotland), 
\texttt{@Boeuz} (Penzberg, Germany), 
Brenna Hamilton (DePere, USA), 
Brian K Bernal (Greeley, USA), 
Carl Groat (Okeechobee, USA), 
Clara Garza (West Covina, USA), 
C. J. A. Dukes (Oxford, United Kingdom), 
Dr. David Collinson (Mentone, Australia), 
Edmund Frank Perozzi (Glen Allen, USA), 
\texttt{@EEZuidema} (Driezum, Netherlands), 
Dr. Elisabeth Chaghafi (Tübingen, Germany), 
\texttt{@graham\_d} (Hemel Hempstead, England), 
Ivan A. Terentev (Petrozavodsk, Russia), 
Juli Fowler (Albuquerque, USA), 
Leah Mulholland (Peoria, USA) 
M. M. Habram-Blanke (Heidelberg, Germany), 
Marvin W. Huddleston (Mesquite, USA), 
Michael Jason Pearson (Hattiesburg, USA), 
Milton K. D. Bosch, MD (Napa, USA), 
\texttt{@mitch} (Chilliwack, Canada), 
Patricia MacMillan (Fredericksburg, USA), 
Petyerák Jánosné (Fót, Hungary), 
R. Banfield (Bad Tölz, Germany), 
Scott Virtes (Escondido, USA), 
Sergey Y. Tumanov (Glazov, Russia), 
Stikhina Olga Sergeevna (Tyumen, Russia), 
Thorsten Eschweiler (Übach-Palenberg, Germany), 
Tiffany Shaw-Diaz (Dayton, USA), 
Timothy Scott (Baddeck, Canada), 
and 
Virgilio Gonano (Udine, Italy). 

\section{Funding}
This material is based upon work supported by the \acl{NSF} \acl{GRFP} under grant No.\ (2018258765). Any opinions, findings, and conclusions or recommendations expressed in this material are those of the author(s) and do not necessarily reflect the views of the National Science Foundation.  C.O.C., H.H.H. and C.A.T. also acknowledge support from the NASA Solar System Observations program (grant 80NSSC19K0869).

This work was supported in part by NSF awards 1461200, 1852478, and 1950901 (\acs{NAU} REU program in astronomy and planetary science). 

\section{General Acknowledgements}
Computational analyses were run on Northern Arizona University's Monsoon computing cluster, funded by Arizona's Technology and Research Initiative Fund. This work was made possible in part through the State of Arizona Technology and Research Initiative Program. World Coordinate System (WCS) corrections facilitated by the \textit{Astrometry.net} software suite \citep{langAstrometryNetBlind2010}.

This research has made use of data and/or services provided by the International Astronomical Union's Minor Planet Center. 
This research has made use of NASA's Astrophysics Data System. 
This research has made use of The Institut de M\'ecanique C\'eleste et de Calcul des \'Eph\'em\'erides (IMCCE) SkyBoT Virtual Observatory tool \citep{berthierSkyBoTNewVO2006}. 
This work made use of the {FTOOLS} software package hosted by the NASA Goddard Flight Center High Energy Astrophysics Science Archive Research Center. 
This research has made use of SAOImageDS9, developed by Smithsonian Astrophysical Observatory \citep{joyeNewFeaturesSAOImage2006}. 
This work made use of the Lowell Observatory Asteroid Orbit Database \textit{astorbDB} \citep{bowellPublicDomainAsteroid1994,moskovitzAstorbDatabaseLowell2021}. 
This work made use of the \textit{astropy} software package \citep{robitailleAstropyCommunityPython2013}.

This project used data obtained with the \acf{DECam}, which was constructed by the \acf{DES} collaboration. Funding for the \acs{DES} Projects has been provided by the US Department of Energy, the US \acl{NSF}, the Ministry of Science and Education of Spain, the Science and Technology Facilities Council of the United Kingdom, the Higher Education Funding Council for England, the National Center for Supercomputing Applications at the University of Illinois at Urbana-Champaign, the Kavli Institute for Cosmological Physics at the University of Chicago, Center for Cosmology and Astro-Particle Physics at the Ohio State University, the Mitchell Institute for Fundamental Physics and Astronomy at Texas A\&M University, Financiadora de Estudos e Projetos, Fundação Carlos Chagas Filho de Amparo à Pesquisa do Estado do Rio de Janeiro, Conselho Nacional de Desenvolvimento Científico e Tecnológico and the Ministério da Ciência, Tecnologia e Inovação, the Deutsche Forschungsgemeinschaft and the Collaborating Institutions in the Dark Energy Survey. The Collaborating Institutions are Argonne National Laboratory, the University of California at Santa Cruz, the University of Cambridge, Centro de Investigaciones Enérgeticas, Medioambientales y Tecnológicas–Madrid, the University of Chicago, University College London, the \acs{DES}-Brazil Consortium, the University of Edinburgh, the Eidgenössische Technische Hochschule (ETH) Zürich, Fermi National Accelerator Laboratory, the University of Illinois at Urbana-Champaign, the Institut de Ciències de l’Espai (IEEC/CSIC), the Institut de Física d’Altes Energies, Lawrence Berkeley National Laboratory, the Ludwig-Maximilians Universität München and the associated Excellence Cluster Universe, the University of Michigan, \acs{NSF}’s \acs{NOIRLab}, the University of Nottingham, the Ohio State University, the OzDES Membership Consortium, the University of Pennsylvania, the University of Portsmouth, \acs{SLAC} National Accelerator Laboratory, Stanford University, the University of Sussex, and Texas A\&M University.

Based on observations at \acl{CTIO}, \acs{NSF}’s \acs{NOIRLab} 

(
NOAO Prop. ID 2012B-0001, PI: Frieman; 
NOAO Prop. ID 2013A-0327, PI: Rest; 
NOAO Prop. ID 2013B-0453, PI: Sheppard; 
NOAO Prop. ID 2013B-0536, PI: Allen; 
NOAO Prop. ID 2014A-0303, PI: Sheppard; 
NOAO Prop. ID 2014A-0479; PI: Sheppard 
NOAO Prop. ID 2014B-0303, PI: Sheppard; 
NOAO Prop. ID 2014B-0404, PI: Schlegel; 
NOAO Prop. ID 2015A-0351, PI: Sheppard;
NOAO Prop. ID 2015A-0620, PI: Bonaca; 
NOAO Prop. ID 2015B-0265, PI: Sheppard; 
NOAO Prop. ID 2016A-0190, PI: Dey; 
NOAO Prop. ID 2016A-0401, PI: Sheppard; 
NOAO Prop. ID 2016B-0288, PI: Sheppard; 
NOAO Prop. ID 2017A-0367, PI: Sheppard; 
NOAO Prop. ID 2017B-0307, PI: Sheppard; 
NOAO Prop. ID 2019A-0272, PI: Zenteno; 
NOAO Prop. ID 2019A-0305, PI: Drlica-Wagner; 
NOAO Prop. ID 2019A-0337, PI: Trilling; 
NOAO Prop. ID 2019B-0323, PI: Zenteno
), 
which is managed by the \acf{AURA} under a cooperative agreement with the \acl{NSF}.

This research has made use of the NASA/IPAC Infrared Science Archive, which is funded by the National Aeronautics and Space Administration and operated by the California Institute of Technology.

The Legacy Surveys consist of three individual and complementary projects: the Dark Energy Camera Legacy Survey (DECaLS; Proposal ID \#2014B-0404; PIs: David Schlegel and Arjun Dey), the Beijing-Arizona Sky Survey (BASS; NOAO Prop. ID \#2015A-0801; PIs: Zhou Xu and Xiaohui Fan), and the Mayall z-band Legacy Survey (MzLS; Prop. ID \#2016A-0453; PI: Arjun Dey). DECaLS, BASS and MzLS together include data obtained, respectively, at the Blanco telescope, Cerro Tololo Inter-American Observatory, NSF's NOIRLab; the Bok telescope, Steward Observatory, University of Arizona; and the Mayall telescope, Kitt Peak National Observatory, NOIRLab. The Legacy Surveys project is honored to be permitted to conduct astronomical research on Iolkam Du'ag (Kitt Peak), a mountain with particular significance to the Tohono O'odham Nation. BASS is a key project of the Telescope Access Program (TAP), which has been funded by the National Astronomical Observatories of China, the Chinese Academy of Sciences (the Strategic Priority Research Program ``The Emergence of Cosmological Structures'' Grant \# XDB09000000), and the Special Fund for Astronomy from the Ministry of Finance. The BASS is also supported by the External Cooperation Program of Chinese Academy of Sciences (Grant \# 114A11KYSB20160057), and Chinese National Natural Science Foundation (Grant \# 11433005). The Legacy Survey team makes use of data products from the Near-Earth Object Wide-field Infrared Survey Explorer (NEOWISE), which is a project of the Jet Propulsion Laboratory/California Institute of Technology. NEOWISE is funded by the National Aeronautics and Space Administration. The Legacy Surveys imaging of the DESI footprint is supported by the Director, Office of Science, Office of High Energy Physics of the U.S. Department of Energy under Contract No. DE-AC02-05CH1123, by the National Energy Research Scientific Computing Center, a DOE Office of Science User Facility under the same contract; and by the U.S. National Science Foundation, Division of Astronomical Sciences under Contract No. AST-0950945 to NOAO.

These results made use of the Lowell Discovery Telescope (LDT) at Lowell Observatory.  Lowell is a private, non-profit institution dedicated to astrophysical research and public appreciation of astronomy and operates the LDT in partnership with Boston University, the University of Maryland, the University of Toledo, Northern Arizona University and Yale University. The Large Monolithic Imager was built by Lowell Observatory using funds provided by the National Science Foundation (AST-1005313). NIHTS was funded by NASA award \#NNX09AB54G through its Planetary Astronomy and Planetary Major Equipment programs.

Based in part on data collected at Subaru Telescope and obtained from the SMOKA, which is operated by the Astronomy Data Center, National Astronomical Observatory of Japan \citep{2002ASPC..281..298B}.

Plots were primarily created using the Python tool \texttt{Matplotlib} \citep{hunterMatplotlib2DGraphics2007}. 
Python mathematical and array operations were frequently executed using the Python \texttt{NumPy} tool \citep{harrisArrayProgrammingNumPy2020}. We made use of Pandas dataframes \citep{mckinneyDataStructuresStatistical2010,rebackPandasdevPandasPandas2022}. We made use of the \texttt{Rebound} dynamical simulator \citep{reinREBOUNDOpensourceMultipurpose2012,reinHybridSymplecticIntegrators2019} with the \texttt{Mercury} \citep{chambersHybridSymplecticIntegrator1999} and \texttt{IAS15} \citep{reinIAS15FastAdaptive2015} integrators. We made use of the \texttt{SciPy} Python suite \citep{virtanenSciPyFundamentalAlgorithms2020}.

We used \texttt{Theli3}\footnote{\url{https://github.com/schirmermischa/THELI}} \citep{schirmerTHELIConvenientReduction2013} for some of our data reduction. 

\clearpage 
\singlespacing
\setcounter{tocdepth}{3} 
\tableofcontents
\doublespacing

\clearpage
\listoftables

\clearpage
\listoffigures

\clearpage
\chapter*{Dedication}

This dissertation is dedicated to my partner, Mark Jesus Mendoza Magbanua, and to my parents, Arthur and Jeanie Chandler; this work would not have been possible without their encouragement and support.

\clearpage
\chapter*{Preface}

The included manuscript chapters have been written to appear in peer-reviewed scientific journals. Redundancy in text is a result of reformatting to conform to University formatting requirements. References have been consolidated into a unified bibliography that appears at the end of this dissertation.

The following chapters have already been published:

Chapter \ref{chap:SAFARI} -- Manuscript I: ``Searching Asteroids for Activity Revealing Indicators (SAFARI)'' \citep{chandlerSAFARISearchingAsteroids2018}

Chapter \ref{chap:Gault} -- Manuscript II: ``Six Years of Sustained Activity from Active Asteroid (6478)~Gault'' \citep{chandlerSixYearsSustained2019}

Chapter \ref{chap:2014OG392} -- Manuscript III: ``Cometary Activity Discovered on a Distant Centaur: A Nonaqueous Sublimation Mechanism'' \citep{chandlerCometaryActivityDiscovered2020a}

Chapter \ref{chap:2005QN173} -- Manuscript IV: ``Recurrent Activity from Active Asteroid (248370) 2005~QN173: A Main-belt Comet'' \citep{chandlerRecurrentActivityActive2021a}

The remaining chapters have been submitted to the American Astronomical Society journals for publication as soon as possible: 

Chapter \ref{chap:282P} -- Manuscript V: ``Migratory Outbursting Quasi-Hilda Object 282P/(323137) 2003 BM80''

\pagenumbering{arabic}
\setcounter{page}{1}
\clearpage
\chapter{Introduction}
\label{chap:intro}
\acresetall

\section{Definitions}
\label{intro:sec:definitions}

It is important to define some terms that will be used throughout this work. ``Comet'' as a class of object is very loosely defined, with many individuals adopting essentially personal definitions of the term. For example, some people will consider anything seen with a tail as being a comet. Other people will only require that an object have an orbit typical of a comet class, such as \ac{JFC}, and the object will be called a comet even if activity has never been seen. It is not uncommon to mix visual and dynamical elements to define a type of comet, for example the Manx comets \citep{meech2013P2Pan2014}, defined as on a comet-like orbit but displaying little to no tail. For this work I define comets as objects that (1) exhibit activity and (2) are on orbits typically associated with dynamical classes with the word ``comet'' in them, including comet, long-period comet, short-period comet, \ac{JFC}, and \ac{QHC}.

As my adopted definition of comet is not based purely on orbital properties, I use the terms ``class'' or ``group'' to refer to collections of minor planets with common traits, most commonly orbital properties, but not necessarily limited to these attributes. I refer to volatile reservoirs as object classes with two or more objects known to harbor volatiles. These are a reservoir in the sense that comets and asteroids are two reservoirs thought to have supplied water to Earth in the past. Classes of so-called ``transition objects'' (defined in this work as objects only temporarily belonging to a particular class of objects) -- for example Centaurs (defined in Section \ref{intro:sec:activeCentaurs}) -- can still be viable reservoirs because these regions are being continually replenished with new objects (Centaurs in this example).

\section{Background}
\label{intro:sec:background}
Volatiles, such as water, are essential for life as we know it, yet fundamental knowledge about these materials, such as their present-day location within our solar system, and even what they are, is incomplete. This information is crucial for future space exploration and to help answer outstanding fundamental questions about the solar system, including Earth. For example, where did Earth's water originate? It is generally accepted that some water was already present when Earth formed, and that some additional quantity was delivered later, but not even a rough ratio is known. Comets were thought to be the only objects that delivered water post-formation, but the consensus today is that comets alone cannot account for the volume of water we find on Earth (see review, \citealt{alexanderOriginInnerSolar2017}). One explanation is that comets represent just one of multiple ``volatile reservoirs,'' classes of solar system bodies that harbor volatiles on or below their surfaces. Today, asteroids are considered likely contributors to the volatile budget on Earth.

\begin{figure*}
    \centering
    \begin{tabular}{cc}
         \includegraphics[width=0.45\columnwidth]{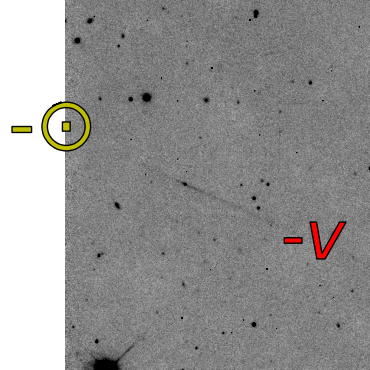} & 
         \includegraphics[width=0.45\columnwidth]{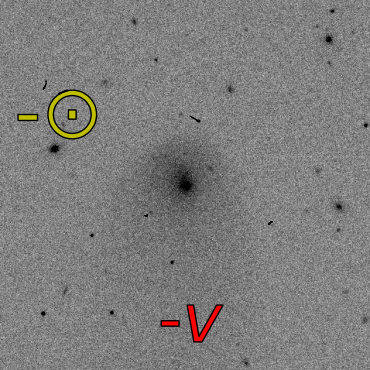}
    \end{tabular}
    \caption{Activity in the form of tails or a coma. Images credit: this work. 
    \textbf{Left:} 2005~QN$_{173}$ imaged on \ac{UT} 2021 December 8 at the \acf{VATT} at the \acf{MGIO} in Arizona (Prop. ID S165, PI Chandler). This image reveals that the object has two tails: one pointing away from the Sun (yellow $-\odot$) and one pointing opposite of the object's apparent direction of motion (red $-v$). These are the two directions most commonly associated with tails. \textbf{Right:} Comet C/2007 F4 (McNaught) displays a prominent coma in this image captured on \acs{UT} 2014 January 13 (Prop. ID 2012B-0001, PI Frieman).}
    \label{intro:fig:activeObjects}
\end{figure*}

Comets are known for their remarkable displays of \textit{cometary activity} (Figure \ref{intro:fig:activeObjects}), like a tail or a shroud of material known as a coma (plural: comae). This activity is typically associated with volatile sublimation, the direct phase transition of a material from solid to gas (e.g., when dry ice turns into carbon dioxide gas) at conditions typical on Earth. However, mechanisms other than sublimation can result in mass loss that takes the form of tails or comae. Here we use the term ``activity'' to describe any situation where a body loses material to space.

Surprisingly, comets are not the only objects known to display comet-like activity. (See review by \cite{jewittAsteroidCometContinuum2022} for a comprehensive discussion on the increasingly blurred lines between comets and asteroids.) As a result, other groups of objects, such as active asteroids and active Centaurs (discussed below), may represent viable volatile reservoirs in their own right. However, in sum fewer than 50 members of these active object groups have been discovered since the first active asteroid was identified in 1949 \citep{harrisCometNotesComet1950} and, as a result, it is virtually impossible to draw robust conclusions about the amount and type of volatiles held by these groups. In order to enable the study of potential volatile reservoirs as populations we set out to create a platform that facilitates discovering many additional active bodies, with a long-term goal to increase the numbers of known active minor planets by a factor of two or more. Here I describe the platform we created, as well as several discoveries we made along the way, including five that resulted in peer-reviewed publications \citep{chandlerSAFARISearchingAsteroids2018,chandlerSixYearsSustained2019,chandlerCometaryActivityDiscovered2020a,chandler2483702005QN2021}. We also include a link\footnote{\url{http://activeasteroids.net}} to the online component of the platform, thereby enabling \textit{you} to participate in this exciting scientific endeavor.

\section{Layout of the Solar System}
\label{intro:sec:layoutOfSolarSystem}

\begin{figure}
    \centering
    \includegraphics[width=0.85\linewidth]{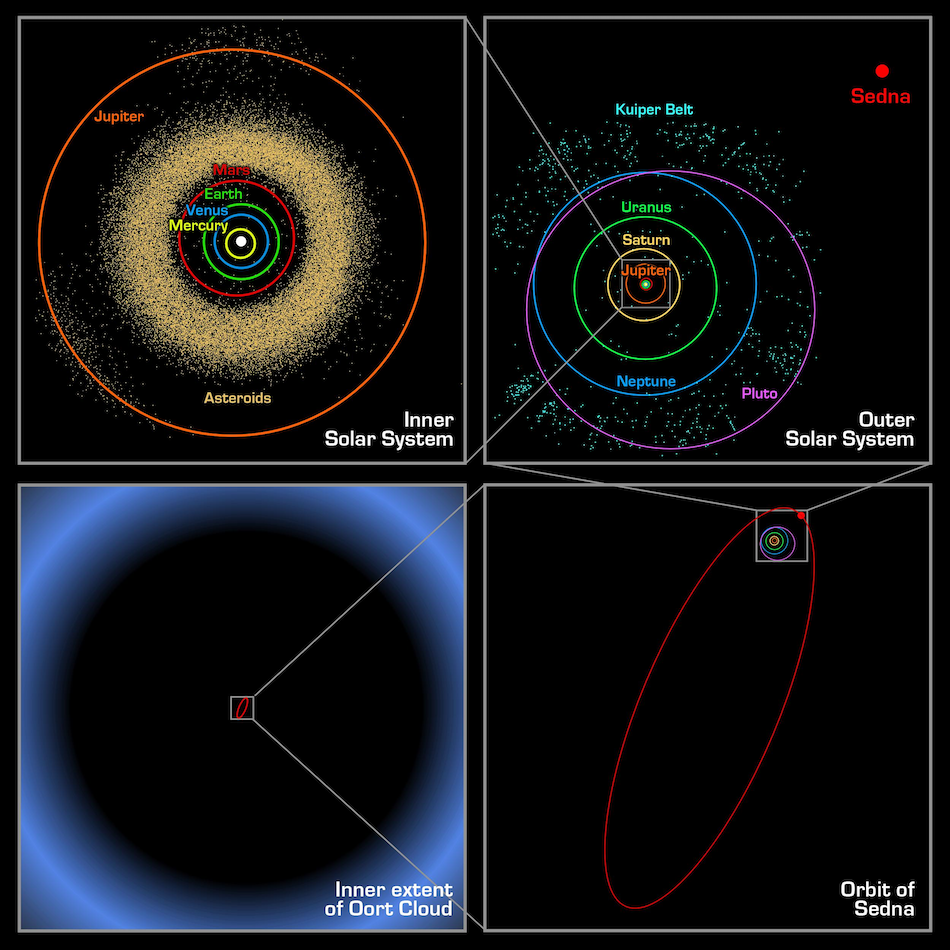}
    \caption{The solar system seen from four different scales. The inner solar system (top-left) is shown with the orbit of Jupiter (red circle) as the outermost orbit. Zooming out (top-right) provides a view of the outer solar system, which includes the orbit of Pluto and the Kuiper Belt. Zooming out again (bottom-right) the orbit of (90377)~Sedna can be seen, which gives context to the view of the inner Oort Cloud (bottom-left), the location from which most comets are thought to originate. Image credit: NASA/CalTech.}
    \label{intro:fig:solarSystem}
\end{figure}

\begin{table}[h]
    \centering
    \begin{tabular}{clrrS[table-format=3.3]}
            Symbol  & Name      & $a$   & $e$   & Mass\\
                    &           & [au]  &       & [M$_\oplus$]\\
                    \hline\hline
           \mercury & Mercury   & 0.5   & 0.21  & 0.06\\
           \venus   & Venus     & 0.7   & 0.01  & 0.81\\
           $\oplus$ & Earth     & 1.0   & 0.02  & 1.00\\
           \mars    & Mars      & 1.5   & 0.09  & 0.11\\
           \jupiter & Jupiter   & 5.2   & 0.05  & 318\\
           \saturn  & Saturn    & 9.5   & 0.05  & 95\\
           \uranus  & Uranus    & 19.2  & 0.05  & 15\\
           \neptune & Neptune   & 30.1  & 0.01  & 17\\
    \end{tabular}
    \caption{Orbital elements for the major planets of the solar system: the semi-major axis $a$, orbital eccentricity $e$, and the mass of the body in Earth masses $M_\oplus$.}
    \label{intro:tab:planets}
\end{table}

Kepler's First Law states that the planets orbit in ellipses (ovals). The average distance from the Sun to an object over one complete orbit is equal to the semi-major axis ($a$), the distance between the center of an ellipse and the farthest point from the center. Astronomers typically measure distances of planets and small solar system bodies in terms of \textit{astronomical units} (au), defined as the average distance between the centers of Earth and the Sun. 
Eccentricity $e$ describes how elongated an orbit is, with $e=0$ being a perfect circle, and $e=1$ a parabola and thus not a closed loop. The planets typically have low eccentricity (Table \ref{intro:tab:planets}), with a median of $e=$0.05, while comets typically have high eccentricity, on average $e=$0.9.

Other objects also orbit the Sun, and these are collectively referred to as minor planets (or small solar system bodies), or dwarf planets (e.g., Sedna). As of 1 July 2022 there are roughly 1.2 million known minor planets, of which fewer than 4,000 are comets. There are two circumstellar ``belts'' containing large numbers of minor planets. The Asteroid Belt (Figure \ref{intro:fig:solarSystem}) is found between the orbits of Mars and Jupiter, roughly between 2~au and 4~au. The Asteroid Belt has a mass of less than one thousandth of Earth's mass (e.g., \citealt{krasinskyHiddenMassAsteroid2002}). The Kuiper Belt (sometimes referred to as the Edgeworth-Kuiper Belt or the Trans-Neptunian Belt) extends from the orbit of Neptune (30~au) to around 50~au. Notably, the Kuiper belt is some 20 times wider in radial extent that the Asteroid Belt, and at one tenth the mass of Earth the Kuiper Belt is 200 times more massive than the Asteroid Belt  \citep{gladmanStructureKuiperBelt2001,pitjevaMassesMainAsteroid2018,diruscioAnalysisCassiniRadio2020}.

To date four active minor planet classes (other than comet classes) have members known to display comet-like activity. These are (1) comets, (2) main-belt asteroids, (3) Centaurs (icy bodies orbiting between 5~au and 30~au), (3) \acfp{NEO}, also known as \acfp{NEA}, 
(4) \acp{QHO}, a type of asteroid with orbits similar to the Hildas (which are in 3:2 orbital resonance with Jupiter), and (4) one interstellar object, designated 2I/Borisov \citep{borisovMPEC2019R106COMET2019}.

Comets are thought to originate from two sources: (1) the Kuiper Belt, and (2) the Oort cloud, a spherical cloud of objects orbiting between 2,000~au to 200,000~au from the Sun \citep{oortStructureCloudComets1950}. Comets are classified by two different means. The first is by recognizing their activity, an approach dating back thousands of years (see catalog by \citealt{kronkCometographyCatalogComets1999}), a technique still valid today. Notably, this definition is not based on orbital characteristics at all, and thus the class ``comet'' is not necessarily dynamically derived. Comets are also identified based on properties of their orbits through essentially two systems of dynamical classification.

(1) The period-based comet classifications are: (i) Hyperbolic comets, which may be interstellar in origin. These have enough momentum to leave the solar system, and so they do not have a period. (ii) Long-period comets have orbits longer than 200 years. (iii) Halley-type comets, named after the famed Halley's Comet, have periods ranging between 20 and 200 years. (4) Short-period comets have periods less than 20 years. These are also sometimes referred to as \acfp{JFC}.

(2) The other system for classifying comets makes use of an orbital metric that describes a body's close approach speed to Jupiter, and can be considered a descriptor of how strongly an orbit is influenced by Jupiter. The metric is known as Tisserand's Parameter with respect to Jupiter ($_\mathrm{J}$), and is described in detail in Chapter \ref{safari:sec:introduction}. Orbits constrained by $T_\mathrm{J}$ include \acp{JFC} (the $T_\mathrm{J}$ definition), having orbits that are strongly influenced by Jupiter.

\section{Active Asteroids}
\label{intro:sec:activeAsteroids}

For a more in-depth discussion of Active Asteroids, see Chapter \ref{safari:sec:introduction}.

\begin{figure}[h]
    \centering
    \begin{tabular}{cc}
         \includegraphics[width=0.45\linewidth]{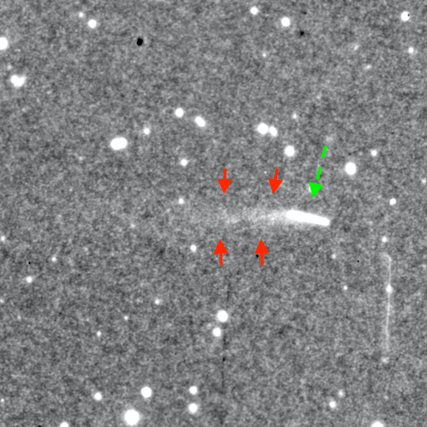} & \includegraphics[width=0.45\linewidth]{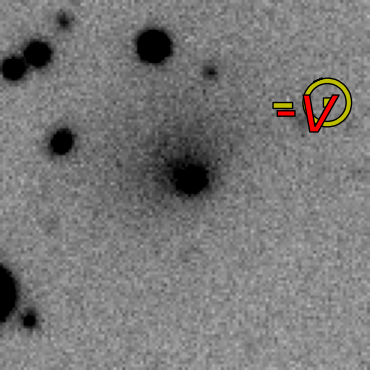}\\
    \end{tabular}
    \caption{\textbf{Left:} This image, taken on 1949 November 19, shows the first asteroid activity discovered, emanating from (4015)~Wilson Harrington (green dashed arrow) in the form of a tail (red arrows). Image Credit: ESO and Palomar Observatory (\url{https://www.eso.org/public/images/eso9212b/}). \textbf{Right:} The first active Centaur discovered (retroactively, following the 1977 discovery of the first identified Centaur, (2060) Chiron \citealt{kowal1977UB1977}), 29P/Schwassmann-Wachmann~1 \citep{schwassmannNEWCOMET1927}, shown here in a 60~s $R$-band image acquired with the \ac{PTF} Mosaic imager on \ac{UT} 2011 February 9 at \ac{PTF} (Prop. ID, PI Kulkarni). 29P was first identified as active in 1927 \citep{schwassmannNEWCOMET1927}, but the object was not considered a Centaur until after the discovery of Centaur (2060)~Chiron in 1977. Image Credit: this work.}
    \label{intro:fig:famousAAs}
\end{figure}

The first asteroid observed with cometary features was \ac{NEO} (4015)~Wilson-Harrington \citep{cunninghamPeriodicCometWilsonHarrington1950}. Astronomers identified a clear tail in images taken in 1949 (Figure \ref{intro:fig:famousAAs}). However, when astronomers were able to observe the object again, no activity was seen. Since then no conclusive evidence of further activity has been detected, despite considerable efforts \citep{degewij1979VAPhysical1980,chamberlin4015WilsonHarrington22011996,licandroSpitzerObservationsAsteroidcomet2009,ishiguroSearchCometActivity2011,urakawaPhotometricObservations107P2011}.

In 1996, nearly five decades later, the modern era of active asteroids was ushered in with the discovery of an active object orbiting within the asteroid belt, 133P/Elst-Pizarro \citep{elstComet1996N21996}. This object has been observed to be repeatedly active, especially near perihelion \citep{hsiehReturnActivityMainbelt2010} -- the point in an object's orbit when it is closest to the sun. Repeated periodic activity when an object is near to the Sun, the warmest period of an object's orbit, is strong evidence that the activity is sublimation-driven, and this object became the first to be designated a \ac{MBC}. The \acp{MBC} are a subset of active asteroids that orbit within the Asteroid Belt and exhibit sublimation-driven activity \citep{hsiehPopulationCometsMain2006}. As of this writing, sublimation is inferred -- activity has been too weak to confirm the presence of volatiles spectroscopically, despite efforts to do so (e.g., \citealt{hsiehObservationalDynamicalCharacterization2012}). Fewer than ten \acp{MBC} have been found, although there are additional candidates suspected of being \ac{MBC} members.

\section{Active Centaurs}
\label{intro:sec:activeCentaurs}

Centaurs are cold bodies that originate from the Kuiper Belt (see review \citealt{morbidelliCometsTheirReservoirs2008}). Confusingly, there are multiple discrepant definitions of Centaur. In this work we adopt the definition of \cite{jewittActiveCentaurs2009} that defines Centaurs as (1) objects with semi-major axes and perihelion distances that both fall between the orbits of Jupiter (5.2~au) and Neptune (30~au), and (2) the object must not be in a resonant orbit with a giant planet. A \textit{resonant orbit} is defined as any situation when two bodies orbiting a parent body (the Sun in this case) share a similar orbit (1:1 ratio) or have mean orbital periods that are in integer ratios of each other (e.g., 3:2). For example, Jupiter Trojans (Figure \ref{intro:fig:solarSystem}) are co-orbital with Jupiter, leading and following Jupiter in its orbit by 60$^\circ$, so these are not classified as Centaurs.

Centaurs, being significantly more distant than main-belt asteroids and \acp{NEO}, are much fainter and, consequently, harder to detect. Unlike active asteroids, active Centaurs were first identified belatedly from objects previously classified as comets. The first known active Centaur is often cited as being 29P/Schwassmann-Wachmann~1 (Figure \ref{intro:fig:famousAAs}), discovered in 1927 \citep{schwassmannNEWCOMET1927} and, at the time, considered a comet. Notably, Centaur 2020~MK$_4$, an object which has an orbit very similar to that of 29P, was recently found to be active \citep{delafuentemarcosActiveCentaur20202021}. In 1977 the prototype Centaur (2060)~Chiron was discovered \citep{kowalSlowMovingObjectKowal1977}, an object that was itself later found to be active \citep{meechAtmosphere2060Chiron1990}.

\section{Dynamical Evolution}
\label{intro:sec:transitionObjects}
Orbits of all bodies in the solar system change continuously because of the influence of gravity imparted by other objects. Minor planets may experience gravitational perturbations that result in their orbit classification changing entirely, for example from Centaur to \ac{JFC}. We define bodies that are in the process of migrating from one dynamical class to another as \textit{transition objects}.

My favorite transition object, 39P/Oterma, was discovered by Liisi Oterma at Turku (Finland) in 1943 \citep{otermaNEWCOMETOTERMA1942}. At the time of discovery, 39P/Oterma had a perihelion distance of 3.4~au and a semi-major axis of 4.0~au, placing it interior to the Centaur region. At the time, 39P/Oterma was either a \ac{QHC} or \ac{JFC}. However, on \ac{UT} 1963 April 12, 39P/Oterma had a close encounter (0.095~au) with Jupiter that dramatically altered its orbit. Ever since, 39P/Oterma has had a perihelion distance and semi-major axis exterior to 5.4~au, placing this object firmly within the Centaurian orbital regime.

A recent example is P/2019 LD2 (ATLAS), an object in an orbit similar to a Jupiter Trojan. Jupiter Trojans lead and trail Jupiter by 60$^\circ$ in orbit, however P/2019 LD2 is not presently in either of those locations. P/2019 LD2 (ATLAS) was most likely a Centaur before it arrived in its current orbit, and will return to a Centaurian orbit in 2028, followed by a \ac{JFC} orbit in 2063 \citep{hsiehTransientJupiterTrojanlike2021}.

In Chapter \ref{chap:282P} we present a study of 282P/(323137) 2003 BM$_{80}$, and object we classified as a \ac{QHO}. Prior to 180 years ago 282P was likely a Centaur or possibly a \ac{JFC}. After numerous close encounters with Jupiter, 282P migrated inward and was captured in a Quasi-Hilda orbit, which is an orbit with properties similar to the Hilda group that is in 3:2 resonance with Jupiter. Over the next 300 years or so 282P will undergo more encounters with Jupiter before it probably migrates to a \ac{JFC} orbit.

\section{Activity Detection Techniques}
\label{intro:sec:activityDiscovery}
Prior to the invention of the telescope, cometary activity was discovered with the naked eye. Documented discoveries date back thousands of years, with written records of lost comets beginning with comet D/-674, and comets still familiar today starting with 1P/Halley, first recorded in -239 B.C. \citep{kronkCometographyCatalogComets1999}. Here I discuss different modalities of activity detection, limiting the discussion to active asteroids and active Centaurs. Two important notes to bear in mind: (1) not all techniques have yielded new active body discoveries, and (2) some techniques have yet to validate activity claims through empirical visible activity identification (i.e., see a tail or coma), though some (but not all) disclaim that the objects highlighted should be considered candidates. These considerations are addressed as appropriate below.

\subsection{Visually Observed Activity}
\label{intro:subsec:visibleActivity}
Visual identification of a tail and/or coma remains the gold standard of activity detection. The \ac{MPC} adds additional requirements for comet discovery, namely that the activity must be visible in at least two images taken during one observing night, and that two sets of images, preferably from adjacent observing dates, be submitted.

There can be a great deal of uncertainty when searching for activity indicators like coma(e) and/or tail(s). For example, background galaxies and image artifacts can masquerade as activity, especially when looking at just one image instead of a sequence where the solar system object moves against the sky. To account for ambiguity I created an informal system to describe the level of apparent activity in an image, ranging from 0 (unable to locate the object at all) to 9 (any individual shown a single image would have no doubt whatsoever that they are looking at cometary activity), described in Section \ref{methods:subset:trainingSet}.

Myriad techniques have been used to enhance images to bring out additional detail in the tails. In the same way modern image or photo editing tools can minimize shadows or enhance contrast, images of activity can be enhanced to bring out more detail. Another common technique is to add multiple images together (co-addition), sometimes referred to as stacking, thereby strengthening the overall image signal.

\begin{figure}
    \centering
    \begin{tabular}{cc}
        \includegraphics[width=0.39\linewidth]{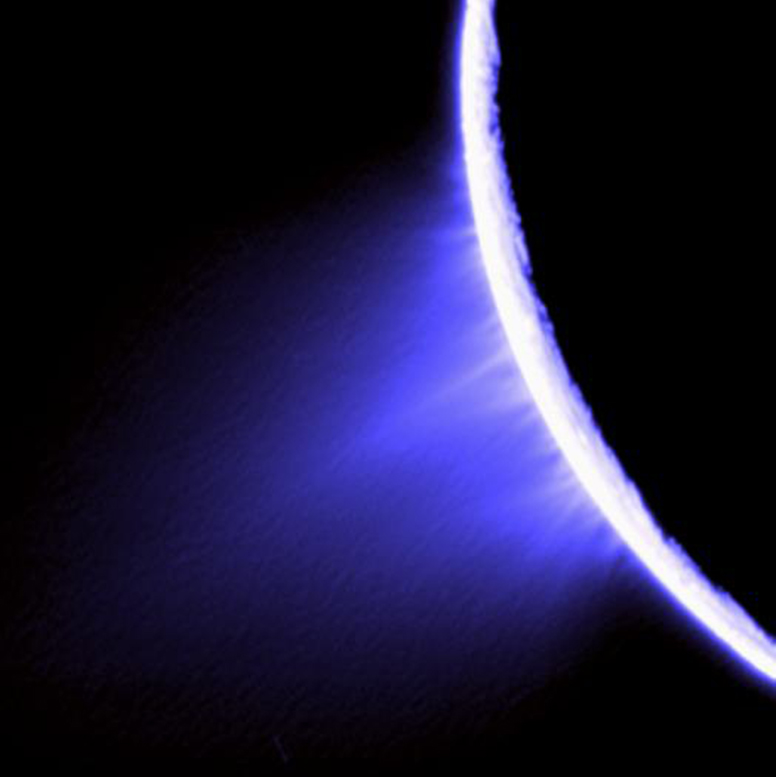} & \includegraphics[width=0.6\linewidth]{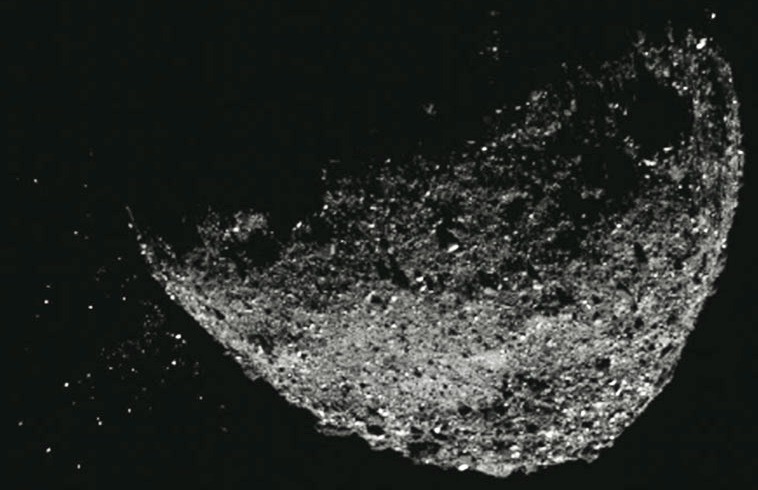}  \\
         & 
    \end{tabular}
    \caption{Activity discovered by fly-by and orbiter spacecraft missions may not be detectable from Earth.  
    (a) Geysers of water reaching high above the surface of Enceladus were imaged by the Cassini spacecraft. Image Credit: NASA/JPL/Space Science Institute. 
    (b) Gravel-sized particles (left) being ejected from the surface of (101955)~Bennu were captured by the cameras aboard the OSIRIS-REx spacecraft. Image Credit: NASA/Goddard/University of Arizona/Lockheed Martin.
    }
    \label{intro:fig:activity}
\end{figure}

Activity detection is strongly influenced by measurement sensitivity. I normally think of activity detection as roughly falling into two categories: remote sensing and \textit{in situ}. This dissertation focuses primarily on the study of activity detected from Earth, however several objects have been found to be active once visited by spacecraft. For example, (101955)~Bennu had not been suspected of being active, but upon arrival of the \ac{OSIRIS-REx} mission spacecraft in 2018, unexpected activity was documented by the cameras aboard the spacecraft (Figure \ref{intro:fig:activity}). Thus Bennu is definitely active, but that activity has never been observed from Earth. Similarly, some planetary moons have been found to be active by spacecraft, for example Enceladus \citep{spencerCassiniEncountersEnceladus2006}, as shown in Figure \ref{intro:fig:activity}. Pluto and Ceres represent special cases where an an atmosphere (or exosphere was detected remotely first, then by spacecraft later. In the case of Pluto, an atmosphere was detected remotely in 1989 \citep{elliotPlutoAtmosphere1989}, then later studied up close by the New Horizons spacecraft flyby in 2015 \citep{sternPlutoSystemInitial2015}. In the case of Ceres, the first asteroid discovered \citep{piazziRisultatiOsservazioniNuova1801}, water vapor was discovered by \cite{kuppersLocalizedSourcesWater2014} with Herschel telescope observations. Following the arrival of the Dawn spacecraft at Ceres, activity in the form of water vapor was detected \citep{nathuesSublimationBrightSpots2015,thangjamHazeOccatorCrater2016}, although a later study by \cite{schroderResolvedSpectrophotometricProperties2017} did not find any evidence of sublimation.

\subsection{Brightness}
\label{intro:subsec:brightness}
Approaches connected with measuring the brightness of an object are often used in conjunction with other evidence to support a claim of activity, to allow for additional analyses, or both. There are three primary techniques involving brightness that have been used to find potentially active asteroids, two of which have yielded proven results.

\paragraph{Discrepant Brightness} This method involves looking for unexpected brightening of objects which could be caused by activity reflecting additional light (see  \citealt{cikotaPhotometricSearchActive2014} for an example of a broad application). This is achievable by measuring how much light the asteroid reflects, and then comparing this result with an expected value. Note, however, that the source for expected values must provide precision photometry for a source with a well-measured phase function, ideally whilst inactive; sources such as \ac{JPL} Horizons, while highly convenient, are widely considered only accurate to within a couple of magnitudes. Objects with activity should reflect more light, and thus should appear brighter than expected.

\paragraph{Point Spread Function Analysis} The shape and size of points in an image is called a \ac{PSF}. By comparing the \ac{PSF} of an object in an image to a comparably bright star in the same image can reveal if the width of a source is wider than expected, indicating that the source is extended (i.e., elongated). Finding asteroids that have unexpected broadening can be used for detecting activity (e.g., \citealt{hsiehMainbeltCometsPanSTARRS12015}). The measurement of excess breadth is often used as direct evidence of activity, and measurements of the excess flux can reveal important information about the activity, such as the amount of material in a coma (this can also be considered image analysis, with the source of data being photons). See Figure \ref{og:fig:sbrp} for an example.

\subsection{Spectroscopic Indicators}
Spectroscopy is the technique that measures markers in refracted light, as with a prism. The general idea is to identify features in spectra that reveal an object is active, whether the material be composed of volatiles or dust. However, to date spectroscopy has yet to identify an active asteroid that has a tail or coma visible from Earth.

This technique has been used successfully before to identify an asteroid as active. A notable example is the case of (1)~Ceres (e.g., \citealt{kuppersLocalizedSourcesWater2014}), albeit  confirmation \citep{nathuesSublimationBrightSpots2015,thangjamHazeOccatorCrater2016} from the visiting Dawn mission spacecraft 
is disputed \citep{schroderResolvedSpectrophotometricProperties2017}. Moreover, spectroscopic study of minor planets has successfully identified surface ices in the past, including on bodies known to be active (e.g., (2060)~Chiron; see review, \citealt{peixinhoCentaursComets402020}). However, sublimation from \acp{MBC} has never been spectroscopically confirmed; see review by \cite{snodgrassXshooterSearchOutgassing2017}.

Recently a group has applied a spectroscopic technique to attempt to identify active asteroids, most recently (24)~Themis and (449)~Hamburga \citep{busarevSimultaneousSublimationActivity2021}. Alas, to date none have been observed to display a visibly identifiable tail or coma, despite archival and observational efforts by astronomers, including archival and observational efforts by our team. In 2018 (162173)~Ryugu, one of the objects identified as active by this technique \citep{busarevNewCandidatesActive2018}, was visited by the \ac{JAXA} spacecraft Hayabusa2 \citep{watanabeHayabusa2MissionOverview2017} for a sample return mission. No activity was reported, though there was evidence that the predominately dehydrated Ryugu \citep{sugitaGeomorphologyColorThermal2019} had spun rapidly in the past \citep{watanabeHayabusa2ArrivesCarbonaceous2019} which could have resulted in mass shedding.

\subsection{Non-gravitational Acceleration}
All objects in the solar system experience acceleration due to the force of gravity. Unexplained acceleration can be caused by the gravity of an unknown body, such as the perturbations that resulted in the discovery of Neptune (see account by \citealt{standage2000neptune}) or, more recently, the hypothesized Planet 9 \citep{trujilloSednalikeBodyPerihelion2014,batyginPlanetNineHypothesis2019}. However, not all unexplained acceleration is caused by gravity.

Sources of non-gravitational acceleration include the Yarkovsky effect (first measured on (6489) Golevka, \citealt{chesleyDirectDetectionYarkovsky2003}), a force resulting from imparted solar radiation received by a body being reemitted later as thermal radiation. 
These types of forces are negligible over short timescales, yet some objects have demonstrably experienced changes in their orbits that could not be readily explained. For example, interstellar object 1I/\omuamua{} evidently experienced non-gravitational acceleration \citep{micheliNongravitationalAccelerationTrajectory2018}, possibly attributable to unseen activity \citep{seligmanAnomalousAcceleration1I2019}.

Activity can provide one potential source of non-gravitational acceleration, for example jets of gas. Thus, in principle, it should be possible to search for activity by scrutinizing the orbits of small solar system bodies and looking for unexplained changes. A recent example tries to link two bodies, 2019~PR$_2$ and 2019~QR$_6$, to cometary activity \citep{fatkaRecentFormationLikely2022}, though no visible activity has been conclusively observed as of this writing.

\subsection{Meteor Showers}
This technique has yet to identify a new active asteroid, but active asteroid (3200)~Phaethon has been identified as the apparent parent of the Geminid meteor showers \citep{whippleIAUC388119831983}. In addition to the Geminid Meteor Stream, Phaethon shares an orbit with 2005~UD \citep{ohtsukaApolloAsteroid20052006} and 1999~YC \citep{ohtsukaApolloAsteroid19992008}, which suggests all co-orbital elements may have originated from the breakup of a single parent body. The case of Phaethon implies that it may be possible to connect other meteor streams back to parent bodies (e.g., \citealt{dumitruAssociationMeteorShowers2017}) that may themselves be active. See also reviews by \cite{babadzhanovExtinctCometsAsteroidmeteoroid2015,yeMeteorShowersActive2018}.

\subsection{Magnetic Anomalies}
This approach has only been used once, to claim activity coming from (2201)~Oljato or an outgassing debris trail in its orbit \citep{russellInterplanetaryMagneticField1984,kerrCouldAsteroidBe1985}. A series of  interplanetary magnetic field enhancements were measured by the Pioneer spacecraft that was orbiting Venus. These events were correlated with the passage of Oljato during 7 of the 11 magnetic anomalies, with the likelihood the anomalies were coincidental given as 4 in $10^4$. However, despite some further evidence supporting Oljato activity \citep{cochranSpectroscopyAsteroidsUnusual1986,mcfaddenEnigmaticObject22011993,chamberlin4015WilsonHarrington22011996}, all activity associated with Oljato to date has been inferred, rather than directly observed in the form of a coma or tail. The event may have been transient as Oljato's orbit is considered chaotic \citep{milaniDynamicsPlanetcrossingAsteroids1989}.


\section{Activity Mechanisms}
\label{intro:sec:mechanisms}

Volatile sublimation is not the only cause of activity we observe. Moreover, the myriad mechanisms potentially responsible for observed activity are not mutually exclusive, and one activity mechanism may trigger another or occur simultaneously. Some events are stochastic (one-off), while other mechanisms are recurrent by nature. Consequently, recurrent activity is an important diagnostic indicator when ascertaining an underlying activity mechanism.

Below is a listing of mechanisms that may result in the kind of activity we observe associated with active asteroids and active Centaurs. See also the reviews of \cite{jewittActiveAsteroids2015a} and \cite{jewittActiveCentaurs2009}, as well as Table \ref{safari:Table:TheAAs}.

\subsection{Volatile Sublimation}
In the same way that dry ice goes directly from a solid to a gas on Earth's surface, ices can sublimate in space to great effect, releasing volatiles and ejecting dust and rocky material. The primary activity mechanism of comets is volatile sublimation, and these bodies have been studied at length from Earth and \textit{in situ} with spacecraft visits (e.g., Rosetta mission to 67P/Churyumov–Gerasimenko, \citealt{glassmeierRosettaMissionFlying2007,sierksNucleusStructureActivity2015}). In addition to water ice, carbon dioxide, carbon monoxide, ammonia, methane, nitrogen and other molecules have been detected on asteroids and Centaurs (see Chapter \ref{og:sec:introduction}).

To be clear, volatiles need not be on the surface in order to sublimate, however remote detection of ices on inactive bodies requires ice to be present on the surface. These bodies may have reservoirs just under their surfaces or buried far below. Any group of bodies that harbors ice, no matter where that material is located on or within a body, represents a volatile reservoir. However, some bodies may not have any ices at all -- especially silica-rich asteroids known as S-type asteroids. Silica, on Earth commonly associated with desiccant and sand, is typically dehydrated and thus not expected to contain volatiles.

Sublimation requires some form of energy change to take place, with energy imparted directly by the Sun being the most ubiquitous. The closer a body gets to the Sun, the more energy it receives and, as a result, activity becomes more likely if ice is present. Many recurrently active objects, especially comets, are observed to be preferentially active as they get closer to the Sun. Energy may also come from other sources, such as tidal heating (e.g., Europa, \citealt{greenbergTectonicProcessesEuropa1998}) or ice phase transitions (see review, \citealt{jewittActiveCentaurs2009}).

Different substances sublimate at different temperatures. Water ice, for example, will not appreciably sublimate at the orbital distances where Centaurs are found, but it can readily sublimate on bodies found in the asteroid belt. As a consequence the lifetime of ices varies by orbital distance so that, for example, it could be expected that water ice could survive on a body orbiting at 5~au but carbon monoxide and methane would have been depleted long ago \citep{schorghoferLifetimeIceMain2008,snodgrassMainBeltComets2017}. As described in Chapter \ref{og:sec:sublimationmodeling}, this knowledge can be leveraged to help identify which material(s) are most likely responsible for observed sublimation-driven activity.

\subsection{Rotational Instability}
All minor planets in our solar system rotate to some extent. Bodies that rotate rapidly can actually break apart or lose loose surface material to space. Small solar system bodies are susceptible to being ``spun up'' over time by the Sun by a process known as the \ac{YORP} effect (see e.g., \citealt{bottkeYarkovskyYorpEffects2006,lowryDirectDetectionAsteroidal2007} for details about \ac{YORP} forces). Consequently, it is possible that rotational instability can lead to recurrent activity that is unrelated to sublimation \citep{jewittEpisodicEjectionActive2015,chandlerSixYearsSustained2019}. Even if activity is recurrent, the onsets of activity would be uncorrelated with perihelion distance.

Rotational instability can lead to sublimation involvement if, for example, a breakup or landslide exposes previously buried volatiles that subsequently sublimate. Activity may cease again if the volatiles become smothered by settling material that was previously ejected.

\subsection{Thermal Fracture}
Heating solid materials can cause fracture. On Earth we see this happen, for example, when pouring boiling water on a frozen car windshield. Fracture is caused when stress induced by temperature change (e.g., expansion of heated material) overcomes the tensile strength of adjoining material \citep{jewittActiveAsteroids2015a}. Depending on the orbit of an object, this could take place repeatedly, especially when the object is close to the Sun. This action alone may eject material into space and result in the appearance of cometary activity. Moreover, these fractures may exposure sequestered volatiles that subsequently sublimate.

In the case of (3200)~Phaethon, an active asteroid that is thought to be responsible for the Geminid meteor stream \citep{whippleIAUC388119831983}, the body's temperatures reaches roughly 1100~K (1520$^\circ$~F) at its 0.14~U perihelion \citep{ohtsukaSolarRadiationHeatingEffects2009}, temperatures that can causes thermal fracture \citep{licandroNatureCometasteroidTransition2007,kasugaObservations1999YC2008} and in turn may result in mass loss \citep{liRecurrentPerihelionActivity2013,huiResurrection3200Phaethon2017}. The \ac{JAXA} mission \ac{DESTINY+}, scheduled to launch in 2024, is designed to provide more insights into Phaethon and its activity \citep{ozakiMissionDesignDESTINY2022}.

\subsection{Impact}
\label{intro:subsubsec:impact}
I divide impact events into two categories: significant events involving one or more large impactors (meter to kilometer scale), and micrometeorite impacts that involve multiple impacts by very small impactors typically of order 1~cm and smaller.

Two important cases of impact are worth mentioning here. (1) (596)~Scheila is widely considered the seminal example of an impact-driven asteroid activity event \citep{bodewitsCollisionalExcavationAsteroid2011,ishiguroObservationalEvidenceImpact2011,moreno596ScheilaOutburst2011}. (2) The first active asteroid discovered, (4015)~Wilson-Harrington (Figure \ref{intro:fig:famousAAs}) is thought to have undergone a significant impact event because the object was never observed to be active again, despite searches spanning over seventy years (e.g., \citealt{chamberlin4015WilsonHarrington22011996}).

Recently the \ac{OSIRIS-REx} spacecraft arrived at asteroid (101955)~Bennu for a sample return mission. Cameras aboard the spacecraft recorded what appeared to be gravel and other particulate leaving and returning to the surface (Figure \ref{intro:fig:activity}). Micrometeorite impacts have been suggested as a potential cause \citep{laurettaEpisodesParticleEjection2019,bottkeMeteoroidImpactsSource2020,hergenrotherIntroductionSpecialIssue2020}, though thermal fracture and other mechanisms are still being investigated.

\textit{Gardening} describes a process by which micrometeorite impacts overturn the outermost layer of a body. First described on the Moon (e.g., \citealt{chapmanLunarCrateringErosion1970} and references therein), gardening theory can be used to place limits on the amount of mass shed by myriad processes, including electrostatic lofting (see below), micrometeorite impacts (impact gardening) and photons (e.g., solar gardening; \citealt{grundySolarGardeningSeasonal2000}). Although impact gardening has been correlated with absolute ages only on the Moon \citep{gaultMixingLunarRegolith1974}, impact gardening serves an important function of bringing ices closer to a body's surface \citep{schorghoferPredictionsDepthtoiceAsteroids2016}, thus increasing the availability of material for sublimation.

\subsection{Cryovolcanism}
Just as volcanoes can eject molten material, cryovolcanoes can eject liquid and/or gaseous material from a cold body. Saturn's moon Enceladus is also thought to undergo cryovolcanism (Figure \ref{intro:fig:activity}) resulting from tidal heating \citep{nimmoShearHeatingOrigin2007}, first observed as plumes by the Cassini spacecraft in 2005 (e.g., \citealt{spencerCassiniEncountersEnceladus2006}). Asteroid (1)~Ceres is thought to undergo cryovolcanic activity (e.g., \citealt{soriVanishingCryovolcanoesCeres2017,nathuesRecentCryovolcanicActivity2020}), most likely driven by radioactive heating \citep{mccordCeresItsOrigin2011}, but evidence of cryovolcanic activity was only confirmed after the arrival of the Dawn spacecraft. To date it is unclear whether or not any activity on active asteroids or active Centaurs is primarily due to cryovolcanism, though cryovolcanism has been reported as responsible for the activity of 29P/Schwassmann-Wachmann~1 \citep{milesDiscreteSourcesCryovolcanism2016}.

\subsection{Radiation Pressure Sweeping}
Solar wind exerts a force that can, in principle, sweep particles off of the surface of an airless body, especially one with a small gravitational field \citep{jewittActiveAsteroids2015a}. This effect may play an important role as a secondary action, carrying away material ejected via other means that would have otherwise settled back on the surface. This is thought to play an important role for (3200)~Phaethon given its close (0.14) perihelion passage where radiation pressure is significant \citep{jewittActivityGeminidParent2010}. To date this effect has not been directly measured at an active asteroid, though the \ac{DESTINY+} mission may help us better understand the mechanisms at play on Phaethon.

\subsection{Electrostatic Lofting}
This mechanism was first observed by Apollo astronauts in the 1960 as a ``lunar horizon glow'' \citep{rennilsonSurveyorObservationsLunar1974,wangDustChargingTransport2016}. The electrostatic forces behind this mechanism may be powerful enough to eject material from the surface of small airless bodies such as asteroids. Should the material be lofted without sufficient energy for escape, a second activity mechanism (e.g., radiation pressure sweeping) could help carry the material away. Electrostatic lofting is a weak phenomenon, so it is unclear if this mechanism could result in activity detectable at distances farther than spacecraft orbit. However, \cite{sonnettLimitsSizeOrbit2011} suggest very low-level activity on a broad scale ($\sim$5\% of main-belt asteroids, based on a study of $\sim1000$ asteroids from \citealt{masieroThousandAsteroidLight2009}) which may involve electrostatic lofting \citep{jewittActiveAsteroids2015a}.

\subsection{Phyllosilicate Dehydration} 
Phyllosilicates are a class of minerals that are characterized by layering, such as mica or smectite clays. Hydrated phyllosilicates have volatiles like water trapped between layers. Laboratory studies of meteorites rich in hydrated phyllosilicates reveal that these volatiles can be released with significant energy when heated sufficiently (e.g., \citealt{gibsonInorganicGasRelease1974}). On large scales this modality may be the underlying cause of thermal fracture, but on small scales this release of energy has the potential to eject material from the surface of an airless body. This mechanism may be at play on asteroid (101955)~Bennu \citep{laurettaEpisodesParticleEjection2019} because the surface has an abundance of hydrated phyllosilicates \citep{hamiltonEvidenceWidespreadHydrated2019}.

\subsection{Binary Rubbing}
The two bodies of a binary asteroid may eventually spiral in and become a contact binary. The physical interaction between the rubbing binaries could cause material to be shed from the surfaces, resulting in a coma or tail. However, this mechanism has yet to be conclusively identified as the cause behind the activity of any known active asteroid, though it has been proposed as one possible explanation for the activity of active asteroid P/2013~P5 \citep{hainautContinuedActivity20132014}.

\section{Citizen Science Project}
\label{intro:sec:citSciProject}

Note: For completeness, this section provides a cursory introduction only. The project is described at length in Chapter \ref{methods:sec:citsci}.

Citizen Science is a paradigm that aims to accomplish scientific goals while simultaneously engaging the public by seeking assistance from volunteers to accomplish tasks that are too numerous for individuals or small groups to complete, and which are also too complex for computers to handle. As described in Chapter \ref{methods:subsec:workflow}, our root method is to ask volunteers whether or not they see a tail or coma in images of known minor planets (e.g., asteroids, Centaurs) extracted from publicly available \ac{DECam} data. Once images are examined we can conduct follow-up investigation and study, as described in Chapter \ref{methods:subsec:archivalInvestigation}.

Zooniverse\footnote{\url{https://zooniverse.org}} is an online Citizen Science platform, known for its highly successful inaugural project \textit{Galaxy Zoo} \citep{lintottGalaxyZooMorphologies2008} that launched in 2007. We selected Zooniverse to host our project because of their proven ability to host and support Citizen Science projects. Our Citizen Science project \textit{Active Asteroids}, a \ac{NASA} Partner, launched on 31 August 2021. The project immediately began yielding results, and volunteers exhausted our original pool of data in just a few days. Since launch, over six thousand volunteers have completed roughly 2.5 million classifications of some two hundred thousand images.

\section{Manuscript Introduction}
\label{sec:intro:manuscripts}
We published numerous discoveries during preparations for the project
. Here I chronologically introduce the manuscripts included in this dissertation, provide a brief synopsis of each, and describe how they relate to the overall dissertation theme of detecting and characterizing active minor planets via astroinformatics and/or Citizen Science. Key points are indicated by \textbf{bold} typeface.


As discussed at the start of this chapter, I identified a need to identify additional active minor planets in order to facilitate their study. From the start we considered launching a Citizen Science project to assist with this tasks, however it seemed logical to first carry out a proof-of-concept to ensure we could, in fact, supply images of known solar system objects to Citizen Scientists who would check for activity like tails and comae. \acf{SAFARI} was the title for our proof-of-concept \citep{chandlerSAFARISearchingAsteroids2018}, provided in Chapter \ref{chap:SAFARI}. We began by creating a software \textbf{pipeline} that produces thumbnail images (small cutouts from a larger image) from \ac{DECam} data, each displaying a known minor planet at the center. This pipeline became the foundation of our \ac{HARVEST} pipeline, discussed at length in Chapter \ref{methods:sec:methods:pipeline}.


From 594 \ac{DECam} images we extracted a total of 35,640 thumbnail images that contained 11,703 unique solar system objects. We examined all of these thumbnails visually (by eye) and identified activity emanating from what turned out to be one already known active asteroid: (62412) 2000~SY$_{178}$. Identifying an active object in our data served as our \textbf{proof-of-concept}, demonstrating that \ac{DECam}, with its wide ($2.2^\circ\times2.2^\circ$) field of view and large 4~m aperture that probes very faintly, is well-suited for finding active bodies. 
We estimated an \textbf{activity occurrence rate} of 1 in 11,000 objects is active, in rough agreement with past studies that found a rate of roughly 1 in 10,000 \citep{jewittActiveAsteroids2015a,hsiehMainbeltCometsPanSTARRS12015}. 
As part of our included background review we constructed a \textbf{comprehensive table} listing all active asteroids, along with details such as orbital distance, number of activity epochs, and diagnosed or suspected activity mechanism(s).


With the proof-of-concept a success we set out to improve upon the \ac{HARVEST} pipeline, for example to query the \ac{DECam} public archive to provide us with a plentitude of data, and to query the archive daily in order to keep our library of minor planet thumbnails up-to-date. Another capability we added to \ac{HARVEST} was the ability to quickly provide us with thumbnail images from our repository of a single solar system object. An opportunity arose for us to make use of this feature after a telegram announced asteroid (6478)~Gault was active \citep{smith6478Gault2019}. For the work we would publish, \textit{Six Years of Sustained Activity from Active Asteroid (6478)~Gault} \citep{chandlerSixYearsSustained2019}, provided in Chapter \ref{chap:Gault}), we produced thumbnails using this \ac{HARVEST} feature and found evidence that Gault had been active during at least two prior orbits, thus Gault experienced \textbf{recurrent activity}. We also introduced \textit{observability}, a metric describing how many hours per night an object is observable from a given location (Figure \ref{Gault:fig:ActivityTimeline}). This metric highlights observational biases (e.g., observer location) that can influence analyses such as activity mechanism diagnosis, as well as allowing us to assess how many images of a given object we have in our archive.

The onset of sublimation-driven activity typically occurs preferentially near perihelion, but we showed that Gault's activity was not correlation with heliocentric distance. 
We carried out simple \textbf{thermal modeling} to estimate the temperatures experienced by Gault over the course of its orbit, and found it consistently too warm for water ice to have survived. In sum, we found activity unlikely to be sublimation-driven because (a) Gault is from a desiccated asteroid family (Phocaea), and (b) we found activity was unrelated to heliocentric distance. We proposed Gault may represent a \textbf{new type of active object}: recurrently active due to rotational spin-up.

Following the aforementioned expansion of the \ac{HARVEST} pipeline to work with all publicly available \ac{DECam} data we set out to examine large collections of thumbnails, on the order of 10,000 or more. The purpose of this exercise was to identify potential problems that would manifest in the thumbnails, such as handling chip gaps (Section \ref{methods:sec:methods:pipeline}) and enhancing contrast (Section \ref{methods:subsec:thumbnailExtraction}). While examining a thumbnail collection composed of Centaurs we noticed visible evidence of activity in images of Centaur 2014~OG$_{392}$, an object not yet known to be active. Our \textbf{discovery of an active Centaur} would become the foundation for our work which culminated in the publication of \textit{Cometary Activity Discovered on a Distant Centaur: A Nonaqueous Sublimation Mechanism} \citep{chandlerCometaryActivityDiscovered2020a}, Chapter \ref{chap:2014OG392} of this dissertation.

Our discovery represented a milestone for our project as our first active object discovery, but first we needed to confirm the activity was real and not an image artifact. To this end, we first acquired follow-up observations with the \ac{DECam} instrument on the Blanco 4~m telescope at \ac{CTIO} in Chile, the same instrument from which our archival images originated. Co-adding the four 250~s exposures revealed evidence of a coma. We next employed the \ac{IMACS} instrument on the 6.5~m Walter Baade Telescope atop the Las Campanas Observatory in Chile, and these images provided strong evidence of activity. Finally we made use of the \ac{LMI} on the 4.3~m Lowell Observatory \ac{DCT}, now called the \ac{LDT}. Here we were able to measure colors of 2014~OG$_{392}$ that revealed it is optically red.

With the images we acquired we were able to detect a coma composed of $\sim2.4\times10^{12}$~kg extending to at least a distance of 400,000~km from 2014~OG$_{392}$. 
We introduced \textbf{a novel approach for estimating which molecular species are most likely responsible for observed activity}. We found carbon dioxide ice and/or ammonia ice the most likely to be sublimating, yet these two materials are optically grey. Thus an as yet unknown reddening agent must must be present to account for the red color of 2014~OG$_{392}$. Upon publication of our work, the \ac{MPC} gave this object the new designation of C/2014~OG392~(\acs{PANSTARRS}). (Note: although provisional names like 2014~OG$_{392}$ are retained as part of comet designation (except for short-period comets), the numeric portion at the end of the designation is no longer represented with subscript text.)

Just prior to launching our Citizen Science project, a telegram announced the discovery of a new active asteroid, (248370) 2005~QN$_{173}$ \citep{fitzsimmons2483702005QN1732021}. We carried out an archival investigation into (248370) 2005~QN$_{173}$ that ultimately led to our publication \textit{Recurrent Activity from Active Asteroid (248370) 2005~QN173: A Main-belt Comet} \citep{chandlerRecurrentActivityActive2021a}, dissertation Chapter \ref{chap:2005QN173}. We found a single \ac{DECam} image from 2016 July 22 (Figure \ref{QN:fig:wedgephot}) that unambiguously showed a tail emanating from (248370) 2005~QN173. Here \textbf{we introduced Wedge Photometry}, a tool that measures tail angle for (a) comparison with anti-Solar and anti-motion angles computed by \ac{JPL} Horizons (Figure \ref{intro:fig:activeObjects}), and (b) activity detection techniques. Our Wedge Photometry tool found the tail orientation to be $251.3\pm1.3^\circ$, in close agreement with the 251.6$^\circ$ and 251.7$^\circ$ orientations computed by the \ac{JPL} Horizons service.

The archival image showing activity we uncovered provided proof that the object had been active during at least one prior orbit. Thus \textbf{(248370) is recurrently active}, having undergone activity during at least two separate orbits. Moreover, we found (248370) was preferentially active near perihelion, only the 8$^\mathrm{th}$ such main-belt asteroid known to exhibit this behavior. Recurrent activity near perihelion is a strong diagnostic indicator of volatile sublimation as the underlying activity mechanism. This combination of recurrent sublimation-driven activity of a main-belt asteroid is evidence that \textbf{(248370) is most likely a member of the \acp{MBC}}. After we announced our discovery via electronic telegram \citep{chandler2483702005QN2021}, (248370) 2005~QN$_{173}$ was assigned an additional designation: comet 433P.

At this point in time (31 August 2022) we successfully launched our \ac{NSF} funded \ac{NASA} Partner Citizen Science project \textit{Active Asteroids} (Section \ref{methods:sec:citsci}) and we started to identify candidate active objects and make discoveries. For example, \textbf{volunteers overwhelming classified two \ac{DECam} images of 282P/(323137) 2003~BM$_{80}$ from 2021 as active}. Additionally, Citizen Scientists classified an image of 282P from 2013, then the only published activity epoch, as active. As described in \textit{Migratory Outbursting Quasi-Hilda Object 282P/(323137) 2003 BM80} (Chapter \ref{chap:282P}, paper submitted to Astrophysical Journal Letters), we carried out a multifaceted study of 282P, also designated 2003~FV$_{112}$, consisting of an archival investigation, telescope follow-up observations, dynamical modeling, and thermodynamical modeling. Notably, this work represents the first peer-reviewed publication to stem from our \textit{Active Asteroids} Citizen Science project.

Our archival investigation yielded additional images of the object that showed evidence of activity. The last images of 282P activity we identified were over a year prior to the Citizen Science project discovery, so we set out to conduct an observational campaign with the goals of (1) determining whether or not 282P was still active and, if so, (2) evaluating ongoing activity for changes in morphology (e.g., shape, extent). Unfortunately, 282P was transiting the Milky Way, which meant there were too many stars to effectively identify activity indicators. We were awarded a Gemini \ac{DDT} observation at Gemini South that we timed to take place during an 11 day window when 282P was passing in front of a less dense region of the Galaxy. The program was successfully executed, yielding 18 images of 282P, and we saw an unmistakable tail (Figure \ref{282P:fig:282P}). These images, in combination with our archival evidence of activity starting in March 2021, indicated that \textbf{282P had been active for at least 15 months}. Furthermore, we found 282P activity preferentially occurs near perihelion passage, typical of sublimation-driven activity, although additional study is needed to confirm this hypothesis.

Our orbital simulations (dynamical modeling) revealed that 282P only recently ($<200$ yr ago) arrived at its present orbit. \textbf{We determined that 282P is a member of the rare ($<$15) active Quasi-Hilda class} (e.g., \citealt{tothQuasiHildaSubgroupEcliptic2006,gil-huttonCometCandidatesQuasiHilda2016}), also referred to as \acfp{QHO}, \acfp{QHC} or \acfp{QHA}. While Hilda asteroids orbit with a 3:2 resonance with Jupiter, \acp{QHC} are only loosely bound to the same region, and not necessarily in 3:2 resonance with Jupiter, as discussed at length in Chapter \ref{chap:282P}. Active Quasi-Hildas like 282P are rare with fewer than 15 active Quasi-Hildas have been found to date. 282P experienced strong interactions with Jupiter and Saturn that dramatically altered its orbit. Of special note, our simulations revealed that 282P used to orbit primarily beyond Jupiter, but now it orbits predominitely interior to Jupiter's orbit. Moreover, 282P will undergo a Jovian interactions in roughly 300 years that will again substantially change its orbit. Giant planet perturbations are so strong, in fact, that dynamical chaos prior to 200 years ago and 300 years in the future prevent exact determination of 282P's origin and future orbital regime. However, when evaluating the possible outcomes of the forward and backward simulations, we identified \acp{JFC} and Centaurs as potential origins of 282P, and that 282P will likely become a \ac{JFC} in the future, thought there is also a chance it may become an active asteroid. Thus \textbf{282P reveals a potential pathway that informs us about the origins of some active asteroids.}

Although outside the scope of this dissertation, we mention here that we are actively working on additional discoveries stemming from the \textit{Active Asteroids} project. These include newfound active asteroids, active Centaurs, \acp{QHC}, \acp{JFC}, and companions to objects such as \acp{TNO}. Anyone interested can partake in this exciting scientific journey by participating in \textit{Active Asteroids}\footnote{\url{http://activeasteroids.net}}.


\clearpage
\singlespacing
\chapter{Comprehensive Discussion of Methods and Materials}
\label{chap:methods}
\acresetall
\doublespacing

Each individual manuscript included in this dissertation includes descriptions of the methods utilized for the work therein. Here I provide updates, additional depth, and unified descriptions.

\section{HARVEST Pipeline}
\label{methods:sec:methods:pipeline}

The pipeline that ultimately produces thumbnails for examination by Citizen Scientists, myself, and my science team, is called \ac{HARVEST}. The pipeline was first described in the \ac{SAFARI} proof-of-concept (\citealt{chandlerSAFARISearchingAsteroids2018}, Chapter \ref{chap:SAFARI}). \ac{HARVEST} has evolved substantially since then into the pipeline described here.

\subsection{Pipeline Overview}

\begin{figure*}
    \centering
    \includegraphics[width=1.0\linewidth]{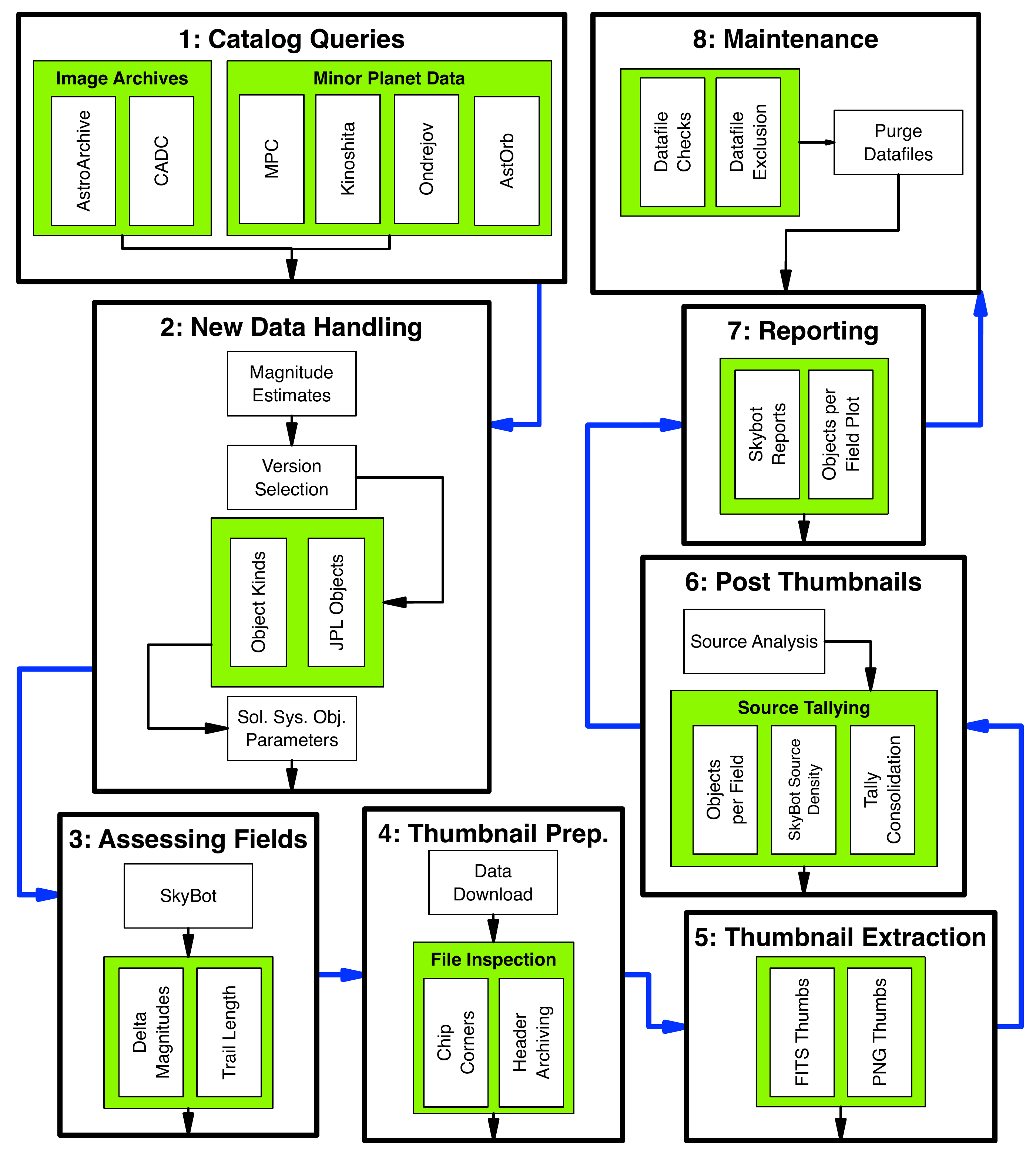}
    \caption{The \acf{HARVEST} daily pipeline. The primary eight steps are sequential. Tasks within each step start from the top of the enclosing rectangle and finish at the bottom. Green shaded regions indicate processes that may be executed in parallel. Each step and the constituent tasks are described in text.}
    \label{methods:fig:harvestFlow}
\end{figure*}

As shown in Figure \ref{methods:fig:harvestFlow}, the \ac{HARVEST} pipeline runs in a series of steps, from Step 1 to Step 8 (described in detail below). In Step 1 we query public astronomical image archives for metadata that describes observations (e.g., \acs{UT} observing date, sky coordinates, exposure time). Concurrently we query external services, such as the \ac{MPC}, for new and updated information about minor planets, such as designations, orbital parameters (e.g., semi-major axis, eccentricity). In Step 2 we estimate the depth (faintness reached) for each image, exclude duplicate datafiles that stemmed from the same observation, assign dynamical classes to objects, and compute parameters such as Tisserand's parameter with respect to Jupiter. In Step 3 we query \texttt{SkyBot} to determine which minor planets may be in each field discovered in Step 1; here we also estimate the detection viability for each object identified by SkyBot. In Step 4 we begin downloading data from the astronomical archives, then determine the sky coordinates for all \ac{CCD} chips. In Step 5 we extract thumbnail images of the desired minor planets from the archival data. During Step 6 we examine the thumbnail images, excluding those that we deem unlikely to yield results via our Citizen Science project. In Step 7 we run automated reporting code that describe, for example, how many thumbnail images per square degree on sky we have in \ac{HARVEST}. Finally, in Step 8, we conduct maintenance tasks such as purging datafiles that have completed all stages of the pipeline.

The \ac{HARVEST} pipeline is run daily and is a blend of serial and parallel processes, visually depicted in Figure \ref{methods:fig:harvestFlow}. All steps are carried out on Monsoon, the \ac{NAU} High Performance Computing cluster. All parallel tasks and most serial tasks are executed using the Slurm scheduler \citep{yooSLURMSimpleLinux2003}.

\ac{HARVEST} is primarily written in Python (v3.x) with parent shell scripts executing code via Slurm scripts. Binary components are entirely supplied by the Anaconda\footnote{\url{https://www.anaconda.com/}} environment and thus require no manual compilation. Most operations involve a \texttt{My\acs{SQL}}\footnote{\url{https://www.mysql.com/}} relational database of my own design, which I describe in the next section.

\subsubsection{Overarching Principles}
Three critical overarching principles apply throughout the entire \ac{HARVEST} pipeline:

\paragraph{A. Task Counting} An optimization that applies to virtually every element of the pipeline is task counting. This determines how much work is necessary in order to (a) evaluate whether or not Slurm is necessary and, if so, (b) assess how many Slurm jobs are needed for each task. Task counting is also used in reporting to ensure the daily pipeline is not falling behind in any area (e.g., catalog queries). The number of jobs launched depends on factors such as the amount of memory needed and the intensity of disk access required. We arrived at these values through trial and error during the development of the \ac{HARVEST} pipeline.

\paragraph{B. Exclusion} Nearly every database table and most tasks involve either excluding data we determine is unfit for use, or ignoring previously excluded data. Exclusion is discussed in in Sections \ref{methods:subsubsec:imageArchives}, \ref{methods:subsubsec:sourceAnalysis}, and \ref{methods:subsubsec:datafileRoutines}.

\paragraph{C. Self-awareness of Elapsed Computing Time} To account for the finite time allotted for each process, all tasks executed through the \ac{HARVEST} pipeline keep track of elapsed time and periodically compute how much time is remaining. This is primarily accomplished through environment variables, but a secondary layer of signaling via the operating system and/or Slurm allows tasks to be systematically warned they are almost out of time. This awareness enables (a) tasks to safely wrap up whatever they are doing (e.g., write data to disk), (b) the pipeline to understand the underlying tasks were not finished, and (c) Slurm scheduler optimization, especially back-fill (making use of leftover compute time on the compute cluster).

\paragraph{D. Self-limiting Service Calls} Tasks that call external services (e.g., \ac{JPL}, \ac{MPC}) may unduly tax host servers. Services typically determine the origin of queries by one or two methods: the originating \ac{IP} address and/or (if supported) an identifier (e.g., email address) embedded in an \ac{API} call. Some external services may ``blacklist'' (block) queries, usually from a specific \ac{IP} address. To address this, we limit the number of queries per minute based on rates we ascertained via documentation and communicate with external service providers as needed.

\paragraph{E. Logging} Extensive logging is crucial for diagnosing problems and, importantly, easily assessing what (if any) tasks need to be repeated. Thus we output information such as how far a task has progressed and what item or objects are being processes.

\paragraph{F. Notifications} Mobile device notifications allow me to monitor the pipeline and attend to problems promptly. For example, a program will monitor jobs submitted for execution and send one of the science team that the jobs are complete.

\paragraph{G. Dynamically Interactive Code} Interactive elements are eliminated from the automated pipeline, but interactive elements play a crucial role in parts of the system that are executed outside of the pipeline, such as choosing a replacement for a bad datafile. We designed and implemented a system that enables tasks to ``know'' whether or not they are being executed in a supervised mode amenable to interaction. Otherwise, tasks running on compute nodes would ask for user input which is impossible to acquire.

\paragraph{H. Permanence} A completely dynamic system can be flexible and change information, for example updating a minor planet's designation. This is possible internally in \ac{HARVEST}, but some properties must remain fixed once set because of Citizen Science involvement. For example, a thumbnail name cannot change once assigned, because the thumbnail name is the identifying feature that allows linkage between the Citizen Science project and \ac{HARVEST}.

\paragraph{I. Minor Planet Identity Management} Managing names, identifications, and designations has proven to be one of the most difficult aspects of the pipeline. Requesting a specific solar system object occurs during the vast majority of pipeline tasks. Challenges include handling identity when an object breaks apart (e.g., P/2016 J1-A and P/2016 J1-B, \citealt{huiSplitActiveAsteroid2017}) or, more commonly, deciding how to handle objects that become consolidated (linked, e.g., P/2022 C$_4$ = P/2010 LK$_{36}$ = P/2016 MD = P/2016 PM$_1$ WISE-PANSTARRS, \citealt{fitzsimmonsCBAT5137COMET2022}) or retracted (e.g., 2011~UH$_{413}$ and 2013~QQ$_{95}$, \citealt{deenMPEC2020N22RETRACTION2020}).

\subsection{HARVEST Step 1: Catalog Queries}

The first step of my pipeline is to query catalog services that provide details about telescopic observations and information about solar system objects.

\subsubsection{Image Archives}
\label{methods:subsubsec:imageArchives} 

We query astronomical image archives for metadata, especially \ac{UT} observation date/time, sky coordinates, broadband filter, and the release date when the data becomes publicly available. At this stage we perform several initial checks and exclude images that do not meet specific criteria:

\begin{enumerate}
    \item \textbf{Airmass:} Pointings must be below 3 airmasses.
    \item \textbf{Moon Separation:} Observations with Moon separations $<4^\circ$ are excluded; we compute the Moon separation ourselves, with the distance computed between the Moon's center and the center of the telescope pointing provided in the image archive metadata.
    \item \textbf{Invalid Coordinates:} We exclude observations with illegal coordinates (e.g., \ac{RA} above 360$^\circ$, e.g., \ac{DECam} archive filename c4d\_201123\_040730\_ooi\_i\_v1.fits.fz).
    \item \textbf{Processing Type:} We exclude raw data, stacked (co-added) images, and non-image data (e.g., data quality masks). We exclude RAW data because (a) activity is harder to detect, and (b) the embedded \ac{WCS} are insufficient for our thumbnail purposes (i.e., the object may not appear near the center of a thumbnail image). Stacked data are excluded because they often produce images that eliminate the object or introduce image artifacts likely to complicate activity detection.
    \item \textbf{Broadband Filter:} We keep broadband filters (e.g., Sloan) and exclude others (e.g., H$\alpha$) that are unlikely to show faint activity. The retained filters are U, B, V, R, I, u, g, r, i, z, Y, J, H, K, M, VR, V+R, g+r+i, g+i, and variants therein (e.g., i variants include DECam-i, SDSS-i, and MegaPrime-i).
\end{enumerate}

We note that it is not uncommon in big data applications, such as with the \ac{HARVEST} processing astronomical image archive metadata, to encounter rare but problematic peculiarities such as sky coordinates outside of the defined range (e.g., RA$>360^\circ$) or representing an impossible elevation for a telescope to reach. Hence we implement safeguards to screen for these data which would otherwise cause problems for software not designed to handle these situations.

\paragraph{NSF's NOIRLab AstroArchive} The \ac{NSF} \ac{NOIRLab} AstroArchive\footnote{\url{https://astroarchive.noirlab.edu}} is the primary source for  data submitted for volunteer examination, \ac{DECam} image data from the Blanco 4~m telescope at the \ac{CTIO} in Chile. We also incorporate the \ac{KPNO} 4~m telescope archival data that AstroArchive hosts. Note: DECam images and metadata were hosted at the \ac{NOAO} prior to their consolidation into \ac{NOIRLab}.

\paragraph{CADC} The \ac{CADC}\footnote{\url{http://www.cadc-ccda.hia-iha.nrc-cnrc.gc.ca}} hosts archival data and provides search tools for many astronomical instruments. At present we query the \ac{CADC} for \ac{CFHT} MegaPrime and \ac{WIRCam} metadata.

\subsubsection{Minor Planet Data}
\label{methods:subsec:objData}

These tasks query external services for minor planet data including orbital parameters, designations, and rotation periods.

\paragraph{Minor Planet Center} The \ac{MPC} provides tabular data describing objects (asteroids and comets) and their orbits. The \ac{MPC} also serves as our first line of inquiry for object numbers, names, and provisional designations.

\paragraph{Kinoshita Comet Pages} The late Kazuo Kinoshita was an amateur astronomer in Japan who computed comet orbits and maintained a website\footnote{\url{https://jcometobs.web.fc2.com}} of comets. These pages included some orbital elements (e.g., perihelion distance) and name permutations we were unable to locate anywhere else. Kazuo Kinoshita passed away in 2021 July.

\paragraph{Ondrêjov Discoveries} This website, hosted at \ac{ASU} \footnote{\url{http://www.asu.cas.cz/~asteroid/news/numbered.htm}}, contains a list of discoveries made at Ond\^{r}ejov Observatory (site code 557). These names may not be present in \ac{MPC} or \ac{JPL} Horizons data, either because the object has a slightly different spelling or the object has a name completely unknown to the \ac{MPC} and Horizons. Importantly, the \ac{IMCCE} Quaero Service \footnote{\url{https://ssp.imcce.fr/webservices/ssodnet/api/quaero/}}, which serves the names returned by \ac{IMCCE} SkyBot (described below in Section \ref{methods:subsubsec:skyBot}), may return names that are found at this website but not necessarily at the \ac{MPC} or \ac{JPL} Horizons (Section \ref{methods:objectSpecificData}).

\paragraph{Astorb} The Lowell Observatory \ac{AstOrb} \footnote{\url{https://asteroid.lowell.edu/main/astorb/}} \citep{bowellPublicDomainAsteroid1994,moskovitzAstorbDatabaseLowell2021} contains object designations, orbital classifications, and related orbital elements (e.g., semi-major axis). This dataset is primarily used to provide orbital elements we were unable to attain elsewhere, and as an additional source for minor planet designations.

\subsection{HARVEST Step 2: New Data Handling}
\label{methods:subsec:newDataHandling}

\subsubsection{Datafile-specific Operations}

\paragraph{Magnitude Estimates} For computationally expedient and consistent estimation of depth we first compute a rough estimated magnitude depth for a given combination of exposure time, telescope/instrument, and broadband filter. The method we use is telescope-specific because whenever possible we use the observatory-supplied \ac{ETC}, sometimes referred to as an \ac{ITC}. In cases where we are unable to use an observatory-supplied \ac{ETC} we make use of the \ac{DECam} \ac{ETC}\footnote{\url{https://noirlab.edu/science/documents/scidoc0494}}, replacing the default mirror effective surface as needed. We compute an estimated magnitude reachable with a \ac{SNR} of 10. For filters that are not already $V$ band we offset the filter-specific depth to a generic $V$-band depth by applying an offset from known solar apparent Vega magnitudes\footnote{\url{http://mips.as.arizona.edu/~cnaw/sun.html}} \citep{willmerAbsoluteMagnitudeSun2018}. The purpose of this step is to match the $V$-band magnitudes supplied by ephemeris services (e.g., \ac{JPL} Horizons) for later use in determining the viability of activity detection (Delta Magnitude, Section \ref{methods:subsec:fieldAssesment}). To be clear, we only need a very rough estimate of depth for the purpose of gauging activity detection viability for each exposure.

\paragraph{Version Selection} involves choosing a single datafile for thumbnail extraction (Section \ref{methods:subsec:thumbnailExtraction}). All others are excluded as they are essentially duplicates. This process is necessary for two reasons: (1) some images have multiple versions of a specific processing application, or (2) multiple viable image processing techniques are available for a single observation.

\subsubsection{Object-Specific Data Handling}
\label{methods:objectSpecificData} 

\paragraph{Object Class} A solar system object can qualify as a member of more than one dynamical class. For example, all Apollo asteroids are types of \acp{NEO} and some (but not all) Mars-crossing asteroids are \acp{NEO}. At present we assign a primary class to each solar system object. Classes in \ac{HARVEST} are Comet, Amor, Apollo, Aten, Mars-crosser, \ac{IMB}, \ac{MMB}, \ac{OMB}, Cybele, Hungaria, \ac{JFC}, Hilda, Trojan, Centaur, Damocloid, \ac{TNO}/\ac{KBO}, or Interstellar Object. The \ac{MPC} and \ac{JPL} \ac{SBDB} do not contain all of these dynamical classes (e.g., Damocloid) so we take additional steps to reclassify objects. For example, we perform a check to reclassify objects labeled as Centaurs with a $T_\mathrm{J}<2$ as Damocloids, and $2<T_\mathrm{J}<3$ as \ac{JFC}s. These distinctions help us to collect thumbnail images from our library and organize them by dynamical class for efficient examination. The default position is to assume the class provided with the \ac{IMCCE} SkyBot results. We generally do not reclassify an object once a thumbnail image is produced because the object class is contained (as a code) in the thumbnail name for convenience, and the thumbnail names cannot be changed once submitted to \textit{Active Asteroids} for Citizen Scientist classification.

\paragraph{NASA JPL Object Data} In order to optimize calculations and reduce query calls to \ac{JPL} we maintain our own internal database of NASA \ac{JPL} provided solar system object-specific parameters, such as rotation period (used for observation planning) and semi-major axis $a$. Our queries to \ac{JPL} come in two forms: the \ac{JPL} Horizons ephemeris service\footnote{\url{https://ssd.jpl.nasa.gov/horizons/}} and the \ac{JPL} Small Body Database Lookup service\footnote{\url{https://ssd.jpl.nasa.gov/tools/sbdb_lookup.html}}, both of which return object-specific information.

\paragraph{Solar System Object Parameters} During this phase we attempt to collect a unified set of object-specific dynamical properties, namely semi-major axis $a$, inclination $i$, eccentricity $e$, perihelion distance $q$, and aphelion distance $Q$. At this stage we compute the Tisserand parameter \citep{tisserandMecaniqueCeleste1896} with respect to Jupiter, $T_\mathrm{J}$, commonly used as a metric to quantify the gravitational interaction between Jupiter and a small body given their respective orbits \citep{kresakJacobianIntegralClassificational1972}. The parameter is given by

\begin{equation}
	T_\mathrm{J} = \frac{a_\mathrm{J}}{a} + 2\sqrt{\left(1-e^2\right)\frac{a}{a_\mathrm{J}}}\cos(i),
	\label{eq:TJ}
\end{equation}

\noindent where $e$ represents osculting eccentricity, $i$ orbital inclination, $a$ the semi-major axis of the body, and $a_\mathrm{J}$ is Jupiter's semi-major axis.

\subsection{HARVEST Step 3: Assessing Fields} 
\label{methods:subsec:fieldAssesment}

These tasks are specific to a Field record, defined as a unique combination of an observing date/time and sky coordinates. Multiple field records may exist for a single observation as sky coordinates are adjusted during the astrometry phase of the pipelines used with, for example, \ac{DECam} data.

\subsubsection{SkyBot}
\label{methods:subsubsec:skyBot}

\paragraph{Query} The \ac{IMCCE} SkyBot service\footnote{\url{https://ssp.imcce.fr/webservices/skybot/}} \citep{berthierSkyBoTNewVO2006} allows a user to determine which solar system objects (if any) may be present in an astronomical image given (1) the \ac{UT} date/time, (2) \ac{RA} and \ac{Dec} sky coordinates, and (3) an angular shape that specifies the size of the area to search. The shape chosen ideally minimizes area not covered by an imaging detector. Given the roughly circular shape of the \ac{DECam} chip arrangement, we primary use the radius (1.1$^\circ$) of the 2.2$^\circ$ \ac{FOV} to conduct a \textit{cone search}. We also use rectangular queries for instruments more suitable to this configuration, for example \ac{CFHT} MegaCam. Queries are carried out using the online SkyBot \ac{API}.

\paragraph{Resubmitting Fields}
For each query submitted, SkyBot returns a unique ticket ID. Along with the date of the query we keep track of the age of each SkyBot Ticket record. We limit the number of SkyBot queries per day, however we try not to allow Ticket dates to exceed 90 days of age. We institute these policies to accommodate new object discovery as well as improvements to object orbital elements. We wrote a program that will flag specific Ticket records for resubmitting to SkyBot based on orbital parameters, however the 90 day age limit has proven far more computationally efficient and fewer overall queries to the SkyBot servers.


\subsubsection{Delta Magnitudes}
\label{methods:subsubsec:deltaMagLim}
During the SkyBot step we compute a simple metric that describes how much brighter (or fainter) an object will appear as compared to the depth of the exposure we computed during Magnitude Estimates (Section \ref{methods:subsec:newDataHandling}. The metric is

\begin{equation}
    \Delta_\mathrm{mag} = V_\mathrm{JPL} - V_\mathrm{ITC},
\end{equation}

\noindent where $V_\mathrm{JPL}$ is the Horizons provided apparent $V$-band magnitude computed for the object at the time of exposure, and $V_\mathrm{ITC}$ is the $V$-band magnitude depth we computed for this exposure, as described in Section \ref{methods:subsec:newDataHandling}. $\Delta_\mathrm{mag}<0$ indicates the object will likely appear bright enough in the image to be detected, whereas $\Delta_\mathrm{mag}>0$ would not. We set a maximum $\Delta_\mathrm{mag}$ of $-1$ for all detections. Note: the apparent magnitude is computed by Horizons, however the object may appear brighter or fainter, for example, in cases of activity outburst or prior mass loss, respectively. At the time of this writing, roughly 57\% ($\sim$21 million) of all potential minor planet detections in \ac{HARVEST} were excluded because of the $\Delta_\mathrm{mag}$ threshold.

\subsubsection{Trail Length}
\label{methods:subsec:deltaPixels}
\begin{figure}
    \centering
    \begin{tabular}{cc}
        \includegraphics[width=0.45\linewidth]{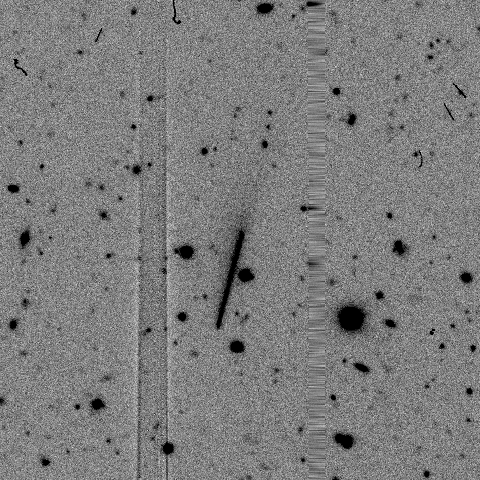} & \includegraphics[width=0.45\linewidth]{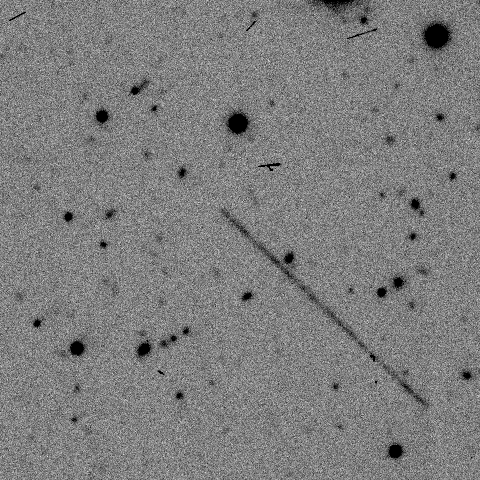}  \\
    \end{tabular}
    \caption{Examples of trails resulting from high rates of apparent motion on the sky. Images credit: this work. 
    \textbf{Left:} Comet C/2014~QU$_2$ (PANSTARRS) as observed by \ac{DECam} on \ac{UT} 2014 September 22 (Prop. ID 2014B-0303, PI Sheppard). Even with a trail length of about 100 pixels, activity is still clearly visible in the upward direction. 
    \textbf{Right:} Apollo class \ac{NEO} 2014~EH$_{45}$ observed by \ac{DECam} on \ac{UT} 2014 March 27 (Prop. ID 2014A-0479, PI Sheppard) with an approximately 280 pixel length trail. The variable brightness seen along the length of the trail is diagnostic of fast rotation.
    }
    \label{methods:fig:trails}
\end{figure}

\textit{Delta pixels} is a metric we compute that describes the number of pixels an object appears to travel across the FOV (Figure \ref{methods:fig:trails}), given the pixel scale of the instrument and the object's apparent rate of motion on the sky. We use this value to screen, for example, images for use in Citizen Science training and the Field Guide, described in Section \ref{methods:sec:citsci}.

This value may be used in the future to exclude thumbnails with long (e.g., $>15$ pixel) trails that may be confusing to volunteers and lead to activity detection false positives. Alternatively, in response to \textit{Active Asteroid} volunteer comments, we may use this value to help search for fast-rotating asteroids known as fast-rotators (Figure \ref{methods:fig:trails}). These trails may appear to have variable width, or even appear dashed, depending on the physical geometry of the object and its rotation speed.

\subsection{HARVEST Step 4: Thumbnail Preparations}
\label{methods:subsec:thumbPrep}
The thumbnail preparation tasks in this section are conducted in parallel because we do not wait for datafiles to download. Instead, we work with the datafiles already on disk, and return to incomplete or missing datafiles during a subsequent pipeline execution. This avoids long delays induced if we wait for large quantities of data to finish downloading from archives.

\subsubsection{Data Download}
During this phase we perform several tasks: (1) check datafiles flagged as ``on disk'' to ensure they are actually present and, if not, flag the datafile as ``to download.'' We  generate bash scripts for daemons (persistent tasks) dedicated to downloading data. Datafiles are typically flagged for download for three reasons: (1) new data was added to an astronomical archive that does not have a proprietary period, (2) datafiles known by \ac{HARVEST} with proprietary periods that have since ended, and (3) a new solar system object was found in a datafile we previously had finished processing and purged from disk.

\subsubsection{File Inspection}
\label{methods:subsec:datafileIntrospection}

\begin{figure}
    \centering
    \includegraphics[width=0.5\linewidth]{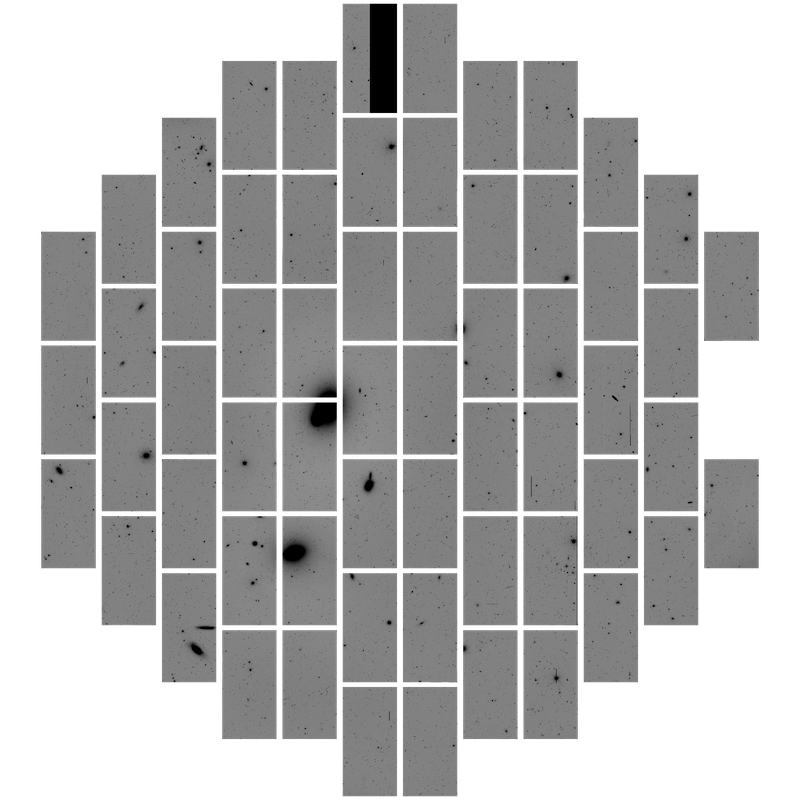}
    \caption{This mosaic shows the \ac{CCD} chip arrangement of the \acf{DECam} instrument. Image data is not recorded anywhere outside of the \ac{CCD} chips (rectangles). Areas between chips are commonly referred to as ``chip gaps.''}
    \label{methods:fig:mosaic}
\end{figure}

\paragraph{Chip Corners} Here we examine in parallel downloaded image data and populate a database table with the sky coordinates (RA, Dec) of each corner of each camera chip. This step provides an order-of-magnitude speed increase (as compared with manually reading the bounds from an image file) when determining which solar system objects do not fall on a chip, either because they are on a chip gap, or on a chip that is not functional at the time of observation (as is the case for some \ac{DECam} images). Moreover, with the corner coordinates recorded, we do not need to download a datafile previously processed and purged to check chip corners because the coordinates are already cached in our database. Chips gaps are the spaces between the \ac{CCD} chips of cameras that use multiple detectors; these gaps do not record any image data.

\paragraph{Header Archiving} As part of Chip Corners we also write a compressed text file to disk containing all (primary and chip-specific) header information for each archival image file. This allow us to make use of data held in headers even after the parent datafile has been purged.

\subsection{HARVEST Step 5: Thumbnail Extraction}
\label{methods:subsec:thumbnailExtraction}

This stage consists of two steps executed in parallel. We process up to 50 datafiles simultaneously, which stays below our estimated \ac{HPC} cluster file system throughput limit.

\subsubsection{FITS Thumbnails}

We extract \ac{FITS} format ``cutouts'' (extracted region of image data) of each unexcluded SkyBot result (Section \ref{methods:subsubsec:skyBot}). We first execute a \ac{SQL} query of our internal database table that stores records of chip corners (Section \ref{methods:subsec:datafileIntrospection}) to determine which chip should contain the image of the object. If the object does not fall within the bounds of any chip, the thumbnail and associated SkyBot records are excluded, so (1) we do not attempt to extract the thumbnail image again, and (2) our database queries remain optimized, and (3) statistical reports reflect current state of \ac{HARVEST} (e.g., number of thumbnail images processed). If the object is located within the bounds of a camera chip then we perform an image cutout that produces a $480\times480$ pixel thumbnail image in \ac{FITS} format. We handle boundary conditions (e.g., the object falls near a chip edge) by filling empty pixels coordinates with \texttt{NULL} values.

Crucial for future analyses, our Thumbnail Extraction Tool adjusts the \ac{WCS} to match the cutout image prior to writing the \ac{FITS} file to disk. \ac{WCS} in \ac{FITS} headers provide a reference that effectively translates pixel coordinates to sky coordinates. For convenience and record completeness we also copy global header values (e.g., observation date and time) and the headers associated with the corresponding chip to the resulting \ac{FITS} format thumbnail image file.

\subsubsection{PNG Thumbnails}
We convert the \ac{FITS} format thumbnail to a \ac{PNG} image file. In order to maximize the likelihood of activity detection, we carry out an iterative rejection technique to determine the best contrast range for the image (see Chapter \ref{safari:procedure} for details).

\subsection{HARVEST Step 6: Post Thumbnails}

Source Analysis, which is carried out in parallel, must conclude prior to the Counting steps, which may all be performed in parallel.

\begin{figure*}
    \centering
    \begin{tabular}{ccc}
         \includegraphics[width=0.32\linewidth]{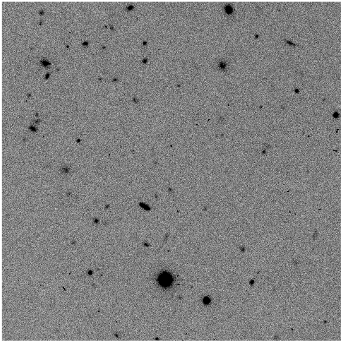} & \includegraphics[width=0.32\linewidth]{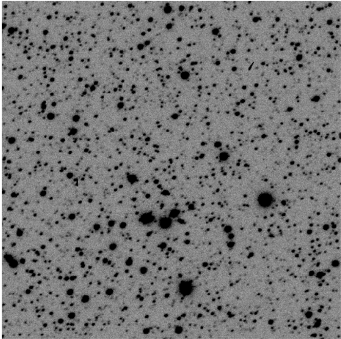} & \includegraphics[width=0.32\linewidth]{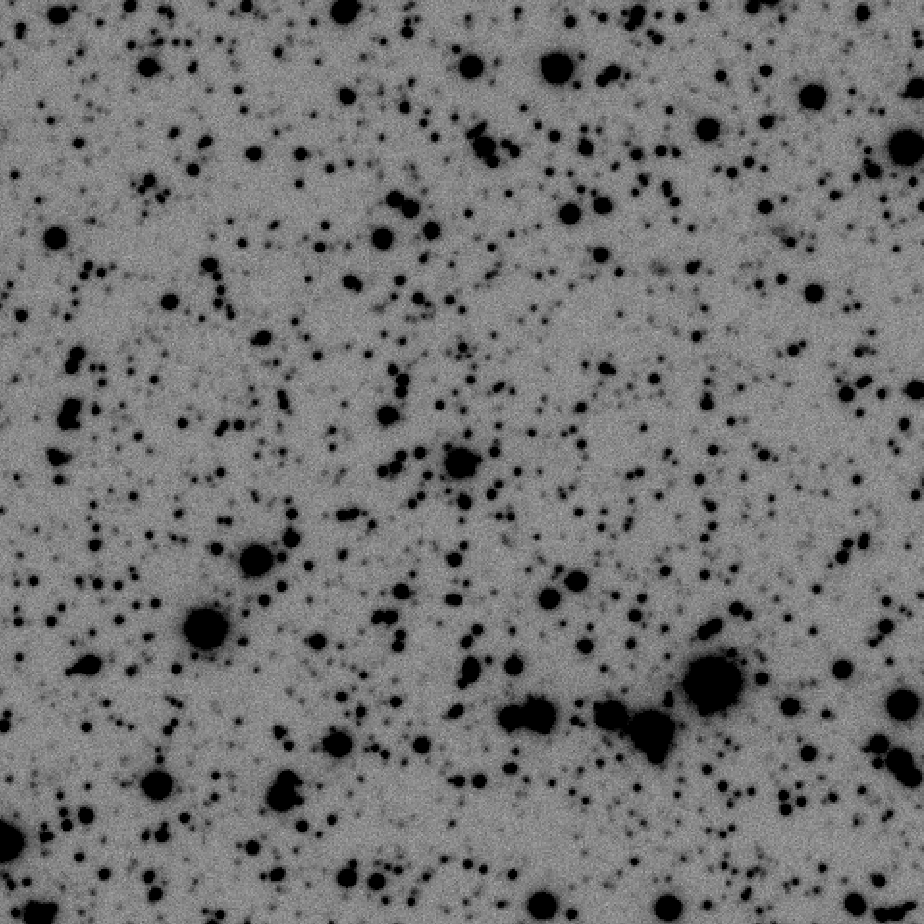} \\
         (a) & (b) & (c)
    \end{tabular}
    \caption{The three causes for excluding a thumbnail during the Source Analysis section of the \acf{HARVEST} pipeline. 
    (a) Empty center: there is no object at the center of the frame to inspect.
    (b) Crowded field: there are too many sources in the center 270$\times$270 pixels, making identification of the target difficult. 
    (c) Blended sources: multiple overlapping sources are seen here, making it difficult to identify which source (if any) is the target.
    }
    \label{methods:fig:exclusionBySourceAnalysis}
\end{figure*}

\subsubsection{Source Analysis}
\label{methods:subsubsec:sourceAnalysis} 
We make use of \texttt{SExtractor} \citep{bertinSExtractorSourceExtractor2010} to determine the pixel coordinates of each point source within each thumbnail image. At this stage we exclude thumbnail images for cases which would be confusing to volunteers or unlikely to lead to an activity detection. The three exclusion causes are given with numbers current as of 10 July 2022, when \ac{HARVEST} had a total of 22,004,739 unexcluded thumbnail images, spanning all instruments. An example from each scenario is provided in Figure \ref{methods:fig:exclusionBySourceAnalysis}.

(1) No source was detected within the center $20\times20$ pixels of the thumbnail image (Figure \ref{methods:fig:exclusionBySourceAnalysis}a). This resulted in 16\% of thumbnails being excluded, 4,248,133 in total. \texttt{SExtractor} detects even faint sources, so if it cannot identify an object at the center of the frame we do not expect volunteers to be able to either. This may be due to poor observing conditions or cases where the object did not appear directly in the center of the frame because, for example, there was a high positional uncertainty.

(2) Too many ($>150$) sources were in the center $270\times270$ pixels of the thumbnail image (Figure \ref{methods:fig:exclusionBySourceAnalysis}b). The 952,289 thumbnails excluded for this reason represents another 4\% reduction in the total number of viable thumbnails. This situation makes identifying which object is the target confusing for volunteers.

(3) When there are too many ($>5$) overlapping (blended) sources at the center of the frame (Figure \ref{methods:fig:exclusionBySourceAnalysis}c) it can be very difficult to determine which source is the object of interest. This is the least common cause for thumbnail exclusion, with just 84,697 (0.4\%) thumbnails excluded. We exclude these images because it can be exceptionally difficult to identify the object of interest in the image.

\subsubsection{Source Tallying}
\label{methods:subsubsec:counting}

\paragraph{Objects per Field} At this point all exclusion routines have finished executing. We tally how many unexcluded solar system objects are present in each field. This step optimizes our analysis of the distribution of on-sky of solar system objects (see Chapter \ref{chap:discussion}).

\paragraph{SkyBot Source Density} We tally unexcluded SkyBot results associated with each SkyBot Ticket record. Tickets include a unique identification for the query performed, and this metric is used for later internal reporting. Tracking tickets is also crucial for troubleshooting with, for example, the \ac{IMCCE}.

\paragraph{Observation Source Tally Consolidation} We consolidate the above tallies into the parent Observation records in our database to optimize analysis tasks. For example, this optimization reduces the compute time for plotting the sky distribution of minor planets in our database (Figure \ref{discussion:fig:asteroidsOnSky}) by orders of magnitude.

\subsection{HARVEST Step 7: Reporting}

\paragraph{SkyBot Reports} Here we determine the age of SkyBot tickets and create an overview of the status of all fields. This provides a status of SkyBot queries within \ac{HARVEST} and can be diagnostic of problems (e.g., if many tickets are much older than 90 days). These reports are periodically reviewed by our science team.

\paragraph{Objects per Field Plot} This plot (Figure \ref{discussion:fig:asteroidsOnSky}) shows how many solar system objects are present for each telescope pointing in our database. These data derive from the tallies executed in Step 6 (Section \ref{methods:subsubsec:counting}). The Objects per Field Plot has many uses, including revealing gaps in sky coverage which, in turn, result in observational biases. For example, this plot shows how far north and to what degree our sky coverage extends from our predominately southern hemisphere originating observations.

\subsection{HARVEST Step 8: Maintenance}

\subsubsection{Datafile Routines}
\label{methods:subsubsec:datafileRoutines} 

\paragraph{Datafile Checks} We inspect downloaded datafiles to ensure datafile integrity; this additional check is a secondary stage, prompted by flags set during \ac{HARVEST} Step 4: Thumbnail Preparations (Section \ref{methods:subsec:thumbPrep}). Here we check \ac{FITS} standards for each image data and header extension of the file, as well as a basic file size check. Datafiles that we determine are corrupt we attempt to download a second time. Should the second datafile also be corrupt, we flag the datafile as ``bad'' in our database and attempt to find a replacement, either a different version or another acceptable image processing type.

\paragraph{Datafile Exclusion by Property} This task excludes datafile records based on specific properties. At present the only task is excluding datafiles with exposure times $<1$~s, including \texttt{NULL}, 0~s, and negative exposure time values.

\paragraph{Purge Datafiles} As we do not have the storage space to keep a copy of all data we download from astronomical data archives we must delete files once we have finished carrying out all steps of the \ac{HARVEST} pipeline.

\subsection{Auxiliary Procedure: Comparison Images}
\label{methods:subsec:comparisonImages}

\begin{figure*}[h]
    \centering
    \begin{tabular}{cc}
        \includegraphics[width=0.45\linewidth]{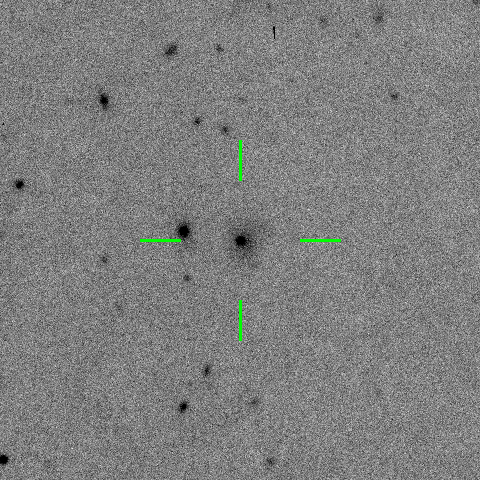} & \includegraphics[width=0.45\linewidth]{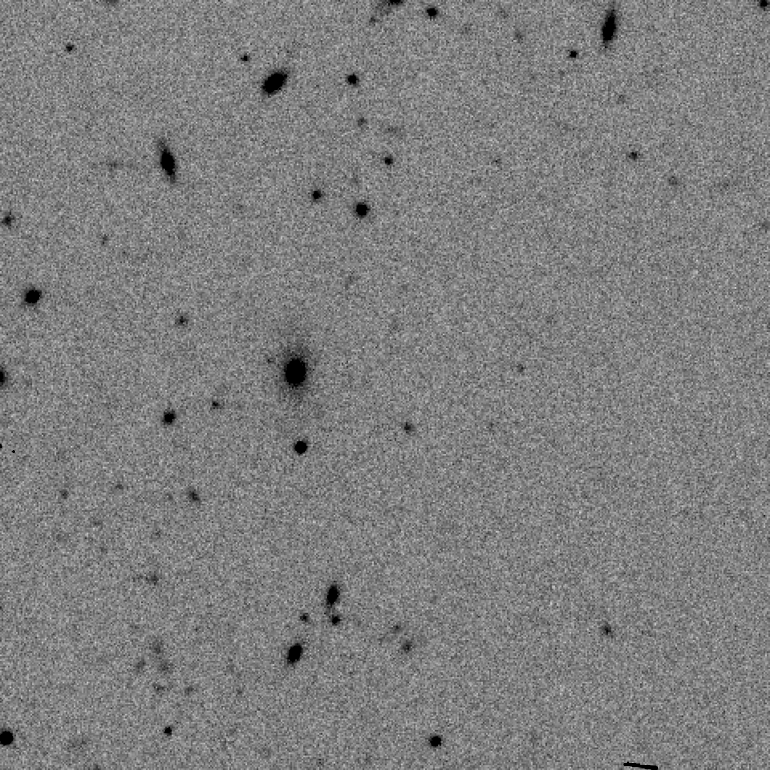} \\
    \end{tabular}
    \caption{Comparison between two identical areas on sky, imaged with similar circumstances (e.g., instrument, exposure time, broadband filter), but taken at times far apart enough that the solar system object would not appear in the comparison image. Images credit: this work. 
    \textbf{Left: } This thumbnail image of \acf{JFC} 2015~TC$_1$ was taken on \ac{UT} 2015 December 19 (Prop. ID 2012B-0001, \acs{PI} Frieman) and submitted for examination by \textit{Active Asteroids} (\url{http://activeasteroids.net}) volunteers; all volunteers classified this image as showing activity, and several users also shared this image in the Talk forums as a potentially active object.
    \textbf{Right: } This image of the same field was captured \ac{UT} 2016 January 1 (Prop. ID 2012B-0001, \acs{PI} Frieman) probes somewhat fainter than the original image on the left. The comparison illustrates that there is no background galaxy or other phenomenon present at the exact coordinate where 2015~TC$_1$ was seen.
    }
    \label{methods:fig:comparison}
\end{figure*}

A common situation arises where a thumbnail shows what appears to be activity, but the activity is ambiguous. For example, Citizen Scientists highlighted a thumbnail image (Figure \ref{methods:fig:comparison}) as having activity, through discussion forums as well as classification. To ascertain whether or not suspected activity is actually a background source (e.g., galaxy), a comparable image (e.g., same area on sky, similar depth) could be useful, so long as enough time has elapsed such that the solar system object is no longer in the field; this could be from minutes to days later, depending on the object's apparent rate of motion.

I created a tool, which we employ on a case-by-case basis, to address this scenario by finding \textit{comparison images} to accompany the image with potential activity. My tool employs a systematic approach to find the closest match as possible, prioritizing (in approximate order) (1) enough time has elapsed to ensure the object is not in the \ac{FOV} based on the object's apparent rate of motion, (2) instrument, (3) delta magnitude limit (Section \ref{methods:subsec:fieldAssesment}), (4) broadband filter, and (5) processing type. (Processing types vary by instrument. For \ac{DECam} we make use of InstCal and Resampled images. The latter incorporates more processing but is not available for all images in the archive.) Crucially, the tool relies on our internal database of chip corner coordinates (Section \ref{methods:subsec:datafileIntrospection}). Once potential comparison image sources have been identified (I set the default to be five comparison images) then the datafiles are flagged for download and download scripts generated. Once downloads have finished we then run a separate tool that produces the thumbnail cutout images.

\section{Citizen Science Project}
\label{methods:sec:citsci}

We chose the Citizen Science platform \textit{Zooniverse}\footnote{\url{https://www.zooniverse.org}} for our project, named \textit{Active Asteroids}\footnote{\url{http://activeasteroids.net}}. Zooniverse has a proven track record of success and, importantly, they provided customization and support that facilitates any hosted project's success.

The overall process, from inception to launch, is as follows:

\begin{enumerate}
    \item Prepare project on Zooniverse (see sections below)
    \item Conduct an initial ``Beta Release'' to test project viability
    \item Formally launch the project (i.e. open to the public)
    \item In a cyclic fashion we continue to
    \subitem a. Interact with volunteers via online forums and Zooniverse messages, the internal Zooniverse inter-user communication system similar to email
    \subitem b. Download and analyze results
    \subitem c. Prepare and upload the next set of images
    \subitem d. Notify volunteers that additional work is available via Twitter and an email newsletter sent by Zooniverse
    \subitem e. Conduct scientific investigation (Section \ref{sec:methods:followup}) into the resulting candidates as they become available
\end{enumerate}

\subsection{Project Components}
\label{methods:subsec:projectComponents}

\begin{figure}
    \centering
    \includegraphics[width=0.8\linewidth]{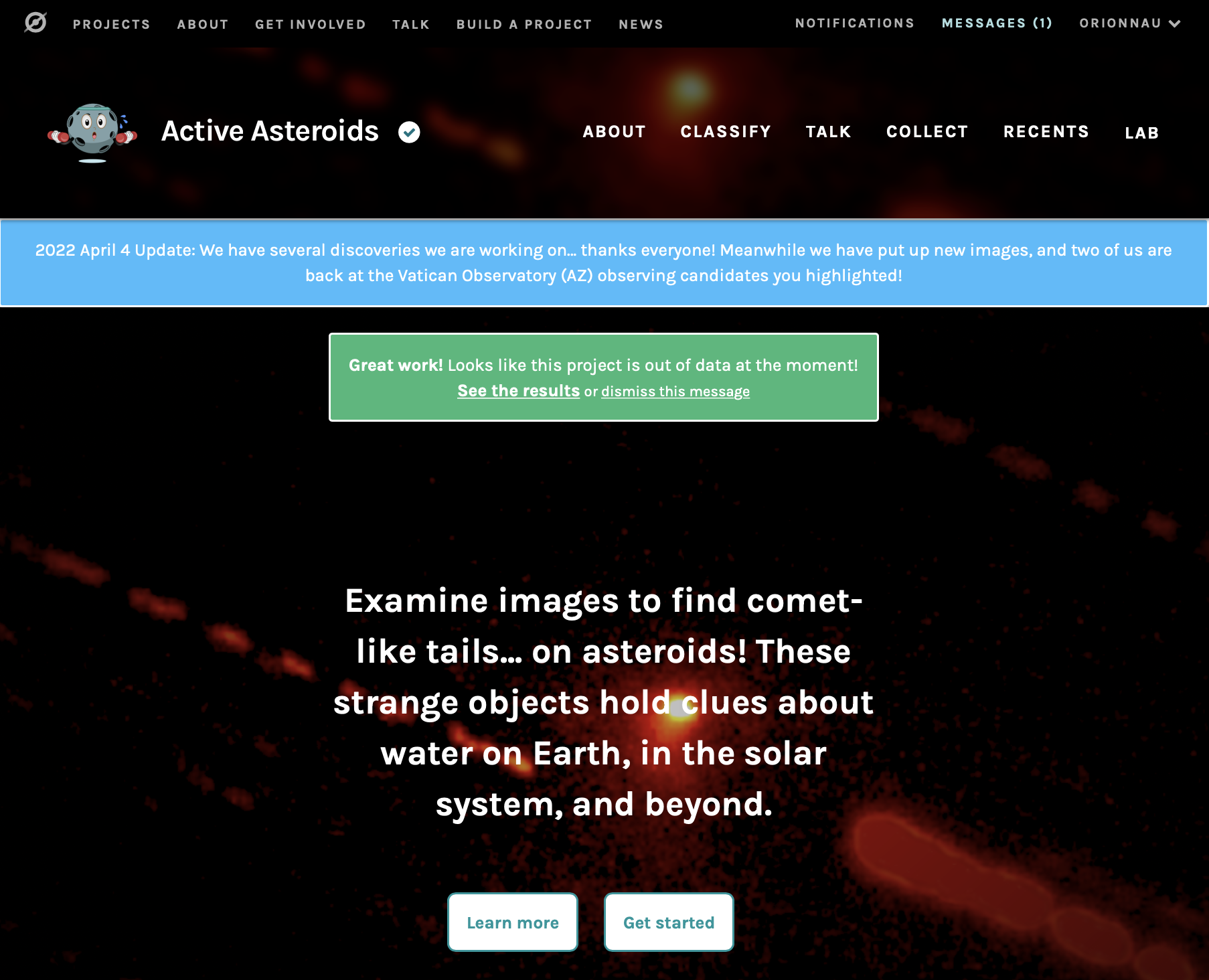}
    \caption{Landing page for the \textit{Active Asteroids} Citizen Science project hosted at Zooniverse. The logo we designed is at the top-left. The horizontal blue text box displays updates we provide to volunteers. The green box indicates the project status; as indicated, the project was complete, so uploading another Subject Set (collection of images) would be the next step in our workflow.}
    \label{methods:fig:citsci:landing}
\end{figure}

Zooniverse provides a standard framework for each project. All users start at the project landing page, shown in Figure \ref{methods:fig:citsci:landing}. From there Citizen Scientists can begin classifying (working) immediately or navigate to one of the other pages, described below.

\subsection{Workflow}
\label{methods:subsec:workflow}

\begin{figure}
    \centering
    \includegraphics[width=0.75\linewidth]{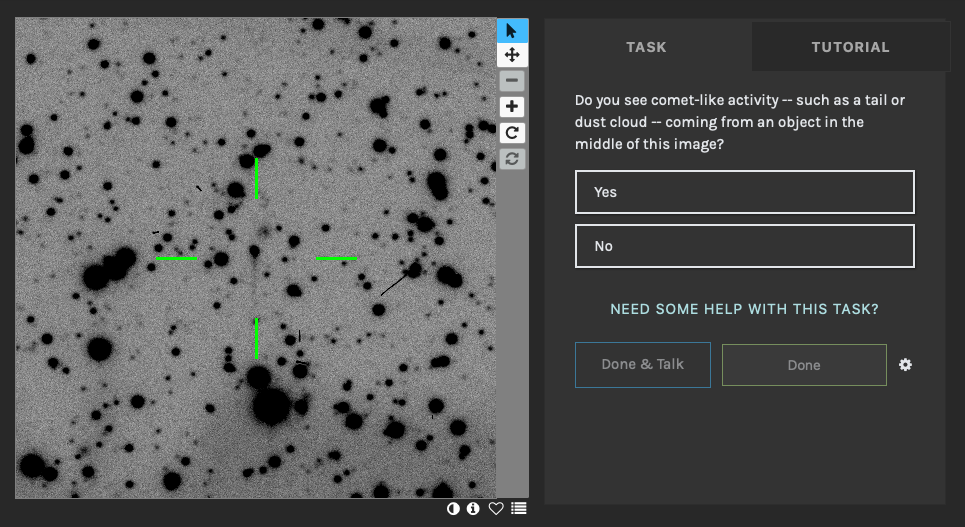}
    \caption{The \textit{Active Asteroids} project workflow is simply to ask whether or not volunteers can see activity emanating from the object at the center of the screen, indicated by the green reticle. Citizen Scientists can also tag thumbnails for discussion in Talk forums or collect them in their own albums.}
    \label{methods:fig:workflow}
\end{figure}

The \textit{Workflow} describes the task we are asking the Citizen Scientists to perform. The \textit{Active Asteroids} workflow is concise: we ask volunteers whether or not they see activity emanating from the object at the center of a thumbnail image, such as the one shown in Figure \ref{methods:fig:workflow}.

\subsubsection{Tutorial}
\begin{figure}
    \centering
    \begin{tabular}{cccc}
        \includegraphics[width=0.22\linewidth]{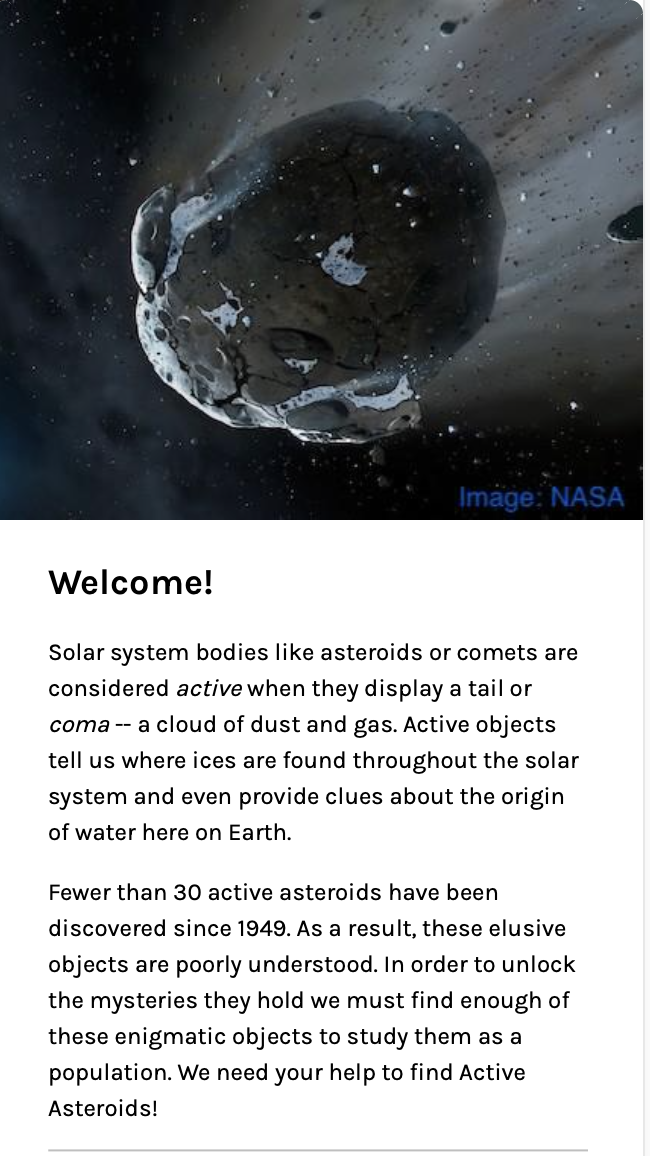} & \includegraphics[width=0.22\linewidth]{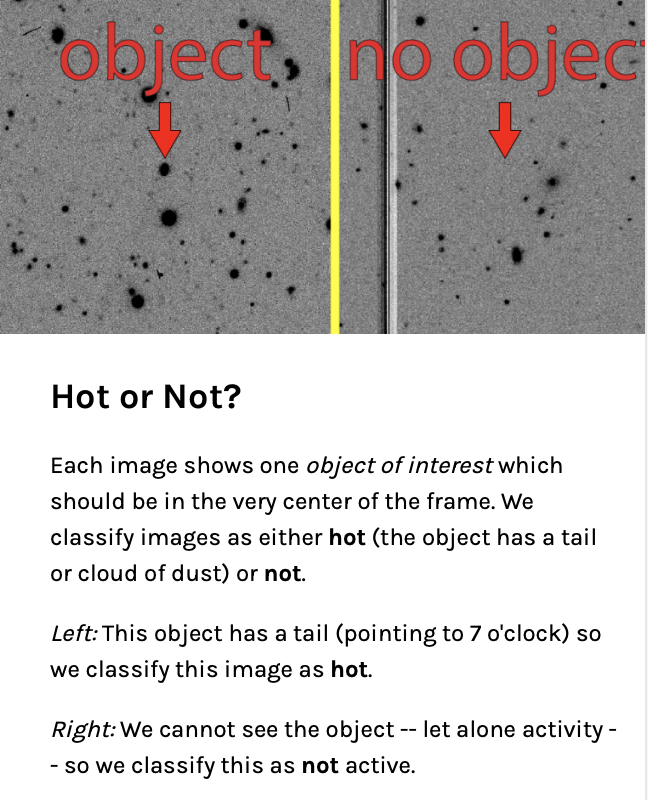} & \includegraphics[width=0.22\linewidth]{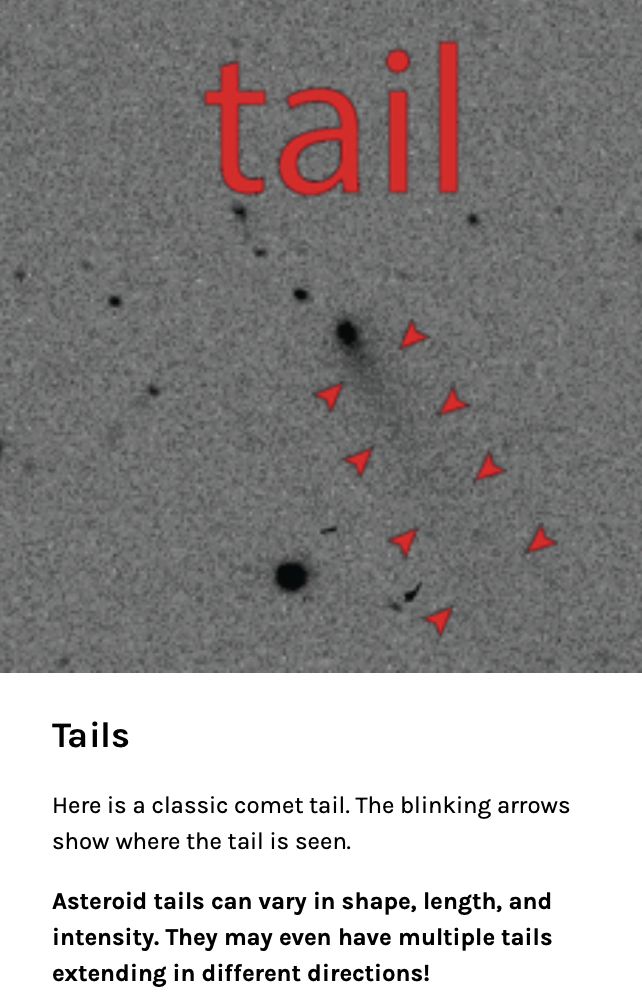} & \includegraphics[width=0.22\linewidth]{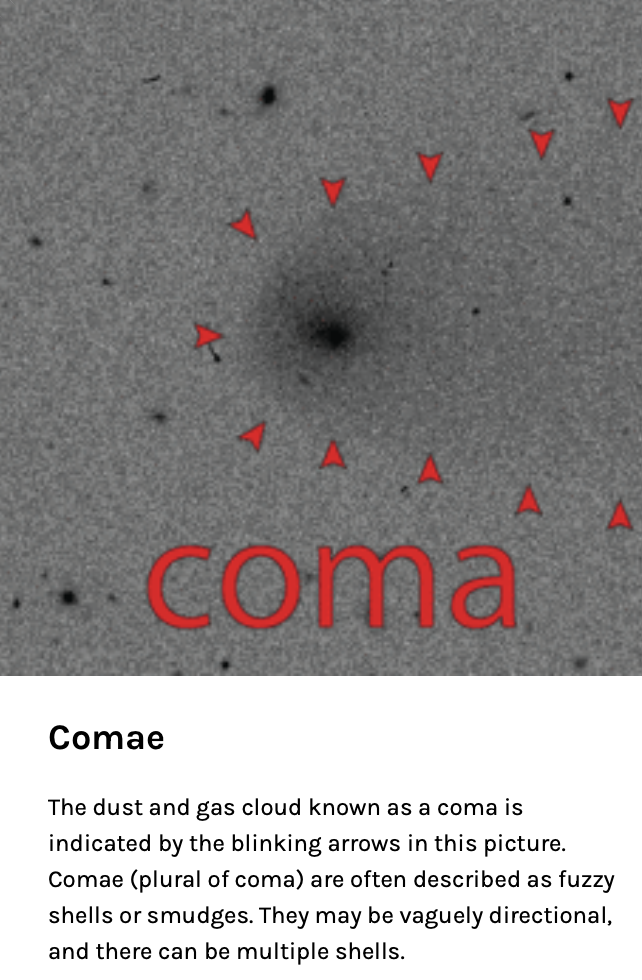}\\
        (1) & (2) & (3) & (4) \\
        \\
        \includegraphics[width=0.22\linewidth]{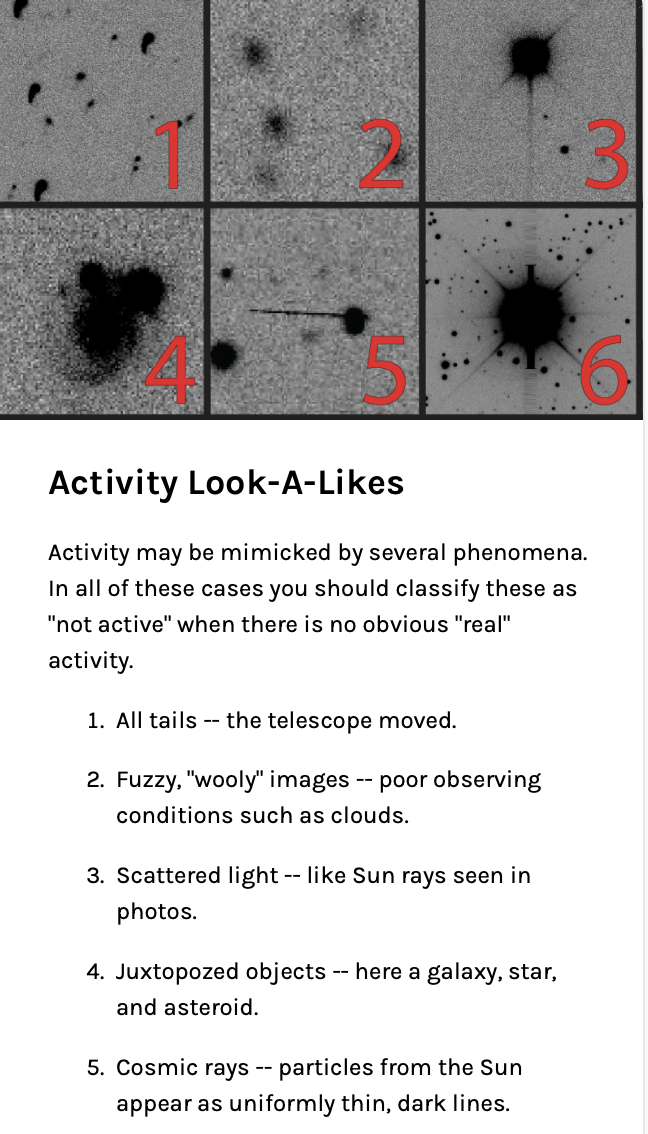} & \includegraphics[width=0.22\linewidth]{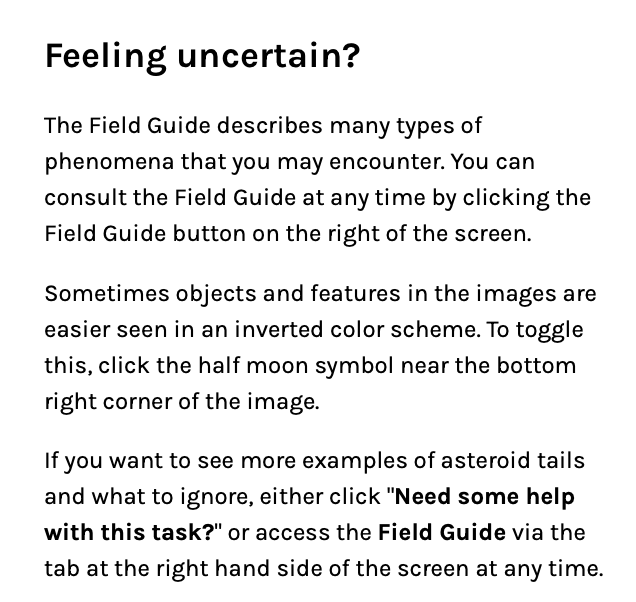} & \includegraphics[width=0.22\linewidth]{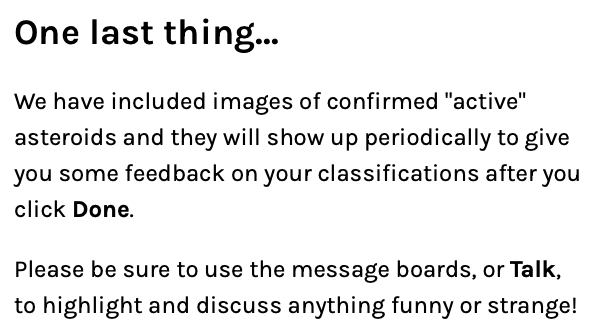} & \includegraphics[width=0.22\linewidth]{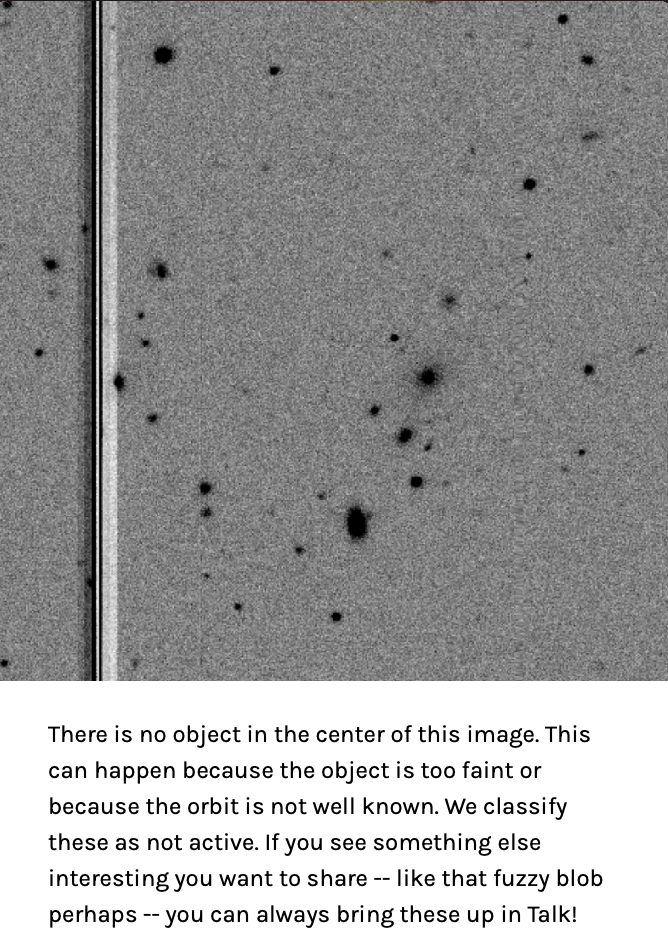}\\
        (5) & (6) & (7) & (8)\\
    \end{tabular}
    \caption{The \textit{Active Asteroids} Tutorial. (1) ``Welcome'' provides a project overview. (2) ``Hot or Not'' describes the workflow. (3) ``Tails'' describes tails with an example. (4) ``Comae'' introduces volunteers to the less familiar coma morphology. (5) ``Activity Look-a-Likes'' describes common false positive scenarios. (6) ``Feeling uncertain?'' explains what to do if a classification is ambiguous. (7) ``One last thing...'' lets users know about the injected training images. (8) This panel, requested by volunteers, describes a common situation where the object cannot be conclusively identified.}
    \label{methods:fig:tutorial}
\end{figure}

Figure \ref{methods:fig:tutorial} shows the eight panels of the project tutorial which is shown to volunteers the first time they participate in the project. The tutorial is also available at all times in a panel of the classification workflow window.

\subsubsection{Field Guide}

\begin{figure}
    \centering
    \begin{tabular}{cccc}
         \includegraphics[width=0.22\linewidth]{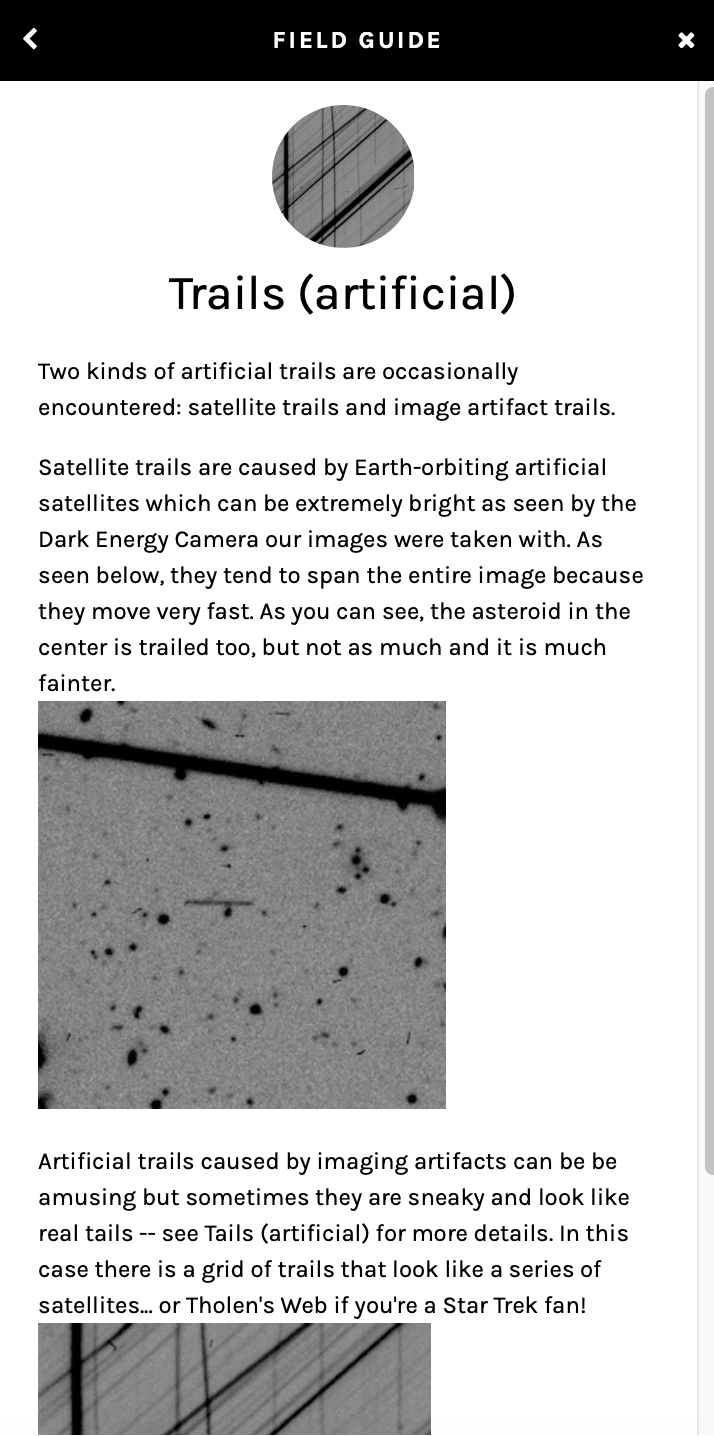} & \includegraphics[width=0.22\linewidth]{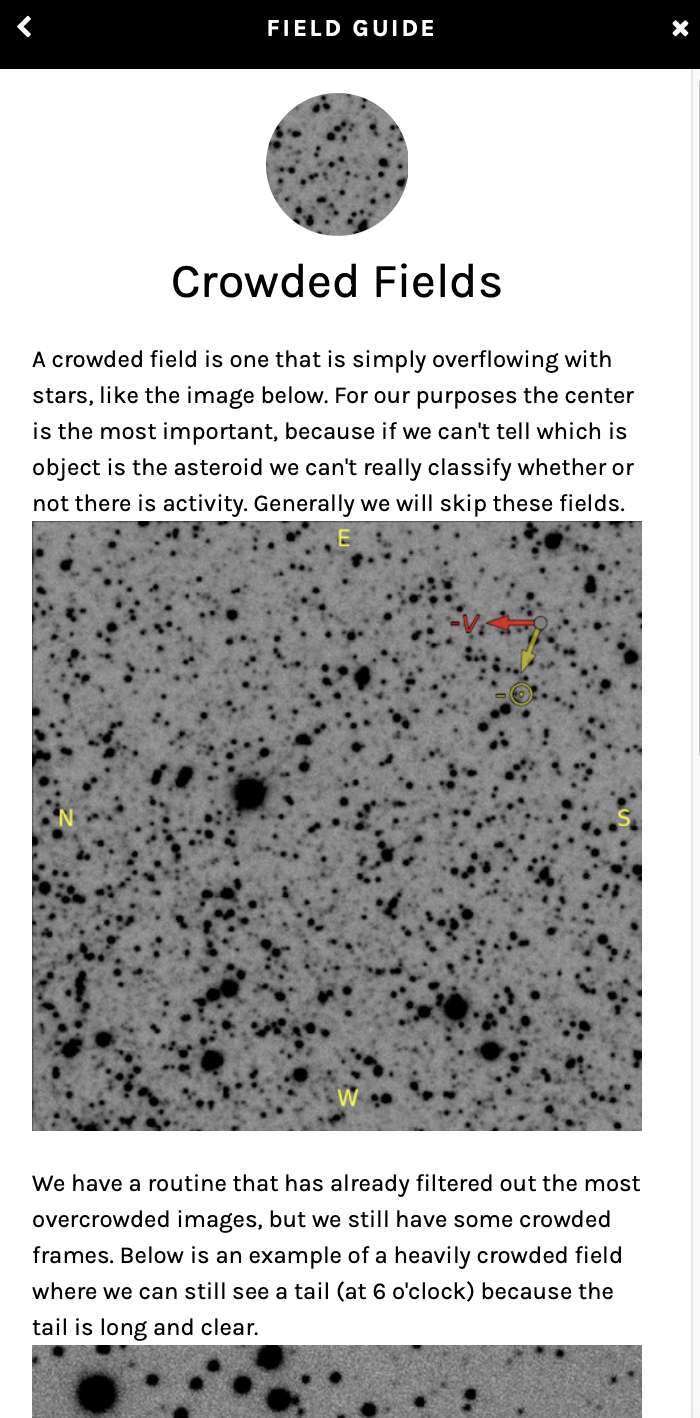} & \includegraphics[width=0.22\linewidth]{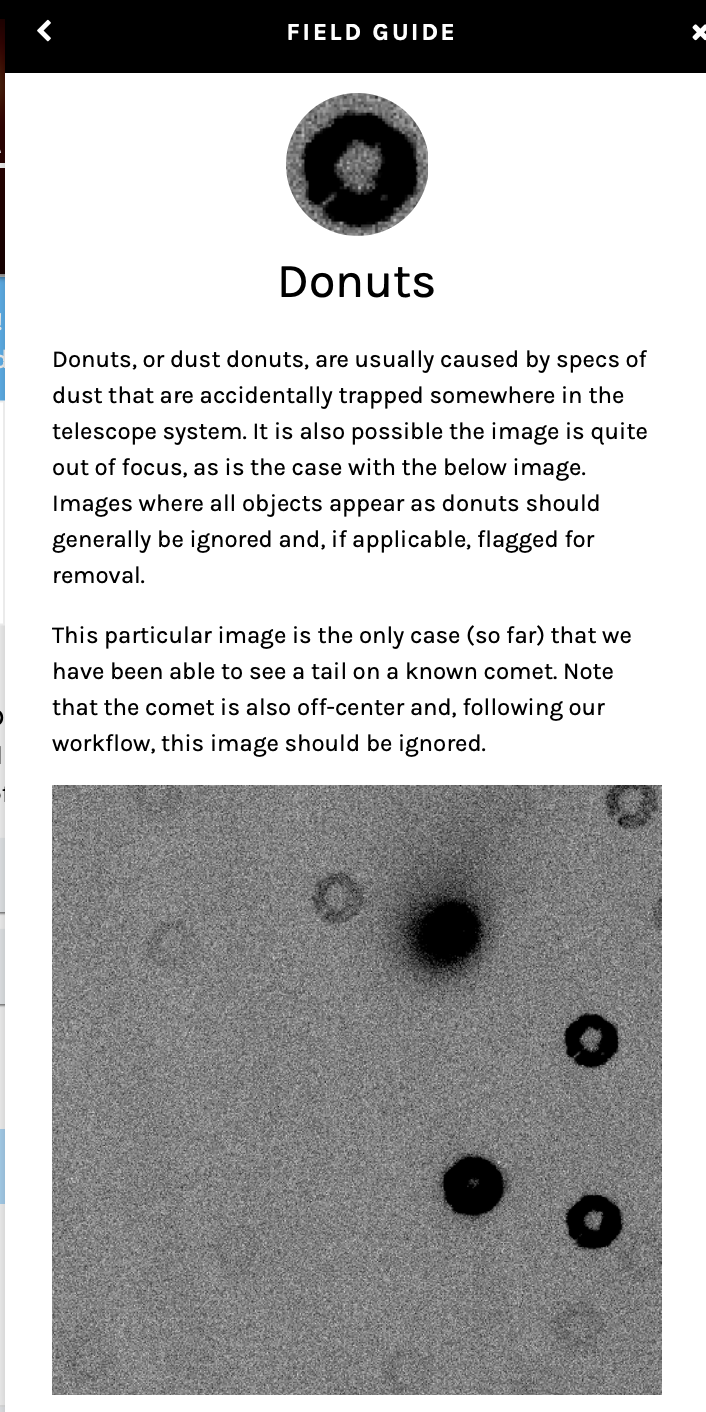} & \includegraphics[width=0.22\linewidth]{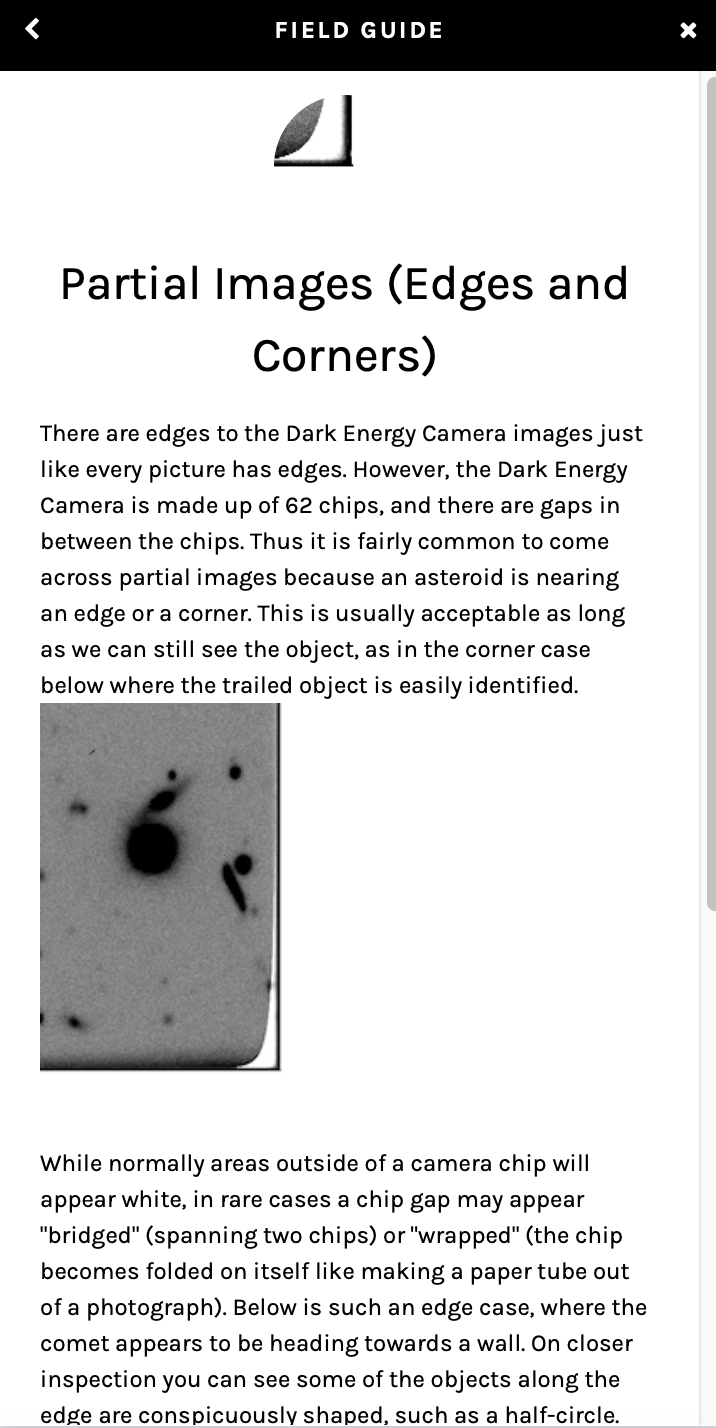} \\
         (1) & (2) & (3) & (4)\\ 
    \end{tabular}
    \caption{Four examples from the Field Guide of the \textit{Active Asteroids} project. (1) ``Trails (artificial),'' includes an example satellite trail and a a example of a telescope tracking problem that results in all sources in a field becoming trailed. (2) ``Crowded Fields'' shows an example of a situation we try to automatically screen out from the project, but which may occasionally make it through our filter. (3) ``Donuts'' discusses this optical issue and includes a donut with activity indicators. (4) ``Partial Images (Edges and Corners)'' explains what happens when an object falls near one or two chip edges and includes an example.}
    \label{methods:fig:fieldGuide}
\end{figure}

The Field Guide is available at all times during the project classification workflow, and can be left open at the side of the window. The guide provides example images and discussion concerning myriad scenarios volunteers may encounter. As of this writing, the topics covered are

\begin{itemize}
    \item \textbf{Asteroids (object of interest)} explains the object should be at center, as a point source or trailed.
    \item \textbf{Trails (natural)} discusses trailing and fast-rotators.
    \item \textbf{Tails (object of interest)} describes possible tail morphology, including multiple tails.
    \item \textbf{Comae} explains how comae are diffuse (as compared to a tail) and provides an example.
    \item \textbf{Missing Object} acknowledges this situation arises and describes how these can make it past our vetting process.
    \item \textbf{Crowded Fields} shows examples of fields with many sources, with and without an active object.
    \item \textbf{Blurry images} states that these low quality images should probably be skipped.
    \item \textbf{Galaxies} shows several galaxy examples, include some juxtaposed with active objects.
    \item \textbf{Cosmic Rays} defines the phenomenon and gives examples of thumbnails with cosmic rays.
    \item \textbf{Trails (artificial)} briefly discusses satellite trails and telescope tracking issues.
    \item \textbf{Saturation and Scattered Light} explains how scattered light could be misinterpreted as activity.
    \item \textbf{Background Object of Note (and size!)} gives examples of interesting clusters and dust clouds that are large in size given the angular size of the thumbnail.
    \item \textbf{Donuts} are displayed, including one showing activity, and we discuss the donut phenomenon.
    \item \textbf{Dead Columns} gives examples of columnar artifacts intersecting with the object at the center of the thumbnail.
    \item \textbf{Partial Images (Edges and Corners)} supplies examples of chip edge and chip corner cases that may be confusing.
    \item \textbf{Object width (object of interest)} discusses why point sources appear to have different widths.
\end{itemize}

\subsubsection{About}

\paragraph{Research} This page provides further discussion of the science justification, a description of the thumbnail production pipeline, explains some of the common challenges for the project, justifies why Citizen Science is needed, and provides a proof-of-concept statement along with links to papers that have resulted from the project and preparations.

\paragraph{The Team} Provides pictures and brief biographical sketches describing team members. At the time of this writing, the \textit{Core Research Team} is made up of (1) this author (Colin Orion Chandler) of \ac{NAU}, (2) Co-founder Chad Trujillo of \ac{NAU}, (3) Co-founder Jay Kueny of Lowell Observatory and \ac{UA}, Project Scientist Will Oldroyd of \ac{NAU}, and Project Scientists Will Burris of \ac{SDSU}. \textit{Contributors} is comprised of Annika Gustaffson of \ac{SwRI}. The \textit{Science Advisory Board} includes Henry Hsieh of the \ac{PSI}, Mark Jesus Mendoza Magbanua of \ac{UCSF}, Michael Gowanlock of \ac{NAU}, David Trilling of \ac{NAU}, and Ty Robinson of \ac{UA} and \ac{NAU}. Our \textbf{Moderator} is Elisabeth Baeten (Belgium).

\paragraph{Results} The Results page currently lists the \ac{SAFARI} proof-of-concept (Chapter \ref{chap:SAFARI}, \citealt{chandlerSAFARISearchingAsteroids2018}), the (6478)~Gault recurrent activity finding (Chapter \ref{chap:Gault}, \citealt{chandlerSixYearsSustained2019}), and our 2014~OG$_{392}$ active Centaur discovery (Chapter \ref{chap:2014OG392}, \citealt{chandlerCometaryActivityDiscovered2020a}). Additional discoveries will be posted on the Results page as they are published and/or announced.

\paragraph{FAQ} The \ac{FAQ} page answers some of the many questions we encountered during project preparations and after project launch.

\subsection{Talk}
\label{methods:subsec:talk}
A facility common to all Zooniverse projects are online forums, known as \textit{Talk}. Here volunteers can, for example, discuss thumbnail images they find interesting, or ask questions. Talk also provides a space where volunteers can build relationships with each other and the scientists. Crucial to the success of Talk are forum Moderators. These individuals help mediate interactions between individuals and answer questions. Our lead forum Moderator is Elisabeth Baeten (see Acknowledgements).

Surprisingly, volunteers posting on Talk became one of two primary paths which facilitate identifying candidate active objects, the other path being our analysis of classification data. Examples include 2017~QN$_{84}$ (Figure \ref{discussion:fig:2017QN84}) and 2015 TC$_1$ (Figure \ref{methods:fig:comparison}). Typically an image that is brought to our attention via Talk are also classified as ``active'' by the majority of volunteers who classified the image.

\subsection{Subject Sets}

\begin{table}
\centering
\caption{Composition of First Citizen Science Subject Set}
\label{methods:tab:CitSciMakeup}
\begin{tabular}{lrr}
\multicolumn{1}{c}{Kind} & \multicolumn{1}{l}{Number} & \multicolumn{1}{l}{Percentage}  \\
\hline\hline
Damocloid                & 343                        & 3.2\%                           \\
Centaur                  & 433                        & 4.1\%                           \\
\ac{JFC}                      & 600                        & 5.6\%                           \\
\ac{KBO}/\ac{TNO}          & 564                        & 5.3\%                           \\
Hungaria                 & 750                        & 7.0\%                           \\
Main-belt Inner                & 1500                       & 14.1\%                          \\
Main-belt Middle               & 1500                       & 14.1\%                          \\
Main-belt Outer                & 1500                       & 14.1\%                          \\
Main-belt Cybele               & 750                        & 7.0\%                           \\
\ac{NEO}                 & 600                        & 5.6\%                           \\
Mars-crosser             & 500                        & 4.7\%                           \\
Trojan                   & 1000                       & 9.4\%                           \\
Hilda                    & 600                        & 5.6\%                           \\
\hline
\textbf{Total}                    & \textbf{10640}                      & \textbf{100.0\%}                        
\end{tabular}
\end{table}

A Subject Set is a Zooniverse element that contains both data (images in our case) and associated metadata for use in the Citizen Science project. By project launch we had uploaded one \textit{Training} subject set (Section \ref{methods:subset:trainingSet}) and one \textit{Test} subject set (comprised of images needing classification) containing roughly 10,000 thumbnail images. Since launch we have submitted an additional 7 Test subject sets. Zooniverse limits the number of subjects a project can contain using a quota system. Quotas are automatically increased with, for example, successful classification of subjects.

\subsubsection{Thumbnail Selection}
\label{methods:subsubsec:thumbnailSelection}
I wrote a Thumbnail Selection Tool to associate thumbnail records in our database with a numbered subject set. The tool allows selection by object class (e.g., outer main-belt asteroid) and delta magnitude (see Section \ref{methods:subsec:fieldAssesment}). There are options to limit the number of thumbnail images per object class, by unique solar system object, or both. These options allow us to provide a variety of objects for Citizen Scientists to look at. Otherwise the thumbnail images would favor bright objects over fainter (and thus more rare) farther objects such as Centaurs.

\paragraph{Subject Set Composition}
In general we limit the number of images per object to one, with the exception of object classes of which we have few observations (e.g., Centaur) which would quickly diminish to zero thumbnails in a batch because of their relative paucity. For the majority of object classes the limit of one image per object stems from my own experience examining thumbnails that indicates the likelihood of a single thumbnail showing activity, and another thumbnail from approximately the same observing \ac{UT} date does not, is very low. Moreover, we limit bias by showing each unique solar system object once, especially main-belt asteroids, prior to showing the same object to volunteers a second time. The ideal scenario would involve all images being examined by volunteers, however this may take years at the current rate of examination (12,000 to 120,000 classifications per day) and the continued \ac{DECam} thumbnail output of $\sim$5,000 thumbnails per day.

I have kept the ratio of object classes roughly the same for each batch, though the exact numbers necessarily varies as certain object classes (e.g., Centaurs) are limited in number because they are, for example, distant and faint. Table \ref{methods:tab:CitSciMakeup} shows the composition of the first test subject set provided for the 2021 August 31 project launch.

\paragraph{Delta Magnitude Limit}
We (the science team) considered prioritizing thumbnails showing objects computed to appear especially bright in the image as computed by our Delta Magnitude metric (Section \ref{methods:subsubsec:deltaMagLim}). However, this would lead to an observational bias favoring objects that appear brighter because of any combination of larger size, closer distance, higher albedo, or favorable phase angle. Additionally, we hoped to avoid volunteer fatigue that could result if we favored the most promising images early in the project, then later only provided images with a lower probability of activity detection. Concern that there may not be enough volunteer interest if the activity occurrence rate was too low was evidently unwarranted, in part because (a) we provide a training set that injects known active images, and (b) there is a significant fraction of images that indeed appear to show activity, be it real or perceived.

\paragraph{Subject Set Size}
A factor that has a significant impact on the overall flow and timeline of the project is the quantity of subjects contained within a given subject set. Because volunteers are shown subjects at random, it takes exponentially longer (in calendar time) for each subject (image) to be \textit{retired} -- the state when an image has been examined by the preset number of volunteers (15) -- for a subject size of 100,000 thumbnails than 10,000 thumbnails. However, even though a smaller subject set will be completed in less time, there is added overhead in preparing subject sets, uploading batches of images, sending out volunteer calls to action, and analyzing the results. At the time of this writing we prefer subject sets containing between 17,000 and 25,000 images.

\subsubsection{Subject Set Preparation}
\label{methods:subsec:subjSetPrep}
I created a \ac{HARVEST} Subject Set Preparation tool to (1) locate images that were flagged as part of a given Citizen Science batch (subject set), (2) copy the thumbnail images to a directory dedicated to permanently storing images uploaded to \textit{Active Asteroids}, and (3) add a green reticle, as shown in Figure \ref{methods:fig:workflow}.

This tool also creates the required manifest that must accompany the images when they are uploaded to Zooniverse. The manifest holds information we later use to identify the thumbnail image in our database, linking it to a specific thumbnail record that represents a unique combination of a SkyBot result and datafile from an astronomical image archive.

\subsubsection{Classification}
\label{methods:citsci:classification}
Once a subject set has been uploaded to Zooniverse the subject set can be assigned to our active Workflow (Section \ref{methods:subsec:workflow}). At this point, the project status will change to reflect the overall completion status. Following Zooniverse advice, we leave all subject sets marked as active; thus when we add additional subjects, the overall completion percentage does not start over at 0\%. Thus far we have not added an additional subject set until all objects have been fully classified (``retired''), so are not amending any subject set that is in the process of being examined.

The general workflow to notify Citizen Scientists that new work is ready is to (1) adjust the announcement banner on the landing page (Figure \ref{methods:fig:citsci:landing}), (2) tweet the event from the project Twitter account (\texttt{@ActiveAsteroids}), and (3) send out an email ``newsletter'' to all past volunteers of the project via the Zooniverse platform. In practice, we have yet to send an email as of this writing. The first two actions have thus far sufficed to accomplish our goals promptly. One other modality, (4) press releases, coincided with project launch\footnote{\url{http://activeasteroids.net}} and seemed effective at drawing new participants to \textit{Active Asteroids}. We note that effectiveness of publicity is impossible to measure as these engagements occurred prior to, during, and following project launch. We intend to issue press releases in the future with new publications and/or announcements in conjunction with our host institutions and NASA.

\subsection{Training Set}
\label{methods:subset:trainingSet}

\begin{table} 
    \centering
    \caption{Expert Activity Scoring Index}
    \begin{tabular}{cl}
         Score & Description \\
         \hline\hline
        0 & missing from center or unidentifiable\\
        1 & visible as a point-source\\
        2 & vaguely fuzzy\\
        3 & fuzzy, but likely not activity-related\\
        4 & suggestive of activity but inconclusive\\
        5 & likely active, but some ambiguity remains\\
        6 & likely activity, not very ambiguous\\
        7 & definitely active, medium-strength indicators\\
        8 & definitely active, strong signs of activity\\
        9 & activity so obvious that a single image suffices\\
    \end{tabular}
    \label{methods:tab:activityScore}
\end{table}

I manually examined over 10,000 images of comets and other active bodies output by the \ac{HARVEST} pipeline. For each image we ranked apparent activity on an integer scale from 0 -- 9, following my own rough classification scheme (Table \ref{methods:tab:activityScore}). We imported these scores into the \ac{HARVEST} database, then prepared a subject set consisting only objects with scores of five or higher (meaning these images definitely showed activity; Section \ref{intro:subsec:visibleActivity}). These became the Training Subject Set for the \textit{Active Asteroids} project.

The primary purpose of the training set is to help teach Citizen Scientists how to recognize activity. As volunteers classify thumbnails in the primary workflow, training images are shown at an interval that is inversely proportional to how many images a volunteer has classified, as determined by Zooniverse. The decaying rate is necessary for (and was requested by) volunteers that become highly proficient and no longer need (or wish) to see training images. The probability a user is shown a training image, $P_\mathrm{T}$, is given by

\begin{equation}
    P_\mathrm{T}(N) = 
    \left\{
    \begin{array}{rr}
        50\% & 1 \le N  \le\; 10\;\\
        20\% & 10 <  N  \le\; 50\\
        10\% & 50 <  N  \le 100\\
        5\% & 100 <  N < \infty\: \: 
    \end{array}
    \right\}
\end{equation}
\noindent where $N$ is the number of classifications a user has carried out in \textit{Active Asteroids}. The training set also serves to validate that the project is working as intended. It was clear during our project's Beta Review that volunteers were adept at identifying activity.

\section{Follow-up Campaign}
\label{sec:methods:followup}

Once we have identified activity candidates we conduct follow-up study, first in the form of an archival investigation. Then, if warranted, we may pursue observations at telescope facilities to further study the object.

\subsection{Archival Investigation}
\label{methods:subsec:archivalInvestigation}

This process is documented extensively in Chapter \ref{QN:sec:secondActivityEpoch}  \citep{chandlerRecurrentActivityActive2021a}. Here, I provide a brief synopsis of the overall process and discuss additional methods and sources we use that were not part of the aforementioned work.

\subsubsection{Step 1: Supplemental Searches}
I carry out searches independent of the \ac{HARVEST} pipeline to find additional archival data for a given solar system object, through (1) three automated supplemental pipelines we wrote and (2) manual query of two sources. This auxiliary pipeline is especially useful for the study of specific objects, where we consider it worth spending extra time to find the maximum amount of data available for an object of interest. However there are many drawbacks when compared with the \ac{HARVEST} pipeline. For example, the overwhelming majority (roughly 90\%, depending on circumstances such as dynamical class) of data are unusable, either not probing faint enough to see the object, or the object is not on a camera chip. Moreover, accessing myriad archives with data from myriad instruments requires us to manually intervene in the pipeline to, for example, address errors generated by \ac{FITS} files that do not conform to the \ac{FITS} standard, or to conduct astrometry in order to embed a valid \ac{WCS} that we need to extract a thumbnail image of the target object. All of these added steps also require additional administrative overhead, from managing downloading data to keeping a separate directory structure organized with results.

\paragraph{CADC SSOIS} This pipeline queries the \ac{CADC} \ac{SSOIS}\footnote{\url{https://www.cadc-ccda.hia-iha.nrc-cnrc.gc.ca/en/ssois/}}. Results include some overlap with \ac{HARVEST} (i.e., \ac{DECam}, MegaPrime, \ac{KPNO} instruments), which are useful for validating \ac{HARVEST} results. Additional instrument archives searched include \ac{SOAR}, \ac{ESO} instruments (e.g., \ac{VST} OMEGACam, \ac{VISTA} {VIRCam}), \ac{NEAT} \ac{GEODSS}, \ac{SDSS}, Subaru SuprimeCam and \ac{HSC}, and \ac{WISE}.

\paragraph{IRSA} This pipeline queries the NASA/CalTech \ac{IRSA}\footnote{\url{https://irsa.ipac.caltech.edu}} archive, which searches both \ac{ZTF} and \ac{PTF} data for moving objects via their \ac{MOST}.

\paragraph{ZTF Alert Stream} This pipeline downloads all \ac{ZTF} alert stream \citep{pattersonZwickyTransientFacility2018} data then prunes out all data unrelated to solar system objects. Alert packets have already been matched to known solar system objects, making these data simple to search. The alerts include image data as well as metadata such as apparent magnitude.

\paragraph{Manual Queries} We manually query the \ac{KOA}\footnote{\url{https://koa.ipac.caltech.edu/cgi-bin/KOA/nph-KOAlogin}} via their \ac{MOST}. We also initiate queries via the \ac{CATCH} tool\footnote{\url{https://catch.astro.umd.edu/}} which provides quick search and image delivery from multiple archives, including \ac{NEAT} \citep{pravdoNearEarthAsteroidTracking1999} and SkyMapper \citep{kellerSkyMapperTelescopeSouthern2007}.

\subsubsection{Step 2: Data Acquisition}
During the aforementioned pipeline processes we automatically generate download scripts that are added to the same queue that executes downloads for the \ac{HARVEST} pipeline. However, several sources require additional steps to download archival image data.

\paragraph{ESO} \ac{ESO} results, including \ac{VST} OMEGACam and \ac{VIRCam}, can be downloaded in an automated fashion.
We generate a separate set of bash scripts we designed to handle downloading, sorting, and preprocessing of \ac{ESO} data.

\paragraph{SMOKA} The \ac{SMOKA}\footnote{\url{https://smoka.nao.ac.jp}} archive  \citep{babaDevelopmentSubaruMitakaOkayamaKisoArchive2002} that serves SuprimeCam and \ac{HSC} data requires an account to request data and carry out downloads. The system is queue-based, with email notifications upon data retrieval readiness. We optimize the process by constructing a custom URL that requests all specific datafiles we need at once. After the request is processed, \ac{SMOKA} sends an email that includes a one-line bash script for downloading the data. Once the data are downloaded, we have separate scripts that prepare the data for analysis. We do not reduce (e.g., flatten) the data, though this is a step we are considering for future work.

\paragraph{CASU} Data hosted at the \ac{CASU} Astronomical Data Centre\footnote{\url{http://casu.ast.cam.ac.uk/casuadc/}}, especially \ac{INT} \ac{WFC}, require data be requested from their queue-based service. Once the data are available an email notice is sent. From this point we produce shell scripts to download the data,  
conduct astrometry, and prepare the data for additional analysis.

\subsubsection{Step 3: Astrometry}
My thumbnail extraction code requires embedded \ac{WCS} of sufficient quality to allow for the target object to appear at or near the center of the thumbnail images we extract. Many archive/instrument combinations (e.g., \ac{PS1}, \ac{ZTF}) provide data with excellent astrometry via embedded \ac{WCS}, however some archives either provide no \ac{WCS} at all, or \ac{WCS} with insufficient or unreliable precision. We perform astrometry as needed via Astrometry.net \citep{langAstrometryNetBlind2010} on the \ac{NAU} Monsoon computing cluster. Astrometry.net makes use of source catalogs, including the Gaia Data Release 2 catalog \citep{gaiacollaborationGaiaDataRelease2018}, and \ac{SDSS} \citep{ahnNinthDataRelease2012}.

\subsubsection{Step 4: Thumbnails}
This tool extracts \ac{FITS} and \ac{PNG} thumbnails with a uniform \ac{FOV} that we specify, the default being 126''$\times$126''. We maintain a database of instrument and telescope parameters so that we can extract thumbnail images (via \ac{HARVEST}-derived code) with North pointing up and East pointing left, standard astronomical orientation. We query \ac{JPL} Horizons and plot symbols indicating the anti-Solar and anti-motion vectors commonly associated with tail direction, such as those shown in Figure \ref{intro:fig:activity}. An additional optional step co-adds \ac{FITS} format thumbnails to help enhance signal for activity searches.

\subsection{New Telescopic Observations}
\label{methods:subsec:telescopicObservations}

\begin{table}
    \caption{Telescopes Utilized}
    \footnotesize
    \begin{tabular}{lllllcc}
        Instrument & Telescope    & Diameter [m]    & Observatory        & Location            & Country & Site Code \\
        \hline
        AltaU-47   & \acs{BLT}          & 0.5         & \acs{ARO}                & Flagstaff, Arizona  & USA     & 687       \\
        \ac{DECam}      & Blanco       & 4.0         & \acs{CTIO}               & Cerro Tololo        & Chile   & 807       \\
        \acs{GMOS}-S     & Gemini South & 8.1         & Gemini             & Cerro Pachon        & Chile   & I11       \\
        \acs{IMACS}      & Baade        & 6.5         & Magellan           & Las Campanas        & Chile   & 304       \\
        \acs{LBCB}, \acs{LBCR} & \acs{LBT}          & 8.5$\times$2 & \acs{MGIO}               & Mt. Graham, Arizona & USA     & G83       \\
        \acs{LMI}, \acs{NIHTS} & \acs{LDT}          & 4.3         & Lowell Observatory & Happy Jack, Arizona & USA     & G37       \\
        VATT4K     & \acs{VATT}         & 1.8         & \acs{MGIO}               & Mt. Graham, Arizona & USA     & 290      
    \end{tabular}
    \raggedright
    \footnotesize
    Definitions: \acf{BLT}, \acf{ARO}, \acf{DECam}, \acf{CTIO}, \acf{GMOS}, \acf{IMACS}, \acf{LBCB}, \acf{LBCR}, \acf{LBT}, \acf{MGIO}, \acf{LMI}, \acf{NIHTS}, \acf{LDT}, \acf{VATT}.
    \label{methods:tab:facilities}
\end{table}

Over the course of the work contained in this dissertation our team made use of telescopes for activity searches and follow-up study. Table \ref{methods:tab:facilities} lists the instruments and associated facilities used for this work.


\clearpage
\setcounter{rownumber}{0}
\singlespacing
\chapter{Searching Asteroids for Activity Revealing Indicators (SAFARI)}
\label{chap:SAFARI}
\acresetall

Colin Orion Chandler\footnote{\label{safari:nau}Department of Physics \& Astronomy, Northern Arizona University, PO Box 6010, Flagstaff, AZ 86011, USA}, Anthony M. Curtis\footnote{Department of Physics, University of South Florida ISA 2019, Tampa, FL 33620, USA}, Michael Mommert$^\mathrm{\ref{safari:nau}}$, Scott S. Sheppard\footnote{Department of Terrestrial Magnetism, Carnegie Institution for Science, 5241 Broad Branch Road. NW, Washington, DC 20015, USA}, Chadwick A. Trujillo$^\mathrm{\ref{safari:nau}}$

\textit{This is the Accepted Manuscript version of an article accepted for publication in Publications of the Astronomical Society of the Pacific.  IOP Publishing Ltd is not responsible for any errors or omissions in this version of the manuscript or any version derived from it. The Version of Record is available online at} \url{https://iopscience.iop.org/article/10.1088/1538-3873/aad03d}\textit{.}

\doublespacing


\section*{Abstract}
\label{safari:Abstract}
Active asteroids behave dynamically like asteroids but display comet-like comae. These objects are poorly understood, with only about 30 identified to date. We have conducted one of the deepest systematic searches for asteroid activity by making use of deep images from the Dark Energy Camera (DECam) ideally suited to the task. We looked for activity indicators amongst \uniquethumbsSAFARI{} unique asteroids extracted from \fitscountSAFARI{} images. We detected three previously-identified active asteroids ((62412), (1) Ceres and (779) Nina), though only (62412) showed signs of activity. Our activity occurrence rate of 1 in \uniquethumbsSAFARI{} is consistent with the prevailing 1 in 10,000 activity occurrence rate estimate. Our proof of concept demonstrates 1) our novel informatics approach can locate active asteroids and 2) DECam data are well-suited to the search for active asteroids.

\textit{Keywords:} minor planets, asteroids: general -- methods: analytical -- techniques -- image processing

\section{Introduction}
\label{safari:sec:introduction}

\begin{figure}
  \centering
	\includegraphics[width=0.5\linewidth]{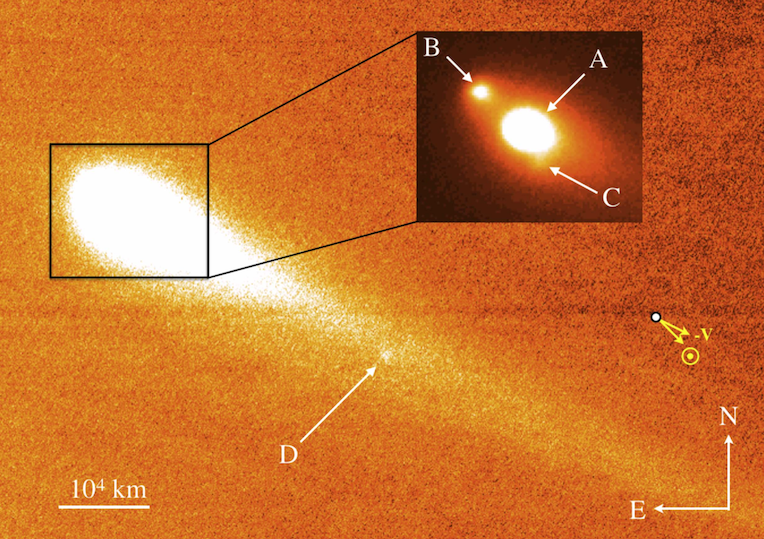}
  \caption{Active asteroid P2013/R3 was imaged in October 2013 while undergoing a breakup (into components A-D) likely caused by rotational instability. The antisolar and negative heliocentric velocity vector arrows are labeled $\odot$ and $-V$, respectively. Reprinted Figure 2 of \cite{jewittAnatomyAsteroidBreakup2017}.}
  \label{safari:fig:ExampleAsteroid}
\end{figure}

\begin{table*}
\footnotesize
\caption{The Active Asteroids (1 of 2)}
\label{safari:Table:TheAAs}
\begin{tabular}{lccrrclcrrclc}
Asteroid Name & $a$ & $e$ & $i$ & Orbit & $T_\mathrm{J}$ &$P$ & $q$ & $R$ & $f$ & $\mathrm{\%}_\mathrm{peri}$ & Act. & Cause\\
& (au) & & ($^\circ$) & & & (years) & (au) & (au) & ($^\circ$) & (\%) & (N) & \\

\rlabel{Ceres}		(1) Ceres 			& 2.77 	& 0.08 	& 10.6 	& MB 		& 3.310 & 4.6 	& 2.60 & 2.72 & 279.3 & 62 	& 3+ 	        & \sublimate{}\ , $\bigwedge$ 					\\
\rlabel{Adeona}		(145) Adeona  		& 2.67 	& 0.14 	& 12.6 	& MB 		& 3.331 & 2.28 	& 2.29 & 2.69 & 258.8 & 47 	& 2 	        & \sublimate 									\\
\rlabel{Constantia}	(315) Constantia 	& 2.24 	& 0.17 	&  2.4 	& MB 		& 3.614 & 3.36 	& 1.86 & 1.94 & 315.9 & 92 	& 0$^\dagger$   & (?) 											\\
\rlabel{Griseldis}	(493) Griseldis 	& 3.12 	& 0.18 	& 15.2  & OMB 		& 3.140 & 5.5 	& 2.57 & 3.34 & 122.4 & 31 	& 1 	        & $\leftrightsquigarrow$ 						\\
\rlabel{Scheila}	(596) Scheila 		& 2.93 	& 0.16 	& 14.7 	& OMB 		& 3.209 & 5.01 	& 2.45 & 3.11 & 239.2 & 90 	& 1 	        & $\leftrightsquigarrow$ 						\\
\rlabel{Interamnia}	(704) Interamnia 	& 3.06 	& 0.15 	& 17.3 	& MB 		& 3.148 & 5.35 	& 2.59 & 2.62 &  19.6 & 97 	& 1 	        & \sublimate 									\\
\rlabel{Nina}		(779) Nina 			& 2.66 	& 0.23 	& 14.6 	& MB 		& 3.302 & 4.35 	& 2.06 & 2.15 &  36.9 & 93 	& 1 	        & \sublimate 									\\
\rlabel{Ingrid}		(1026) Ingrid 		& 2.25 	& 0.18 	&  5.4 	& MB 		& 3.597 & 3.38 	& 1.85 & 2.23 &  97.5 & 16 	& 0$^\dagger$   & (?) 											\\
\rlabel{Beira}		(1474) Beira 		& 2.73 	& 0.49 	& 26.7 	& Mars 		& 3.033 & 4.52 	& 1.39 & 1.57 & 310.9 & 93 	& 1 	        &  \sublimate 									\\
\rlabel{Oljato}		(2201) Oljato 		& 2.17 	& 0.71 	&  2.5	& Apollo 	& 3.298 & 3.21 	& 0.62 & 0.88 &  73.1 & 92 	& 1 	        & (?) 											\\
\rlabel{Phaethon}	(3200) Phaethon 	& 1.27 	& 0.89 	& 22.2 	& Apollo 	& 4.510 & 1.43 	& 0.14 & 0.14 &   5.1 & 87 	& 3 	        & $\bigodot$ 									\\
\rlabel{DonQuixote}	(3552) Don Quixote 	& 4.26 	& 0.71 	& 31.1 	& Amor 		& 2.315 & 8.78 	& 1.24 & 1.23 & 343.6 & 100	& 2 	        & \sublimate,(?) 								\\
\rlabel{Aduatiques}	(3646) Aduatiques 	& 2.75 	& 0.11 	&  0.6 	& MB 		& 3.336 & 4.57 	& 2.46 & 2.56 & 309.0 & 90 	& 0$^\dagger$   & (?) 											\\
\rlabel{WilHar}		(4015) Wil.-Har. 	& 2.63 	& 0.63 	&  2.8 	& Apollo 	& 3.082 & 4.26 	& 0.97 & 1.17 &  51.0 & 95 	& 2$^\ddagger$  & \sublimate{}\ , (?) 							\\
\rlabel{24684}		(24684) 1990 EU4 	& 2.32 	& 0.08 	&  3.9	& MB 		& 3.572 & 3.53 	& 2.13 & 2.28 & 277.9 & 77 	& 0$^\dagger$   & (?) 											\\
\rlabel{35101}		(35101) 1991 PL16 	& 2.60 	& 0.18 	& 12.3 	& MB 		& 3.365 & 4.17 	& 2.12 & 2.86 & 227.0 & 21 	& 0$^\dagger$   & (?) 											\\
\rlabel{62412}		(62412) 			& 3.15 	& 0.08 	&  4.7	& OMB 		& 3.197 & 5.6 	& 2.90 & 3.06 &  74.5 & 68 	& 1 	        & $\circlearrowleft$ 							\\
\rlabel{Ryugu}		(162173) Ryugu 		& 1.19 	& 0.19 	&  5.9	& Apollo 	& 5.308 & 1.3 	& 0.96 & 1.08 & 288.4 & 8 	& 1 	        & \sublimate 									\\
\rlabel{GO98}		(457175) 			& 3.96 	& 0.28 	& 15.6 	& OMB 		& 2.926 & 7.89 	& 2.85 & 3.28 &  66.0 & 81 	& 1 	        & (?) 											\\
\rlabel{ElstPizarro} 133P/Elst--Pizarro & 3.16 	& 0.16 	&  1.4	& OMB 		& 3.184 & 5.63 	& 2.66 & 2.65 &  21.7 & 100	& 4 	        & \sublimate 									\\
\rlabel{176P}		176P/LINEAR 		& 3.20 	& 0.19 	&  0.2	& OMB 		& 3.166 & 5.71 	& 2.58 & 2.59 &  10.1 & 1 	& 1 	        & (?) 											\\
\rlabel{233P}		233P/La Sagra 		& 3.04 	& 0.41 	& 11.3 	& Encke 	& 3.081 & 5.28 	& 1.78 & 2.01 & 309.1 & 91 	& 1 	        & (?) 											\\
\rlabel{238P}		238P/Read 			& 3.16 	& 0.25 	& 1.3 	& OMB 		& 3.154 & 5.64 	& 2.37 & 2.42 &  26.5 & 97 	& 3 	        & \sublimate 									\\
\rlabel{259P}		259P/Garradd 		& 2.73 	& 0.34 	& 15.9 	& MMB 		& 3.217 & 4.51 	& 1.81 & 1.85 &  27.6 & 99 	& 2 	        & \sublimate 									\\
\rlabel{288P}		288P (300163) 		& 3.05 	& 0.20 	&  3.2 	& OMB 		& 3.204 & 5.32 	& 2.44 & 2.45 &  12.2 & 99 	& 2 	        & \sublimate 									\\
\rlabel{311P}		311P/PS 			& 2.19 	& 0.12 	&  5.0	& IMB 		& 3.661 & 3.24 	& 1.94 & 2.15 & 272.8 & 58 	& 2 	        & $\circlearrowleft$\ , \large{\textbf{:}} 		\\
\rlabel{313P}		313P/Gibbs 			& 3.16 	& 0.24 	& 11.0 	& Encke 	& 3.132 & 5.62 	& 2.42 & 2.40 &   8.0 & 100	& 2 	        & \sublimate 									\\
\rlabel{324P}		324P/La Sagra 		& 3.10 	& 0.15 	& 21.4 	& OMB 		& 3.100 & 5.45 	& 2.62 & 2.64 &  20.0 & 98 	& 2 	        & \sublimate 									\\
\rlabel{331P}		331P/Gibbs 			& 3.00 	& 0.04 	&  9.7	& OMB 		& 3.229 & 5.21 	& 2.88 & 3.10 & 140.4 & 11 	& 2 	        & $\leftrightsquigarrow$, $\circlearrowleft$ 	\\
\rlabel{348P}		348P/PS 			& 3.17 	& 0.30 	& 17.6 	& OMB 		& 3.062 & 5.63 	& 2.18 & 2.51 &  60.8 & 83 	& 1 	        & (?) 											\\
\rlabel{354P}		354P/LINEAR 		& 2.29 	& 0.12 	&  5.3	& OMB 		& 3.583 & 3.47 	& 2.00 & 2.01 &  12.2 & 99 	& 1 	        & $\circlearrowleft$,$\circledast$ 				\\
\rlabel{358P}		358P 				& 3.15 	& 0.24 	& 11.1 	& Encke 	& 3.135 & 5.59 	& 2.39 & 2.42 &   7.5 & 99 	& 2 	        & \sublimate{}\ , (?) 							\\
\rlabel{P2013R3}	P/2013 R3 			& 3.03 	& 0.27 	& 0.9 	& OMB 		& 3.184 & 5.28 	& 2.20 & 2.22 &  14.0 & 99 	& 1 	        & $\circlearrowleft$, \sublimate{} 				\\
\rlabel{P2015X6}	P/2015 X6 			& 2.75	& 0.17 	&  4.6	& MMB 		& 3.318 & 4.57 	& 2.28 & 2.64 & 274.5 & 62 	& 1 	        & $\circlearrowleft$ 							\\
\rlabel{P2016G1}	P/2016 G1 			& 2.58	& 0.21	& 11.0	& MMB 		& 3.367 & 4.15 	& 2.04 & 2.52 & 264.7 & 56 	& 1 	        & $\leftrightsquigarrow$ 						\\
\rlabel{P2016J1}	P/2016 J1 			& 3.17 	& 0.23 	& 14.3 	& OMB 		& 3.113 & 5.65 	& 2.45 & 2.46 & 345.9 & 99 	& 1 	        & $\circlearrowleft$, \sublimate{} 				\\
\end{tabular}\\
\\
	Orbital parameters retrieved from the Minor Planet Center and \ac{JPL} Horizons. \\
	$a$: semimajor axis; %
	$e$: eccentricity; %
	$i$: inclination; %
	Orbit: Inner, Mid, Outer, \& Main Belt (IMB, MMB, OMB, MB); %
	$T_\mathrm{J}$:Tisserand parameter with respect to Jupiter; %
	$P$:Orbital Period; %
	$q$:Perihelion distance; %
	$R$: Heliocentric discovery distance. 
    $f$: True anomaly. %
    $\mathrm{\%}_\mathrm{peri}$:Percentage towards perihelion. %
    Act.: Number of times object reported active. 
	$^\dagger$Authors declare object a candidate (activity not yet confirmed). %
	$^\ddagger$\cite{ferrin2009ApparitionMethuselah2012} argue (4015) was also active in 1992, 1996, 2008, and 2009-2010. %
	
	\sublimate{}\hspace{0.5mm}Sublimation; %
	$\circlearrowleft$\hspace{0.25mm}Rotational Breakup; %
	$\leftrightsquigarrow$\hspace{0.25mm}Impact; %
	$\bigwedge$\hspace{0.25mm}Cryovolcanism; %
	\protect{\large{\textbf{:}}}\hspace{0.25mm}Binary; %
	$\bigodot$\hspace{0.25mm}Thermal Fracture; %
	$\circledast$\hspace{0.25mm}Dust Model; %
	(?)\hspace{0.25mm}Unknown %
\end{table*}
\clearpage


\setcounter{table}{0} 

\begin{table*}
\footnotesize
\centering
\caption{The Active Asteroids (2 of 2) \label{safari:Table:TheAAs2}}
\begin{tabular}{llclcccc}

Asteroid Name & Family & $1^\mathrm{st}$Act & Facility & Method & Last & Visit & Refs\\
& & (years) & & & (years) & &\\

(1) Ceres 			& None 		    & 2014 & Herschel 	& Spec.  & 2017 	        & Yes	& [\ref{Ceres}]\\
(145) Adeona  		& Adeona		& 2017 & Terksol 	& Spec.  & 2016 	        & No 	& [\ref{Adeona}]\\
(315) Constantia 	& Flora		    & 2013 & MPCAT 		& Phot.  & 2013 	        & No 	& [\ref{Constantia}]\\
(493) Griseldis 	& Eunomia	    & 2015 & Subaru 	& Visual & 2015 	        & No 	& [\ref{Griseldis}]\\
(596) Scheila 		& None		    & 2010 & CSS 		& Visual & 2010 	        & No 	& [\ref{Scheila}]\\
(704) Interamnia 	& None		    & 2017 & Terksol 	& Spec.  & 2012 	        & No 	& [\ref{Interamnia}]\\
(779) Nina 			& $\cdots$	    & 2017 & Terksol 	& Spec.  & 2012 	        & No 	& [\ref{Nina}]\\
(1026) Ingrid 		& Flora 		& 2013 & MPCAT 		& Phot.  & 2013 	        & No 	& [\ref{Ingrid}]\\
(1474) Beira 		& $\cdots$	    & 2017 & Terksol 	& Spec.  & 2012 	        & No 	& [\ref{Beira}]\\
(2201) Oljato 		& $\cdots$ 	    & 1984 & Pioneer 	& Mag.   & 1984 	        & No 	& [\ref{Oljato}]\\
(3200) Phaethon 	& Pallas 	    & 2009 & STEREO 	& Visual & 2017 	        & Yes 	& [\ref{Phaethon}]\\
(3552) Don Quixote 	& $\cdots$	    & 2009 & Spitzer 	& Visual & 2018 	        & No 	& [\ref{DonQuixote}]\\
(3646) Aduatiques 	& $\cdots$	    & 2013 & MPCAT 		& Phot.  & 2013 	        & No 	& [\ref{Aduatiques}]\\
(4015) Wil.-Har. 	& $\cdots$	    & 1949 & Palomar 	& Visual & 1979$^\ddagger$  & No 	& [\ref{WilHar}]\\
(24684) 1990 EU4 	& $\cdots$	    & 2013 & MPCAT 		& Phot.  & 2013 	        & No 	& [\ref{24684}]\\
(35101) 1991 PL16 	& Eunomia	    & 2013 & MPCAT 		& Phot.  & 2013 	        & No 	& [\ref{35101}]\\
(62412) 			& Hygiea		& 2015 & DECam 		& Visual & 2014 	        & No 	& [\ref{62412}]\\
(162173) Ryugu 		& Clarissa	    & 2017 & MMT 		& Spec.	 & 2017 	        & Yes 	& [\ref{Ryugu}]\\
(457175) 			& Hilda		    & 2017 & CSS 		& Visual & 2017 	        & No 	& [\ref{GO98}]\\
133P/Elst--Pizarro 	& Themis		& 1996 & ESO 		& Visual & 2014 	        & No 	& [\ref{ElstPizarro}]\\	
176P/LINEAR 		& Themis		& 2009 & HTP 		& Visual & 2011 	        & No 	& [\ref{176P}]\\
233P/La Sagra 		& $\cdots$	    & 2009 & LSSS 		& Visual & 2009 	        & No 	& [\ref{233P}]\\
238P/Read 			& Gorchakov	    & 2005 & SW 		& Visual & 2016 	        & No 	& [\ref{238P}]\\
259P/Garradd 		& $\cdots$	    & 2008 & SS 		& Visual & 2017 	        & No 	& [\ref{259P}]\\
288P (300163) 		& Themis		& 2011 & PS 		& Visual & 2017 	        & No 	& [\ref{288P}]\\
311P/PS 			& Behrens 	    & 2013 & PS 		& Visual & 2014 	        & No 	& [\ref{311P}]\\
313P/Gibbs 			& Lixiaohua     & 2014 & CSS 		& Visual & 2015 	        & No 	& [\ref{313P}]\\
324P/La Sagra 		& Alauda		& 2011 & LSSS 		& Visual & 2015 	        & No 	& [\ref{324P}]\\
331P/Gibbs 			& Gibbs		    & 2012 & CSS 		& Visual & 2014 	        & No 	& [\ref{331P}]\\
348P/PS 			& $\cdots$	    & 2017 & PS 		& Visual & 2017 	        & No 	& [\ref{348P}]\\
354P/LINEAR 		& Baptistina    & 2010 & LINEAR 	& Visual & 2017 	        & No 	& [\ref{354P}]\\
358P 				& Lixiaohua	    & 2012 & PS 		& Visual & 2017 	        & No 	& [\ref{358P}]\\
P/2013 R3 			& Mandragora    & 2013 & PS 		& Visual & 2013 	        & No 	& [\ref{P2013R3}]\\
P/2015 X6 			& Aeolia		& 2015 & PS 		& Visual & 2015 	        & No 	& [\ref{P2015X6}]\\
P/2016 G1 			& Adeona		& 2016 & PS 		& Visual & 2016 	        & No 	& [\ref{P2016G1}]\\
P/2016 J1 			& Theobalda	    & 2016 & PS 		& Visual & 2016 	        & No 	& [\ref{P2016J1}]\\
\end{tabular}

	
    $1^\mathrm{st}$Act: Year activity discovered. %
    Facility: Facility originally reporting activity. 
    Last: As of January 2018 submission. %
    Refs: Object-specific references in Appendix \ref{safari:ObjectReferences}. %
	CSS:Catalina Sky Survey; %
	ESO:European Space Observatory 1-metre Schmidt %
	HTP:Hawaii Trails Project; %
	LINEAR:LIncoln Near-Earth Asteroid pRogram; %
	LSSS:La Sagra Sky Survey; %
	MPCAT:Minor Planet Catalog %
	PS:Pan-STARRS; %
	SS:Siding Spring; %
	SW:Spacewatch %
	
\end{table*}
\clearpage

\begin{table}
	\caption{AA Mass-loss Mechanisms}
	\centering
	\begin{tabular}{lrr}
	\hline\hline
		Suspected Mechanism & \ N$^*$ & \%\\
		\hline
		Sublimation & 15 & 44\\
		Rotational Breakup & 7 & 21\\
		Impact\ /\ Collision & 4 & 12\\
		Thermal Fracturing & 1 & 3\\
		Cryovolcanism & 1 & 3\\
		Binary Interaction & 1 & 3\\
		Unknown & 5 & 15\\
	\end{tabular}\\
	\raggedright
	$^*$ Objects with multiple mechanism are counted more than once; objects listed in Table \ref{safari:Table:TheAAs} as candidates were not included in this computation.
	\label{safari:Table}
\end{table}

Active asteroids appear to have tails like comets (Figure \ref{safari:fig:ExampleAsteroid}) but follow orbits predominately within the main asteroid belt. Although the first active asteroid (Wilson--Harrington) was discovered in 1949 \citep{cunninghamPeriodicCometWilsonHarrington1950}, 27 of the 31 objects (87\%) were identified as active in the last decade (Table \ref{safari:Table:TheAAs}). Asteroid activity is thought to be caused by several different mechanisms, combinations of which are undoubtedly at work (e.g., an impact event exposing subsurface ice to sublimation). The number of times (i.e., orbits) an object has displayed activity (Table 1: Act.) is especially diagnostic of the mechanism (Table 1: Cause). A singular (non-recurring) event likely originates from an impact event, e.g., (596)~Scheila. Rotational breakup, as in P/2013~R3 of Figure \ref{safari:fig:ExampleAsteroid}, may be a one-time catastrophic event, or a potentially repeating event if, for example, only a small piece breaks free but the parent body remains near the spin breakup limit. Ongoing or recurrent activity has been observed $\sim$15 times, e.g., 133P/Elst--Pizarro, and is suggestive of sublimation or, in the case of (3200)~Phaethon, thermal fracture. These last two mechanism (sublimation and thermal fracture) should be more likely to occur when an object is closer to the Sun, i.e. perihelion (Table 1:$q$). The Sun-object distance (Table 1: $R$) indicates the absolute distance, but it is can be simpler to consider how close (Table 1: \%$_\mathrm{peri}$) to perihelion the object was when activity was first observed (Table 1: 1$^\mathrm{st}$Act), where 100\% represents perihelion ($q$) and 0\% indicates aphelion:

\begin{equation}
	\mathrm{\%}_\mathrm{peri} = \left[1-\left(\frac{d_\mathrm{disc}-d_\mathrm{peri}}{d_\mathrm{ap}-d_\mathrm{peri}}\right)\right]\cdot 100\mathrm{\%}
\end{equation}

\noindent where $d_\mathrm{disc}$ is the heliocentric object distance at the activity discovery epoch, $d_\mathrm{peri}$ the perihelion distance, and $d_\mathrm{ap}$ the aphelion distance.

While the term ``\ac{MBC}'' often refers to this sublimation-driven subset of active asteroids, we use the more inclusive ``active asteroid'' term throughout this paper. We aimed to include all objects termed ``active asteroids'' in the literature for completeness, but we only include objects which have provided observable signs of activity. Objects known to host surface water ice but which have yet to shown signs of activity, such as (24) Themis \citep{rivkinDetectionIceOrganics2010, campinsWaterIceOrganics2010}, are outside the scope of this paper.

Orbital characteristics also provide insight into the dynamical evolution and even the composition of an object. Objects with conspicuously similar orbital properties may have originated from a catastrophic disruption event that created a family (Table 1:Family) of asteroids \citep{hirayamaGroupsAsteroidsProbably1918}. More generally, asteroids can be categorized (Table 1:Orb.) as interior to the Main Asteroid Belt, within the Main Asteroid Belt (and further subdivided into inner, mid, and outer main belt as IMB, MMB, and OMB respectively), or exterior to the Main Asteroid Belt (e.g., Kuiper belt). Objects interior to the Main Asteroid Belt, including Near Earth Objects (NEOs), include Earth-crossing (Apollo), Earth-orbit nearing (Amor), and Mars-crossing asteroids. Objects whose orbits are similar to Comet 2P/Encke are said to be Encke-type.
 
The Tisserand parameter $T_\mathrm{J}$ (Table \ref{safari:Table:TheAAs}$T_\mathrm{J}$) describes the degree to which an object's orbit is influenced by Jupiter:

\begin{equation}
	T_\mathrm{J} = \frac{a_\mathrm{J}}{a} + 2\sqrt{\left(1-e^2\right)\frac{a}{a_\mathrm{J}}}\cos(i).
	\label{safari:eq:TJ}
\end{equation}

\noindent The orbital elements are given by $a_\mathrm{J}$ the orbital distance of Jupiter (5.2 AU), plus the semi-major axis $a$, eccentricity $e$, $i$ the inclination (Table 1). For the case where $a=a_\mathrm{J}$ you can see $T_\mathrm{J}=3$. Asteroids in the main-belt are typically inside the orbit of Jupiter (i.e. $a<a_\mathrm{J}$) and usually have $T_\mathrm{J}>3$ \citep{jewittActiveAsteroids2014}; however, as Equation \ref{safari:eq:TJ} indicates, it is the combination of all three free parameters ($a$, $e$, $i$) which describes the magnitude of Jovian influence on the object's orbit. One active asteroid definition also constrains membership to objects whose orbits are interior to Jupiter but whose Tisserand parameters are $>$ 3.08 \citep{jewittActiveAsteroids2014}.

Objects not identified in the literature as active asteroids, yet still appear orbitally asteroidal (e.g., Comet 2P/Enke), are not included in this paper, but objects with $T_\mathrm{J}<3$ which are identified in the literature as active asteroids (e.g., (3552) Don Quixote), are included; see e.g., \cite{hsiehPopulationCometsMain2006,tancrediCriterionClassifyAsteroids2014} for further discussion on distinguishing objects within this regime.

We would like to understand active asteroids in part because they may hold clues about solar system formation and the origin of water delivered to the terrestrial planets. The recent discovery of interstellar asteroid \omuamua{} \citep{bacci2017U12017} intensifies interest in understanding our own indigenous asteroid population in order to better understand and characterize ejectoids we encounter in the future, an estimated decadal occurrence \citep{trillingImplicationsPlanetarySystem2017}. There has also long been an interest mining asteroids for their metals, and water could prove an invaluable resource providing, for example: energy, rocket fuel, breathable oxygen, and sustenance for plant and animal life \citep{olearyMiningApolloAmor1977,dicksonCongressApprovesSolar1978,kargelMetalliferousAsteroidsPotential1994,forganExtrasolarAsteroidMining2011,hasnainCapturingNearEarthAsteroids2012,lewickiPlanetaryResourcesAsteroid2013,andrewsDefiningSuccessfulCommercial2015}.

Our knowledge of active asteroids has been limited due to small sample size: only $\sim$20 active asteroids have been discovered to date \citep{jewittActiveAsteroids2015a}. As such, the statistics presented in Table \ref{safari:Table} are poorly constrained (e.g., the thermal fracturing rate is based upon a single object: (3200) Phaethon). Spacecraft visits have been carried out or planned to a number of the active asteroids (Table 1: Visit), and while we may learn a great deal from these individual objects, spacecraft visits will not substantially increase the number of known active asteroids. While spectroscopy has recently shown potential for discovering activity, the overwhelming majority of activity detections have been made by visual examination (Table 1:Method). One notable exception was the 1984 (2201) Oljato outburst first detected by magnetic field disturbances (2201) Oljato outburst \citep{russellInterplanetaryMagneticField1984}.

\begin{table}
	\small
	\centering
	\caption{Surveys which have discovered AAs.}
	\begin{tabular}{lccc}
		AA Discovered by & AAs & Limit & Operation\\
		\hspace{4mm}(survey name) & (N) & (mag) & (years)\\
		\hline
		Catalina Sky Survey & 5 & 22$^a$ & 1998$^f$ --\\
		La Sagra Survey & 2 & 17$^b$ & 2008$^g$--\\
		LINEAR & 1 & 19.6$^c$ & 1997$^h$ --\\
		Pan-STARRS & 8 & 22.7$^d$ & 2008$^i$ --\\
		Spacewatch & 2 & 21.7$^e$ & 1981$^j$ --\\
		Total & 18 & $\cdots$ & 98\\
		\hline
	\end{tabular}
	
	\raggedright
	$^a$\cite{drakeFirstResultsCatalina2009}; 
	$^b$estimated from aperture; 
	$^c$\cite{sesarExploringVariableSky2011,stokesLincolnNearEarthAsteroid2000}; 
	$^d$\cite{chambersPanSTARRS1Surveys2016}; 
	$^e$\cite{larsenSearchDistantObjects2007}; 
	$^f$\cite{larsonCatalinaSkySurvey1998}; 
	$^g$\cite{stossJ75SagraSky2011}; 
	$^h$\cite{stokesLincolnNearEarthAsteroid2000}; 
	$^i$\cite{jedickePanSTARRSFirstSolar2008}; 
	$^j$\cite{gehrelsFaintCometSearching1981}
	
	\label{safari:Surveys}
\end{table}

\begin{table*}
	\centering
	\caption{Active Asteroid Hunting Surveys \& Occurrence Rate Estimates}
	\footnotesize
	\begin{tabular}{llclclrl} 
		Survey & Source & Zone & Activity & $N^\dag$ & Limit & Objects & Method\\
		& & & ($N$ per $10^6$) & (mag) & &\\
		\hline
			\cite{cikotaPhotometricSearchActive2014} & MPC & MBA & \hspace{6mm}$\cdots$ & 1 & 16.7 & 330K & Photometric Excess\\
			\cite{gilbertUpdatedResultsSearch2010} & CFHT & MBA & \hspace{6mm}$40\pm 18$ & 3 & 22.5$^a$ & 25K & By-Eye\\
			\cite{hsiehHawaiiTrailsProject2009} & HTP & OMB & \hspace{6mm}$\cdots$ & 1 & 26 & 600 & By-Eye\\
			\cite{hsiehMainbeltCometsPanSTARRS12015} & \footnotesize{Pan-STARRS} & OMB & \hspace{6mm}$96$ & 4 & 22.6 & 300K & PSF\\
			SAFARI (this work) & DECam & MBA & \hspace{6mm}$80$ & 1 & 24.3 & 11K & By-Eye\\
			\cite{sonnettLimitsSizeOrbit2011} & TALCS & MBA & \hspace{2mm}$<2500$ & 0 & 24.3 & 1K & Excess Sky Flux\\
			\cite{waszczakMainbeltCometsPalomar2013} & PTF & MBA & \hspace{2mm}$<30$ & 0 & 20.5 & 220K & Extendedness\\
		\hline
	\end{tabular}
	\raggedright
	CFTS: Canada-France-Hawaii Telescope; DECam: Dark Energy Camera; HTP: Hawaii Trails Project; MPC: Minor Planet Center; PTF: Palomar Transient Factory; Pan-STARRS: Panoramic-Survey Telescope And Rapid Response System; TALCS: Thousand Asteroid Light Curve Survey; MBA: Main Belt Asteroids; OMB: Outer Main Belt; $^\dag$Includes known AAs;
	$^a$\cite{gilbertSearchingMainbeltComets2009}; PSF: Point Spread Function
	\label{safari:aahunts}
	\label{safari:ActivityRates}
\end{table*}

We chose to visually examine (``by-eye'') images of active asteroids because this technique has so far produced the greatest yield. Other methods have been applied (Table \ref{safari:Surveys}) but with varied degrees of success. \cite{cikotaPhotometricSearchActive2014} examined a large number of objects and searched for unexpected deviations in object brightness; this technique positively identified one known active asteroid, but (so far) the other candidates (\ref{safari:Table:TheAAs}) have not been observed to be active. \cite{sonnettLimitsSizeOrbit2011} examined the regions immediately surrounding asteroids, searching for photometric excess (i.e., a photon count above the sky background level). \cite{waszczakMainbeltCometsPalomar2013} formulated a way to quantify ``extendedness'' of  Palomar Transient Factory objects, with a 66\% comet detection rate and a 100\% Main Belt Comet detection efficiency. \cite{hsiehMainbeltCometsPanSTARRS12015} compared point spread function (PSF) widths between background stars and other objects and flagged exceptionally large PSF radii for further follow-up. All of the aforementioned techniques rely upon visual inspection for confirmation of activity. Spectroscopic detection of activity has also been carried out (Table \ref{safari:Table:TheAAs2}), but so far only (1) Ceres has been observed to be visually active in follow-up, and, in that case, in situ by the Rosetta spacecraft orbiting it. Hayabusa 2 recently arrived at (162173) Ryugu but as of yet no tail or coma has been observed.

Conservative activity occurrence rates of $>$1 in 10,000 are constrained by the magnitude limits of prior surveys \citep{jewittActiveAsteroids2015a}. We reached past the 17-22.7 magnitude limits of previous large-sky surveys (Table \ref{safari:Surveys}) by making use of existing \ac{DECam} data \citep{sheppardNewExtremeTransNeptunian2016} probing a magnitude fainter than other large-sky active asteroid survey. Note that while we are sensitive to more distant populations (e.g., Centaurs, Trans-Neptunian Objects), 99.7\% of our population is from the main asteroid belt.

We set out to determine the viability of \ac{DECam} data for locating active asteroids. We aimed to create a novel, streamlined pipeline for locating known asteroids within our dataset. We planned to examine our new library of asteroid thumbnails to find active asteroids and to test published asteroid activity occurrence rates (Table \ref{safari:ActivityRates}). We applied our technique to \fitscountSAFARI{} \ac{DECam} images ($\sim$5 Tb) to produce \allthumbsSAFARI{} thumbnail images comprising \uniquethumbsSAFARI{} unique objects. We examined the asteroid thumbnails by-eye to identify signs of activity. We show our technique can be applied to an orders-of-magnitude larger publicly-available dataset to elevate active asteroids to a regime where they can be studied as a population.

\section{Methods}
\label{safari:methods}

\begin{figure}
  \centering
	\includegraphics[height=8in]{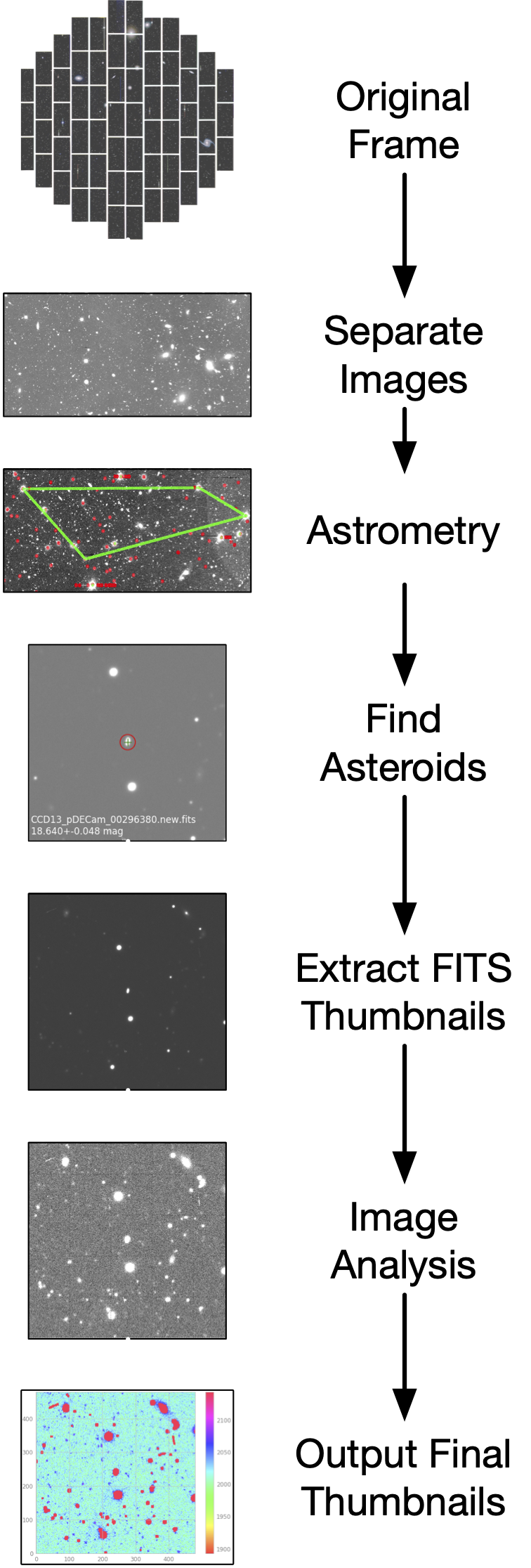}
  \caption{The SAFARI workflow.}
  \label{safari:flow}
\end{figure}

\subsection{Dark Energy Camera} We made use of data taken by the \ac{DECam} instrument on the 4-meter Blanco telescope at the Cerro Tololo Inter-American Observatory in Chile. The instrument has a $\sim$3 square degree field of view, capturing data via a mosaic of 62 \ac{CCD} chips, each $2048\times 4096$ with a pixel scale of 0\farcsec263/pixel \citep{darkenergysurveycollaborationDarkEnergySurvey2016}. Our data consisted of $594\times 2.2$ Gb frames in the VR filter ($500\pm10$ nm to $760\pm10$ nm), each containing $62\times 33$ Mb subsets of data, one per \ac{CCD}. The mean seeing across all images was 1\farcsec.14$\pm$ 0\farcsec13. We made use of software which required each multi-extension \ac{FITS} file be split into its 62 constituent parts, which we refer to as images for the remainder of this paper. Note: some files contained only 61 chips due to an instrument hardware malfunction.

\subsection{High Performance Computing}

We utilized \textit{Monsoon}, the Northern Arizona University (NAU) High Performance Computing (HPC) computing cluster. \textit{Monsoon} uses the \textit{Slurm Workload Manager} \citep{yooSLURMSimpleLinux2003} software suite to manage the 884 Intel Xeon processors to deliver up to 12 teraflops of computing power. The majority of our tasks each utilized 8 cores and 48 Gb of memory. The online supplement contains the complete listing of requirements necessary for each task.

\subsection{photometrypipeline} We utilized the \textit{photometrypipeline} \citep{mommertPHOTOMETRYPIPELINEAutomatedPipeline2017} software package to carry out source extraction via \textit{Source Extractor} \citep{bertinSExtractorSoftwareSource1996,bertinSExtractorSourceExtractor2010}, photometry and astrometry via \textit{SCAMP} \citep{bertinAutomaticAstrometricPhotometric2006,bertinSExtractorSourceExtractor2010}, and asteroid identification via \textit{SkyBot} \citep{berthierSkyBoTNewVO2006} and \textit{Horizons} \citep{giorginiStatusJPLHorizons2015}. We chose the \textit{Anaconda}\footnote{\url{www.anaconda.com}} \textit{Python} programming language distributions (versions 2.7 and 3.5) and the \textit{Python} package \textit{AstroPy} \citep{robitailleAstropyCommunityPython2013}.

\subsection{Procedure}
\label{safari:procedure}

\begin{figure}
	\centering
	\begin{tabular}{cc}
		\includegraphics[width=0.46\linewidth]{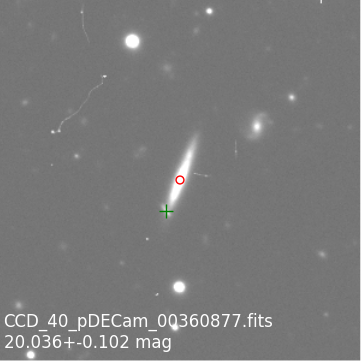} & \includegraphics[width=0.46\linewidth]{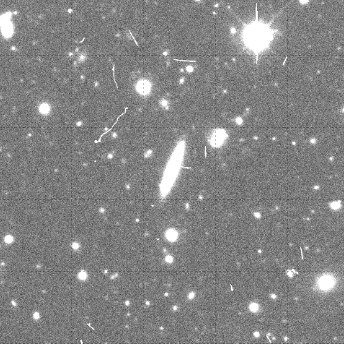}\vspace{-2mm}\\
		\vspace{1mm}
		(\textbf{a}) & (\textbf{b})\\
		\includegraphics[width=0.46\linewidth]{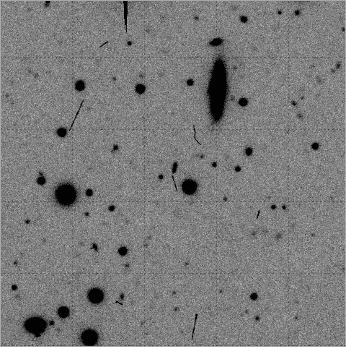} & \includegraphics[width=0.46\linewidth]{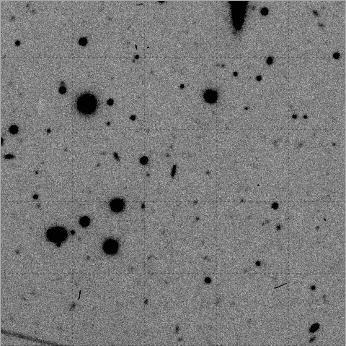}\vspace{-2mm}\\
		(\textbf{c}) & (\textbf{d})\\
	\end{tabular}
	\caption{Two asteroid thumbnail contrast selection approaches are shown in \textbf{a} and \textbf{b}. \textbf{a}) The \textit{photometrypipeline} thumbnail shows increased dynamic range. \textbf{b}) Iterative Rejection sacrifices some dynamic range (notice especially the edges of the center galaxy and the spiral galaxy to its upper-right) in favor of recovering more objects, many of which are not visible in \textbf{a} that can be easily seen in \textbf{b}. \textbf{c} \& \textbf{d}) Asteroid 2012 YU2 is shown in two frames comprising one animated \ac{GIF} file.}
	\label{safari:AnimatedGif}
	\label{safari:ContrastComparison}
	\label{safari:ExposureTimes}
\end{figure}

\begin{enumerate}[itemsep=2pt,leftmargin=10pt]
	\item \textit{Image Reduction}-- We employed standard image reduction techniques where each frame was bias subtracted, then flat-fielded using a combination of twilight flats and a master flat; full details of our imaging techniques can be found in \cite{sheppardNewExtremeTransNeptunian2016}.
	\item \textit{Splitting Multi-Extension FITS Files}-- \ac{DECam} produces multi-extension FITS files, where each extension contains data from one \ac{CCD}; because \textit{photometrypipeline} was incompatible with this format, we split each file into 62 separate \ac{FITS} files via the \textit{FTOOLS} \citep{blackburnFTOOLSFITSData1995} software package. We replicated global and extension headers for each output file to preserve metadata required for our image processing.
	\item \textit{Coordinate Correction}-- Each DECam image came pre-encoded with right ascension (RA) and declination (Dec) information indicating the coordinates of the telescope pointing center. We shifted the RA \& Dec of each remaining \ac{CCD} to their true coordinate values. The RA \& Dec offsets used for each \ac{CCD} are provided with the online supplement.
	\item \textit{World Coordinate System Purging}-- We discovered World Coordinate System (WCS) headers encoded in the \ac{FITS} files were preventing \textit{photometrypipeline} and/or \textit{astrometry.net} from performing astrometry. We were able to resolve the issue by purging all WCS header information as part of our optimization process. The header record names are listed in the online supplement.
	\item \textit{WCS Population via astrometry.net}-- We installed the \textit{astrometry.net} \citep{langAstrometryNetBlind2010} v0.72 software suite on \textit{Monsoon}. We processed all \fitscountSAFARI{} \ac{FITS} files to retrieve coordinate information for each image by matching the image to one or more index files (catalogs of stars for specific regions of sky, designed for astrometric solving).
	\item \textit{photometrypipeline Image Processing}-- We performed source extraction, photometry, astrometry and image correction via the \textit{photometrypipeline} software suite.
	\item \textit{Identifying Known Asteroids} We identified known asteroid in our data by making use of \textit{pp\_distill}, a module of \textit{photometrypipeline}.
	\item \textit{FITS Thumbnail Generation}-- We extracted the RA, Dec, and $(x,y)$ pixel coordinates of each object. We then produced $480\times480$ pixel, lossless, \ac{FITS} format asteroid thumbnails, each a small image centered on an asteroid. For cases where the object was too close ($<$240 pixels) to one or more image edges, we found it best to use the \textit{NumPy}\footnote{\url{www.numpy.org}} \textit{Python} routine to ``roll'' the image array; the technique shifts an array as if it were wrapped around a cylinder. For example: array [0, 1, 2, 3] rolled left by 1 would result in array [1, 2, 3, 0].
	\item \textit{Create PNG Thumbnails}-- We used an iterative-rejection technique to compute contrast parameters, then produced \ac{PNG} image files via \textit{MatPlotLib}\footnote{\url{www.matplotlib.org}}.
	\item \textit{Animated GIF Creation} We combined thumbnails of asteroids observed more than once (Figure \ref{safari:DataStats}c) to create animated \ac{GIF} files (Figure \ref{safari:AnimatedGif}) using the \textit{Python Image Library}\footnote{\url{www.pythonware.com/products/pil/}} software package. There are a number of advantages to this inspection approach, including 1) the opportunity to inspect one asteroid at multiple epochs, 2) activity may not occur at every epoch, and 3) activity may be easier to spot if the inspector has the opportunity to become familiar with an object (e.g., the general shape or streak pattern), even if only briefly.
	\item \textit{Examination of Image Products} -- Three authors served as asteroid thumbnail inspectors. Each inspector conducted a procedure consisting of rapid by-eye examination of asteroid thumbnails and animated \acp{GIF}, covering each thumbnail at least once. We flagged thumbnails and animations containing potential active asteroids for a later en masse review.
\end{enumerate}

\section{Results}
\label{safari:results}

\begin{figure*}
	\centering
	\begin{tabular}{cc}
		\begin{tabular}{cc}
			\includegraphics[width=0.2\linewidth]{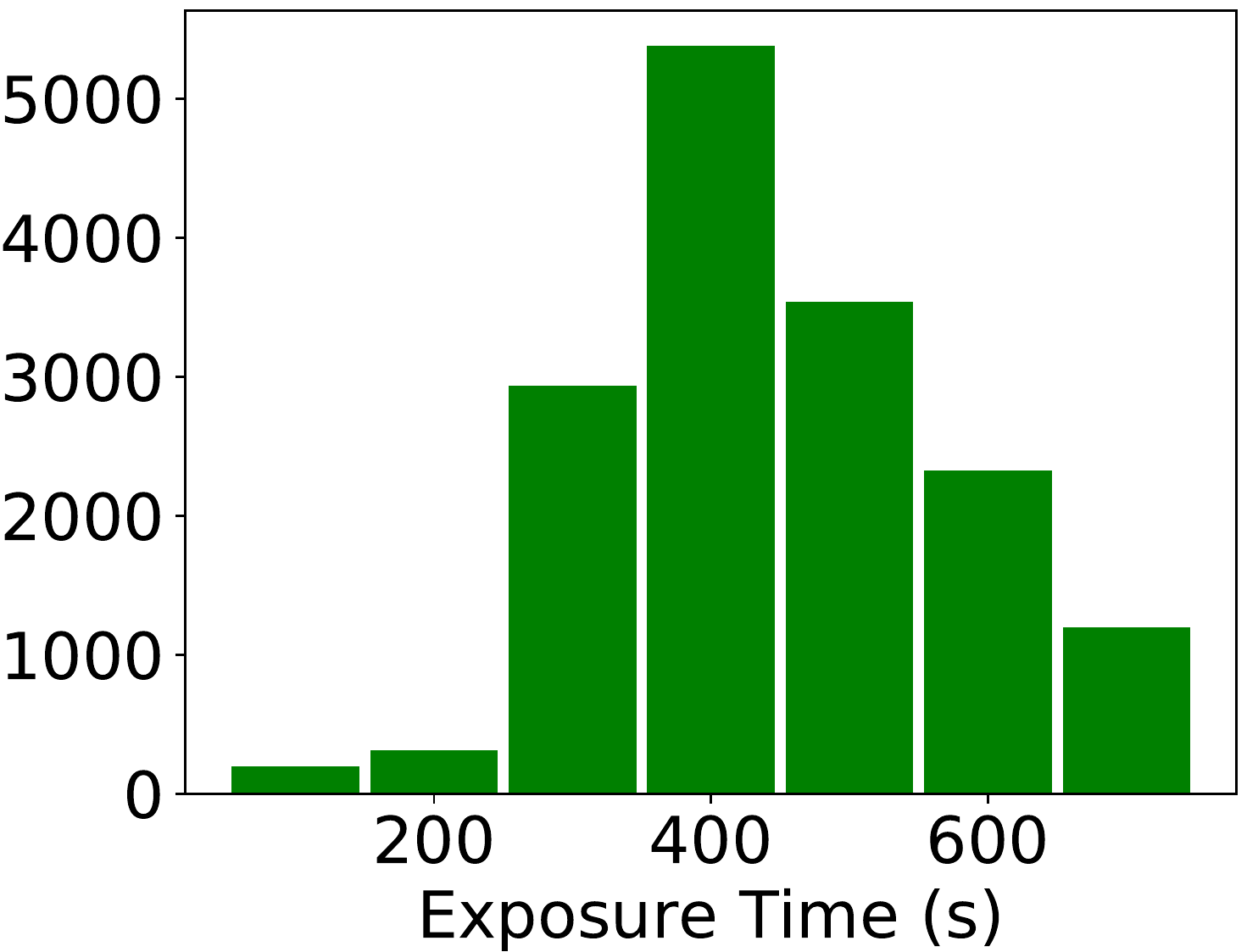} & \includegraphics[width=0.2\linewidth]{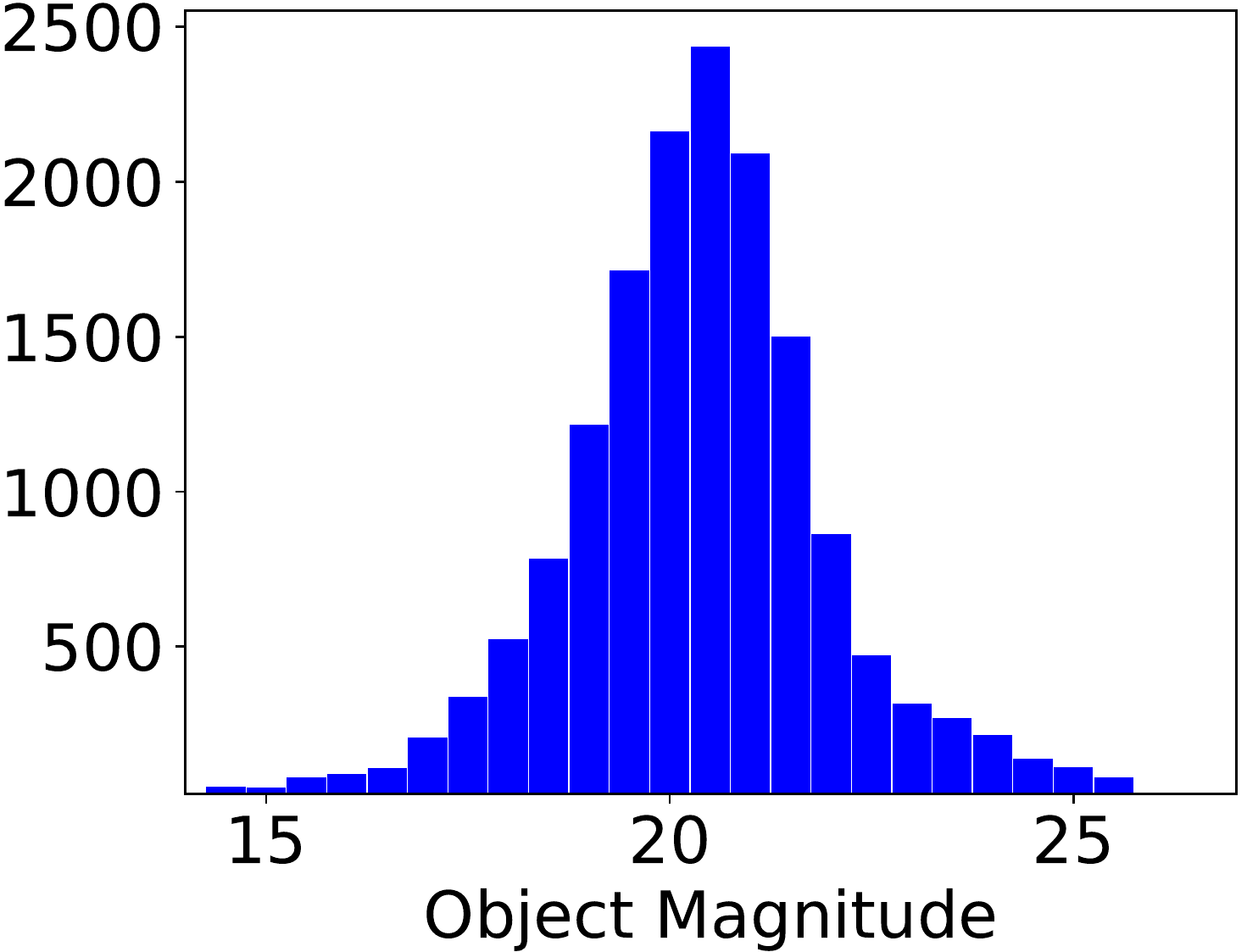}\vspace{-2mm}\\
			(\textbf{a}) & (\textbf{b})\vspace{1mm}\\
			\includegraphics[width=0.19\linewidth]{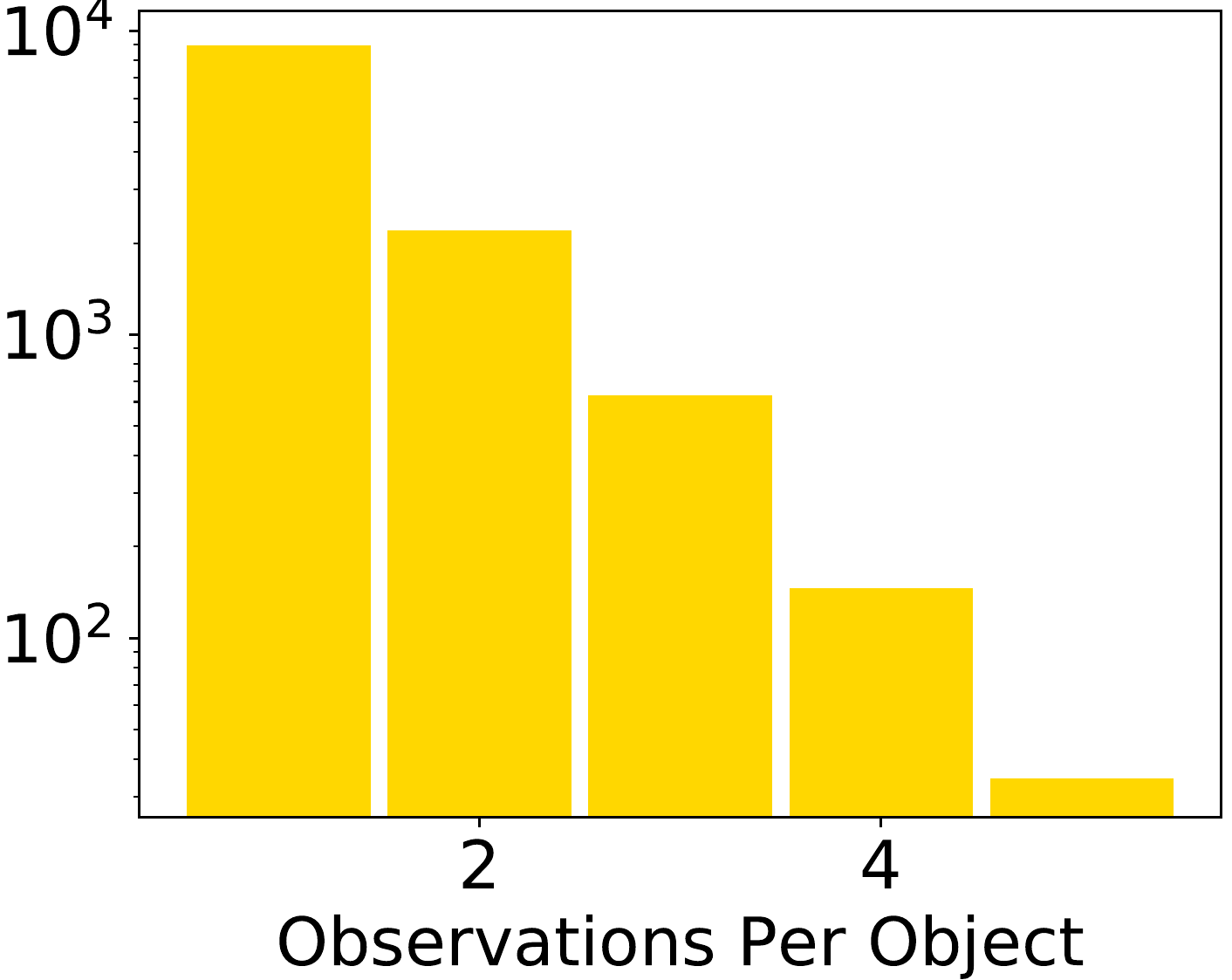} & \includegraphics[width=0.21\linewidth]{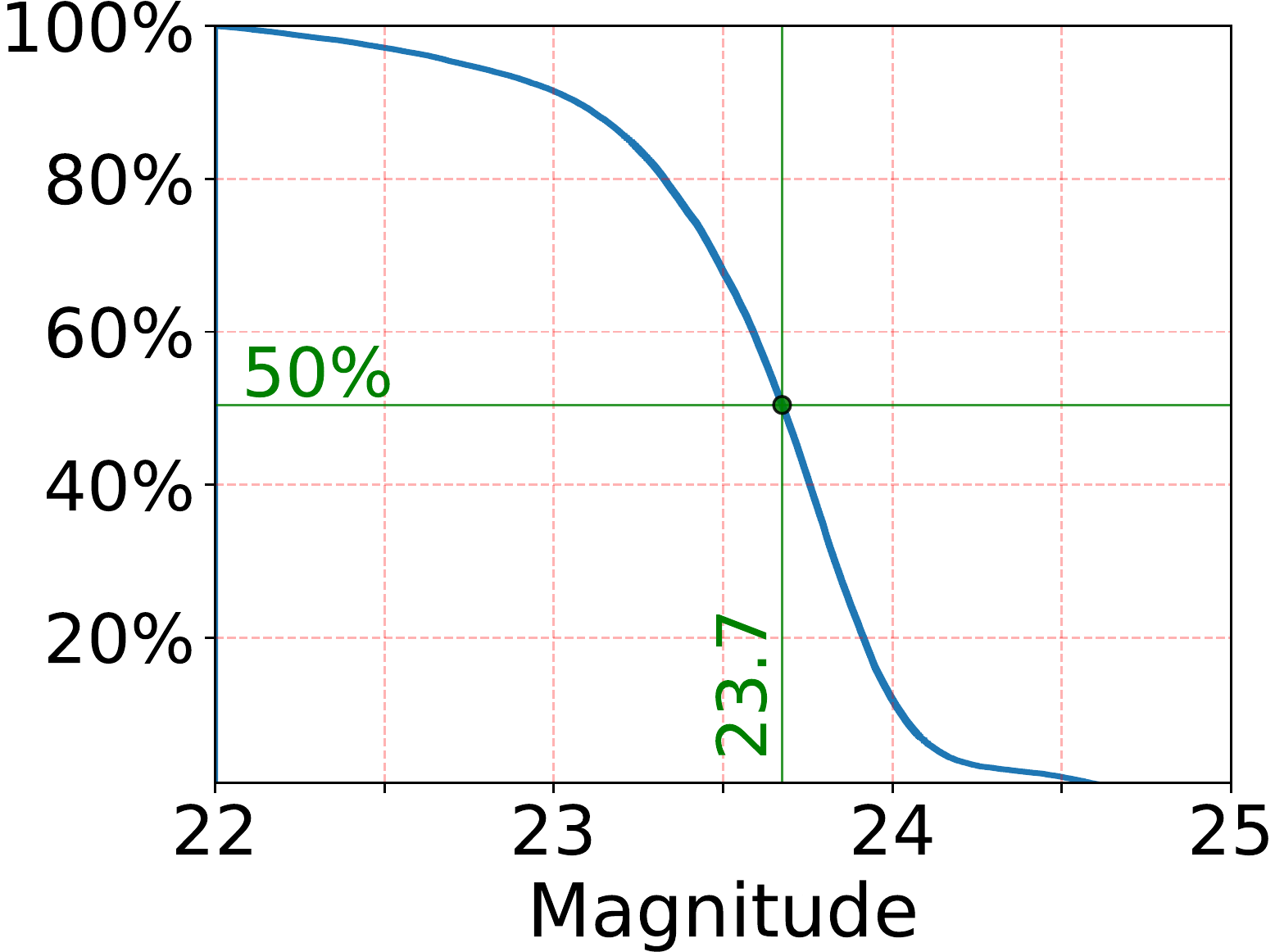}\vspace{-2mm}\\
			(\textbf{c}) & (\textbf{d})\\
		\end{tabular} & \hspace{-10mm}
		\begin{tabular}{c}
			\includegraphics[width=0.5\linewidth]{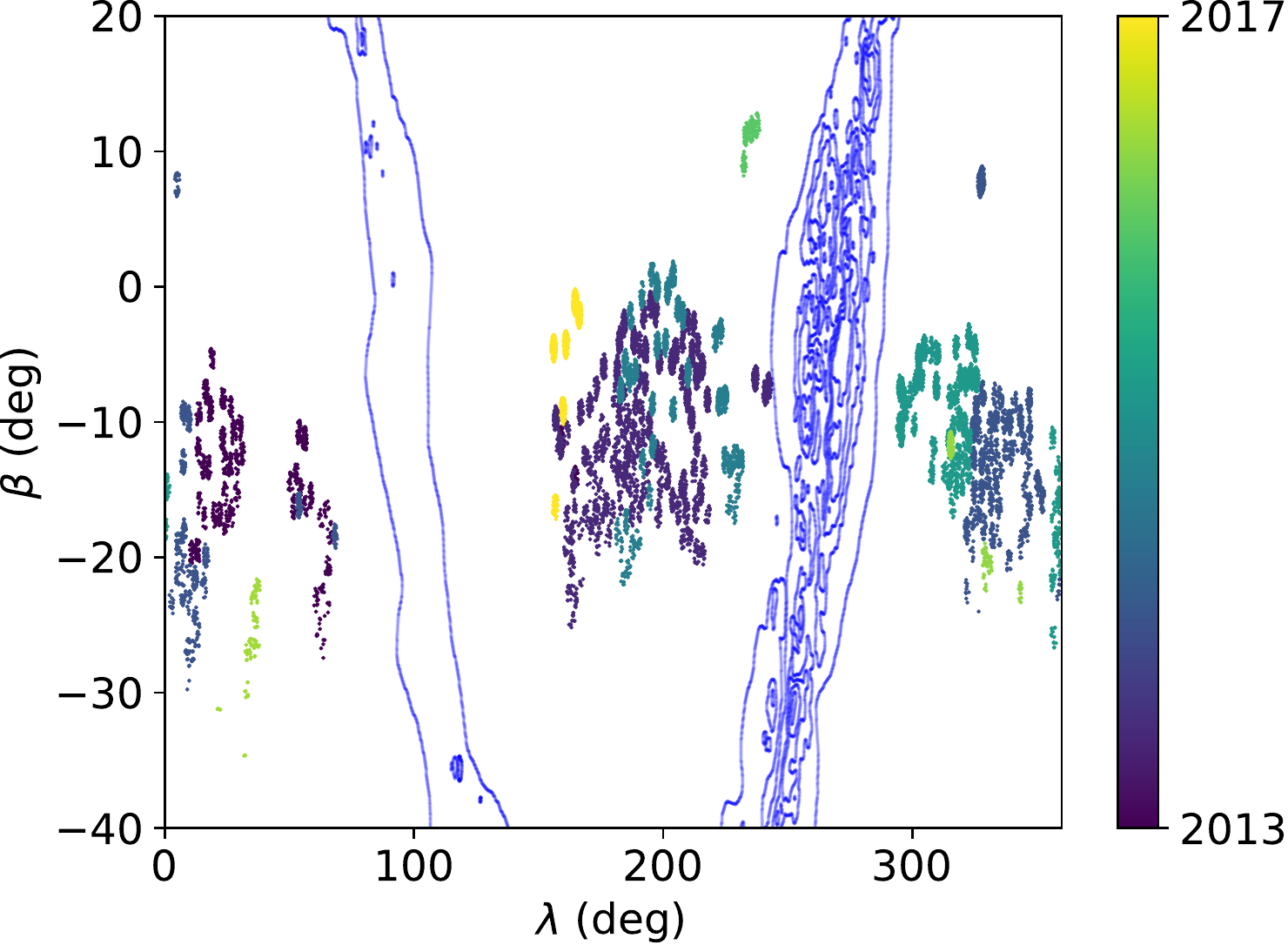}\vspace{-2mm}\\
			(\textbf{e})\\
		\end{tabular}
	\end{tabular}
	\caption{\textbf{a}) Exposure time distribution in our data. \textbf{b}) Histogram of apparent magnitudes for known asteroids we identified in our dataset. \textbf{c}) Observations per object; the \allthumbsSAFARI{} asteroid thumbnails contained \uniquethumbsSAFARI{} unique objects, \gifcountSAFARI{} of which were observed more than once. \textbf{d}) Cumulative histogram showing the depth of magnitudes (stars and asteroids) found in our dataset. 50\% of our images reached a magnitude of $m_\mathrm{R}=23.7$. Sources with a signal-to-noise ratio of $<$5:1 were not included. \textbf{e}) Asteroids encountered shown in geocentric ecliptic space, where $\lambda$ and $\beta$ are the ecliptic longitude and latitude, respectively. Distinct patches sum to $\sim$1000 $\mathrm{deg}^2$, as described in the text. Milky Way coordinates were retrieved from the \textit{D3-Celestial} (\url{http://ofrohn.github.io}) software suite.}
	\label{safari:DataStats}
\end{figure*}

\textit{Pipeline} -- We created a pipeline (Figure \ref{safari:flow}) that takes as its input DECam multi-extension \ac{FITS} files, and returns individual asteroid thumbnails and animated \ac{GIF} files. The initial total compute time requested across all tasks was 13,000 hours (1.5 compute-years), but after optimization (see Optimization section below) only $\sim$500 compute hours were required. See the online supplement for a comprehensive table of resources utilized during this project.

\textit{Image Products} We extracted \allthumbsSAFARI{} asteroid thumbnails from \fitscountSAFARI{} DECam images ($\sim$2 Tb total). Most of our data consisted of exposure times $>$300s (Figure \ref{safari:DataStats}a). These longer integration times allowed us to probe deeper (fainter), with asteroids captured down to $25^\mathrm{th}$ magnitude (Figure \ref{safari:DataStats}b). Each of the \uniquethumbsSAFARI{} unique objects identified in our data were observed between 1 and 5 times, with \gifcountSAFARI{} objects imaged more than once (Figure \ref{safari:DataStats}c). 

To compute our coverage area on sky (depicted in Figure \ref{safari:DataStats}e) we employed a nearest neighbor algorithm to identify the distinct (non-overlapping) regions of our dataset. Two fields were considered overlapping if their center-to-center distance was $<1.8$ degrees, the width of one DECam field. We computed our coverage to be $\sim200$ distinct 3 $\mathrm{deg}^2$ patches comprising $\sim$1000 square degrees.

\begin{figure}
	\centering
	\includegraphics[width=0.5\linewidth]{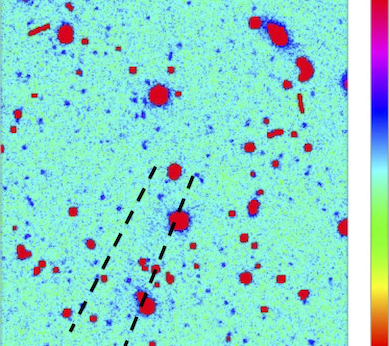}
	\caption{Asteroid (62412) shown with the ``hsv'' colormap and Mitchell interpolation. The asteroid is at the center of the frame and the tail can be seen between the dashed lines.}
	\label{safari:fig:62412}
\end{figure}

\textit{Active Asteroids} -- We imaged one asteroid previously discovered to be active \citep{sheppardDiscoveryCharacteristicsRapidly2015}: (62412). The object shows activity in our image (Figure \ref{safari:fig:62412}; see the online supplement for additional image color map and interpolation permutations) and we were able to identify activity in two other DECam frames that were not part of this work. \cite{sheppardDiscoveryCharacteristicsRapidly2015} confirmed activity with Magellan Telescope follow-up observations. We also imaged two other objects listed as active: (1) Ceres and (779) Nina but neither showed signs of activity.

\textit{Optimization} --  The final pipeline resulted from a series of iterative optimizations carried out with a subset of our large dataset. These optimizations produced order-of-magnitude reductions in compute time, and improved successful pipeline completion from the initial $\sim$35\% to the final 94\%. The implemented optimizations and their results are broken down below by number (matched to the corresponding procedure number of Section \ref{safari:procedure}). The final optimized \textit{Slurm} parameters used on \textit{Monsoon} can be found in the online supplement.

\begin{enumerate}[itemsep=2pt,leftmargin=10pt]
	\item \textit{Image Reduction}: No optimization needed.
	\item \textit{File-Splitting}: Splitting each multi-extension \ac{FITS} file into 62 separate \ac{FITS} files resulted in a larger number of smaller tasks which were better suited for parallel processing.
	\item \textit{Coordinate Correction}: Coordinate corrections proved cumbersome and inefficient, so we added \textit{astrometry.net} to our pipeline.
	\item \textit{WCS Purging}: We identified mismatched distortion coefficients as the primary culprit behind roughly 1/3 of our images failing \textit{photometrypipeline} analysis. We purged all World Coordinate System (WCS) headers, allowing us to employ \textit{astrometry.net} which increased our overall throughput and output.
	\item \textit{astrometry.net Astrometry}: We cached all ($\sim$32 Gb) astrometry index files (described in Section \ref{safari:procedure} item 5) locally so \textit{astrometry.net} would not be dependent on the speed of the internet connection and file host. We optimized the \textit{astrometry.net} computation by supplying the following parameters we extracted from our \ac{FITS} files. Providing a pixel scale range ($\sim$0\farcsec25/pixel to $\sim$0\farcsec28/pixel) and R.A./decl. values narrowed the range of indices that required searching. We found a 15$''$ search radius further reduced computation time without impacting image recognition efficacy. We disabled \textit{astrometry.net} plotting due to a \textit{Slurm} incompatibility, and computation time decreased further still. We found submitting \textit{astrometry.net} ``solve-field'' tasks directly to \textit{Slurm} was much faster. All but 41 images successfully matched for astrometry on first pass, and we improved \textit{astrometry.net} image recognition speed roughly tenfold.
 	\item \textit{photometrypipeline}: Proper configuration of prerequisite software and \textit{photometrypipeline} proved crucial; the online supplement contains the necessary parameters we used. We made minor modifications to the \textit{photometrypipeline} code, described in the online supplement. We found out \textit{astropy} was using home directory temporary storage space, a fatal error for systems with enforced quotas; the home storage space was also slower than the scratch space. Proper configuration reduced computation time and increased the pipeline success rate.
 	\item \textit{Known Asteroid Identification} We added an initial \textit{SkyBot} query to identify the asteroids within each image. We then populated the requisite \texttt{OBJECT} \ac{FITS} header keyword in each of our images, thereby enabling us to call \textit{Horizons} to locate asteroids in our images and provide accurate astrometry. Prepending the \textit{SkyBot} query and populating the \texttt{OBJECT} keyword enabled us to run asteroid identification tasks in parallel, reducing processing time by three orders-of-magnitude.
	\item \textit{FITS Thumbnails}: We ``rolled'' images (described in Section \ref{safari:procedure} item 8) so we could create full-sized ($480\times480$ pixel) thumbnails. While thumbnails sometimes looked peculiar when rolled, this method preserved image statistics used to compute the narrow range of contrast achieved in the next section.
	\item \textit{PNG Thumbnails}: While \textit{photometrypipeline} does output thumbnails by default, we were unable to see enough detail with the default scaling. Therefore, we employed an iterative rejection technique. Figures \ref{safari:ContrastComparison} a and \ref{safari:ContrastComparison} b compare the two contrast ranges. For each of the \allthumbsSAFARI{} asteroid thumbnails, we chose to output different colormap/interpolation combinations: two modes of interpolation (Mitchell--Netravali balanced cubic spline filter and one set unfiltered), each in 11 color schemes (afmhot, binary, bone, gist\_stern, gist\_yarg, gray, hot, hsv, inferno, Purples, and viridis), examples of which are shown in the online supplement. 
The optimized dynamic ranges allowed faint trails to become more visible. These colormap/interpolation schemes gave us, as thumbnail inspectors, the ability to choose a comfortable theme for use while searching thumbnails for asteroid activity, thereby increasing our productivity.
	\item \textit{Animated GIFs}: We produced animated \acp{GIF} enabling an alternative inspection format.
	\item \textit{Examination}: We uncovered common sources of false positives (discussed in Section \ref{safari:FalsePositives}) and incorporated their presence into our visual examination procedures, resulting in a streamlined examination process while simultaneously reduced the number of false-positives.
\end{enumerate}

\section{Discussion}
\label{safari:discussion}

We set out to determine if DECam data would provide a suitable pool from which to search for active asteroids. We crafted a method to extract asteroid thumbnails from DECam data, and the large number of asteroids encountered (\uniquethumbsSAFARI{}) along with the exceptional depth our images probed (Figures \ref{safari:DataStats}b and \ref{safari:DataStats}d) indicate our data are well-suited to locating active asteroids. 

\subsection{Population Traits}

\begin{table}
	\centering
	\small
	\caption{SAFARI Asteroid Populations}
	\begin{tabular}{lclccrr}
		\hline\hline
		Zone 		& $R_i$ & $a_{p_i}$ & $a_{p_o}$ & $R_o$ & \multicolumn2c{SAFARI}\\
		      		& \footnotesize{(A:J)} & \footnotesize{(au)}    & \footnotesize{(au)} & \footnotesize{(A:J)} & (N) & (\%)\ \\
		\hline
		Int.  		& $\cdots$ 	& 0     & 2.064    & 4:3 	  &	115   &  1\\
		IMB   		& 4:3 		& 2.064 & 2.501    & 3:1 	  & 3,605 & 26\\
		MMB   		& 3:1 		& 2.501 & 2.824    & 5:2 	  & 5,358 & 39\\
		OMB   		& 5:2 		& 2.824 & 3.277    & 2:1 	  & 4,599 & 33\\
		Ext.		& 2:1 		& 3.277 & $\infty$ & $\cdots$ & 162   &  1\\
		Total$^*$	& $\cdots$	& $\cdots$ & $\cdots$ & $\cdots$ & 13,839 & 100\\
		\hline
	\end{tabular}
	\label{safari:tab:zones}
	
	\raggedright
	\hspace{1.7in}\footnotesize{Int., Ext.,: Interior, Exterior to the main belt}\\
	\hspace{1.7in}\footnotesize{IMB, MMB, OMB: Inner, Mid, Outer Main Belt}\\	
	\hspace{1.7in}\footnotesize{$a_{p_i}$,$a_{p_o}$: inner, outer proper semi-major axis}\\
	\hspace{1.7in}\footnotesize{A:J Asteroid:Jupiter; $R_i$, $R_o$: inner/outer resonances}\\
	\hspace{1.7in}\footnotesize{$^*$Not included: 791 objects with unknown parameters.}
\end{table}

As indicated by Figures \ref{safari:DataStats}a-d, the population imaged during our survey were subject to selection effects caused by the depth ($\bar{m}_\mathrm{R}=23.7$) of our survey (e.g., closer objects would have appeared as long trails which would have been difficult to identify with our pipeline). We classified the objects following the procedure of \cite{hsiehAsteroidFamilyAssociations2018}; we categorized our population as Inner Main Belt (IMB), Mid Main Belt (MMB), and Outer Main Belt (OMB), plus two additional regions: ``Interior'' (to the IMB) and ``Exterior'' (to the OMB). Table \ref{safari:tab:zones} indicates the boundaries, along with their Asteroid:Jupiter (A:J) resonances.

The synthetic proper semi-major axis $a_p$ aims to minimize the influence of transient perturbations \citep{knezevicSyntheticProperElements2000}. We made use the \textit{AstDyn-2}\footnote{\url{http://hamilton.dm.unipi.it/astdys}} online catalog service \citep{knezevicProperElementCatalogs2003} in determining proper orbital parameters for asteroids in our dataset (Table \ref{safari:tab:zones}).



Our target (object) aperture photometry was computed with a fixed diameter of 10 pixels, though photometric calibration was performed with an aperture radius determined by curve-of-growth analysis (see \citet{mommertPHOTOMETRYPIPELINEAutomatedPipeline2017} for details). To determine the surface brightness limit of our catalog we first computed the limit $SB$ of each image

\begin{equation}
	SB_\mathrm{lim} = \frac{\sum_{k=1}^{k=N} \left(m_{0_k} - 2.5 \log_{10}\left(n \sigma_{\mathrm{bg}_k}\sqrt{1/A}\right)\right)}{N},
\end{equation}

\noindent where $m_0$ is the photometric zero point (determined by \textit{PhotometryPipeline}), $n$ the order of detection level for background noise standard deviation $\sigma_\mathrm{bg}$, and $A$ is the area of one pixel in square arcseconds (\citealt{hsiehSurfaceBrightnessLimits2018}, personal communication). The DECam camera had a pixel scale of 0\farcsec263/pixel, give a pixel area

\begin{equation}
	A = (0\farcsec 263)^2 = 0.069169\ \mathrm{arcseconds}^2.
\end{equation}


For our surface brightness analysis we made use of $N=32,790$ chips for which we had been able to determine a photometric zero point. We computed the $3\sigma$ mean surface brightness limit of our dataset to be ${SB}_\mathrm{lim}=26.44\pm 0.24$ mag/$\mathrm{arcsec}^2$.

\subsection{Occurrence Rates}
We also aimed to validate the published asteroid activity occurrence rates of Table \ref{safari:ActivityRates}. Occurrence rates have been conservatively set at 1 in 10,000 (for all main belt asteroids), with the limiting magnitude of surveys the primary bottleneck. As shown in Figure \ref{safari:DataStats}d, the DECam instrument reaches an average magnitude of 24 \citep{sheppardNewExtremeTransNeptunian2016}, an unprecedented depth for large area active asteroid surveys. While our complete dataset was consistent with the 1:10,000 activity occurrence rate estimate, it is somewhat surprising we did not discover additional asteroidal activity. 

\cite{hsiehMainbeltCometsPanSTARRS12015} postulated many active asteroids could be continuously active throughout their orbits (not just at perihelion), but with weaker activity. We expected then to find active asteroids more frequently in our search, given the objects we observed were indeed of a fainter magnitude (Figure \ref{safari:DataStats}b), though our outer main belt occurrence rate ($\sim$1:4000) was slightly higher than that reported by \cite{hsiehMainbeltCometsPanSTARRS12015} which is in line with their prediction. Small number statistics may have contributed to the possible discrepancy, and it is plausible we missed activity indications due to the limitations of visual inspection which were further compounded by an increased prevalence of background sources compared to shallower surveys. The use of a point spread function (PSF) comparison technique (e.g., \citet{hsiehMainbeltCometsPanSTARRS12015} or a photometric search (e.g., \citet{cikotaPhotometricSearchActive2014}) could help us identify candidates, features we plan to investigate in future work.

\subsection{False Positives}
\label{safari:FalsePositives}

\begin{figure}
	\centering
			\includegraphics[width=0.5\columnwidth]{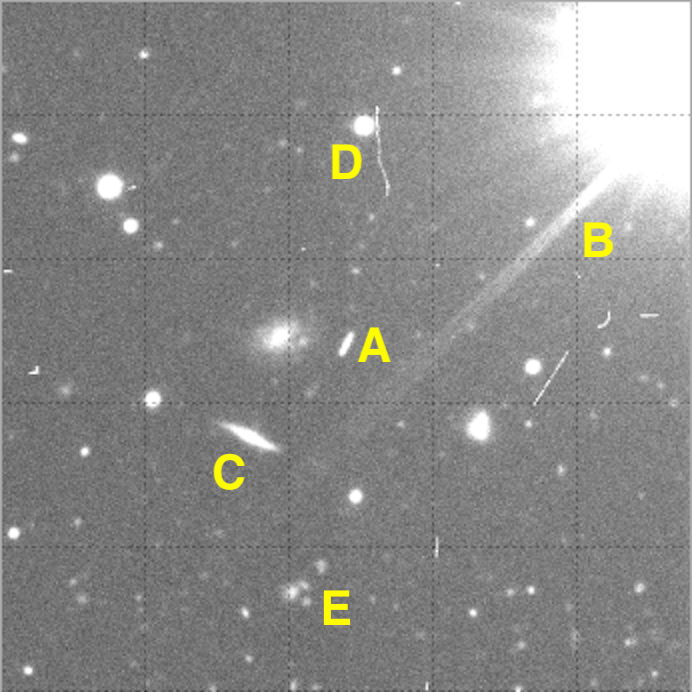}\\
	\caption{Common potential false-positives encountered in an asteroid thumbnail. (\textbf{a}) This thumbnail includes 4 potential false-positive sources: (\textbf{A}) Asteroid (432345). \textbf{B}) \textit{Scattered light} from a bright star trails toward the lower-left corner. (\textbf{C}) An \textit{extended source}, such as this edge-on galaxy, can present itself as coma if close to an asteroid. (\textbf{D}) \textit{Cosmic rays} with variable morphology are common throughout our images; they can look like trails if they align with a star as in this case. (\textbf{E}) \textit{Juxtaposed} objects can masquerade as active asteroids, especially when a bright object is near one or more progressively dimmer objects along the direction of apparent motion.}
	\label{safari:FalsePositives1}
\end{figure}

We found false-positive management to be a formidable task, with specific mechanisms responsible for creating false-positives recurring throughout the project. For the rare cases where one of the authors involved in inspecting thumbnails found potential activity in an asteroid thumbnail, we checked other interpolation and color schemes, other thumbnails of the same asteroid, and the animated \ac{GIF} if available. We checked frames showing the same region on the sky, including original \ac{CCD} images, for background sources or image artifacts. What follows is a discussion of the primary culprits in order to convey the challenges faced during by-eye inspection (which is subjective by nature).

\textit{Juxtaposition} -- Figure \ref{safari:FalsePositives1}A marks asteroid (432345); the object is in close proximity to a galaxy, which, if juxtaposed in a confusing manner, could give the appearance of a coma. \ref{safari:FalsePositives1}D shows how a cosmic ray can be juxtaposed with a star. Figure \ref{safari:FalsePositives1}E demonstrates how multiple objects may appear to be an extended source.

\textit{Extended Sources} -- Extended sources, especially galaxies, were present in a myriad of orientations and configurations. They can appear like active asteroids, as in the edge-on galaxy shown in Figure \ref{safari:FalsePositives1}C. For a given brightness, galaxies occupied more sky area in a frame than other types of natural (i.e., non-artifact) objects and were more likely to be juxtaposed with other objects.

\textit{Scattered Light} -- Figure \ref{safari:FalsePositives1}B is scattered light associated with an especially bright star; the flare originates from the star and tapers off the further the ``tail'' gets from the source. While obvious in Figure \ref{safari:FalsePositives1}, the ``tail'' can be more difficult to identify as scattered light if the source is outside of the thumbnail.

\textit{Cosmic Rays} -- Cosmic rays (e.g. Figure \ref{safari:FalsePositives1}D) are common throughout our images, most of which have exposure times of 300 seconds or longer (see Figure \ref{safari:ExposureTimes}a). Figure \ref{safari:FalsePositives1} D demonstrates how cosmic rays may not appear as straight lines, and they may seem to connect two or more objects together.

\textit{Poor Seeing} -- Images with poor and/or rapidly varying seeing conditions suffered from fuzziness (potentially coma-like) and elongation implying a trailed object (e.g., an asteroid).

\subsection{Limitations of By-eye Inspection}
\label{safari:HumanLimitations}

As proof-of-concept for future projects making use of larger datasets, we sought a general understanding of our throughput as thumbnail inspectors. It is worth noting we did not impose time limits upon ourselves. We noted markedly different inspection rates, with the time required to inspect all thumbnails ranging from 2 to 6 hours. Furthermore, our attention spans varied, with inspection sessions lasting roughly between 10 minutes to 3 hours before requiring a break. The false positive handling described above undoubtedly impacted our image examination efficacy to some degree. Given these challenges, it is evident a computational approach to screen for potential active asteroids (through e.g., PSF comparison) would improve our detection rate.

\subsection{Asteroid Selection}
\label{safari:AsteroidSelection}
We examined only known asteroids during this work, but certainly many unknown asteroids are present within our data. Future efforts involving Citizen Scientists could locate these objects and quantify previously unrecognized biases inherent to locating activity among known asteroids. We used observations from a southern observatory, and while there may be little to no effect on observed activity occurrence rates, we acknowledge this selection effect nonetheless.

\subsection{Future Work}
\label{safari:FutureWork}

A broader study of the efficacy of human inspectors should be carried out if employing a larger number of inspectors. Injecting artificial active asteroids into the datasets would enable quantifying detection rates. The enormous datasets (2M+ thumbnails) we plan to generate will necessitate the deployment of a Citizen Science project, an endeavor that would thoroughly flush out these detection rates.

Citizen Science endeavors enable scientists to analyze otherwise prohibitively large datasets, with the added benefit of providing the scientific community with invaluable outreach opportunities proven to engage the public and spark far-reaching interest in science. \textit{Zooniverse}\footnote{\url{www.zooniverse.org}}, designed with the average scientist in mind, facilitates deployment of crowd-sourcing science projects. Volunteers are enlisted to interpret data too complex for machines, but accomplishable by anyone with minimal training. \textit{Zooniverse} has a proven track record, with notable successes such as \textit{Galaxy Zoo} which, within 24 hours of launch, reached 70,000 identifications/hour \citep{coxDefiningMeasuringSuccess2015}. While traditional and social media coverage undoubtedly boosted the performance of \textit{Galaxy Zoo} and other exemplary Citizen Science projects, the platform is designed to facilitate such exposure, especially through social media connectivity.

Our aim is to expand our survey to a second, comparably sized dataset already in-hand. We will first explore strategies to quantify active asteroid candidacy through computational techniques such as PSF comparison. We will then use the combined datasets to design, implement and test a Citizen Science project. We plan to start with a moderate ($\sim$ 10 member) group of thumbnail inspectors consisting of undergraduate and graduate students, whose feedback will inform the documentation and training system which is crucial to the success of a Citizen Science project. We subsequently intend to expand our dataset to the entire DECam public archive, at which point we would open our analysis system to public participation. We hope to incorporate machine learning into our pipeline as a means of reducing the number of thumbnails sent to the Citizen Science project or to help locate candidates missed at any point in the process.

\section{Summary}
\label{safari:summary}

We have developed an approach for finding active asteroids, rare objects visually like comets but dynamically like asteroids. We show DECam data are suitable for active asteroid searches. The approach involved processing \fitscountSAFARI{} \ac{FITS} files and extracting \allthumbsSAFARI{} asteroid thumbnails (small images centered on an asteroid) consisting of \uniquethumbsSAFARI{} unique objects. Upon visual examination of all thumbnails, we identified one previously known active asteroid (62412); our discovery rate of 1 in \uniquethumbsSAFARI{} is consistent with the currently accepted active asteroid occurrence rate of 1 in 10,000. We did observe (1) Ceres and (779) Nina, though the former is a special case of \textit{a priori} activity knowledge \citep{ahearnWaterVaporizationCeres1992,kuppersLocalizedSourcesWater2014}, and neither object has ever shown signs of activity visible from Earth; as we did not observe activity in either object, we did not include them in our activity occurrence rate estimate. From our proof-of-concept study, we conclude a significantly larger survey should be carried out to locate active asteroids, finally placing them into a regime where they may be studied as a population.

\section{Acknowledgements}
\label{safari:acknowledgements} 

 The authors thank the referee, Henry Hsieh (Planetary Science Institute), whose thorough and thoughtful feedback vastly improved the quality of this work. Prof. Ty Robinson of Northern Arizona University (NAU) helped us keep this project a priority and provided fresh perspectives on scientific dilemmas. Dr. Mark Jesus Mendoza Magbanua (University of California San Francisco) whose insights greatly improved the quality of the paper. Annika Gustaffson (NAU) provided frequent input and encouraged us to move this project forward. The enthusiastic support provided by Monsoon supercomputer system administrator Christopher Coffey (NAU) was essential in overcoming countless technical challenges. The Trilling Research Group (NAU) provided insight and feedback about our data visualization techniques. Prof. Mike Gowanlock (NAU) inspired numerous computational techniques which reduced our analysis compute time.

Computational analyses were run on Northern Arizona University's Monsoon computing cluster, funded by Arizona's Technology and Research Initiative Fund. This work was made possible in part through the State of Arizona Technology and Research Initiative Program. Michael Mommert was supported in part by NASA grant NNX15AE90G to David E. Trilling.

This research has made use of the VizieR catalogue access tool, CDS, Strasbourg, France. The original description of the VizieR service was published in A\&AS 143, 23 \citep{ochsenbeinVizieRDatabaseAstronomical2000}. This research has made use of data and/or services provided by the International Astronomical Union's Minor Planet Center. This research has made use of NASA's Astrophysics Data System. This research has made use of the The Institut de M\'ecanique C\'eleste et de Calcul des \'Eph\'em\'erides (IMCCE) SkyBoT Virtual Observatory tool \citep{berthierSkyBoTNewVO2006}. This work made use of the {FTOOLS} software package hosted by the NASA Goddard Flight Center High Energy Astrophysics Science Archive Research Center.

This project used data obtained with the Dark Energy Camera (DECam), which was constructed by the Dark Energy Survey (DES) collaboration. Funding for the DES Projects has been provided by the U.S. Department of Energy, the U.S. National Science Foundation, the Ministry of Science and Education of Spain, the Science and Technology Facilities Council of the United Kingdom, the Higher Education Funding Council for England, the National Center for Supercomputing Applications at the University of Illinois at Urbana-Champaign, the Kavli Institute of Cosmological Physics at the University of Chicago, Center for Cosmology and Astro-Particle Physics at the Ohio State University, the Mitchell Institute for Fundamental Physics and Astronomy at Texas A\&M University, Financiadora de Estudos e Projetos, Funda\c{c}\~{a}o Carlos Chagas Filho de Amparo, Financiadora de Estudos e Projetos, Funda\c{c}\~ao Carlos Chagas Filho de Amparo \`{a} Pesquisa do Estado do Rio de Janeiro, Conselho Nacional de Desenvolvimento Cient\'{i}fico e Tecnol\'{o}gico and the Minist\'{e}rio da Ci\^{e}ncia, Tecnologia e Inova\c{c}\~{a}o, the Deutsche Forschungsgemeinschaft and the Collaborating Institutions in the Dark Energy Survey. The Collaborating Institutions are Argonne National Laboratory, the University of California at Santa Cruz, the University of Cambridge, Centro de Investigaciones En\'{e}rgeticas, Medioambientales y Tecnol\'{o}gicas–Madrid, the University of Chicago, University College London, the DES-Brazil Consortium, the University of Edinburgh, the Eidgen\"ossische Technische Hochschule (ETH) Z\"urich, Fermi National Accelerator Laboratory, the University of Illinois at Urbana-Champaign, the Institut de Ci\`{e}ncies de l'Espai (IEEC/CSIC), the Institut de Física d'Altes Energies, Lawrence Berkeley National Laboratory, the Ludwig-Maximilians Universit\"{a}t M\"{u}nchen and the associated Excellence Cluster Universe, the University of Michigan, the National Optical Astronomy Observatory, the University of Nottingham, the Ohio State University, the University of Pennsylvania, the University of Portsmouth, SLAC National Accelerator Laboratory, Stanford University, the University of Sussex, and Texas A\&M University.

Based on observations at Cerro Tololo Inter-American Observatory, National Optical Astronomy Observatory (NOAO Prop. IDs 2015A-0351
2016B-0288, 2017A-0367, 2015B-0265, 2013B-0453, 2014B-0303, 2016A-0401 and 2014A-0479; PI: Scott Sheppard), which is operated by the Association of Universities for Research in Astronomy (AURA) under a cooperative agreement with the National Science Foundation.


\clearpage

\section{Appendix}

\subsection{Object-specific References}

\label{safari:ObjectReferences}

SPK-ID are found at the JPL Horizons Small Bodies Database (\url{https://ssd.jpl.nasa.gov/sbdb.cgi}).

\begin{itemize}
\setlength\itemsep{0.01mm}
	\item[\ref{Ceres}.] (1) Ceres, 1943 XB, A899 OF, SPK-ID=2000001; Activity Discovered:\cite{ahearnWaterVaporizationCeres1992,kuppersLocalizedSourcesWater2014}; Mechanism: \cite{kuppersLocalizedSourcesWater2014}; Activity Obs.: 1 (1992) -- \cite{ahearnWaterVaporizationCeres1992}, 2 (2011-2013) -- \cite{kuppersLocalizedSourcesWater2014, nathuesSublimationBrightSpots2015}, 3 (2015-2016) -- \cite{thangjamHazeOccatorCrater2016,nathuesEvolutionOccatorCrater2017,landisConditionsSublimatingWater2017,rothConstraintsWaterVapor2018}; Visit: Dawn \citep{russellDawnArrivesCeres2016}; Absence of Family Association: \cite{rivkinCaseMissingCeres2014,hsiehAsteroidFamilyAssociations2018}; Additional: \cite{tuSublimationdrivenExosphericModel2014, witzeBrightSpotsCeres2015, hayneThermalStabilityIce2015, nathuesSublimationBrightSpots2015, liSurfaceAlbedoSpectral2016, rothConstraintsExosphereCeres2016, prettymanExtensiveWaterIce2017, mckayObservationalConstraintsWater2017, nathuesOxoCraterCeres2017, landisConditionsSublimatingWater2017}

	\item[\ref{Adeona}.] (145) Adeona, SPK-ID=2000145; Activity Discovery: \cite{busarevMaterialCompositionAssessment2016}; Mechanism: \cite{busarevMaterialCompositionAssessment2016}; Activity Obs.: 1 (2012) -- \cite{busarevMaterialCompositionAssessment2016}$^\ast$; Visit: Dawn (cancelled)\footnote{\scriptsize{\url{https://www.nasa.gov/feature/new-horizons-receives-mission-extension-to-kuiper-belt-dawn-to-remain-at-ceres}}}; Additional: \cite{busarevNewCandidatesActive2018}

	\item[\ref{Constantia}.] (315) Constantia, SPK-ID=2000315; Candidacy: \cite{cikotaPhotometricSearchActive2014}; Flora family association: \cite{alfvenAsteroidalJetStreams1969}

	\item[\ref{Griseldis}.] (493) Griseldis, 1902 JS, A915 BB, SPK-ID=2000493;  Activity Discovery: \cite{tholenEvidenceImpactEvent2015};  Activity Obs.: 1 (2015) -- \cite{tholenEvidenceImpactEvent2015,seargentWeirdCometsAsteroids2017}; Unknown impactor size: \cite{huiNongravitationalAccelerationActive2017}; Absence of Family Association: \cite{hsiehAsteroidFamilyAssociations2018}

	\item[\ref{Scheila}.] (596) Scheila, 1906 UA, 1949 WT, SPK-ID=2000596;  Activity Discovery: \cite{larson596Scheila2010};  Mechanism: \cite{jewittHubbleSpaceTelescope2011, bodewitsCollisionalExcavationAsteroid2011, yangNearinfraredObservationsCometlike2011a, moreno596ScheilaOutburst2011, ishiguroObservationalEvidenceImpact2011, ishiguroInterpretation596Scheila2011, hsiehOpticalDynamicalCharacterization2012, husarikRelativePhotometryPossible2012, neslusanDustProductivityImpact2016};  Activity Obs.: 1 (2010-2011) -- \cite{jewittHubbleSpaceTelescope2011, bodewitsCollisionalExcavationAsteroid2011, yangNearinfraredObservationsCometlike2011a, ishiguroObservationalEvidenceImpact2011, hsiehOpticalDynamicalCharacterization2012, husarikRelativePhotometryPossible2012,neslusanDustProductivityImpact2016}; Absence of Family Association: \cite{hsiehAsteroidFamilyAssociations2018}

	\item[\ref{Interamnia}.] (704) Interamnia, 1910 KU, 1952 MW, SPK-ID=2000704; Activity Discovery, Mechanism: \cite{busarevMaterialCompositionAssessment2016}; Activity Obs.:  1 (2012) -- \cite{busarevMaterialCompositionAssessment2016}$^\ast$; Absence of Family Association: \cite{rivkinCaseMissingCeres2014}; Shape Model: \cite{sato3DShapeModel2014}; Additional: \cite{busarevNewCandidatesActive2018}

	\item[\ref{Nina}.] (779) Nina, 1914 UB, A908 YB, A912 TE, SPK-ID=2000779;  Activity Discovery, Mechanism: \cite{busarevMaterialCompositionAssessment2016}; Activity Obs.:  1 (2012) -- \cite{busarevMaterialCompositionAssessment2016}$^\ast$, 2 (2016) -- \cite{busarevNewCandidatesActive2018}

	\item[\ref{Ingrid}.] (1026) Ingrid, 1923 NY, 1957 UC, 1963 GD, 1981 WL8, 1986 CG2, 1986 ES2, SPK-ID=2001026; Candidacy: \cite{cikotaPhotometricSearchActive2014}; Follow-up Observation (negative): \cite{betzlerPhotometricObservations10262015}; Flora family association: \cite{alfvenAsteroidalJetStreams1969}; Additional: \cite{nakano1026Ingrid1986,busarevNewCandidatesActive2018}

	\item[\ref{Beira}.] (1474) Beira, 1935 QY, 1950 DQ, SPK-ID=2001474;  Activity Discovery: \cite{busarevMaterialCompositionAssessment2016};  Mechanism: \cite{busarevMaterialCompositionAssessment2016};  Activity Obs.:  1 (2012) -- \cite{busarevMaterialCompositionAssessment2016}$^\ast$; Chaotic Cometary Orbit: \cite{hahnAsteroidsCometaryOrbits1985}; Additional: \cite{busarevNewCandidatesActive2018}

	\item[\ref{Oljato}.] (2201) Oljato, 1947 XC, 1979 VU2, 1979 XA, SPK-ID=2002201;  Activity Discovery: \cite{russellInterplanetaryMagneticField1984}; Activity Obs.: 1 (1984) -- \cite{russellInterplanetaryMagneticField1984}, Negative (1996) -- (\cite{chamberlin4015WilsonHarrington22011996};  Visit: \cite{perozziBasicTargetingStrategies2001}; Additional: \cite{kerrCouldAsteroidBe1985, mcfaddenEnigmaticObject22011993, connorsUnusualAsteroid22012016}

	\item[\ref{Phaethon}.] (3200) Phaethon, 1983 TB, SPK-ID=2003200; Activity Discovery: \cite{battams3200Phaethon2009};  Mechanism: ;  Activity Obs.: Negative -- \cite{chamberlin4015WilsonHarrington22011996,hsiehSearchActivity32002005}, 1 (2009) --  \cite{battams3200Phaethon2009,jewittActivityGeminidParent2010} 2 (2012) -- \cite{liRecurrentPerihelionActivity2013,jewittDustTailAsteroid2013}, 3 (2016) -- (\cite{huiResurrection3200Phaethon2017};  Visit: Destiny Plus \citep{iwataStudiesSolarSystem2016};  Pallas Family Association: \cite{todorovicDynamicalConnectionPhaethon2018}; Additional: \cite{jewittActivityGeminidParent2010, ryabovaPossibleEjectionMeteoroids2012, liRecurrentPerihelionActivity2013, jewittDustTailAsteroid2013, ansdellRefinedRotationalPeriod2014, jakubikMeteorComplexAsteroid2015, hanusNearEarthAsteroid32002016, sarliDESTINYTrajectoryDesign2017}

	\item[\ref{DonQuixote}.] (3552) Don Quixote, 1983 SA, SPK-ID=2003552; Activity Discovery, Mechanism: \cite{mommertDiscoveryCometaryActivity2014}; Activity Obs.: 1 (2009) -- \cite{mommertDiscoveryCometaryActivity2014}, (2018) -- \cite{mommertCBET4502201803292018}; Chaotic Cometary Orbit (as 1983 SA): \cite{hahnAsteroidsCometaryOrbits1985}

	\item[\ref{Aduatiques}.] (3646) Aduatiques, 1985 RK4, 1979 JL, 1981 WZ6, SPK-ID=2003646; Candidacy: \cite{cikotaPhotometricSearchActive2014}; Follow-up (inconclusive): \cite{sosaoyarzabalPhotometricSearchActivity2014}

	\item[\ref{WilHar}.] (4015) Wilson--Harrington, 1979 VA, 107P,  SPK-ID=2004015; Activity Discovery: \cite{cunninghamPeriodicCometWilsonHarrington1950};  Activity Obs.: 1 (1949) -- \cite{cunninghamPeriodicCometWilsonHarrington1950}, 2 (1979) -- \cite{degewij1979VAPhysical1980}, Negative (1992) -- \cite{bowell40151979VA1992a}, Negative (1996) \cite{chamberlin4015WilsonHarrington22011996}, Negative (2008) -- \cite{licandroSpitzerObservationsAsteroidcomet2009}, Negative (2009-2010) -- \cite{ishiguroObservationalEvidenceImpact2011,urakawaPhotometricObservations107P2011}, 3-6 (1992, 1996, 2008, 2009-2010) \cite{ferrin2009ApparitionMethuselah2012};  Visits: Failed \citep{raymanDeepSpaceExtended2001}, Concept \citep{sollittMissionConcepts40152009};  Chaotic Cometary Orbit (as 1979 VA): \cite{hahnAsteroidsCometaryOrbits1985}; Additional: \cite{harrisCometNotesComet1950, vanbiesbroeckCometNotesComet1951, helin1979VaPossibleCarbonaceous1980, helinDiscoveryUnusualApollo1981, osipRotationState4015WilsonHarrington1995, fernandezAnalysisPOSSImages1997}

	\item[\ref{24684}.] (24684), 1990 EU4, 1981 UG28, SPK-ID=2024684; Candidacy: \cite{cikotaPhotometricSearchActive2014}

	\item[\ref{35101}.] (35101) 1991 PL16, 1998 FZ37, SPK-ID=2035101; Candidacy: \cite{cikotaPhotometricSearchActive2014}; Eunomia Family Association: \cite{cikotaPhotometricSearchActive2014}

	\item[\ref{62412}.] (62412), 2000 SY178, SPK-ID=2062412;  Activity Discovery: \cite{sheppardDiscoveryCharacteristicsRapidly2015}; Activity Obs.: 1 (2014) \cite{sheppardDiscoveryCharacteristicsRapidly2015}; Hygiea Family Association: \cite{sheppardDiscoveryCharacteristicsRapidly2015,hsiehAsteroidFamilyAssociations2018}

	\item[\ref{Ryugu}.] (162173) Ryugu, SPK-ID=2162173;  Activity Discovery, Mechanism, Activity Obs.: 1 (2007) -- \cite{busarevNewCandidatesActive2018}$^\ast$;  Visit: Hayabusa 2 \citep{tsudaSystemDesignHayabusa2013a}; Clarissa Family Association \cite{campinsOriginAsteroid1621732013,lecorreGroundbasedCharacterizationHayabusa22018}; Thermal Inertia: \cite{liang-liangInvestigationThermalInertia2014}; Additional: \cite{suzukiInitialInflightCalibration2018, pernaSpectralRotationalProperties2017}

	\item[\ref{GO98}.] (457175), 2008 GO98, 362P, SPK-ID=2457175;  Activity Discovery: \cite{kimNewObservationalEvidence2017};  Activity Obs.: 1 (2017) \cite{masi4571752008GO2017};  Hilda Family Association: \cite{warnerLightcurveAnalysisHilda2018}; Additional: \cite{sato4571752008GO2017, yoshimoto4571752008GO2017, birtwhistle4571752008GO2017, bacci4571752008GO2017, bell4571752008GO2017, bryssinck4571752008GO2017}

	\item[\ref{ElstPizarro}.] 133P/Elst--Pizarro, (7968), 1979 OW7, 1996 N2, SPK-ID=2007968;  Activity Discovery: \cite{elstComet1996N21996};  Mechanism: \cite{hsiehStrangeCase133P2004, jewittHubbleSpaceTelescope2014};  Activity Obs.: 1 (1996) \cite{elstComet1996N21996}, 2 (2002) \cite{hsiehStrangeCase133P2004}, Negative (2005) \cite{tothSearchCometlikeActivity2006}, 2 (2007) \cite{hsiehReturnActivityMainbelt2010, bagnuloPolarimetryPhotometryPeculiar2010, rousselotNearinfraredSpectroscopy133P2011}, 3 (2013) \cite{jewittHubbleSpaceTelescope2014};  Visit: Castalia \citep{snodgrassCastaliaMissionMain2017};  Themis Family Association: \cite{boehnhardtImpactInducedActivityAsteroidComet1998}; Additional: \cite{tothImpactgeneratedActivityPeriod2000, ferrinSecularLightCurves2006, prialnikCanIceSurvive2009}

	\item[\ref{176P}.] 176P/LINEAR, (118401), P/1999 RE$_{70}$, 2001 AR7, SPK-ID=2118401;  Activity Discovery: \cite{hsiehComet1999RE2006,hsiehHawaiiTrailsProject2009};  Mechanism: \cite{hsiehSearchReturnActivity2014}; Activity Obs.: 1 (2005) \cite{hsiehComet1999RE2006}, Negative (2006-2009) \cite{hsiehPhysicalPropertiesMainbelt2011}, Negative (2011) \cite{hsiehSearchReturnActivity2014}; Themis Family Association: \cite{hsiehHawaiiTrailsProject2009,hsiehAsteroidFamilyAssociations2018} Additional: \cite{hsiehAlbedosMainBeltComets2009,licandroTestingCometNature2011,deval-borroUpperLimitWater2012}

	\item[\ref{233P}.] 233P (La Sagra), P/2009 W$_{50}$, 2005 JR71, SPK-ID=1003062; Activity Discovery: \cite{mainzerComet2009WJ2010}, Activity Obs.: 1 (2009) \cite{mainzerComet2009WJ2010}; Absence of Family Association: \cite{hsiehAsteroidFamilyAssociations2018}

	\item[\ref{238P}.] 238P/Read, P/2005 U1, 2010 N2, SPK-ID=1001676;  Activity Discovery: \cite{readComet2005U12005};  Activity Obs.: 1 (2005) \cite{readComet2005U12005}, 2 (2010) \cite{hsiehMainbeltComet238P2011}, 3 (2016) \cite{hsiehReactivationsMainBeltComets2017}; Gorchakov Family Association: \cite{hsiehAsteroidFamilyAssociations2018}; Former Themis Family Association: \cite{haghighipourDynamicalConstraintsOrigin2009}; Additional: \cite{hsiehPhysicalPropertiesMainBelt2009, pittichovaComet2005U12010}

	\item[\ref{259P}.] 259P/Garradd, 2008 R1, SPK-ID=1002991; Activity Discovery: \cite{garraddComet2008R12008}; Mechanism: \cite{jewittMainBeltComet20082009}; Activity Obs.: 1 (2008) \cite{garraddComet2008R12008}, 2 (2017) \cite{hsiehComet259PGarradd2017,hsiehReactivationsMainBeltComets2017}; Absence of Family Association: \cite{hsiehAsteroidFamilyAssociations2018}; Additional: \cite{kossackiMainBeltComet2012, maclennanNucleusMainbeltComet2012, kleynaFaintMovingObject2012}

	\item[\ref{288P}.] 288P, (300163), 2006 VW139, SPK-ID=2300163;  Activity Discovery: \cite{hsiehComet2006VW2011}; Activity Obs.: 1 (2011) \cite{hsiehComet2006VW2011}, 2 (2016-2017) \cite{agarwalBinaryMainbeltComet2017,hsiehReactivationsMainBeltComets2017}; Themis Family Association: \cite{hsiehDiscoveryMainbeltComet2012,hsiehAsteroidFamilyAssociations2018}; Additional: \cite{hsiehDiscoveryMainbeltComet2012, novakovic2006VW139Mainbelt2012, agarwalHubbleKeckTelescope2016}

	\item[\ref{311P}.] 311P/Pan-STARRS, P/2013 P5, SPK-ID=1003273; Activity Discovery: \cite{micheliComet2013P52013};  Mechanism: \cite{jewittExtraordinaryMultitailedMainbelt2013, morenoIntermittentDustMass2014, hainautContinuedActivity20132014, jewittEpisodicEjectionActive2015};  Activity Obs.: 1 (2013-2014) \cite{micheliComet2013P52013,jewittEpisodicEjectionActive2015}; Behrens Family Association: \cite{hsiehAsteroidFamilyAssociations2018}

	\item[\ref{313P}.] 313P/Gibbs, P/2014 S4, 2003 S10, SPK-ID=1003344;  Activity Discovery: \cite{gibbsComet2014S42014};  Mechanism, Activity Obs.: 1 (2003) \cite{nakanoComet2003S102014,skiffComet2014S42014,huiArchivalObservationsActive2015}, 2 (2015) \cite{jewittNucleusMassLoss2015}; Lixiaohua Family Association: \cite{hsiehMainbeltComet20122013,hsiehSublimationDrivenActivityMainBelt2015,hsiehAsteroidFamilyAssociations2018}; Additional: \cite{jewittNewActiveAsteroid2015, hsiehSublimationDrivenActivityMainBelt2015, pozuelosDustEnvironmentMainBelt2015}

	\item[\ref{324P}.] 324P/La Sagra, P/2010 R2, 2015 K3, SPK-ID=1003104; Activity Discovery: \cite{nomenComet2010R22010};  Activity Obs.: 1 (2010-2011) \cite{nomenComet2010R22010,hsiehObservationalDynamicalCharacterization2012}, Negative (2013) \cite{hsiehNucleusMainbeltComet2014}, 2 (2015) \cite{hsiehReactivationMainbeltComet2015,jewittHubbleSpaceTelescope2016}; Alauda Family Association: \cite{hsiehAsteroidFamilyAssociations2018};  Additional: \cite{morenoDustEnvironmentMainBelt2011,hsiehObservationalDynamicalCharacterization2012,hsiehNucleusMainbeltComet2014,hsiehReactivationMainbeltComet2015}

	\item[\ref{331P}.] 331P/Gibbs, P/2012 F5, SPK-ID=1003182;  Activity Discovery: \cite{gibbsComet2012F52012};  Mechanism: \cite{stevensonCharacterizationActiveMain2012, drahusFastRotationTrailing2015};  Activity Obs.: 1 (2012) \cite{gibbsComet2012F52012}, 2 (2015) \cite{drahusFastRotationTrailing2015}; Gibbs family association: \cite{novakovicDiscoveryYoungAsteroid2014}; Additional: \citep{stevensonCharacterizationActiveMain2012,morenoShortdurationEventCause2012}

	\item[\ref{348P}.] 348P, P/2017 A2, P/2011 A5 (PANSTARRS), SPK-ID=1003492; Activity Discovery: \cite{wainscoatComet2017A22017}; Activity Obs.: 1 (2017) \cite{wainscoatComet2017A22017}; Absence of Family Association: \cite{hsiehAsteroidFamilyAssociations2018}

	\item[\ref{354P}.] 354P/LINEAR, P/2010 A2, 2017 B5, SPK-ID=1003055;  Activity Discovery: \cite{birtwhistleComet2010A22010};  Activity Obs.: 1 (2010) \cite{birtwhistleComet2010A22010,jewittComet2010A22010}; Baptistina Family Association: \cite{hsiehAsteroidFamilyAssociations2018}; Additional: \cite{morenoWatericedrivenActivityMainBelt2010,jewittRecentDisruptionMainbelt2010,snodgrassCollision2009Origin2010,jewittPreDiscoveryObservationsDisrupting2011,hainaut2010A2LINEAR2012,kimMultibandOpticalObservation2012,agarwalDynamicsLargeFragments2012,kleyna2010A2LINEAR2013,jewittLargeParticlesActive2013,agarwalDynamicsLargeFragments2013,kimNewObservationalEvidence2017,kimAnisotropicEjectionActive2017}

	\item[\ref{358P}.] 358P/PanSTARRS, P/2012 T1, 2017 O3, SPK-ID=1003208;  Activity Discovery: \cite{wainscoatComet2012T12012}; Activity Obs.: 1 (2012) \cite{wainscoatComet2012T12012}, 2 (2017) \cite{kimNewObservationalEvidence2017};  Mechanism: \cite{hsiehMainbeltComet20122013}; Lixiaohua Family Association: \citep{hsiehMainbeltComet20122013,hsiehAsteroidFamilyAssociations2018}; Additional: \cite{morenoDustEnvironmentMainBelt2013, orourkeDeterminationUpperLimit2013, snodgrassXshooterSearchOutgassing2017}

	\item[\ref{P2013R3}.] P/2013 R3 (Catalina-Pan-STARRS), SPK-ID=1003275 (P/2013 R3-A SPK-ID=1003333, P/2013 R3-B SPK-ID=1003334); Activity Discovery: \cite{bolinComet2013R32013,hillComet2013R32013}; Activity Obs.: 1 (2013-2015) \cite{bolinComet2013R32013,hillComet2013R32013,jewittAnatomyAsteroidBreakup2017}; Mandragora Family Association: \cite{hsiehAsteroidFamilyAssociations2018}; Additional: \cite{jewittDisintegratingAsteroid20132014, hirabayashiConstraintsPhysicalProperties2014}

	\item[\ref{P2015X6}.] P/2015 X6 (Pan-STARRS), SPK-ID=1003426;  Activity Discovery: \cite{lillyComet2015J22015};  Activity Obs.: 1 (2015) \cite{lillyComet2015J22015,tubbioloComet2015X62015,morenoDustLossActivated2016}; Aeolia Family Association: \cite{hsiehAsteroidFamilyAssociations2018}

	\item[\ref{P2016G1}.] P/2016 G1 (Pan-STARRS), SPK-ID=1003460;  Activity Discovery: \cite{werykCOMET2016G12016};  Mechanism: \cite{morenoEarlyEvolutionDisrupted2016}; Activity Obs.: 1 (2016) \cite{werykCOMET2016G12016,morenoDisruptedAsteroid20162017}; Adeona Family Association: \cite{hsiehAsteroidFamilyAssociations2018}

	\item[\ref{P2016J1}.] P/2016 J1 (Pan-STARRS), P/2016 J1-A (SPK-ID=1003464), P/2016 J1-B (SPK-ID=1003465);  Activity Discovery: \cite{wainscoat2016J1Panstarrs2016}; Activity Obs.: 1 (2016) \cite{wainscoat2016J1Panstarrs2016,huiSplitActiveAsteroid2017}; Theobalda Family Association: \cite{hsiehAsteroidFamilyAssociations2018}

\end{itemize}
$^\ast$: Under review.

\clearpage
\singlespacing
\chapter{Manuscript II: Six Years of Sustained Activity from Active Asteroid (6478)~Gault}
\chaptermark{Six Years Sustained Activity from Active Asteroid (6478)~Gault}
\label{chap:Gault}
\acresetall

Colin Orion Chandler\footnote{\label{GaultNAU}Department of Physics \& Astronomy, Northern Arizona University, PO Box 6010, Flagstaff, AZ 86011, USA}, Jay Kueny$^\mathrm{\ref{GaultNAU}}$, Annika Gustafsson$^\mathrm{\ref{GaultNAU}}$, Chadwick A. Trujillo$^\mathrm{\ref{GaultNAU}}$, Tyler D. Robinson$^\mathrm{\ref{GaultNAU}}$, David E. Trilling$^\mathrm{\ref{GaultNAU}}$

\textit{This is the Accepted Manuscript version of an article accepted for publication in Astrophysical Journal Letters.  IOP Publishing Ltd is not responsible for any errors or omissions in this version of the manuscript or any version derived from it.  The Version of Record is available online at}  \url{https://iopscience.iop.org/article/10.3847/2041-8213/ab1aaa}\textit{.}








\doublespacing


\section{Abstract}
\label{Gault:Abstract}
We present archival observations demonstrating that main belt asteroid (6478)~Gault has an extensive history of comet-like activity. Outbursts have taken place during multiple epochs since 2013 and at distances extending as far as ~2.68 au, nearly aphelion. (6478)~Gault is a member of the predominately S-type (i.e., volatile-poor) Phocaea family; no other main belt object of this type has ever shown more than a single activity outburst. Furthermore, our data suggest this is the longest duration of activity caused by a body spinning near the rotational breakup barrier. If activity is indeed unrelated to volatiles, as appears to be the case, (6478) Gault represents a new class of object, perpetually active due to rotational spin-up.

\textit{Keywords:} minor planets, asteroids: individual ((6478) Gault) --- comets: individual ((6478) Gault) 

\section{Introduction}
\label{Gault:introduction}

\begin{figure}[H]
	\centering
	\includegraphics[width=0.5\linewidth]{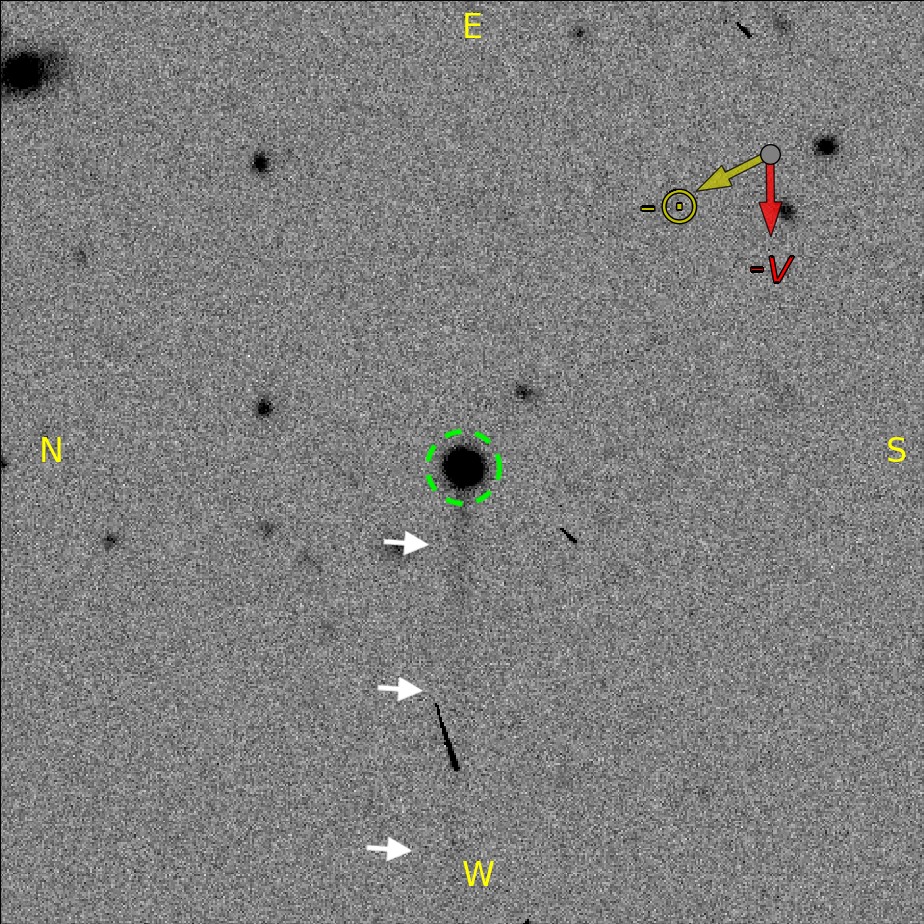}
    \caption{(6478)~Gault (dashed green circle) displays a prominent tail (indicated by white arrows) during this 2013 September 28  apparition when (6478)~Gault was halfway between perihelion and aphelion. This 90 s $g$-band exposure reached $\sim$7 mag fainter than (6478)~Gault. The anti-Solar direction ($-\odot$; yellow) and negative heliocentric velocity vector ($-\vec{v}$; red) are shown.}
    \label{Gault:fig:gault2013a}
\end{figure}

Active asteroids like (6478)~Gault (Figure~\ref{Gault:fig:gault2013a}, this work) are dynamically asteroidal objects but they uncharacteristically manifest cometary features such as tails or comae \citep{hsiehActiveAsteroidsMystery2006}. With only $\sim$20 known to date (see Table 1 of \citealt{chandlerSAFARISearchingAsteroids2018}), active asteroids remain poorly understood, yet they promise insight into solar system volatile disposition and, concomitantly, the origin of water on Earth \citep{hsiehPopulationCometsMain2006}.

Active asteroids are often defined as objects with (1) comae, (2) semimajor axes interior to Jupiter, and (3) Tisserand parameters with respect to Jupiter $T_\mathrm{J}>3$; $T_\mathrm{J}$ describes an object's orbital relationship to Jupiter by 

\begin{equation}
    T_\mathrm{J} = \frac{a_\mathrm{J}}{a} + 2 \sqrt{\frac{a\left(1-e^2\right)}{a_\mathrm{J}}}\cos\left(i\right)
\end{equation}

\noindent where $a_\mathrm{J}$ is the semimajor axis of Jupiter, $e$ is the eccentricity, and $i$ is the inclination; see \cite{jewittActiveAsteroids2012} for a thorough treatment. Main belt comets are a subset of active asteroids dynamically constrained to the main asteroid belt and thought to have volatile-driven activity (see e.g., \citealt{snodgrassMainBeltComets2017} for an in-depth discussion). It is worth pointing out that some objects have had multiple classifications, for instance (3552) Don Quixote has an asteroid designation due to its low activity and has been called a near-Earth asteroid \citep{mommertDiscoveryCometaryActivity2014} but has a $T_\mathrm{J}$ of 2.3 which indicates it is more properly a Jupiter family comet.

Discovering these objects has proven observationally challenging. The first active asteroid, (4015)~Wilson-Harrington, was discovered in 1949 \citep{harrisCometNotesComet1950}. In the mid-1980s a connection between bow-shock magnetic field disturbances detected by the \textit{Pioneer} spacecraft suggested (2201)~Oljato was leaving behind a distant comet-like gas trail \citep{kerrCouldAsteroidBe1985}, even if not detected at the object itself \citep{russellInterplanetaryMagneticField1984}. Despite many efforts (see e.g., \citealt{chamberlin4015WilsonHarrington22011996}) it was not until the 1996 discovery of activity in (7968)~Elst-Pizarro that another active asteroid was visually identified \citep{elstComet1996N21996}. Though initially impact appeared a possible cause (e.g., \citealt{tothImpactgeneratedActivityPeriod2000}), when activity recurred \citep{hsiehReturnActivityMainbelt2010} it was more indicative of volatile sublimation.

A significant complication hindering our understanding of active asteroids arises when assessing underlying activity mechanisms: causes are neither few nor mutually exclusive (see \citealt{jewittActiveAsteroids2012} for a comprehensive overview). Responsible primary processes include volatile sublimation (e.g., 133P/Elst-Pizarro,  \citealt{hsiehReturnActivityMainbelt2010}), impact events (e.g., (596) Scheila,  \citealt{bodewitsCollisionalExcavationAsteroid2011,morenoDustEnvironmentMainBelt2011}), rotational breakup (e.g., 311P/PanSTARRS, \citealt{jewittExtraordinaryMultitailedMainbelt2013,morenoIntermittentDustMass2014}), thermal fracture (e.g., (3200) Phaethon, discussed below), and cryovolcanism (e.g., (1) Ceres, \citealt{kuppersLocalizedSourcesWater2014,witzeBrightSpotsCeres2015}). Physical interaction, or ``rubbing binary,'' has been proposed as a primary mechanism in the case of 311P (\citealt{hainautContinuedActivity20132014}, cf. \citealt{jewittNucleusActiveAsteroid2018}). Secondary mechanisms such as electrostatic gardening, physical properties like chemical makeup, and geometric effects (e.g., the opposition effect) may influence our ability to reliably detect and quantify outbursts.

One crucial diagnostic indicator of the underlying activity mechanism is whether or not activity recurs. If activity is observed on only one occasion (i.e., a single apparition), then the object may have experienced a recent impact event. Expelled material and/or exposed volatiles sublimating may both cause comae or tails to appear. Activity would then cease once ejecta dissipated or the volatile supply is exhausted, reburied, or refrozen.

Recurrent activity is typically associated with volatile sublimation. For example, Geminid Meteor Shower parent body (3200) Phaethon is thought to undergo thermal fracture during the rapid temperature changes accompanying its perihelion passages \citep{liRecurrentPerihelionActivity2013} where it experiences temperatures $>800$ K \citep{ohtsukaSolarRadiationHeatingEffects2009}. Fracture events may directly expel material in addition to exposing volatiles for sublimation.

Thermally induced activity is thought to increase with decreasing heliocentric distance; that is, the closer a body is to the Sun, the more likely an outburst is to occur. 
Active asteroids are more likely to exhibit activity during perihelion passage (see Table 1 of \citealt{chandlerSAFARISearchingAsteroids2018}). Notable exceptions where activity was discovered at distances far from perihelion include 311P/PanSTARRS \citep{jewittExtraordinaryMultitailedMainbelt2013} and (493) Griseldis \citep{tholenEvidenceImpactEvent2015}. Activity in ``traditional'' comets has been reported at distances that are substantially farther than the main asteroid belt, for instance Comet C/2010 U3 (Boattini) at 27~au \citep{hui2010U3Boattini2019}. Of the $\sim$20 active asteroids known to date, 16 are carbonaceous (i.e., C-type) but only four are believed to be composed of silicate-rich non-primitive material (i.e., \citealt{demeoExtensionBusAsteroid2009} S-type taxonomy): (2201) Oljato (Apollo-orbit), 233P/La Sagra (Encke-orbit), 311P/PanSTARRS (inner main belt), and 354P/LINEAR (outer main belt). 

(6478)~Gault activity was first reported in 2019 January \citep{smith6478Gault2019}. Ensuing follow-up observations \citep{maury6478Gault2019} confirmed activity with subsequent reports \citep{leeEarlyDetection64782019,yeContinuedActivity64782019} providing evidence of ongoing activity. \cite{yeMultipleOutburstsAsteroid2019} reported that multiple outbursts actually began in 2018 December. Analysis of dust emanating from (6478)~Gault via Monte Carlo tail brightness simulations indicate the current apparition, comprised of two outbursts, could have begun as early as 2018 November 5 \citep{morenoDustPropertiesDoubletailed2019}. Simultaneous to our own work, \cite{jewittEpisodicallyActiveAsteroid2019} reported three tails with independent onsets, the earliest being 2018 October 28.

We set out to determine if any data in our local repository of National Optical Astronomy Observatory (NOAO) Dark Energry Camera (DECam) images showed signs of activity. The $\sim$500 megapixel DECam instrument on the Blanco 4 m telescope situated on Cerro Tololo, Chile, probes faintly ($\sim$24 mag) and, as we demonstrated in \cite{chandlerSAFARISearchingAsteroids2018}, it is well-suited to detect active asteroids. We produced novel tools taking into account (1) orbital properties of (6478)~Gault (summarized in Appendix \ref{Gault:subsec:GaultData}) and (2) observational properties (e.g., apparent magnitude, filter selection, exposure time) to find ideally suited candidate images.


\section{Methods}
\label{Gault:sec:methods}

We searched our own in-house database of archival astronomical data (e.g., observation date, coordinates) in order to locate images that are likely to show (6478)~Gault. Our database includes the entire NOAO DECam public archive data tables along with corresponding data from myriad sources (e.g., NASA JPL Horizons \citealt{giorginiJPLOnLineSolar1996}; see also the Acknowledgements).

\section{Locating Candidate Images}
\label{Gault:subsec:candidates}

We began our search for (6478)~Gault by making use of a fast grid query in R.A. and decl. space. We then passed these results through a more accurate circular filter prescribed for the DECam image sensor arrangement. Lastly, we computed image sensor chip boundaries precisely to ensure that the object fell on a sensor rather than, for example, gaps between camera chips. This progressively more precise query approach cut down image search time by orders of magnitude.

\section{Observability Assessment}
\label{Gault:subsec:observability}

We created a reverse exposure time calculator to estimate how faintly (i.e., the magnitude limit) candidate images probed. After applying color coefficient corrections (see \citealt{willmerAbsoluteMagnitudeSun2018} for procedure details) we transformed the color-corrected magnitudes to the absolute bolometric system used by the DECam exposure time calculator\footnote{\url{http://www.ctio.noao.edu/noao/node/5826}}. These steps enabled us to compute differences between apparent magnitude and the specific magnitude limit of the DECam exposure so that we could produce a list of images where (6478)~Gault could be detected.

\section{Thumbnail Extraction}
\label{Gault:subsec:acquisition}

We downloaded the image files containing (6478)~Gault from the NOAO archive and, following the procedures of \citet{chandlerSAFARISearchingAsteroids2018}, we extracted flexible image transport system (FITS) thumbnails of (6478)~Gault. We then performed image processing to enhance contrast before finally producing portable network graphics (PNG) image files for inspection.

\section{Image Analysis}
\label{Gault:subsec:imageanalysis}

We visually inspected our (6478)~Gault thumbnail images to check for signs of activity. PNG thumbnails with activity indicators were examined in greater detail via the corresponding FITS thumbnail image.

To assess the influence of heliocentric distance on activity level we employed a simple metric (see \cite{chandlerSAFARISearchingAsteroids2018} ``$\mathrm{\%}_\mathrm{peri}$'' for motivation) describing how far from perihelion $q$ the target $T$ was located (at distance $d$) relative to its aphelion distance $Q$ by

\begin{equation}
		\%_{T\rightarrow q} = \left(\frac{Q - d}{Q-q}\right)\cdot 100\mathrm{\%}.
		\label{Gault:eq:percentperi}
\end{equation}

\vspace{2mm}


\section{Results}
\label{Gault:sec:results}

\begin{table*}
\caption{(6478)~Gault Archival DECam Observations Examined}
\footnotesize
\centering
\begin{tabular}{ccrlrcccccrr}
Activity & UT Date & Time & Processing  & $t_\mathrm{exp}$ & Filter & $m_\mathrm{lim}$ & $m_V$ & $\Delta m$ & $r$ & $\%_{T\rightarrow q}$ & $\angle_\mathrm{STO}$  \\
&                        &                       &           &           (s)            &           &           &                   &   &   (au) &            &   ($^\circ$)\\
\hline\hline
 & 2013 Sep 22               & 3:03                     & R, I, Re    & 45                               & Y          & 21.0                                 & 17.2                     & -4                           & 2.27                          & 54                        & 3.41                            \\
$*$ & 2013 Sep 28               & 2:23                     & R, I, I, Re & 90                               & g          & 24.2                               & 17.0                       & -7                           & 2.28                          & 52                        & 0.45                            \\
$*$ & 2013 Oct 13               & 3:06                     & I, I, I, Re & 90                               & i          & 23.3                               & 17.1                     & -6                           & 2.32                          & 49                        & 8.21                            \\
& 2013 Oct 13               & 3:08                     & I, I, Re    & 90                               & z          & 22.7                               & 17.1                     & -6                           & 2.32                          & 49                        & 8.21                            \\
$*$ & 2016 Jun 09               & 4:45                     & R, I, Re    & 96                               & r          & 23.9                               & 16.8                     & -7                           & 1.86                          & 100                           & 22.38                           \\
$*$ & 2016 Jun 10               & 4:40                     & R, I, Re    & 107                              & g          & 24.2                               & 16.8                     & -7                           & 1.86                          & 100                           & 22.48                           \\
 & 2017 Oct 23               & 8:57                     & R, I, Re    & 80                               & z          & 22.7                               & 18.8                     & -4                           & 2.66                          & 10                         & 14.79                           \\
 & 2017 Nov 11               & 7:13                     & R, I, Re    & 80                               & z          & 22.7                               & 18.5                     & -5                           & 2.68                          & 8                        & 10.92                           \\
$*$ & 2017 Nov 12               & 5:14                     & R, I, Re    & 111                              & g          & 24.3                               & 18.5                     & -6                           & 2.68                          & 8                        & 10.81                          \\
\hline
\end{tabular}

\raggedright
\footnotesize
\vspace{1mm}
\textbf{Note.} Process types: Raw (R), InstCal (I), Resampled (Re); $r$: Sun-target distance; $\%_{T\rightarrow q}$: target distance toward perihelion from aphelion (Equation \ref{Gault:eq:percentperi}); $t_\mathrm{exp}$: exposure time; $m_\mathrm{lim}$: estimated exposure magnitude limit; $m_V$: (6478)~Gault apparent $V$-band magnitude; $\Delta m$: $m - m_\mathrm{lim}$; $\angle_\mathrm{STO}$: Sun-Target-Observer (phase) angle. 
 Thumbnails are included in Appendix \ref{Gault:subsec:ThumbnailGallery}.
\label{Gault:tab:observations}
\end{table*}

\begin{figure*}
    \centering
    \begin{tabular}{cc}
    \includegraphics[width=0.45\linewidth]{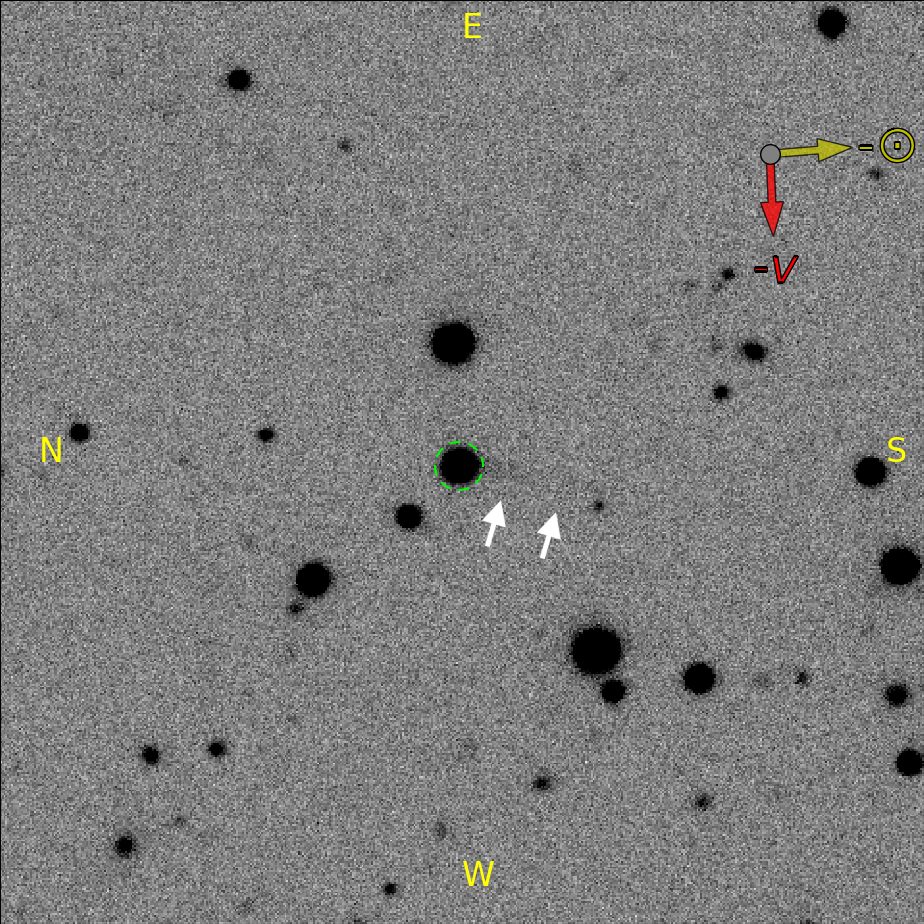}  &  \includegraphics[width=0.45\linewidth]{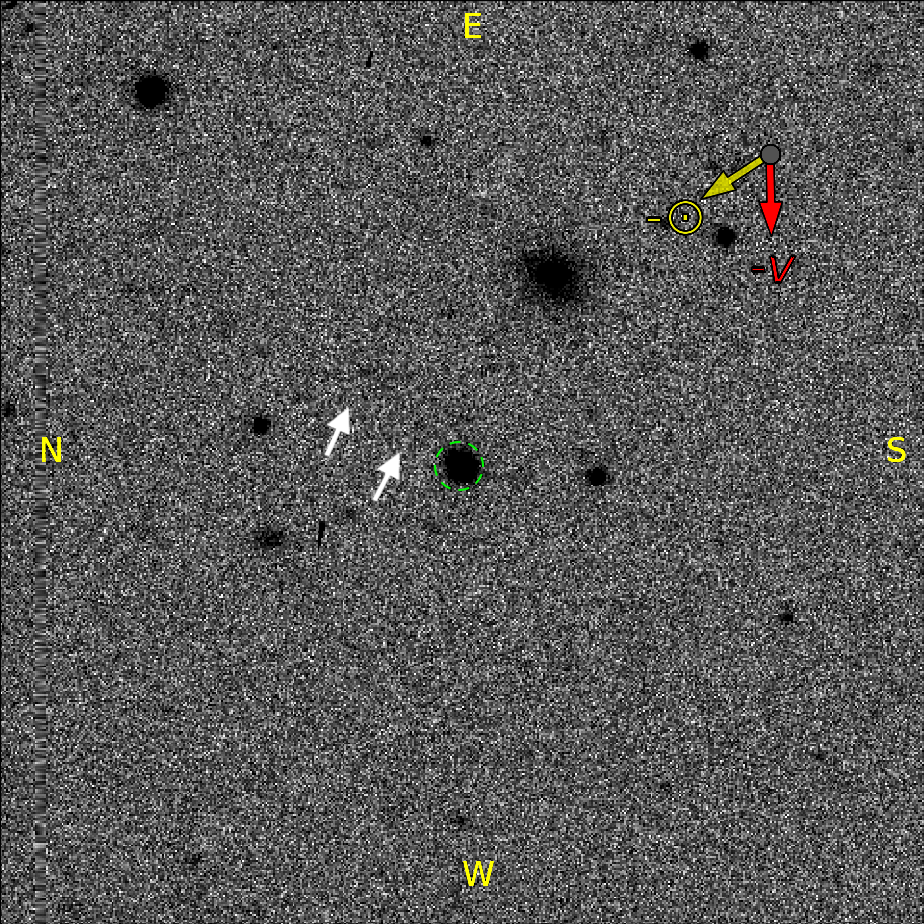}\\
    \textbf{(a)}     & \textbf{(b)}\\
    \end{tabular}
    \caption{\textbf{(a)} A tail (indicated by white arrows) at $\sim$8 o'clock is seen in this 107-second $g$-band exposure imaged June 10, 2016. The yellow arrow indicates anti-Solar ($-\odot$) direction and red the negative heliocentric velocity vector ($-\vec{v}$). \textbf{(b)} (6478)~Gault top seen on November 12, 2017. The 111 second exposure in the $g$-band delivered a flux limit 6 magnitudes fainter than (6478)~Gault, revealing a faint tail ($\sim$2 o'clock, indicated by white arrows) and coma. The yellow arrow indicates anti-Solar ($-\odot$) direction and the red arrow negative heliocentric velocity ($-\vec{v}$). Dashed green circles outline (6478)~Gault and white arrows have been placed perpendicular to any observed activity.}
    \label{Gault:fig:thumb2016}
\end{figure*}

We successfully extracted thumbnails from 9 archival observations of (6478)~Gault; see Table \ref{Gault:tab:observations} for details. Most data were available in raw and calibrated form (``InstCal'' and ``Resampled'' are described by \citealt{darkenergysurveycollaborationDarkEnergySurvey2016}) allowing us to extract $\sim$30 thumbnail images in total. Figure~\ref{Gault:fig:gault2013a} shows (6478)~Gault in 2013 with a pronounced tail in the 6 o'clock direction. Figure~\ref{Gault:fig:thumb2016} \textbf{(a)} and Figure~\ref{Gault:fig:thumb2016} \textbf{(b)} show (6478)~Gault in 2016 and 2017, respectively. Additional images may be found in Appendix \ref{Gault:subsec:ThumbnailGallery}.

\begin{figure*}[ht]
	\centering
	\includegraphics[width=1.0\linewidth]{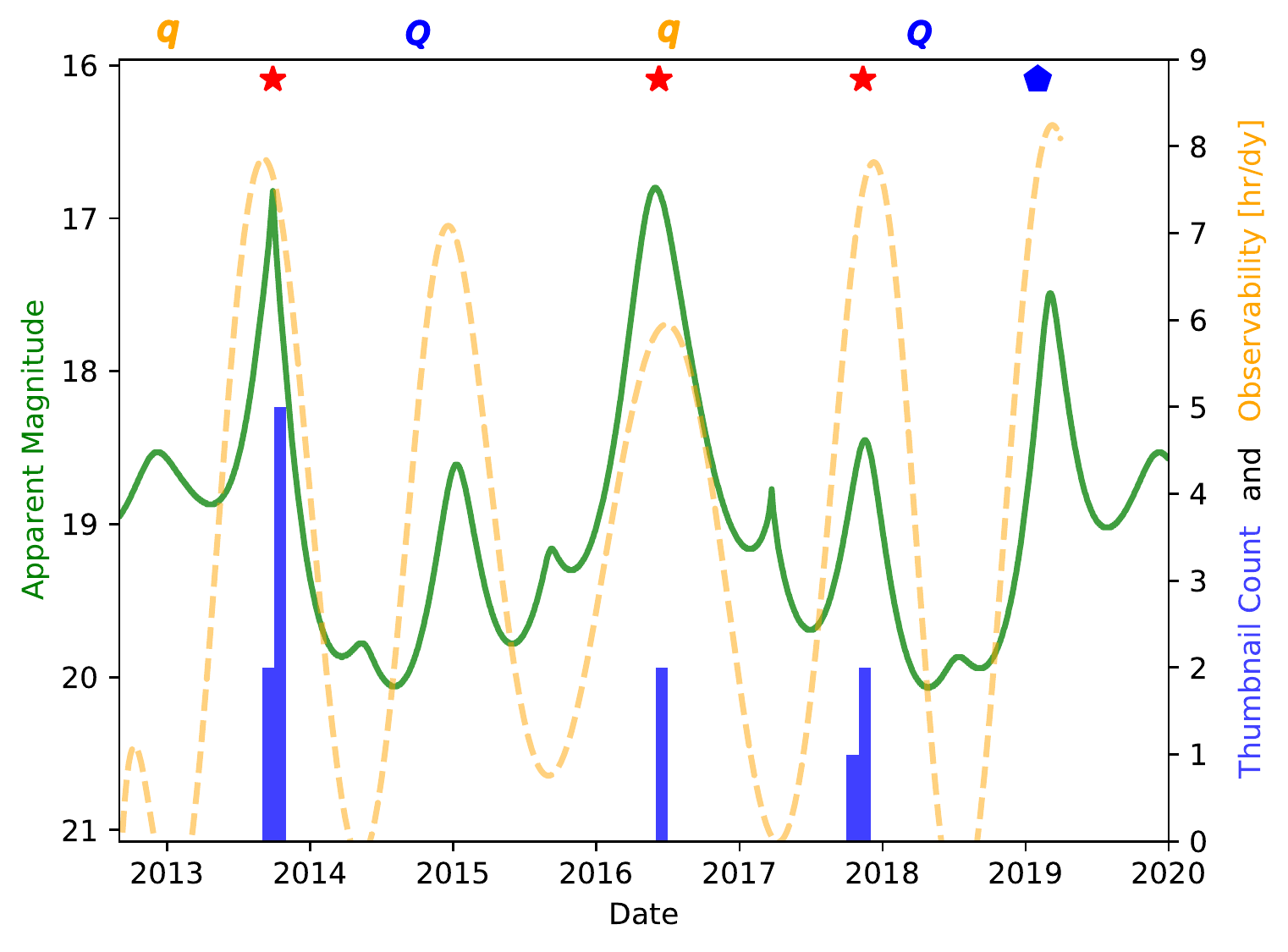}
	\caption{(6478)~Gault activity timeline beginning with DECam operation commencement (2012 September) to present. Red stars show when we found visible activity; the blue pentagon represents the current apparition where prominent activity has been seen. Above the top axis are marked perihelion ($q$) and aphelion ($Q$) events. The solid green line indicates the apparent $V$-band magnitude of (6478)~Gault as viewed from Earth. The dashed yellow line shows our ``observability'' metric, defined as the number of hours per UT observing date meeting both of the following conditions possible for DECam: (1) elevation $> 15^\circ$, and (2) the refracted solar upper-limb elevation was $< 0^\circ$ (i.e., nighttime). Peaks in apparent magnitude coinciding with peaks in observability indicate opposition events; conversely, secondary magnitude peaks aligned with observability troughs highlight solar conjunctions, i.e., when (6478)~Gault was ``behind'' the Sun as viewed from Earth. All activity has been observed near opposition events. Also, activity was seen at every epoch in our data. The histogram (vertical blue bars) indicate the number of thumbnails that we extracted for a given observing month.}
	\label{Gault:fig:ActivityTimeline}
\end{figure*}

\begin{figure}
    \centering
    \includegraphics[width=1.0\linewidth]{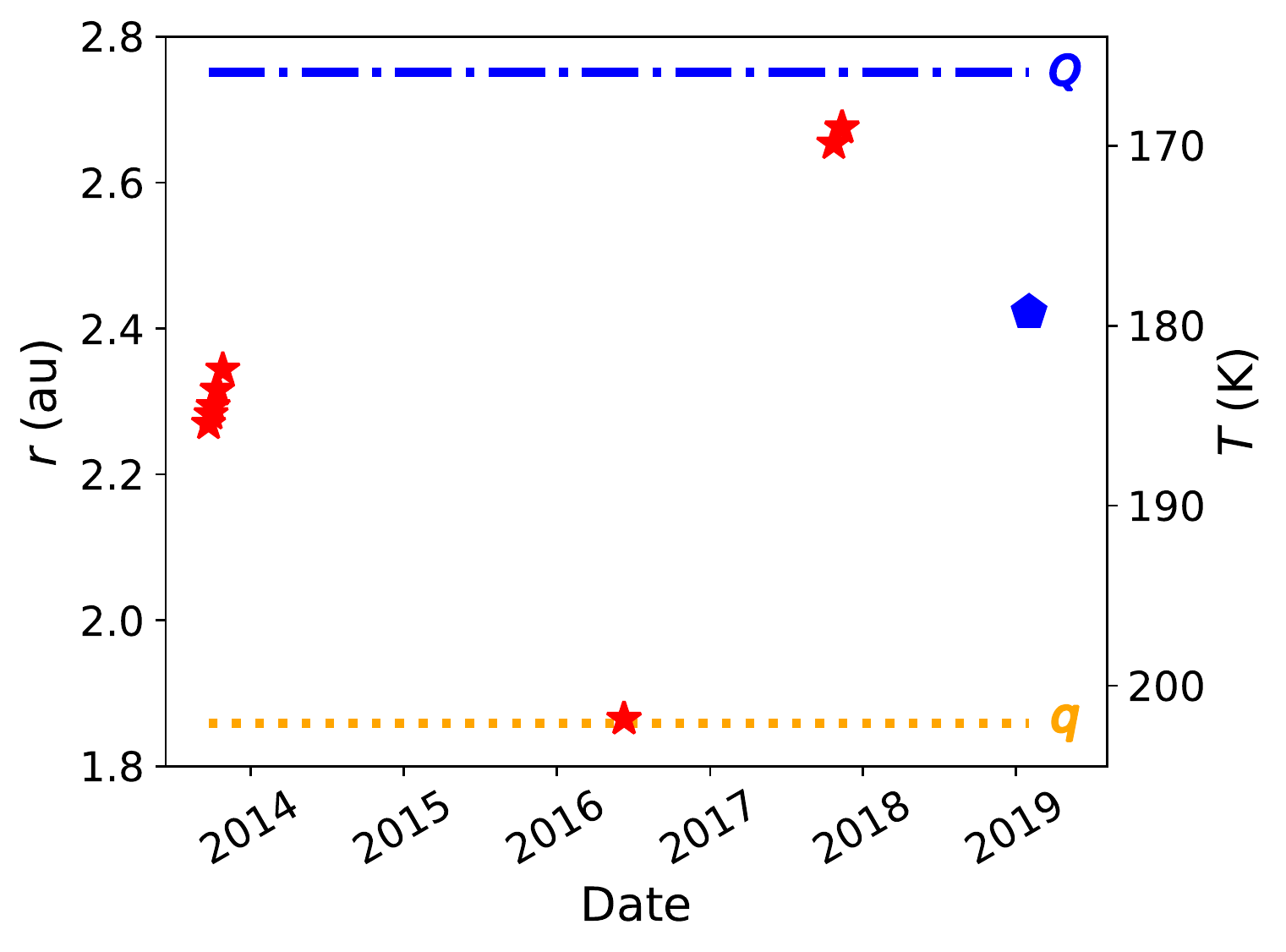}
    \caption{Positive detections of (6478)~Gault activity with DECam as a function of heliocentric distance $r$ (au) and surface temperature $T$ (K). Our activity observations are indicated by red stars, whereas the current apparition is represented by the blue pentagon. Distance and temperature of (6478)~Gault  perihelion $q$ (orange dashed line) and aphelion $Q$ (blue dashed-dotted line) events are shown. During the course of one full orbit, (6478)~Gault is exposed to temperatures greater than 165~K. As a result, (6478)~Gault is consistently subjected to temperatures that are too high for water ice to form at the 5 au ice formation distance \citep{snodgrassMainBeltComets2017}.}
    \label{Gault:fig:tempsanddistance}
\end{figure}

Figure~\ref{Gault:fig:ActivityTimeline} summarizes our observed activity. We found activity at least once in every set of observations and no correlation with distance. We plotted apparent $V$-band magnitude (solid green line) and found that all periods of activity were observed near opposition events. We further define ``observability'' (dashed yellow line) as when the object was (1) above 15$^\circ$ elevation, and (2) visible outside of daylight hours. This allowed us to assess potential observational biases specific to the southern hemisphere where our data were collected. As demonstrated by the coinciding of apparent magnitude maxima with spikes in apparent magnitude, the primary observability factor was solar elongation.

Figure~\ref{Gault:fig:tempsanddistance} shows how (6478)~Gault varies in both temperature $T$ and distance $r$ through time. Indicated are our activity observations (red stars) and the current apparition (blue pentagon). Temperature varies between $\sim$165 K at aphelion $Q$ (blue dash-dotted line) and $\sim$200K at perihelion $q$ (orange dotted line).

We define persistent activity as activity detectable across contiguous sets of observations spanning at least two epochs, even if activity is not visible in every image (due to, for example, exposure time and/or filter selection). We also expect to see activity at all positions throughout the orbit where (6478)~Gault may be detectable by DECam, given appropriate observing parameters (e.g., exposure time, filter selection).


\section{Discussion} 
\label{Gault:sec:discussion}

Most active asteroids, like comets, are composed of low-albedo (i.e.,~dark), primitive material allowing for sublimation or release of volatiles to occur when the body is heated during close passages with the Sun \citep{hsieh2016ReactivationsMainbelt2018}. 

Of the $\sim$20 known active asteroids, four belong to the S-type asteroid taxonomy defining non-primitive silicate-rich material \citep{demeoExtensionBusAsteroid2009}. For these four objects, the causes of activity, when identifiable, are thought to be rotational breakup or impact. Rotational breakup and impact events are consistent with single apparitions or short-lived activity. Furthermore, \cite{hsiehAsteroidFamilyAssociations2018} found that processed material bodies, such as S-types, are more likely to become active due to disruption, while primitive material bodies, such as C-types, can become active via multiple mechanisms due to their volatile abundances. 

(6478)~Gault has been identified as a core member of the Phocaea Family \citep{knezevicProperElementCatalogs2003}. The Phocaea family is dominated by 75\% S-types, followed by 15\% C-types, and 10\% a mix of other asteroid taxonomies \citep{carvanoSpectroscopicSurveyHungaria2001}. While this work was in review, \cite{smith6478Gault2019, jewittEpisodicallyActiveAsteroid2019} reported color measurements suggesting that (6478)~Gault is closer in taxonomic class to a C-type body, rather than an S-type. However, gases were not detected in their spectra, suggesting that sublimation may not be the underlying cause.


Sustained activity near perihelion normally can point to sublimation driven activity, but we observe activity nearly at aphelion during opposition. We do see variability in activity intensity, but we observe activity in (6478)~Gault in at least one image in each set of observations in our DECam data set, suggesting that the target is perpetually active. As a result, we conclude there is no correlation between distance and activity for (6478)~Gault.

Because we observe persistent activity, impact-driven disruption seems unlikely as we would expect the timescale for the activity to be relatively short. The most probable cause for activity has been presented as disruption due to rotational breakup of (6478)~Gault \citep{morenoDustPropertiesDoubletailed2019, yeMultipleOutburstsAsteroid2019, yeContinuedActivity64782019} due to the Yarkovsky--O'Keefe--Radzievskii--Paddack (YORP) effect spin-up \citep{kleynaSporadicActivity64782019}; see \cite{mcneillBrightnessVariationDistributions2016}, \cite{lowryDirectDetectionAsteroidal2007}, and \cite{bottkeYarkovskyYorpEffects2006} for detailed explanations of YORP forces.

Rotational breakup holds for an S-type composition where we would anticipate landslides or surface material redistribution caused by rapid rotation near the 2.2 hr spin rate barrier, which is consistent with the measured $\sim$2 hr light curve period reported by \cite{kleynaSporadicActivity64782019}. We predict that (6478)~Gault will continue to show signs of activity as it has for the last 6 years in a relatively steady state. 
We do not expect catastrophic disruption of (6478) Gault (cf.\ \citet{morenoDustPropertiesDoubletailed2019}). 

The activity observed in (6478)~Gault over multiple epochs and throughout its orbit make (6478)~Gault the first known sustained-activity active asteroid in the main asteroid belt. As a likely S-type asteroid, this is also the first time that we have observed a sustained active body at the rotational barrier for such an extended duration. If activity is in fact not volatile-related, then Gault is a new class of object, perpetually active due to spin-up. We encourage continued monitoring of both the lightcurve and activity level of (6478) Gault, as well as photometric color observations or spectra to further explore its composition.


\section{Acknowledgements}
\label{Gault:subsec:GaultAcknowledgements}

The authors thank the anonymous referee whose comments greatly improved the quality of this Letter.

We wish to thank Nick Moskovitz (Lowell Observatory), who strongly encouraged us to pursue this report. We thank Dr.\ Mark Jesus Mendoza Magbanua (University of California San Francisco) for his frequent and timely feedback on the project. Many thanks to David Jewitt and Man-To Hui (Univeristy of California Los Angeles) for their helpful comments. The authors express their gratitude to Jonathan Fortney (University of California Santa Cruz), Mike Gowanlock (NAU), Cristina Thomas (NAU), and the Trilling Research Group (NAU), all of whom provided invaluable insights which substantially enhanced this work. The unparalleled support provided by Monsoon cluster administrator Christopher Coffey (NAU) and his High Performance Computing Support team facilitated the scientific process wherever possible.

This material is based upon work supported by the National Science Foundation Graduate Research Fellowship Program under Grant No.\ (2018258765). Any opinions, findings, and conclusions or recommendations expressed in this material are those of the author(s) and do not necessarily reflect the views of the National Science Foundation.

Computational analyses were run on Northern Arizona University's Monsoon computing cluster, funded by Arizona's Technology and Research Initiative Fund. This work was made possible in part through the State of Arizona Technology and Research Initiative Program. ``GNU's Not Unix!'' (GNU) Astro \textit{astfits} \citep{akhlaghiNoisebasedDetectionSegmentation2015} provided command-line FITS file header access.

This research has made use of data and/or services provided by the International Astronomical Union's Minor Planet Center. This research has made use of NASA's Astrophysics Data System. This research has made use of the The Institut de M\'ecanique C\'eleste et de Calcul des \'Eph\'em\'erides (IMCCE) SkyBoT Virtual Observatory tool \citep{berthierSkyBoTNewVO2006}. This work made use of the {FTOOLS} software package hosted by the NASA Goddard Flight Center High Energy Astrophysics Science Archive Research Center. This research has made use of SAO ImageDS9, developed by Smithsonian Astrophysical Observatory \citep{joyeNewFeaturesSAOImage2003}. This work made use of the Lowell Observatory Asteroid Orbit Database \textit{astorbDB} \citep{bowellPublicDomainAsteroid1994,moskovitzAstorbLowellObservatory2018}. This work made use of the \textit{astropy} software package \citep{robitailleAstropyCommunityPython2013}.

This project used data obtained with the Dark Energy Camera (DECam), which was constructed by the Dark Energy Survey (DES) collaboration. Funding for the DES Projects has been provided by the U.S. Department of Energy, the U.S. National Science Foundation, the Ministry of Science and Education of Spain, the Science and Technology Facilities Council of the United Kingdom, the Higher Education Funding Council for England, the National Center for Supercomputing Applications at the University of Illinois at Urbana-Champaign, the Kavli Institute of Cosmological Physics at the University of Chicago, Center for Cosmology and Astro-Particle Physics at the Ohio State University, the Mitchell Institute for Fundamental Physics and Astronomy at Texas A\&M University, Financiadora de Estudos e Projetos, Funda\c{c}\~{a}o Carlos Chagas Filho de Amparo, Financiadora de Estudos e Projetos, Funda\c{c}\~ao Carlos Chagas Filho de Amparo \`{a} Pesquisa do Estado do Rio de Janeiro, Conselho Nacional de Desenvolvimento Cient\'{i}fico e Tecnol\'{o}gico and the Minist\'{e}rio da Ci\^{e}ncia, Tecnologia e Inova\c{c}\~{a}o, the Deutsche Forschungsgemeinschaft and the Collaborating Institutions in the Dark Energy Survey. The Collaborating Institutions are Argonne National Laboratory, the University of California at Santa Cruz, the University of Cambridge, Centro de Investigaciones En\'{e}rgeticas, Medioambientales y Tecnol\'{o}gicas–Madrid, the University of Chicago, University College London, the DES-Brazil Consortium, the University of Edinburgh, the Eidgen\"ossische Technische Hochschule (ETH) Z\"urich, Fermi National Accelerator Laboratory, the University of Illinois at Urbana-Champaign, the Institut de Ci\`{e}ncies de l'Espai (IEEC/CSIC), the Institut de Física d'Altes Energies, Lawrence Berkeley National Laboratory, the Ludwig-Maximilians Universit\"{a}t M\"{u}nchen and the associated Excellence Cluster Universe, the University of Michigan, the National Optical Astronomy Observatory, the University of Nottingham, the Ohio State University, the University of Pennsylvania, the University of Portsmouth, SLAC National Accelerator Laboratory, Stanford University, the University of Sussex, and Texas A\&M University.

Based on observations at Cerro Tololo Inter-American Observatory, National Optical Astronomy Observatory (NOAO Prop. IDs 2012B-0001, PI: Frieman; 2014B-0404, PI: Schlegel), which is operated by the Association of Universities for Research in Astronomy (AURA) under a cooperative agreement with the National Science Foundation.

\section{Appendix}

\subsection{Gault Data}
\label{Gault:subsec:GaultData}

For reference we provide essential information regarding (6478) below.

\begin{center}
	\textbf{Properties of (6478)~Gault}
	\begin{tabular}{lll}
		Parameter & Value & Source\\
		\hline\hline
		Discovery Date & 1988 May 12 & \citet{schmadelDictionaryMinorPlanet2003}\\
		Discovery Observers & C. S. \& E. M. Shoemaker & \citet{schmadelDictionaryMinorPlanet2003}\\
		Discovery Observatory & Palomar & \citet{schmadelDictionaryMinorPlanet2003}\\
		Activity Discovery Date & & \citet{smith6478Gault2019}\\
		Alternate Designations & 1988 JC1, 1995 KC1 & NASA JPL Horizons\\
		Orbit Type & Inner Main Belt & NASA JPL Horizons\\
		Family & Phocaea (Core Member) & {\citet{knezevicProperElementCatalogs2003}}\\
		Taxonomic Class & S & via Phocaea association\\
		Diameter & $D=4.5$ km & {\citet{harrisAsteroidsThermalInfrared2002}}\\ 
		Absolute $V$-band Magnitude & $H=14.4$ & NASA JPL Horizons\\
		Slope Parameter & $G=0.15$ & NASA JPL Horizons\\
		Orbital Period & $P=3.5$ yr & NASA JPL Horizons \\
		Semimajor Axis & $a=2.305$ au & NASA JPL Horizons\\
		Eccentricity & $e=0.1936$ & NASA JPL Horizons\\
		Inclination & $i=22.8113^\circ$ & NASA JPL Horizons\\
		Longitude of Ascending Node & $\Omega=183.558$ & Minor Planet Center\\
		Mean Anomaly & $M=289.349^\circ$ & Minor Planet Center\\
		Argument of Perihelion & $\omega=83.2676^\circ$ & NASA JPL Horizons\\
		Perihelion Distance & $q=1.86$ au & NASA JPL Horizons\\
		Aphelion Distance & $Q=2.75$ au & NASA JPL Horizons\\
		Tisserand Parameter w.r.t. Jupiter & $T_J=3.461$ & astorbDB\\
	\end{tabular}
\end{center}

\subsection{Thumbnail Gallery}
\label{Gault:subsec:ThumbnailGallery}

For all thumbnails, red arrows indicate the negative motion vector $-\vec{v}$ of (6478)~Gault; yellow arrows point away from the Sun ($-\odot$). When possible, (6478)~Gault has been circled with a dashed green line and white arrows placed perpendicular to any observed activity. Areas outsize of chip boundaries appear black in color.

\begin{center}
\begin{tabular}{cc}
     \includegraphics[width=0.45\linewidth]{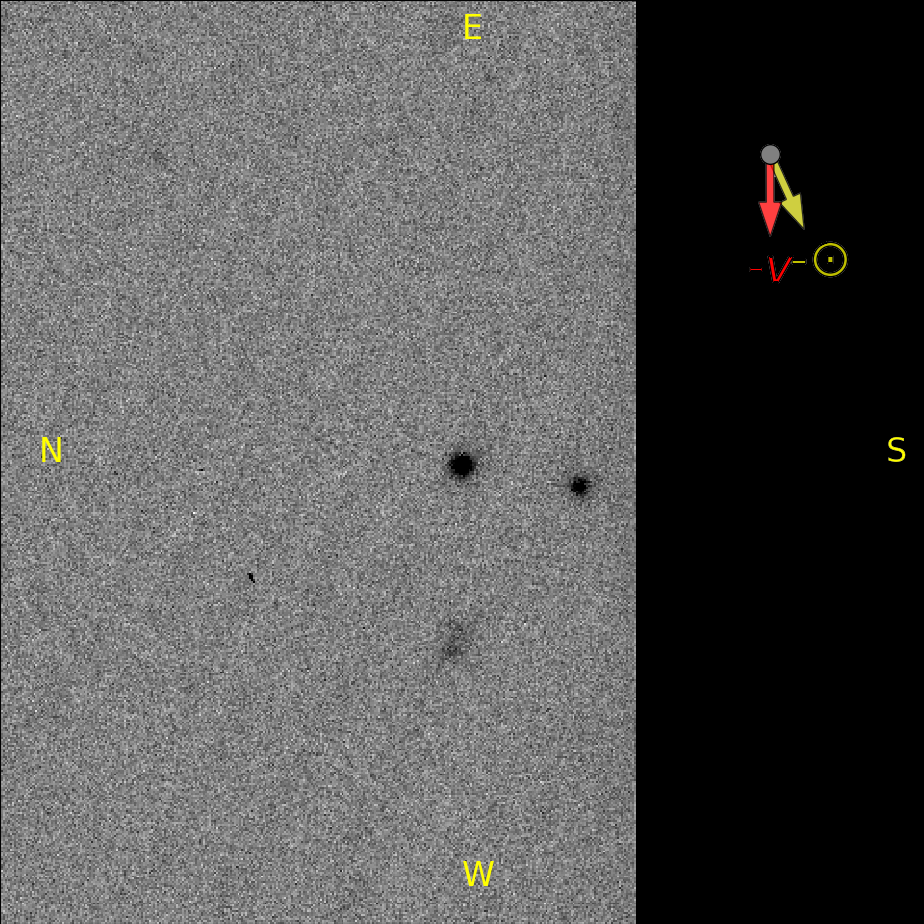} &  \includegraphics[width=0.45\linewidth]{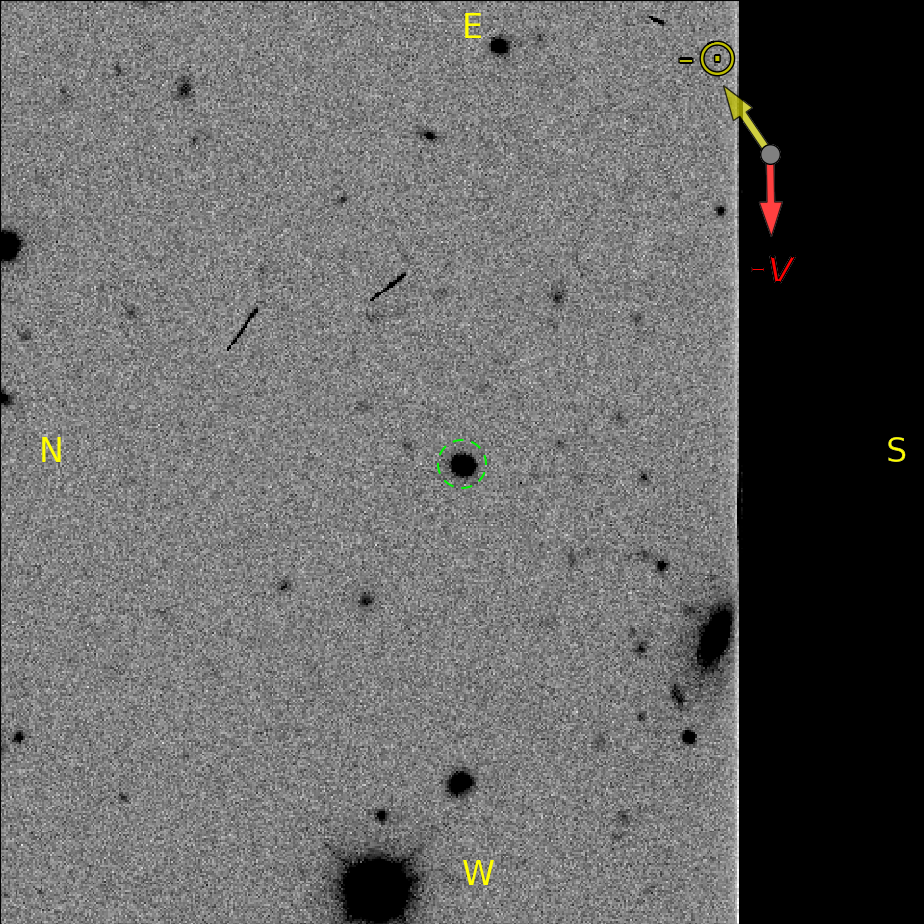}\\
\end{tabular}
\footnotesize
    Left panel: 2013 September 22 3:03 (UT); 45 s $Y$-band. Right panel: 2013 October 13 03:06 (UT); 90 s $i$-band.
\end{center}


\begin{center}
\begin{tabular}{cc}
    \includegraphics[width=0.45\linewidth]{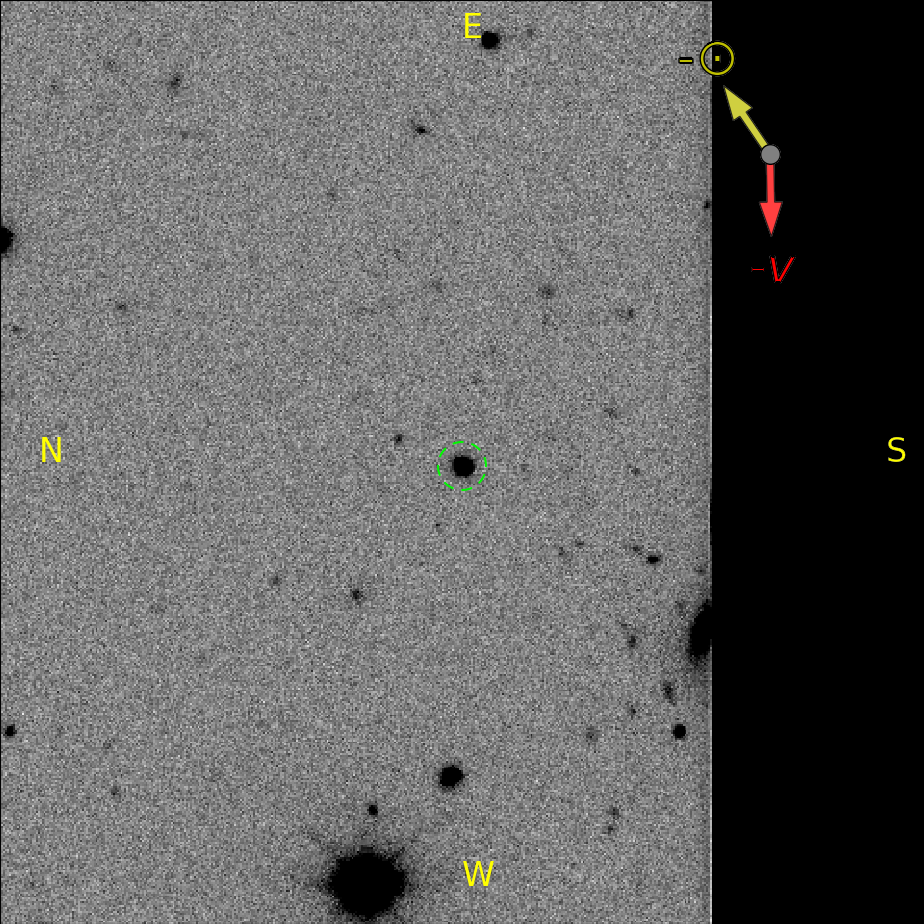} &
    \includegraphics[width=0.45\linewidth]{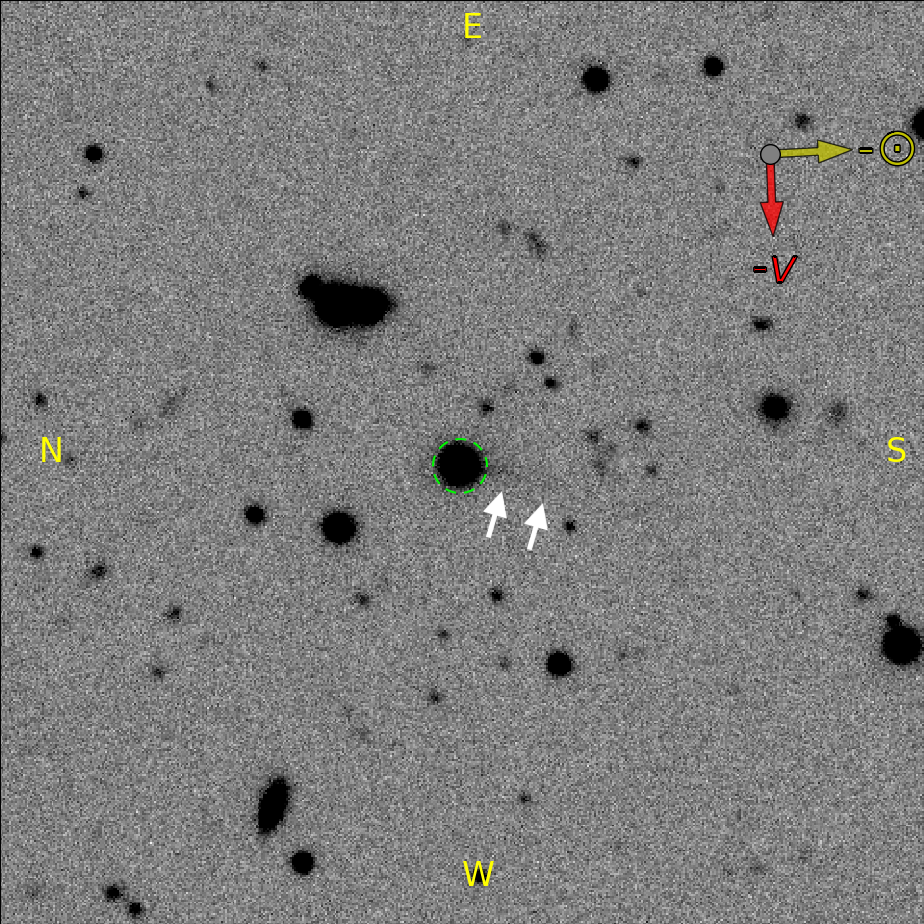} \\
\end{tabular}
\footnotesize
    Left panel: 2013 October 13 03:08 (UT); 90 s $z$-band. Right panel: 2016 June 09 04:45 (UT); 96 s $r$-band.
\end{center}

\begin{tabular}{cc}
     \includegraphics[width=0.45\linewidth]{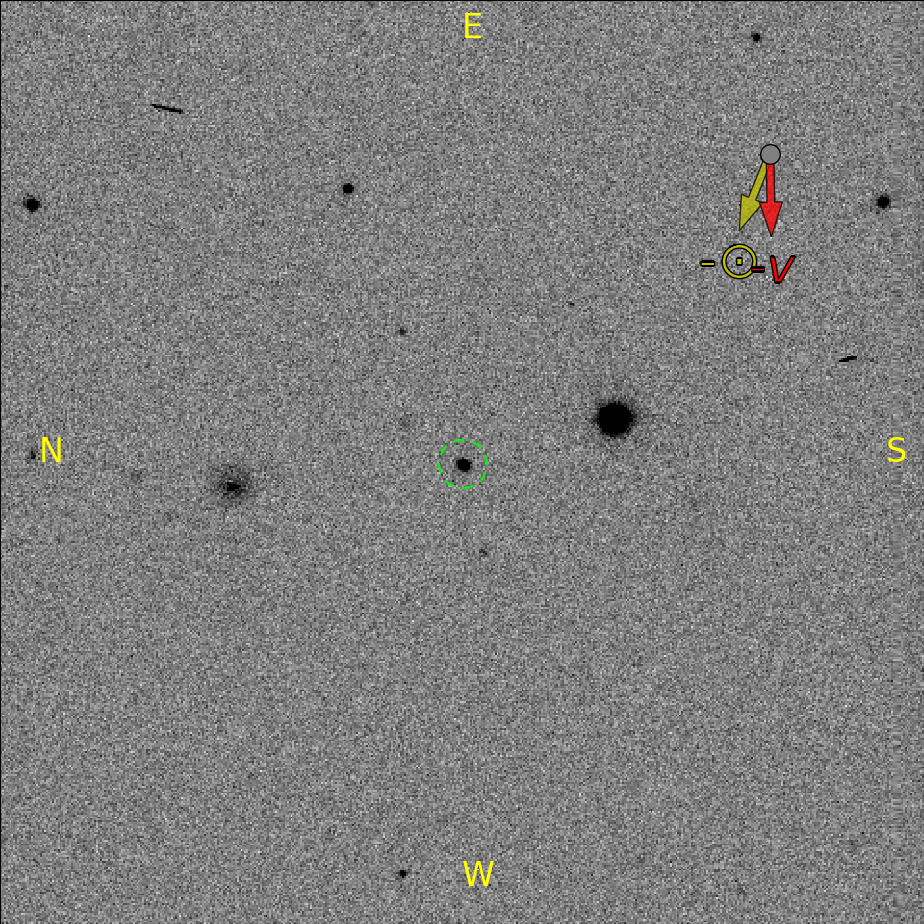} &
     \includegraphics[width=0.45\linewidth]{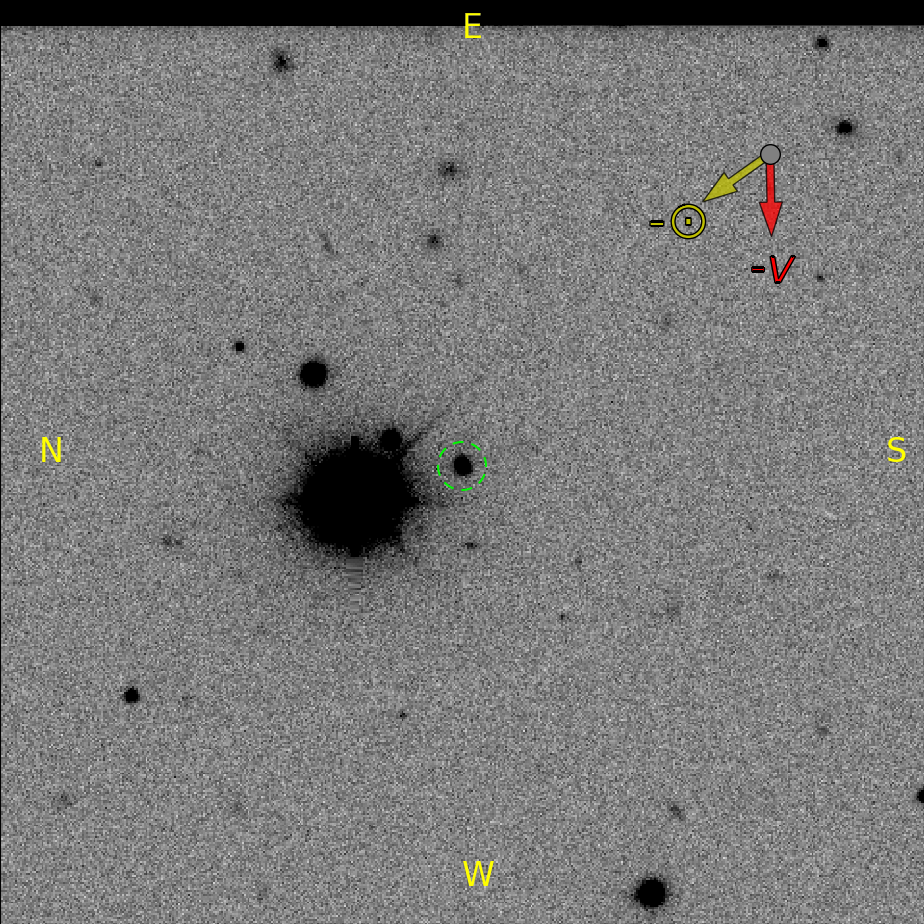} \\
\end{tabular}
\begin{center}
\footnotesize
    Left panel: 2017 October 23 08:57 (UT); 80 s $z$-band. Right panel:  2017 November 11 07:13 (UT); 80 s $z$-band.
\end{center}

\clearpage
\singlespacing
\chapter{Manuscript III: Cometary Activity Discovered on a Distant Centaur: A Nonaqueous Sublimation Mechanism}
\chaptermark{Nonaqueous Cometary Activity Discovered on a Distant Centaur}
\label{chap:2014OG392}

\acresetall

Colin Orion Chandler\footnote{\label{og:nau}Department of Physics \& Astronomy, Northern Arizona University, PO Box 6010, Flagstaff, AZ 86011, USA}, Jay K. Kueny$^\mathrm{\ref{og:nau},}$\footnote{Lowell Observatory, 1400 W Mars Hill Rd, Flagstaff, AZ 86001, USA}, Chadwick A. Trujillo$^\mathrm{\ref{og:nau}}$, David E. Trilling$^\mathrm{\ref{og:nau}}$, William J. Oldroyd$^\mathrm{\ref{og:nau}}$

\textit{This is the Accepted Manuscript version of an article accepted for publication in Astrophysical Journal Letters. IOP Publishing Ltd is not responsible for any errors or omissions in this version of the manuscript or any version derived from it. The Version of Record is available online at }\url{https://iopscience.iop.org/article/10.3847/2041-8213/ab7dc6}\textit{.}

\doublespacing

\section{Abstract}
Centaurs are minor planets thought to have originated in the outer solar system region known as the Kuiper Belt. Active Centaurs enigmatically display comet-like features (e.g., tails, comae) even though they orbit in the gas giant region where it is too cold for water to readily sublimate. Only 18 active Centaurs have been identified since 1927 and, consequently, the underlying activity mechanism(s) have remained largely unknown up to this point. Here we report the discovery of activity emanating from Centaur \og{}, based on archival images we uncovered plus our own new observational evidence acquired with the Dark Energy Camera (Cerro Tololo Inter-American Observatory Blanco 4~m telescope), the Inamori-Magellan Areal Camera \& Spectrograph (Las Campanas Observatory 6.5~m Walter Baade Telescope), and the Large Monolithic Imager (Lowell Observatory 4.3~m Discovery Channel Telescope). We detect a coma as far as 400,000 km from \og{}, and our novel analysis of sublimation processes and dynamical lifetime suggest carbon dioxide and/or ammonia are the most likely candidates for causing activity on this and other active Centaurs. We find \og{} is optically red, but CO$_2$ and NH$_3$ are spectrally neutral in this wavelength regime so the reddening agent is as yet unidentified.

\textbf{Keywords:} Centaurs (215), Comae (271), Comet tails (274), Astrochemistry (75)

\section{Introduction}
\label{og:sec:introduction}

Prior to the mid-20th century, comets were thought to be the only astronomical objects with tails or comae. Unsurprisingly, then, the first two active Centaur discoveries---29P/Schwassman--Wachmann~1 \citep{schwassmannNEWCOMET1927} and 39P/Oterma \citep{otermaNEWCOMETOTERMA1942}---were initially classified as comets.

\begin{figure}[H]
    \centering
	\includegraphics[width=0.5\columnwidth]{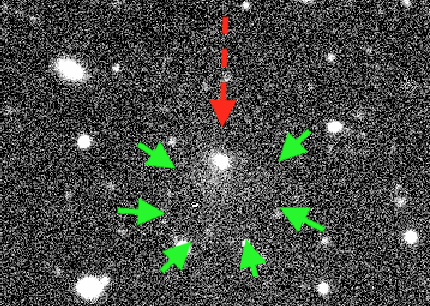}
	\caption{\og{} (dashed arrow) displays a coma (short arrows) during our 2019 August 30 observations. Stack of $4\times250$~s DECam exposures.}
	\label{og:fig:stacked}
\end{figure}

\noindent In 1949 the discovery of the first active asteroid, (4015)~Wilson--Harrington (also designated 107P), blurred the dividing line between asteroid and comet \citep{cunninghamPeriodicCometWilsonHarrington1950}. In 1977 (2060)~Chiron was discovered \citep{kowalSlowMovingObjectKowal1977}, the first member of the population now known as Centaurs. (2060)~Chiron was later found to be active, making it the first object to be identified as a Centaur prior to activity discovery \citep{meechAtmosphere2060Chiron1990}.

We adopt the Centaur classification system \citep{jewittActiveCentaurs2009} that defines Centaurs as objects (1) with perihelia and semi-major axes between the orbits of Jupiter ($\sim$5 au) and Neptune ($\sim$30~au) and (2) not in 1:1 mean-motion resonance with a giant planet (as is the case for the Trojans). We distinguish between Centaurs and Jupiter-Family Comets (following \citealt{levisonLongTermDynamicalBehavior1994}) via the Tisserand parameter with respect to Jupiter, given by

\begin{equation}
	T_\mathrm{J} = \frac{a_\mathrm{J}}{a} + 2\sqrt{\left(1-e^2\right)\frac{a}{a_\mathrm{J}}}\cos(i),
	\label{og:eq:TJ}
\end{equation}

\noindent with eccentricity $e$, inclination $i$, and the semi-major axes of the body and Jupiter $a$ and $a_\mathrm{J}$, respectively. Centaurs have $T_\mathrm{J}\geqslant3$ whereas Jupiter-Family Comets range between $2<T_\mathrm{J}<3$.

\begin{table*}
    \centering
    \footnotesize 
	\caption{Active Centaurs}
	\label{og:tab:activecentaurs}
    \hspace{40mm}Orbital Elements\hspace{25mm}Activity Discovery\\
	\begin{tabular}{l|rrrr|crccr}
		Object Name or Designation & $P$  & $a$  & $q$  & $Q$  & $r$  & $\%_{T\rightarrow q}$	& $M_V$	& Date & Ref.\\
									& (yr) & (au) & (au) & (au) & (au) & & & (UT) & \\
		\hline
		Chiron~(95P)				& 50.5	& \ 6.0 &  8.5 & 18.9	& 11.8  & 68 	& 17.0 & 1989 Apr 10 & 1\\
		Echeclus~(174P)				& 35.3	&  10.8 &  5.9 & 15.6	& 13.1	& 25	& 21.1	& 2005 Dec 04 & 2\\
		29P/Schwassmann--Wachmann 1	& 14.8	& \ 6.0 &  5.5$^\dagger$ & \ 6.6    & \ 6.0	& 53 & 15.3	& 1927 Nov 15 & 3\\
		39P/Oterma					& 19.5	& \ 7.2 &  3.4$^\dagger$ & \ 9.0	& \ 3.5	& 99 & 15.1 & 1942 Feb 12 & 4\\
		165P/LINEAR					& 76.4	&  18.0 &  6.8 & 29.3	& \ 6.9	& 100	& 19.4	& 2000 Jan 09 & 5\\
		166P/NEAT					& 51.9	&  13.9 &  8.6 & 19.2	& \ 8.6	& 100	& 19.6	& 2001 Oct 15 & 6\\
		167P/CINEOS					& 64.8	&  16.1 & 11.8 & 20.5	&  12.2 & 96	& 20.7	& 2004 Jun 07 & 7\\
		P/2005~S${2}$ (Skiff)		& \ 22.5 	& \ 8.0 & \ 6.4 & \ 9.5	& \ 6.5	& 98	& 19.7	& 2005 Sep 16 & 8\\
		P/2005~T$_{3}$ (Read)		& \	20.6	& \ 7.5 & \ 6.2 & \ 8.8	& \ 6.2	& 100	& 20.7	& 2005 Aug 07 & 9\\
		P/2011~C${2}$ (Gibbs)		& \ 20.0	& \ 7.4 & \ 5.4 & \ 9.3	&  \ 5.5	& 97	&   20.3	&2011 Feb 12 & 10\\
		C/2011~P${2}$ (PanSTARRS)	& \ 30.6	& \ 9.8 & \ 6.2 & 13.4	&  \ 6.3	&  98	&   20.3	& 2011 Aug 03 & 11\\
		P/2011~S${1}$ (Gibbs)		& \ 25.4	& \ 8.6 & \ 6.9 & 10.4	&  \ 7.5	&   82	&   21.0	&2011 Sep 18 & 12\\
		C/2013~C${2}$ (Tenagra)	& \ 64.4	&  16.1 & \ 9.1 & 23.0	& \ 9.8	&   96	&   19.1	&2013 Feb 14 & 13\\
		C/2013~P${4}$ (PanSTARRS)	& \ 56.8	&  14.8 & \ 6.0 & 23.6	&  \ 6.3	&   98	&   19.5	&2013 Aug 15 & 14\\
		P/2015~M${2}$ (PanSTARRS)	& 19.3     & \ 7.2  & \ 5.9& \ 8.5	& \ 5.9    &   100 &   19.5	& 2015 Jun 28 & 15\\
		C/2015~T${5}$ (Sheppard--Tholen)& 147.9	&  28.0 & \ 9.3 &   46.6	&  \ 9.4	&   100	&   22.3	&2015 Oct 13 & 16\\
		C/2016~Q${4}$ (Kowalski)	& \ 69.0	&  16.8 & \ 7.1 & 26.5	&  \ 7.5	&   98	& 20.1	& 2016 Aug 30 & 17\\
		2003~QD$_{112}$				& \ 82.8	&  19.0 & \ 7.9 & 30.1	&   12.7 	&   57	&   21.7     & 2004 Oct 10 & 18\\
		\og{}				& \ 42.5	&  12.2 & \ 10.0 & 14.4	& 10.6	&   86 &21.1  & 2017 Jul 18 & 19\\
	\end{tabular}
	
	\raggedright
	\footnotesize
	\vspace{1mm}
	\textbf{Notes.} $P$: orbital period; $a$: semi major axis; $q$: perihelion--distance; $Q$: aphelion distance; $r$: heliocentric distance; $\%_{T\rightarrow q}$: fractional perihelion-aphelion distance (Equation \ref{og:eq:percentperi}); $M_V$: apparent $V$-band magnitude. $Q$ computed via $Q=a(1+e)$ when otherwise unavailable. Asteroid parameters provided by the Minor Planet Center. Heliocentric distance and apparent magnitude courtesy of JPL Horizons \citep{giorginiJPLOnLineSolar1996}.
	
	$^a$ Original value(s) from activity discovery epoch adopted where available; otherwise, values adopted from more recent epoch(s). Reference points to a source that discusses activity of the object.
    
    \textbf{References.} 1:\cite{meechAtmosphere2060Chiron1990}, 2:\cite{choi605582000EC2006}; 3:\cite{schwassmannNEWCOMET1927}, 4:\cite{otermaNEWCOMETOTERMA1942}, 5:\cite{greenComets165P20002005}, 6:\cite{pravdoComet2001T42001}, 7:\cite{romanishinCOMET2004PY422005}, 8:\cite{gajdosComet2005S22005}, 9:\cite{readComet2005T32005}, 10:\cite{gibbsComet2011C22011}, 11:\cite{wainscoatComet2011P22011a}, 12:\cite{gibbsComet2011S12011}, 13:\cite{holvorcemComet2013C22013}, 14:\cite{wainscoatComet2013P42013}, 15:\cite{bacciComet2015M22015}, 16:\cite{tholenComet2015T52015}, 17:\cite{kowalskiCOMET2016Q42016}, 18:\cite{jewittActiveCentaurs2009}, 19:this work
\end{table*}

Centaurs are thought to have migrated inward from the Kuiper Belt (see review; \citealt{morbidelliCometsTheirReservoirs2008}), a region that spans 30~au (Neptune's orbital distance) to 50~au. Neptune Trojans may also serve as a Centaur reservoir \citep{hornerNeptuneTrojansNew2010}. Centaurs all orbit exterior to the 3~au water ice line so they cannot readily undergo sublimation. Surprisingly, though, 18 Centaurs ($\sim$ 4\% of known Centaurs) have been found to display prominent comet-like features such as comae (e.g., Fig. \ref{og:fig:stacked}) or tails; these are the active Centaurs. Table \ref{og:tab:activecentaurs} lists the known active Centaurs along with key physical parameters and discovery circumstances.

Our understanding of active Centaurs has been limited because of their faint apparent magnitudes (the mean apparent magnitude $m_V$ at discovery is $\sim$20; Table \ref{og:tab:activecentaurs}), since it is necessary to probe several magnitudes fainter in order to reliably detect activity via telescopic imaging. Spectroscopy has been used with some success to identify cometary activity originating from asteroids \citep{busarevNewCandidatesActive2018} but this method requires even brighter targets than detection by imaging. Discovering activity on Centaurs is observationally challenging because they are faint, telescope time-intensive, and because they are rare. Active centaurs are discovered, on average, within $\sim$10\% of their perihelion distance (Table \ref{og:tab:activecentaurs}) where they are significantly brighter and, importantly, warmer.

Another significant obstacle to understanding active Centaurs stems from the extreme cold found at their orbital distances. Water and methanol ices have been detected on the surfaces of $\sim$10 Centaurs, but only one of these, (2060)~Chiron, has also been visibly active (see review, \citealt{peixinhoCentaursComets402020}). At surface temperatures less than 150~K and pressures below $\sim10^{-12}$ bar many thermodynamical properties (e.g., enthalpy of sublimation) of volatile ices are not well known from laboratory experiments \citep{fraySublimationIcesAstrophysical2009a}. Moreover, ices may exist in two or more different structural forms; energy from the H$_2$O crystalline--amorphous state transition may even play a role in generating activity \citep{jewittActiveCentaurs2009}.

\section{Mining Archival Data}
\label{og:sec:archivaldata}

In order to overcome the observational challenges discussed in Section \ref{og:sec:introduction} we began by searching archival images captured with the 0.5~gigapixel \acf{DECam} on the Blanco 4~m telescope at the Cerro Tololo Inter-American Observatory in Chile. Archival data from this facility allow the detection of faint activity because of the relatively large aperture and because a large number of objects serendipitously imaged by the instrument can be searched.

We identified Centaurs in our own proprietary database cataloging the NSF’s National Optical-Infrared Astronomy Research Laboratory (NSF’s OIR Lab, formerly NOAO) public \ac{DECam} archive following the methodology outlined in \cite{chandlerSAFARISearchingAsteroids2018}. Our general approach was to correlate image celestial coordinate and temporal data with object ephemeris services such as NASA JPL Horizons \citep{giorginiJPLOnLineSolar1996} and IMCCE SkyBot (\citet{berthierSkyBoTNewVO2006}; see also the acknowledgements).

We (1) extracted event information from the entire \ac{DECam} public archive database, (2) submitted objects to SkyBot or matched against ephemerides produced via the Minor Planet Center and/or Horizons, and then (3) carried out a database query to identify potential images containing Centaurs. 

After (4) downloading the data, we (5) checked each chip for the presence of the Centaur to ensure the object was visible and free of imaging complications (e.g., gaps between chips, scattered light from bright stars, cosmic rays). Finally, we (6) adhered to the routine outlined in \cite{chandlerSAFARISearchingAsteroids2018} where, following image file retrieval of \og{} from the archive, we extracted Flexible Image Transport System (FITS) and Portable Network Graphics (PNG) thumbnails (480$\times$480 pixel images). We subjected these thumbnails to image processing techniques in order to assist by-eye analysis.

While examining each Centaur PNG thumbnail image by eye we flagged any with apparent activity for later analysis. FITS thumbnail images corresponding to those flagged were subjected to additional image processing techniques in an effort to enhance image quality, especially comae contrast.

To ascertain potential heliocentric distance effects we made use of a simple metric \citep{chandlerSAFARISearchingAsteroids2018}, $\%_{T\rightarrow q}$, which describes how close to perihelion ($q$) an object's distance ($d$) is relative to its aphelion distance ($Q$):

\begin{equation}
		\%_{T\rightarrow q} = \left(\frac{Q - d}{Q-q}\right)\cdot 100\mathrm{\%}.
		\label{og:eq:percentperi}
\end{equation}

\begin{figure*}
    \centering
	\includegraphics[width=1.0\linewidth]{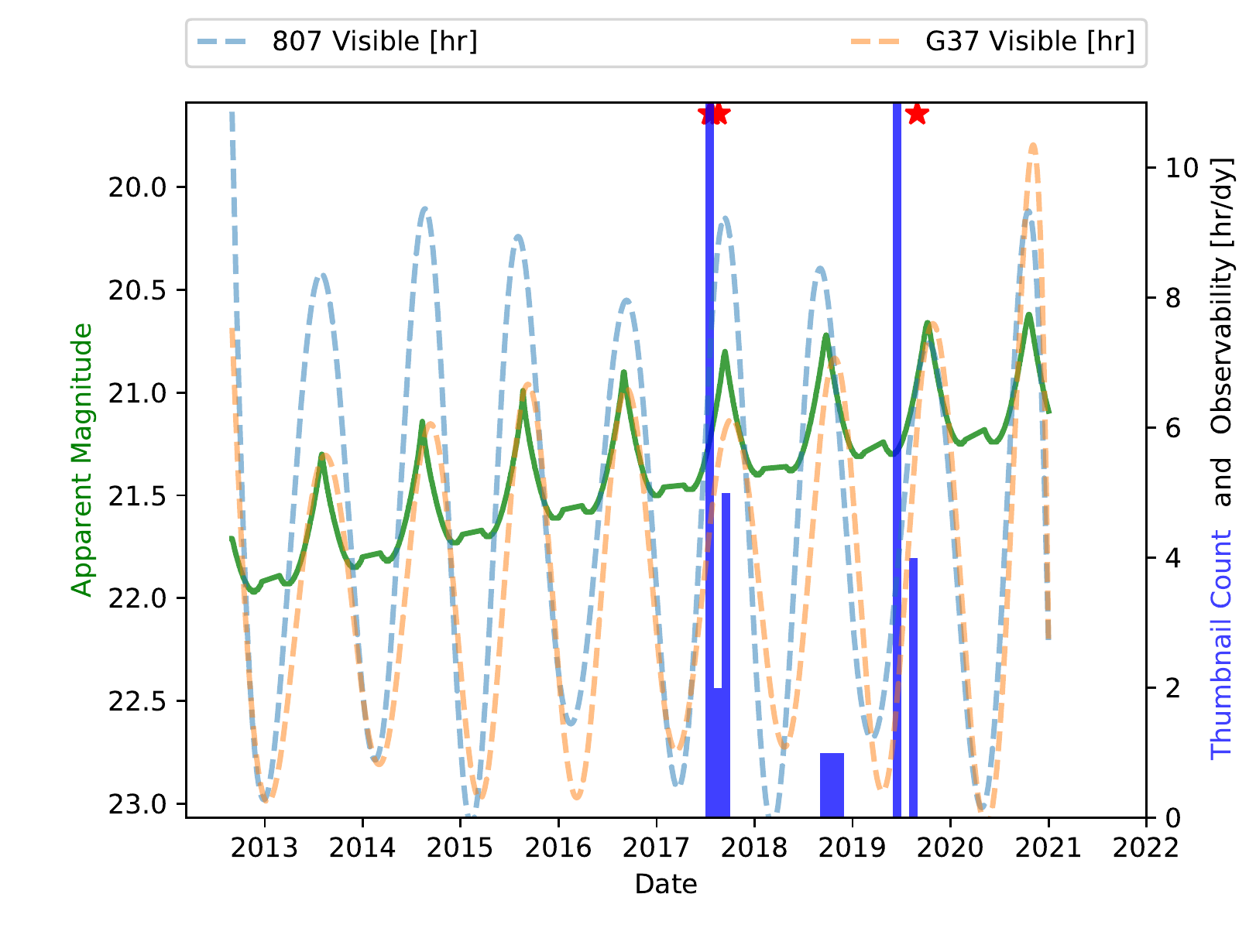}
	\footnotesize\caption{\og{} activity timeline beginning 2012 September (\ac{DECam} first light) to present. Red stars show when we found visible activity. The orbital period is $\sim$42 yr so neither perihelion (2021 December 3) nor aphelion are visible on this plot. The solid green line (left vertical axis) shows the geocentric apparent $V$-band magnitude of \og{}. Dashed lines (right vertical axis) indicate the number of nighttime hours with elevation $> 15^\circ$ for the southern hemisphere \ac{DECam} (blue; site code: 807) and for the northern hemisphere DCT (orange; site code: G37).  The overlaid histogram (vertical blue bars and right axis) shows the number of thumbnail images captured during one calendar month. Note that in all instances when observability was high and many thumbnails were present, activity was observed.
    }
	\label{og:fig:ActivityTimeline}
\end{figure*}

From \ac{DECam} archival data we extracted $\sim20$ thumbnail images of \og{}; Figure \ref{og:fig:ActivityTimeline} shows the number of thumbnails obtained along with the predicted apparent $V$-band magnitude and observability of \og{}. In images from 2017, July and August, we spotted what appeared to be activity emanating from \og{} (see gallery; Figure \ref{og:fig:archivalimages}); at that time the object was 10.60~au from the Sun.

\section{Follow-up Observing}
\label{og:sec:observations}

To confirm the presence of activity we used the same \ac{DECam} instrument and made additional observations on UT 2019 August 30. Fig.\ 1 shows \og{} with a telltale coma revealed by a combined 1000~s exposure. Figure \ref{og:fig:newobserations} contains a gallery showing the four constituent 250 s \ac{DECam} exposures, plus two images where isophotal contours were overplotted to help identify coma extent for each of the first two exposures (Figure \ref{og:fig:isophotalcontours}).

We made use of three observatories for follow-up observations of \og{}: (1) NSF's OIR Labs \ac{DECam} with $VR$ filter on the Blanco 4~m telescope at the Cerro Tololo Inter-American Observatory in Chile (2) WB4800-7800 filtered imaging with the Magellan 6.5~m Walter Baade  Telescope equipped with the Inamori-Magellan Areal Camera \& Spectrograph (IMACS) at the Las Campanas Observatory on Cerro Manqui, Chile, and (3) $g$, $r$, and $i$ filter images taken with the \ac{LMI} at the Lowell Observatory 4.3~m Discovery Channel Telescope (DCT) in Arizona, USA.  Galleries showing our Magellan images and DCT images are shown in Figure \ref{og:fig:magellanobservations} \ref{og:fig:dctobservations}.
respectively. A log of observations is provided in Appendix \ref{og:sec:observationdetails}. Astrometric calibration was performed using the \textit{astrometry.net} \citep{langAstrometryNetBlind2010} and/or \textit{PhotometryPipeline} \citep{mommertPHOTOMETRYPIPELINEAutomatedPipeline2017} software packages.

\section{Simulating Dynamical Lifetime}
\label{og:sec:dynamicallifetimesimulation}
Determining the total mass loss possible for different volatiles requires knowledge of the dynamical lifetime of \og{} in the Centaur region (where both perihelion distance and semi-major axis are between 5 and 30~au). To this end we made use of the REBOUND $N$-body integrator to model the orbits of \og{} and giant planets Jupiter, Saturn, Uranus, and Neptune \citep{reinHybridSymplecticIntegrators2019}. We also carried out 25 simulations of \og{}, each with an orbital clone derived from the orbital uncertainties published by the Minor Planet Center. From these dynamical integrations, we found that the lifetime of \og{} spans the range of 13,000--1.8 million years, roughly in agreement with prior work \citep{liuInvestigationOriginCentaurs2019}.

\section{Sublimation Modeling}
\label{og:sec:sublimationmodeling}

In order to better assess potential processes responsible for \og{} activity, we computed equilibrium temperatures and modeled mass-loss rates for seven astrophysically relevant ices: ammonia (NH$_3$), carbon dioxide (CO$_2$), carbon monoxide (CO), methane (CH$_4$), methanol (CH$_3$OH), nitrogen (N$_2$), and water (H$_2$O). 

Object distance is the primary factor in determining potential ice sublimation effects. We began with a simple sublimation model \citep{hsiehMainbeltCometsPanSTARRS12015} well suited to gaining broad insight into the observed activity from \og{}; we expanded the procedure to apply more generally to other volatile ices. As we do not know the composition of \og{} we cannot make use of a more comprehensive model which includes effects of, for example, porosity, tortuosity, or crystal structure \citep{schorghoferLifetimeIceMain2008a}. Moreover, \og{} is undoubtedly not composed of a single ice, and mixtures of ices can exhibit behavior uncharacteristic for any lone constituent \citep{grundySolarGardeningSeasonal2000}. For the limiting case of an inert gray body orbiting at a distance $R$ from the Sun (measured in au)

\begin{equation}
    \frac{F_\odot}{R^2}(1-A)=\chi \epsilon \sigma T_\mathrm{eq}^4
\end{equation}

\noindent where the fiducial solar flux $F_\odot$ is 1360 W m$^{-2}$, $A$ is the Bond albedo (we choose 0.1 as representative for Centaurs; \citealt{peixinhoCentaursComets402020}), $\epsilon$ is the infrared emissivity of the ice (set here as 0.9), $T_\mathrm{eq}$ is the equilibrium temperature of the body, and $\sigma$ is the Stefan--Boltzmann constant ($5.670\times10^{-8}$W m$^{-2}$ K$^{-4}$). Here $\chi$ is a factor that describes the rotational and axial tilt effects on how much flux is received from the Sun: $\chi=1$ indicates the maximum heating scenario where the body is a ``slab'' facing the Sun at all times; $\chi=\pi$ describes a body that rotates quickly with no axial tilt with respect to the Sun; and $\chi=4$, which we adopt here, is used for a fast-rotating (on the order of a few hours) isothermal body in thermodynamic equilibrium.
Here ``fast-rotating'' means that the rotation period of the object is short compared to the thermal wave propagation time \citep{schorghoferLifetimeIceMain2008a,hsiehMainbeltCometsPanSTARRS12015}.

We next consider an energy balance that incorporates sublimation in addition to blackbody radiation \citep{hsiehMainbeltCometsPanSTARRS12015}:

\begin{equation}
    \frac{F_\odot}{R^2} (1-A) = \chi\left[\epsilon\sigma T^4 + L f_\mathrm{D}\dot{m}_\mathrm{S}(T)\right]
    \label{og:eq:sublimationfull}
\end{equation}

\noindent where $f_\mathrm{D}$ is the ``diffusion barrier factor'' that describes how much emission is blocked by overlaying material (e.g., regolith), and $L$ the latent heat of sublimation. The mass-loss rate $\dot{m}_\mathrm{S}(T)$ is given by

\begin{equation}
    \dot{m} = P_\mathrm{v}(T)\sqrt{\frac{\mu}{2\pi k T}}
\end{equation}
\noindent with $\mu$ the SI mass of one molecule, and $k$ the Boltzmann constant of $1.38069\times10^{-23}$ J K$^{-1}$. The vapor pressure (in Pa) of the substance can be related to temperature by the Clausius--Clapeyron relationship

\begin{equation}
    P_\mathrm{v}(T) = e_\mathrm{S} \exp \left[\frac{\Delta H_\mathrm{subl}}{R_\mathrm{g}}\left(\frac{1}{T_\mathrm{triple}}-\frac{1}{T}\right)\right]
\end{equation}

\noindent in which $e_\mathrm{S}$ is the saturation vapor pressure (in Pa) of the substance at the triple-point temperature $T_\mathrm{triple}$, $\Delta H_\mathrm{subl}$ is the heat of sublimation of the substance (in kJ mol$^{-1}$), and $R_\mathrm{g}$ is the ideal gas constant ($\rm 8.341~J/mol\cdot K$).

Solving Equation \ref{og:eq:sublimationfull} for heliocentric distance $R$ (in au) yields

\begin{equation}
    R(T) = \sqrt{\frac{F_\odot (1-A)}{\chi\left[\epsilon\sigma T^4 + L f_\mathrm{D}\dot{m}_\mathrm{S}(T)\right]}}.
\end{equation}

Energy of sublimation values \citep{lunaNewExperimentalSublimation2014b} and triple-point temperatures and pressures \citep{fraySublimationIcesAstrophysical2009a} were incorporated as needed. To validate our model we computed the mass-loss rate for (2060)~Chiron assuming $\chi=4$, an albedo of $0.057$, a diameter of 206 km, and an orbit ranging from 8.47 au at perihelion to 18.87~au at aphelion. Our (2060)~Chiron model validation results were in rough agreement with the 0.5--20 kg s$^{-1}$ mass-loss rate reported by \citet{womackCODistantlyActive2017}.

\begin{figure*}
    \centering
    \includegraphics[width=0.85\linewidth]{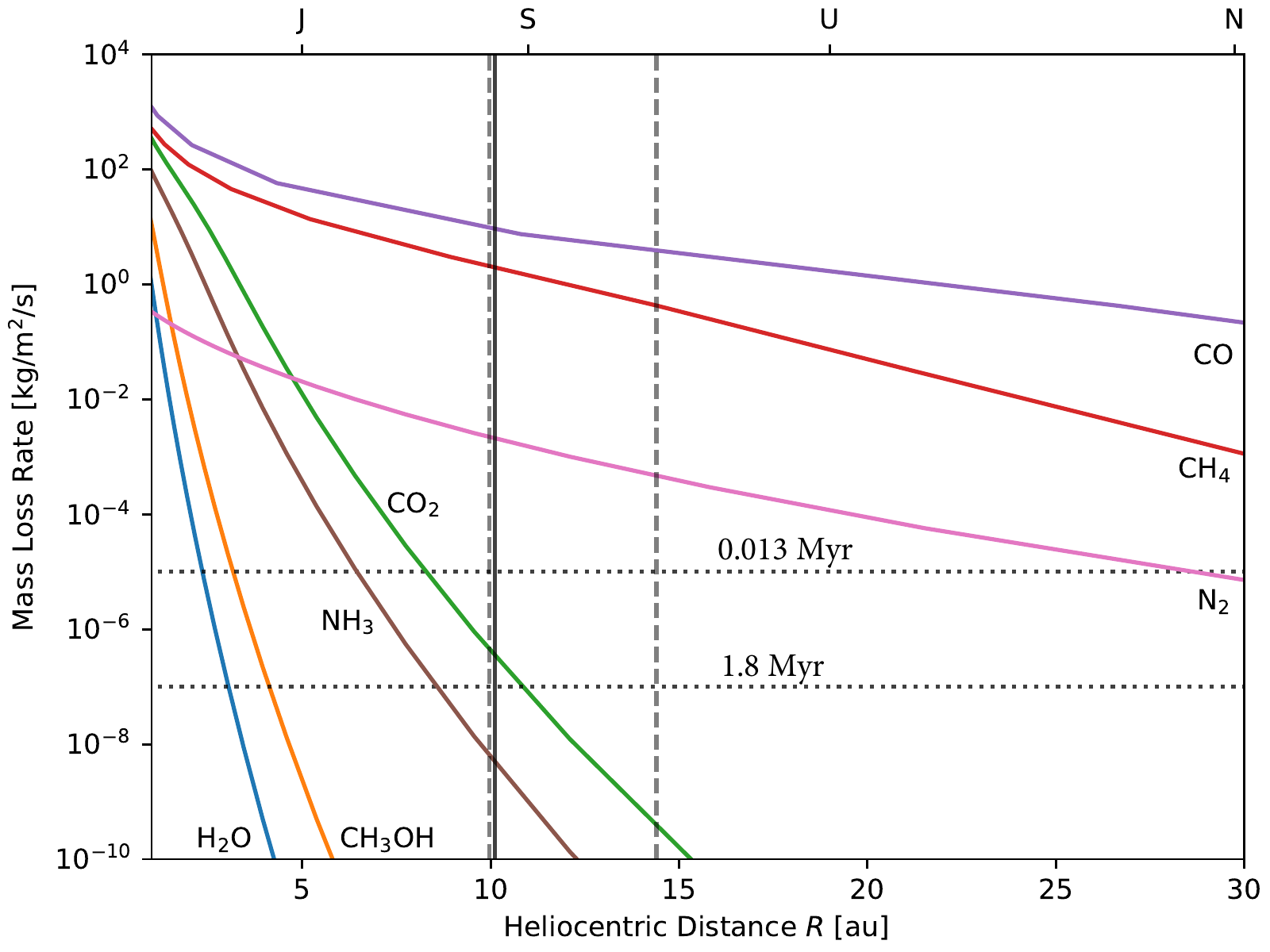}
    \footnotesize \caption{ Mass-loss rates for seven different astrophysically relevant ices on an isothermal ($\chi=4$) body; water (H$_2$O) and methanol (CH$_3$OH) ices have been detected on Centaurs. Orbital distances of Jupiter, Saturn, Uranus, and Neptune are indicated about the top axis. The current 10.11~au heliocentric distance of \og{} is indicated by a vertical black bar, bracketed by perihelion (9.97 au) and aphelion (14.40~au) distances (leftmost and rightmost dashed vertical lines, respectively). Over the course of one orbit (between the vertical dashed lines), water and methanol never appreciably sublimate and carbon monoxide (CO), methane (CH$_4$), and molecular nitrogen (N$_2$) sublimate at high and relatively constant rates; we rule out all of these molecules as potential causes of activity. (The shallow slopes of CO, CH$_4$, and N$_2$ extend beyond 50~au [not shown], which informs us the mass loss would have begun long before \og{} became a Centaur.) However, over the course of one orbit the sublimation rates for CO$_2$ and NH$_3$ vary substantially, presumably producing  significant variation in visible activity. Order-of-magnitude estimates of mass-loss-rate upper limits for the dynamical lifetime of \og{} are shown as horizontal dotted lines. Only CO$_2$ and NH$_3$ have sublimation rates near these limits.}
    \label{og:fig:masslossrates}
\end{figure*}

We use our computed dynamical lifetime to circumstantially constrain the molecule(s) responsible for the sublimation of \og{}. Fig. \ref{og:fig:masslossrates} shows, over the orbit of \og{}, the mass-loss rates for the different ices determined via modeling and validated through laboratory measurements. If \og{} has an albedo of 10\%, similar to that measured for other Centaurs (see review; \citealt{peixinhoCentaursComets402020}), then the body is about 20~km in diameter (see Section \ref{og:sec:absmaganddiameter}). Assuming a spherical body of low density in the range of 1--3~g cm$^{-1}$ suggests a reasonable body mass of $4.2--12.6\times 10^{15}$~kg and a surface area of $3.1 \times 10^{8} \mbox{ m}^2$. Thus, the 13,000--1.8 Myr dynamical lifetime of \og{} suggests a maximum orbit-averaged mass-loss rate in the range of $7.1\times10^{-7}$ to $\rm 3.3\times 10^{-5}\ kg/m^{2}/s$ (horizontal dashed lines in Figure \ref{og:fig:masslossrates}) before the body would be entirely lost due to sublimation.

\section{Colors}
\label{og:sec:colors}
The archival data and our confirmation observations did not contain enough information to determine colors, so we obtained six 300~s exposures of \og{} in a $g$-$r$-$i$ filter sequence at the DCT (Section \ref{og:sec:observations}). We made use of the \textit{PhotometryPipeline} software package \citep{mommertPHOTOMETRYPIPELINEAutomatedPipeline2017} to automate astrometry using SCAMP \citep{bertinAutomaticAstrometricPhotometric2006} which made use of the Vizier catalog service \citep{ochsenbeinVizieRDatabaseAstronomical2000} Gaia Data Release 2 catalog \citep{collaborationGaiaDataRelease2018}, and photometric image calibration using solar stars from the Sloan Digital Sky Survey Data Release 9 (SDSS-DR9) catalog \citep{ahnNinthDataRelease2012}. We carried out manual aperture photometry using the Aperture Photometry Tool \citep{laherAperturePhotometryTool2012a}.

\begin{figure}
    \centering
    \includegraphics[width=0.5\columnwidth]{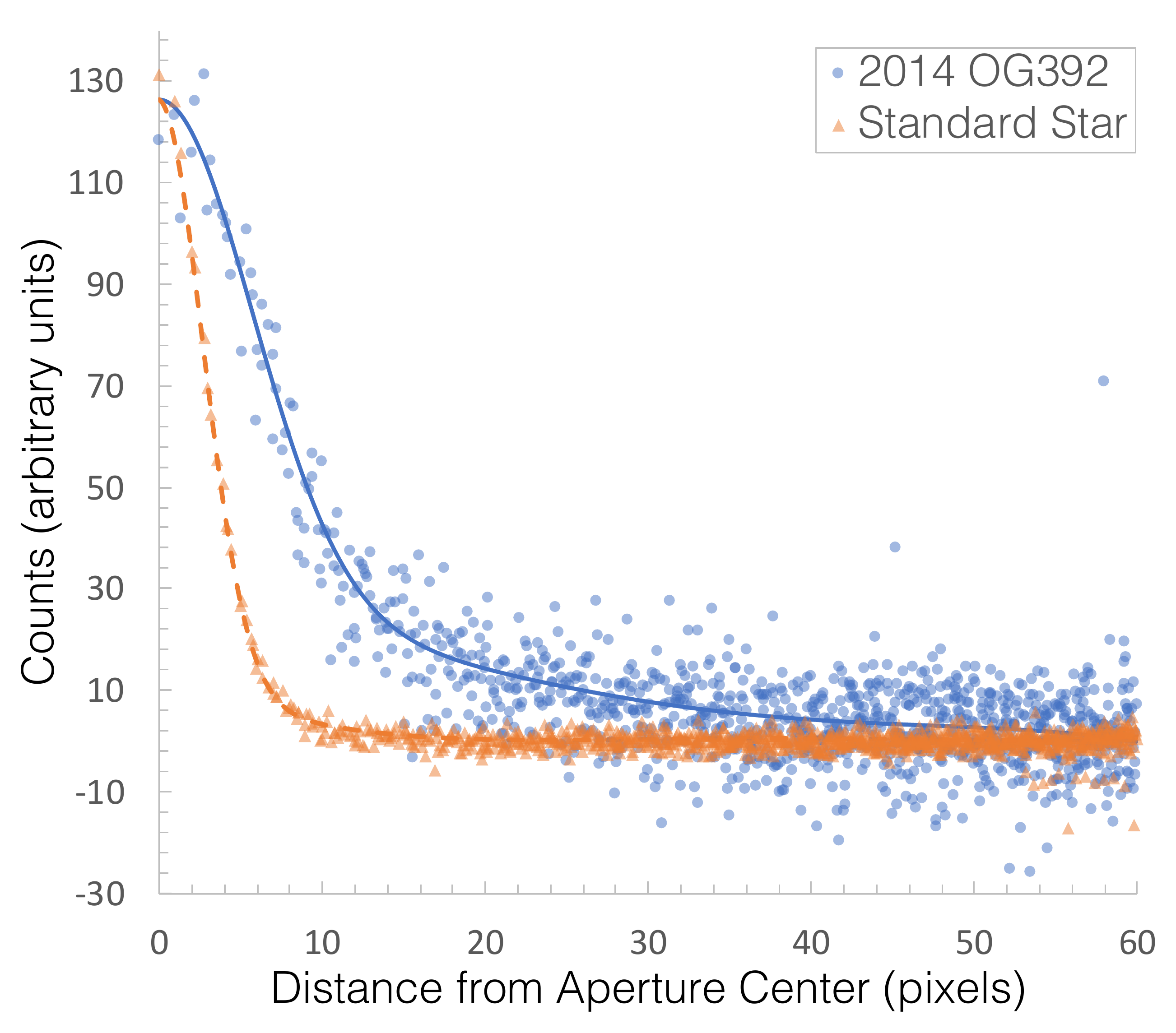}
    \caption{\footnotesize Surface brightness radial profiles of \og{} and a nearby SDSS-DR9 catalog solar-type star (J004840.66-022335.6) are plotted along with a model fit for each object. After subtracting the background flux from the two profiles we normalized the standard star profile to the peak of the \og{} profile. The coma flux tapers from 125 counts to background (0 counts) at $\rho\simeq60$ pixels, or $4.3\times10^5$~km. We estimate there are $\sim5.8\times10^{17}$ particles in the coma assuming a grain radius of 1~mm; for a density of 1~g cm$^{-3}$ the total mass is $2.4\times10^{15}$~g. Data from our 300~s $g$-band exposure taken on UT 2019 December 30 2:29 using the LMI on the Lowell Observatory 4.3~m DCT.} 
    \label{og:fig:sbrp}
\end{figure}

Prior to analysis we examined all thumbnail images showing activity emanating from \og{} to ensure no significant background sources were blended with the nucleus. To help us identify unseen contaminators we measured and modeled surface brightness radial profiles of \og{} (Figure \ref{og:fig:sbrp}) and a nearby solar-type star, using the Aperture Photometry Tool. The radial profile itself (i.e., not the model) was used to identify flux contribution by unseen background sources; we rejected images in which the nucleus or nearby coma was significantly contaminated. We note that we identified at least one background source within the coma in all of our images, although for color measurement we were able to use an aperture small enough (5 pixel radius) to exclude all resolvable background objects. 

We measured \og{} apparent magnitudes to be $g=21.99\pm0.018$, $r=21.19\pm0.016$, and $i=20.81\pm0.018$. We compared our colors of $g-r=0.80\pm0.024$ and $r-i=0.39\pm0.024$ to SDSS reported solar colors of $g-r=0.44\pm0.02$ and $r-i=0.11\pm0.02$\footnote{\url{http://www.sdss.org/dr12/algorithms/ugrizvegasun}}. Centaur colors are often reported in Johnson $B$ -- $R$ colors (see, e.g., \citealt{teglerTWOCOLORPOPULATIONS2016}), so we computed the $B$ -- $R$ color for \og{} via \cite{jesterSloanDigitalSky2005a} transformations. We found $B$ -- $R$ = \ogColor{}, which is about one magnitude redder than the Sun, and red according to the classification system of \cite{teglerTWOCOLORPOPULATIONS2016} (see discussion in Section \ref{og:sec:discussion}).

\section{Absolute Magnitude and Diameter Estimation}
\label{og:sec:absmaganddiameter}

To gauge the overall spatial extent of the coma we examined the radial surface brightness profiles of \og{} and nearby solar-type star J004840.66-022335.6 (see Section \ref{og:sec:colors}). We fit the profiles to the model

\begin{equation}
    S(r) = A + Br + Cr^2 + Dr^3 + Er^4 + F e^{-\frac{r^2}{2\sigma^2}}
\end{equation}

\noindent as described in \cite{gwynSSOSMovingObjectImage2012a}.

After subtracting the sky flux from each profile and each model we scaled the star to the peak flux of the \og{} radial profile. Figure \ref{og:fig:sbrp} shows the radial profiles and their corresponding models plotted; we estimate the coma returns to sky background flux levels at $\sim$60 pixels from the aperture center, thus the coma extent is $\sim4.3\times10^5$ km. The FWHM of \og{} was $13.62\pm0.37$ pixels (3\farcsec2$\pm$0\farcsec09), whereas the star FWHM was $6.05\pm0.05$ pixels ($1.45\pm0\farcsec012$).

As reported in Section \ref{og:sec:sublimationmodeling}, the coma is likely present throughout the orbit of \og{}. As a result, prior absolute ($H$) magnitude estimates would have included the excess flux caused by the coma, as evinced in Figure \ref{og:fig:sbrp}. To estimate the absolute nuclear magnitude of \og{} we compared the ratio of the total (nucleus + coma) flux (blue line and circles, Figure \ref{og:fig:sbrp}) to the scaled stellar flux (orange line and triangles, Figure \ref{og:fig:sbrp}). We estimate the coma accounts for 0.75 and 1.1 magnitudes of the observed $r$-band and $g$-band fluxes, respectively, implying the nucleus apparent magnitudes are $m_r=21.9$ and $m_g=23.1$.

The absolute magnitude of an asteroid, $H$, is commonly used to estimate the size of small bodies in the solar system . $H$ is defined as equal to the apparent $V$-band magnitude of an object observed at a heliocentric distance $R=1$ au, a geocentric distance $\Delta=1$ au, and a phase angle $\alpha=0^\circ$. Here we employ the International Astronomical Union defined \citep{swingsTransactionsInternationalAstronomical1986} $H$ -- $G$ magnitude system approximated from \cite{bowellApplicationPhotometricModels1989}:

\begin{equation}
	V = 5 \log \left(R \Delta\right) + H - 2.5 \log \left[\left(1-G\right)\Phi_1 + G \Phi_2\right]
	\label{og:eq:HG}
\end{equation}

\noindent where the phase function $\Phi$ is given by

\begin{equation}
	\Phi_i = \exp\left[-A_i\tan\left(\alpha/2\right)^{B_i}\right]; i=1,2
\end{equation}

\noindent with constants $A_1=3.33$, $A_2=1.87$, $B_1=0.63$, and $B_2=1.22$.

We make use of the relationships put forth by \cite{jesterSloanDigitalSky2005a} to derive Johnson $V=22.4$ from our $g$ and $r$ nuclear magnitudes. The JPL Horizons ephemerides service \citep{giorginiJPLOnLineSolar1996} provided $G=0.150$ (the standard assumed slope for dark surfaces), $r=10.10$~au, $\Delta=10.01$~au, and $\alpha=5\fdegree 58$ for UT 2019 December 30. Via Equation~\ref{og:eq:HG} we find $H=11.3$, $0.5$ magnitudes fainter than reported by the Minor Planet Center and JPL Horizons.

\cite{harrisRevisionRadiometricAlbedos1997a} provide a convenient method to approximate object diameter $D$,

\begin{equation}
    D = \frac{1329}{\sqrt{G}}\times 10^{-H/5},
\end{equation}

\noindent which, for \og{}, gives $D\approx20$~km.

\section{Coma Dust Analysis}
\label{og:sec:dustproductionmodeling}

To facilitate comparing our \og{} dust-related metrics with other works we adopt the instrument and aperture-independent cometary dust production parameter described by \cite{ahearnCometBowell1980b1984a}. The metric, $Af\rho$ (units of cm), combines the mean albedo $A$ of ejecta grains within an aperture of radius $\rho$ (in cm), scaled by the filling factor $f$ (unitless), which describes how much of the aperture area ($\pi\rho^2$) is filled by $N$ grains of cross section area $\sigma$ (in cm$^2$),

\begin{equation}
    f = \frac{N(\rho)\sigma}{\pi\rho^2}.
    \label{og:eq:f}
\end{equation}

We measured $Af\rho$ (following the method outlined by \citealt{shiResearchActivityMain2019b}) via

\begin{equation}
    A f \rho = 4 R^2 \Delta^2 10^{0.4(m_{\odot,F} - m_{\mathrm{OG},F})} \rho^{-1}
\end{equation}

\noindent where $R$ is the \og{} heliocentric distance in au, $\Delta$ is the geocentric distance of \og{} in cm, and, for filter $F$, $m_{\odot,F}$ and $m_{\mathrm{OG},F}$ are the magnitudes of the Sun and \og{}, respectively. For $m_{\odot,F}$ we made use of solar apparent Vega magnitudes\footnote{\url{http://mips.as.arizona.edu/~cnaw/sun.html}} (see \citet{willmerAbsoluteMagnitudeSun2018} for details) in Table \ref{og:tab:AppMagFilter}: 

\begin{table} 
    \centering
    \caption{Solar Apparent Magnitude by Filter}
    \begin{tabular}{cc}
        Filter & $m_{\odot,F}$\\
        \hline
        SDSS-$g$ & -26.34\\
        SDSS-$r$ & -27.04\\
        SDSS-$i$ & -27.38\\
    \end{tabular}
    \label{og:tab:AppMagFilter}
\end{table}

To estimate the number of particles $N$ within our measured $Af\rho$ we can substitute Equation \ref{og:eq:f} into the equality $Af\rho=Af\rho$

\begin{equation}
    Af=A\frac{N(\rho)\sigma}{\pi\rho^2}\rho
\end{equation}

\noindent and solve for $N(\rho)$,

\begin{equation}
    N(\rho) = Af\rho \frac{\pi\rho}{A\sigma}.
    \label{og:eq:Nrho}
\end{equation}

To quantify the total number of particles in the coma $N_\mathrm{tot}$ we can scale the aperture of Equation \ref{og:eq:Nrho} to the 60 pixel aperture containing the entire coma, $\rho_\mathrm{max}$,

\begin{equation}
    N(\rho_\mathrm{max}) = Af\rho \frac{\pi\rho_\mathrm{max}^2}{A \sigma \rho}.
    \label{og:eq:Nmax}
\end{equation}

\noindent Recall the quantity $Af\rho$, here, is a measured value, so the quantities $A\rho$ do not cancel in Equation \ref{og:eq:Nmax}.

Four of our observations, Images 15-18 (details in Appendix \ref{og:sec:observationdetails}) were suitable for directly measuring $Af\rho$. We found $Af\rho=487\pm 12$~cm with an aperture of $4.3\times10^5$~km. With the albedo adopted for our sublimation modeling ($A=0.1$) and a 1~mm radius grain, the coma around \og{} is composed of roughly $5.8\times10^{17}$ particles. Assuming a grain density of 1~g~cm$^{-3}$ the total coma mass is $\sim2.4\times10^{15}$~g. 

\section{Discussion}
\label{og:sec:discussion}

The activity we observed spans more than two years, which rules out impact-driven activity. We determined that the two ices previously detected on Centaurs, water and methanol, would not appreciably sublimate at any point in \og{}'s orbit and so should still be present in solid form on the surface (Figure \ref{og:fig:masslossrates}). Moreover, CO, N$_2$ and CH$_4$ are highly volatile and sublimate at temperatures low enough that their supply is likely depleted, though reservoirs could still be trapped below the surface. We reiterate our model encompasses single-species ices subjected to the thermodynamic conditions outlined in Section \ref{og:sec:sublimationmodeling}; heterogeneous ice environments may alter sublimation chemistry (see, e.g., \citealt{grundySolarGardeningSeasonal2000}), as can single-species state transitions (e.g., energy released during crystallization of amorphous water ice; see, e.g., \citealt{jewittActiveCentaurs2009}).

We find that the molecule(s) most likely to drive the observed activity is either $\rm CO_2$ and/or possibly $\rm NH_3$. Neither would have sublimated appreciably at Kuiper Belt distances prior to \og{} becoming a Centaur. Interestingly, both of these substances sublimate at rates that vary by over two orders of magnitude over the course of a \og{} orbit, peaking at perihelion. As a result we predict \og{} will become less active post-perihelion. This further implies that all other active Centaurs should follow this trend, with peak sublimation near perihelion and a significant drop in outgassing for most of their orbits.

We determined \og{} is at present roughly one magnitude redder than the Sun at visible wavelengths. However, we were only able to obtain two images in each filter, so uncertainty could be improved upon with additional observations. Our color measurements inexorably included the coma; future observations during a quiescent period (should one exist) would allow for color measurements of the bare nucleus. We did, however, attempt to better estimate the $H$ magnitude by subtracting the coma measured in the radial surface brightness profiles. We found \og{} has $H\approx11.3$, 0.5 magnitudes fainter than previously reported. The $H$ magnitude implies a radius of about 20~km when assuming a slope parameter $G=0.15$ as is typical for a dark surface.

In our images of \og{} background sources were typically present in the coma and/or blended with the nucleus, but from four images we were able to directly measure dust properties. Assuming a 10\% albedo and a grain radius of 1~mm we estimate the coma contains roughly $5.8\times10^{17}$ particles. If the grain density is 1~g cm$^{-3}3$, the total mass is $\sim2.4\times10^{15}$~g, or $\sim0.01$\% the total mass of \og{}. If the coma mass is indeed of this scale, \og{} must be eroding very quickly, undergoing new activity, or the ejecta is accumulating faster than it is escaping. Our measured $Af\rho$ of $487\pm12$~cm is comparable to other Centaurs active at the same orbital distance as \og{}: C/2011~P2~(PANSTARRS) with $Af\rho=161\pm 4$~cm at $\sim 9$~au \citep{epifaniNucleusActiveCentaur2017}, and for 166P~(NEAT) $Af\rho=288\pm 19$~cm at $\sim 12$~au \citep{shiCCDPhotometryActive2015}. 

Centaurs are sometimes classified as either gray or red depending on whether the object has a $B$-$R$ color closer to $\sim$1.2 or $\sim$1.7, respectively (see \citealt{teglerColorsCentaurs2008a} and \citealt{peixinhoCentaursComets402020} reviews for in-depth discussions). We find our derived $B$-$R$ color of \ogColor{} consistent with the red classification. Notably both molecules we find viable for sublimation are spectrally neutral in visible wavelengths so the reddening agent is as yet unidentified. 
\og{} will remain observable through 2020 February and will again be observable beginning around 2020 August. We anticipate imaging and spectroscopy will yield further insight into the nature of these rare objects. We wish to emphasize further lab work is needed to characterize sublimation processes of volatiles under low pressure and temperature regimes.

\section{Acknowledgments}

We thank Dr.\ Mark Jesus Mendoza Magbanua (University of California San Francisco) for his frequent and timely feedback on the project. J.K.K. acknowledges support from Northern Arizona University through a startup award administered by TD Robinson. Mark Loeffler (NAU) and Patrick Tribbett (NAU) helped us interpret lab results. Stephen Tegler (NAU) provided extensive insight into the nuances of Centaur colors. Kazuo Kinoshita provided cometary elements that saved us considerable time and energy. The authors express their gratitude to Mike Gowanlock (NAU), Cristina Thomas (NAU), and the Trilling Research Group (NAU), all of whom provided insights that substantially enhanced this work. Thank you to Stephen Kane (University of California Riverside), Dawn Gelino (NASA Exoplanet Science Institute at California Institute of Technology), and Jonathan Fortney (University of California Santa Cruz) for encouraging the authors to pursue this work. The unparalleled support provided by Monsoon cluster administrator Christopher Coffey (NAU) and his High Performance Computing Support team greatly facilitated the work presented here.

\ This material is based upon work supported by the National Science Foundation Graduate Research Fellowship Program under grant No.\ 2018258765. Any opinions, findings, and conclusions or recommendations expressed in this material are those of the author(s) and do not necessarily reflect the views of the National Science Foundation.

\ Computational analyses were carried out on Northern Arizona University's Monsoon computing cluster, funded by Arizona's Technology and Research Initiative Fund.
\ This work was made possible in part through the State of Arizona Technology and Research Initiative Program.

\ This research used the facilities of the Canadian Astronomy Data Centre operated by the National Research Council of Canada with the support of the Canadian Space Agency. We also employed their solar system Object Search \citep{gwynSSOSMovingObjectImage2012a}.
\ This research has made use of data and/or services provided by the International Astronomical Union's Minor Planet Center.
\ This research has made use of NASA's Astrophysics Data System.
\ This research has made use of the The Institut de M\'ecanique C\'eleste et de Calcul des \'Eph\'em\'erides (IMCCE) SkyBoT Virtual Observatory tool \citep{berthierSkyBoTNewVO2006}.
\ Simulations in this Letter made use of the REBOUND code, which is freely available at http://github.com/hannorein/rebound.
\ This work made use of the {FTOOLS} software package hosted by the NASA Goddard Flight Center High Energy Astrophysics Science Archive Research Center.
\ This research has made use of SAO Image DS9, developed by Smithsonian Astrophysical Observatory \citep{joyeNewFeaturesSAOImage2003}. This work made use of the Lowell Observatory Asteroid Orbit Database \textit{astorbDB} \citep{moskovitzModernizingLowellObservatory2019a}.
\ This work made use of the \textit{astropy} software package \citep{robitailleAstropyCommunityPython2013}.

\ This project used data obtained with the \acf{DECam}, which was constructed by the \acf{DES} collaboration. Funding for the \ac{DES} Projects has been provided by the U.S. Department of Energy, the U.S. National Science Foundation, the Ministry of Science and Education of Spain, the Science and Technology Facilities Council of the United Kingdom, the Higher Education Funding Council for England, the National Center for Supercomputing Applications at the University of Illinois at Urbana-Champaign, the Kavli Institute of Cosmological Physics at the University of Chicago, Center for Cosmology and Astro-Particle Physics at the Ohio State University, the Mitchell Institute for Fundamental Physics and Astronomy at Texas A\&M University, Financiadora de Estudos e Projetos, Funda\c{c}\~{a}o Carlos Chagas Filho de Amparo, Financiadora de Estudos e Projetos, Funda\c{c}\~ao Carlos Chagas Filho de Amparo \`{a} Pesquisa do Estado do Rio de Janeiro, Conselho Nacional de Desenvolvimento Cient\'{i}fico e Tecnol\'{o}gico and the Minist\'{e}rio da Ci\^{e}ncia, Tecnologia e Inova\c{c}\~{a}o, the Deutsche Forschungsgemeinschaft and the Collaborating Institutions in the Dark Energy Survey. The Collaborating Institutions are Argonne National Laboratory, the University of California at Santa Cruz, the University of Cambridge, Centro de Investigaciones Energ\'{e}ticas, Medioambientales y Tecnol\'{o}gicas–Madrid, the University of Chicago, University College London, the DES-Brazil Consortium, the University of Edinburgh, the Eidgen\"ossische Technische Hochschule (ETH) Z\"urich, Fermi National Accelerator Laboratory, the University of Illinois at Urbana-Champaign, the Institut de Ci\`{e}ncies de l'Espai (IEEC/CSIC), the Institut de Física d'Altes Energies, Lawrence Berkeley National Laboratory, the Ludwig-Maximilians Universit\"{a}t M\"{u}nchen and the associated Excellence Cluster Universe, the University of Michigan, NSF’s National Optical-Infrared Astronomy Research Laboratory, the University of Nottingham, the Ohio State University, the University of Pennsylvania, the University of Portsmouth, SLAC National Accelerator Laboratory, Stanford University, the University of Sussex, and Texas A\&M University.

\ This work is based in part on observations at Cerro Tololo Inter-American Observatory, National Optical Astronomy Observatory (Prop. IDs 2019A-0337, PI: Trilling; 2014B-0404, PI: Schlegel), which is operated by the Association of Universities for Research in Astronomy (AURA) under a cooperative agreement with the National Science Foundation.

These results made use of the Discovery Channel Telescope at Lowell Observatory. Lowell is a private, non-profit institution dedicated to astrophysical research and public appreciation of astronomy and operates the DCT in partnership with Boston University, the University of Maryland, the University of Toledo, Northern Arizona University and Yale University. The Large Monolithic Imager was built by Lowell Observatory using funds provided by the National Science Foundation (AST-1005313). This Letter includes data gathered with the 6.5 m Magellan Telescopes located at Las Campanas Observatory, Chile.


\section{Appendix} 
\subsection{Activity Observation details} 
\label{og:sec:observationdetails}

\begin{table}
    \centering
    \caption{Activity Observations}
	\begin{tabular}{cccrc}
		\# & Instrument  & Date/Time			& Exp.	& Filter\\
		 &       & (UT)				& [s]\ 	&			 \\
		\hline
		 1   & DECam$^1$   & 2017-07-18 09:27	& 137	& $z$	\\
		 2   & DECam$^1$   & 2017-07-18 10:20	& 250	& $z$	\\
		 3   & DECam$^1$   & 2017-07-22 05:37	&  79	& $g$	\\
		 4   & DECam$^1$   & 2017-07-25 06:25	&  60	& $r$	\\
		 5   & DECam$^1$   & 2017-07-25 06:32	&  52	& $r$	\\
		 6   & DECam$^1$   & 2017-08-20 04:48	&  67	& $r$	\\
		 7   & DECam$^2$   & 2019-08-30 09:54	& 250	& \textit{VR}\\
		 8   & DECam$^2$   & 2019-08-30 09:58	& 250	& \textit{VR}\\
		 9   & DECam$^2$   & 2019-08-30 10:03	& 250	& \textit{VR}\\
		10   & DECam$^2$   & 2019-08-30 10:08	& 250	& \textit{VR}\\
		11   & IMACS   & 2019-12-27 00:54  & 300   & WB4800-7800\\
		12   & IMACS   & 2019-12-27 01:01  & 300   & WB4800-7800\\
		13   & IMACS   & 2019-12-27 01:36  & 600   & WB4800-7800\\
		14   & LMI     & 2019-12-30 02:08    & 300   & $g$\\
		15   & LMI     & 2019-12-30 02:17    & 300   & $r$\\
		16   & LMI     & 2019-12-30 02:23    & 300   & $i$\\
		17   & LMI     & 2019-12-30 02:29    & 300   & $g$\\
		18   & LMI     & 2019-12-30 02:35    & 300   & $r$\\
		19   & LMI     & 2019-12-30 02:41    & 300   & $i$\\
	\end{tabular}
	
	$^1$Program 2014B-0404 (PI: Schlegel)\\
	$^2$Program 2019A-0337 (PI: Trilling)\\
	\label{og:tab:observations}
\end{table}

Table \ref{og:tab:observations} provides a listing of the observations used in this work.

\subsection{Thumbnail Gallery} 
\label{og:sec:thumbnailgallery}

\begin{figure}
	\begin{tabular}{ccc}
	\includegraphics[width=0.30\linewidth]{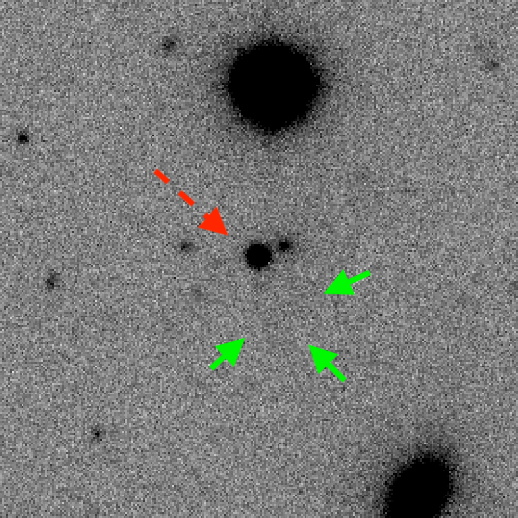} &
	\includegraphics[width=0.30\linewidth]{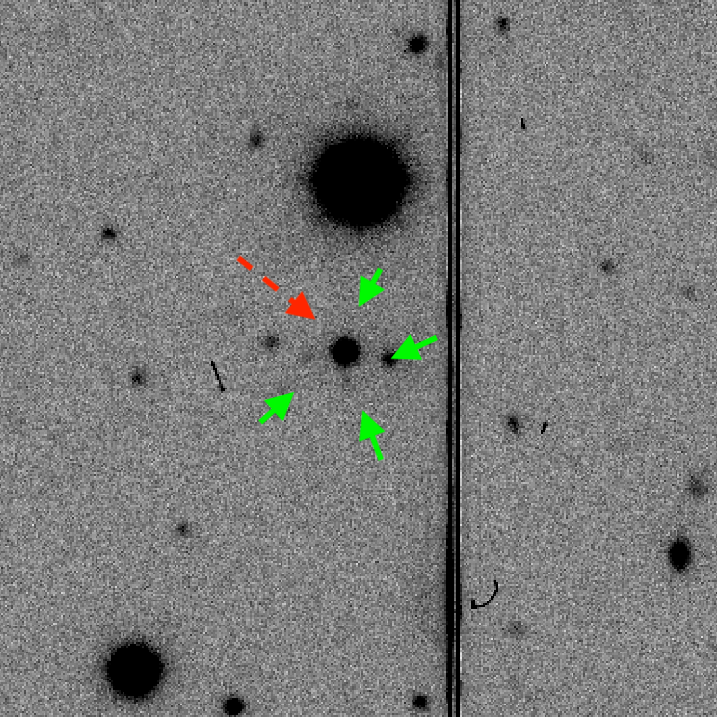} &
	\includegraphics[width=0.30\linewidth]{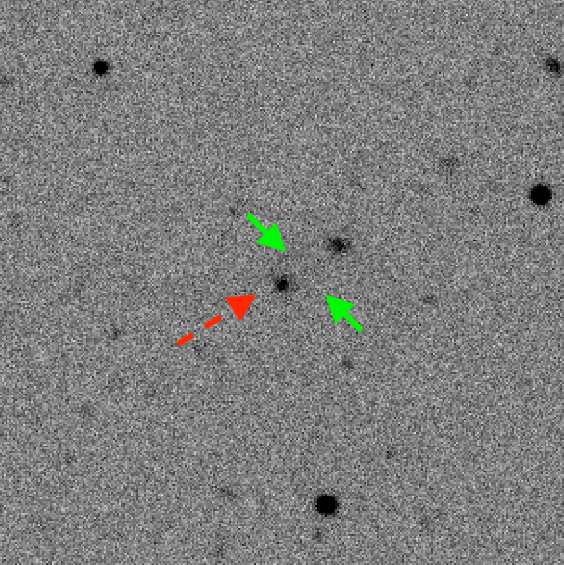}\\
 	\includegraphics[width=0.30\linewidth]{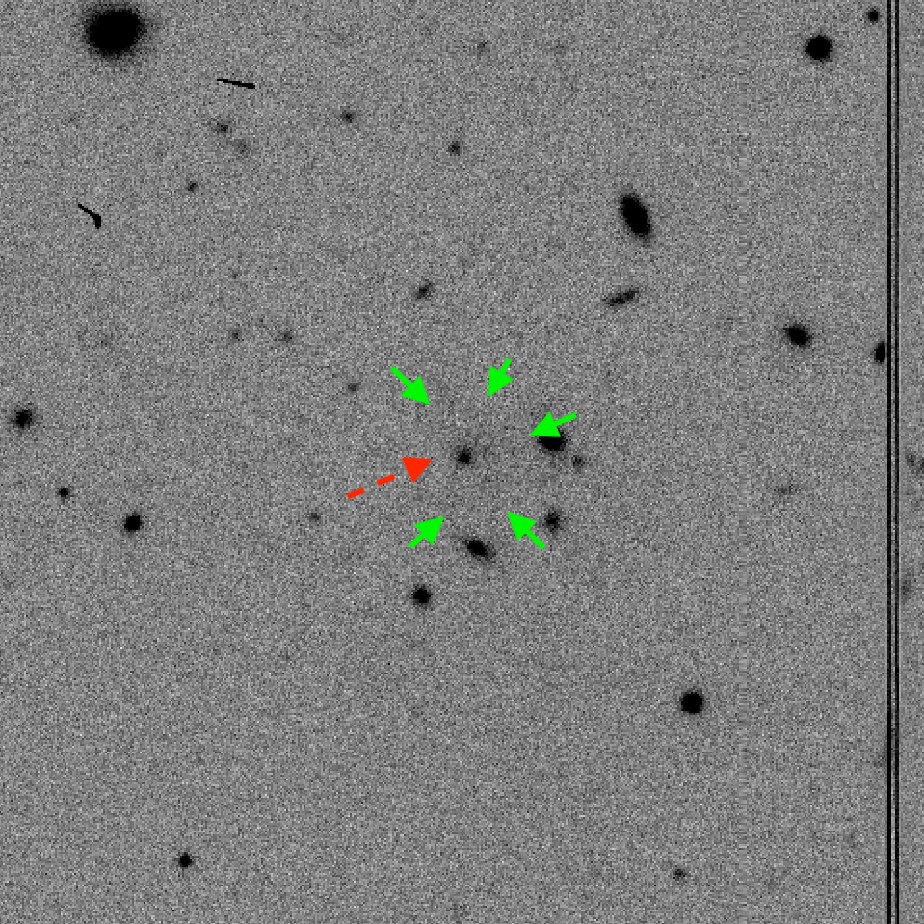} &
 	\includegraphics[width=0.30\linewidth]{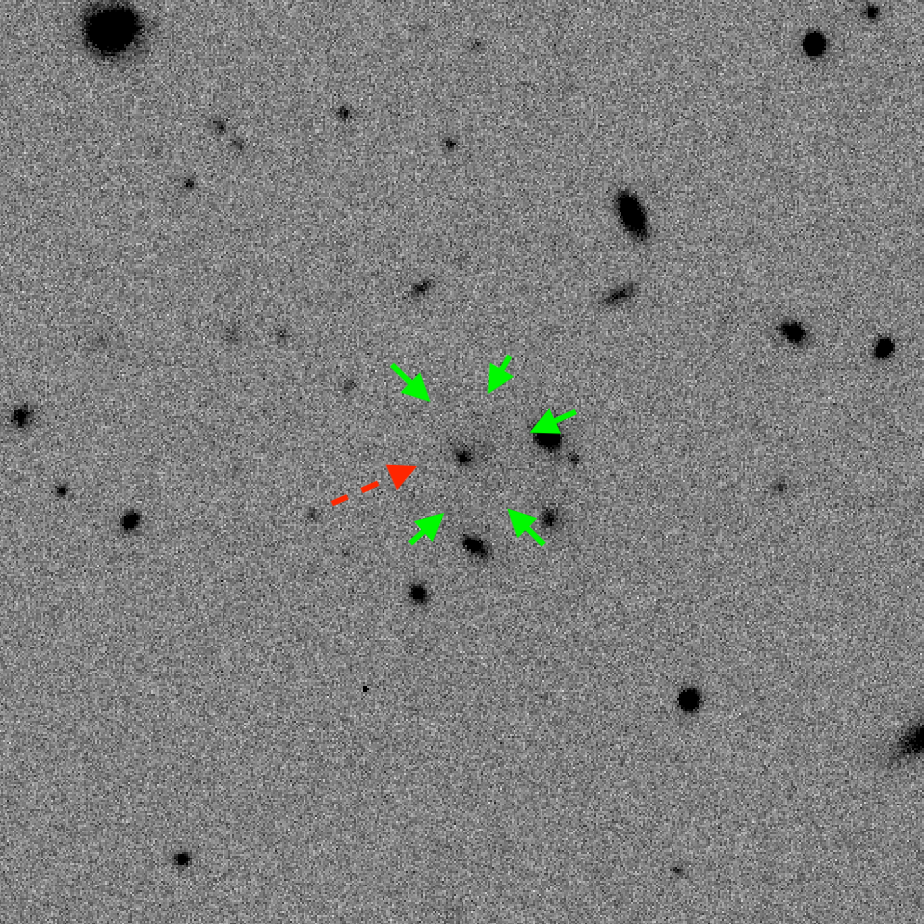} &
 	\includegraphics[width=0.30\linewidth]{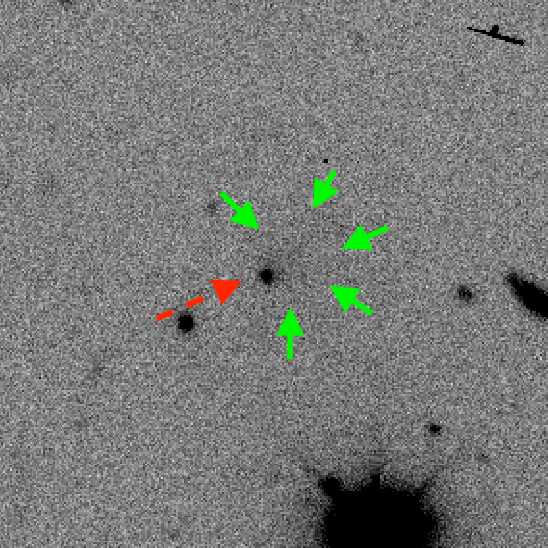} \\
	\end{tabular}
	\caption{\textit{DECam Archival Images}. Top left: UT 2017-Jul-18 09:27 -- 137~s $z$-band. Top center: UT 2017-Jul-18 10:20 -- 250~s $z$-band. Top right: UT 2017-Jul-22 05:37 -- 79~s $g$-band. Bottom left: UT 2017-Jul-25 06:25 -- 60~s $r$-band. Bottom center: UT 2017-Jul-25 06:32 -- 52~s $r$-band. Bottom right: UT 2017-Aug-20 04:48 -- 67~s $r$-band. All Images: The coma (green arrows) was exceptionally faint in all of these DECam archival images of \og{} (indicated by dashed red arrows) but nevertheless they prompted us to obtain follow-up observations.}
	\label{og:fig:archivalimages}
\end{figure}

\begin{figure}
    \centering
	\begin{tabular}{cc}
    \includegraphics[width=0.4\linewidth]{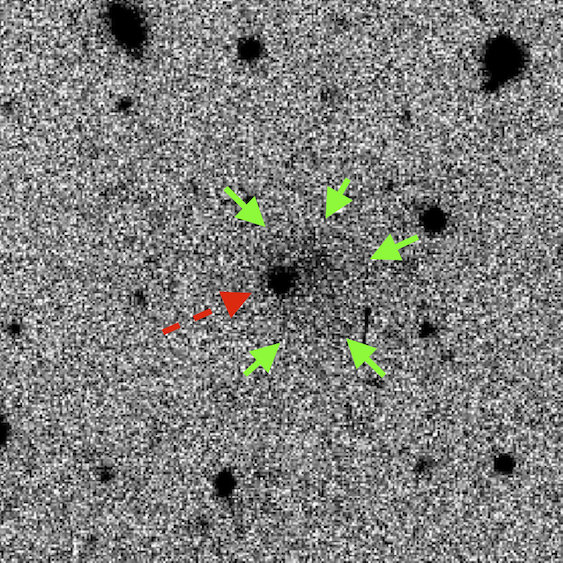} & \includegraphics[width=0.4\linewidth]{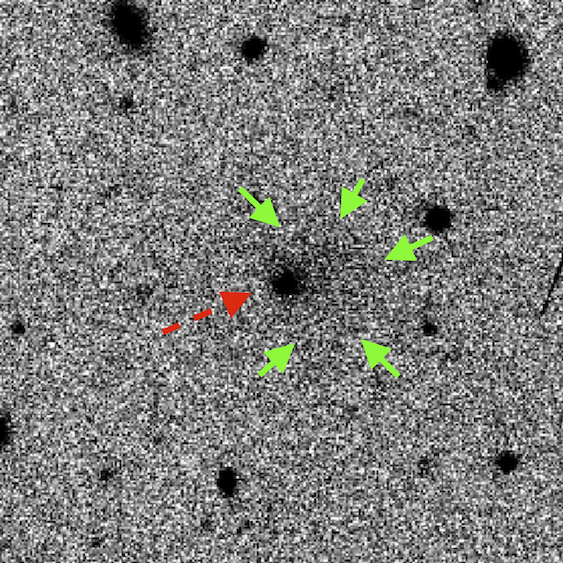}\\
    \includegraphics[width=0.4\linewidth]{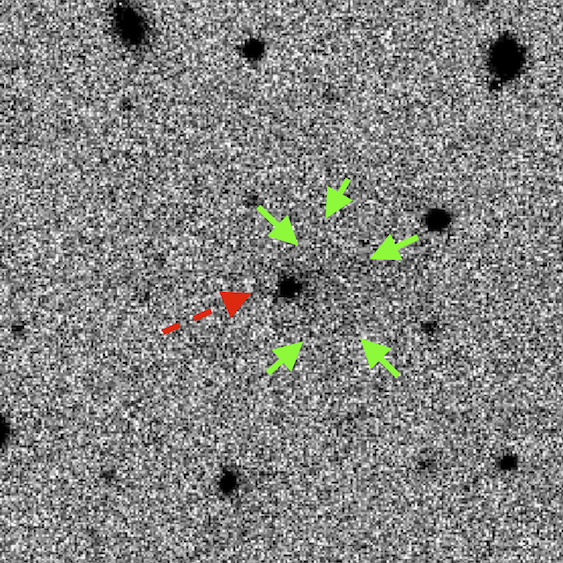} & \includegraphics[width=0.4\linewidth]{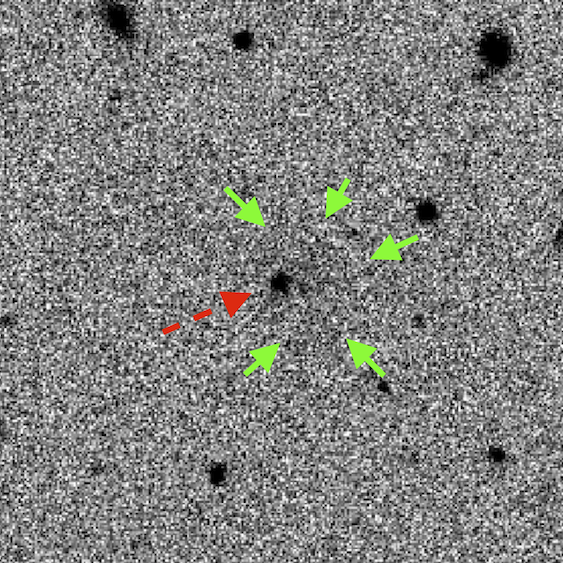}
	\end{tabular}
	\caption{\textit{New DECam Observations Gallery}. Top left: UT 9:54. Top right: UT 9:58. Bottom left: UT 10:03; Bottom right: UT 10:08. All images: (1) dashed red arrow points to \og{}, (2) green arrows highlight the comae if visible, (3) observing date was UT 2019 August 30, (4) filter was \textit{VR}, (5) exposure time was 250~s. The apparent decrease in coma prominence was the result of increasing background noise as images were taken into twilight.}
	\label{og:fig:newobserations}
\end{figure}

\begin{figure}
    \centering
	\begin{tabular}{cc}
		\includegraphics[width=0.45\linewidth]{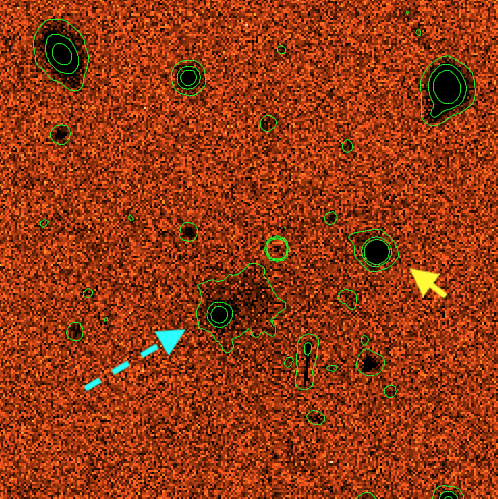} & \includegraphics[width=0.45\linewidth]{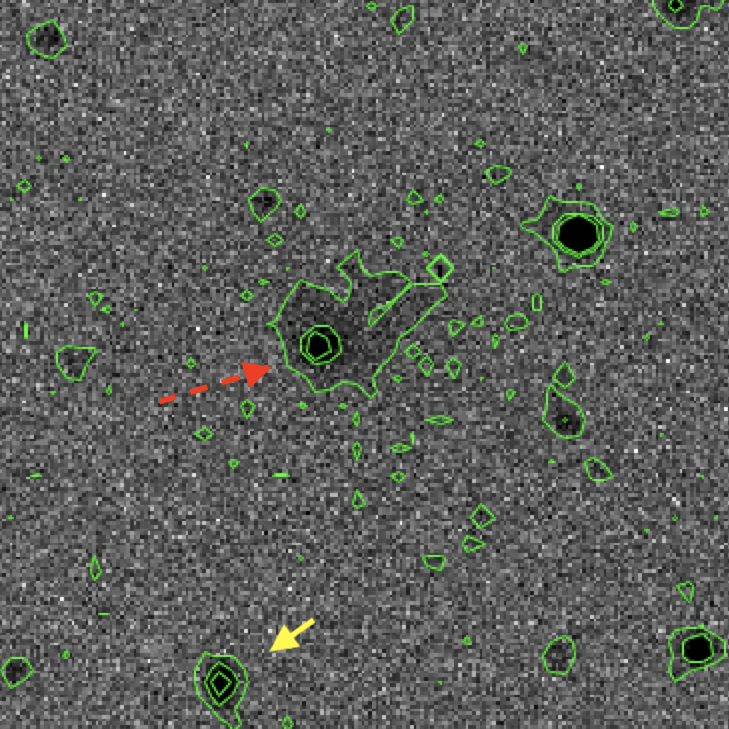}\\
	\end{tabular}
	\caption{\textit{Isophotal Contours.} Isophotal contours indicate the extent and irregularity of the coma emanating from \og{} (dashed arrows), especially when contrasted with background objects (yellow arrows) presenting relatively symmetric radial profiles. These two 250 s \textit{VR}--band exposures were taken at 9:54 (left) and 9:58 (right) during our 2019 August 30 follow-up campaign.}
	\label{og:fig:isophotalcontours}
\end{figure}

\begin{figure}
    \centering
    \begin{tabular}{ccc}
        \includegraphics[width=0.30\linewidth]{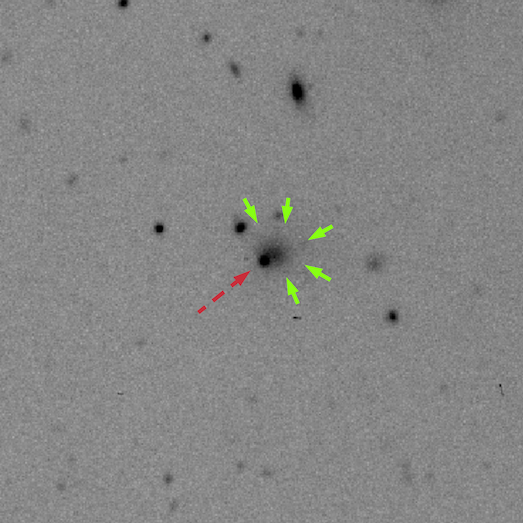} & \includegraphics[width=0.30\linewidth]{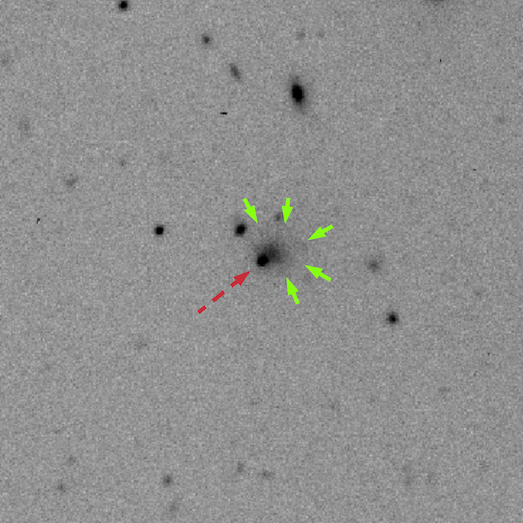} & \includegraphics[width=0.30\linewidth]{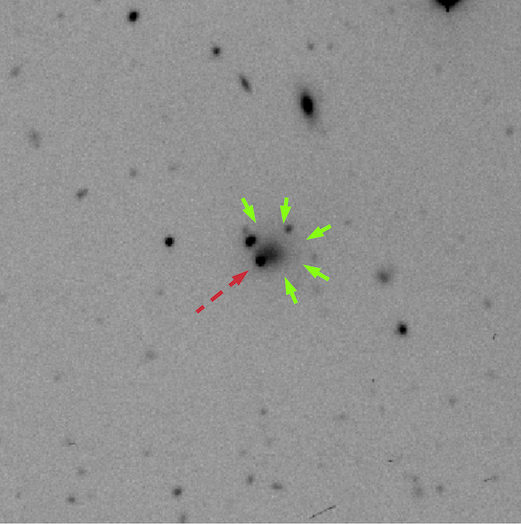} \\
    \end{tabular}
    \textit{New Magellan Observations Gallery}. \caption{\og{} imaged December 27, 2019 via the Magellan 6.5m Baade Telescope using the WB4800-7800 filter on the Inamori-Magellan Areal Camera \& Spectrograph (IMACS) at Las Campanas Observatory on Cerro Manqui, Chile. The three images reveal an apparent coma (green arrows) emerging from the object (red dashed arrow) and were taken at 300~s (left, center) exposures and one 600~s exposure (right).}
    \label{og:fig:magellanobservations}
\end{figure}

\begin{figure}
    \begin{tabular}{ccc}
        \includegraphics[width=0.30\linewidth]{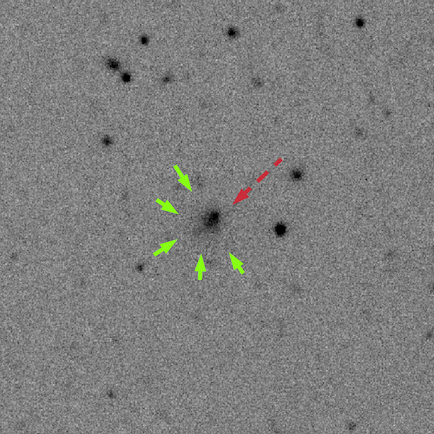} & \includegraphics[width=0.30\linewidth]{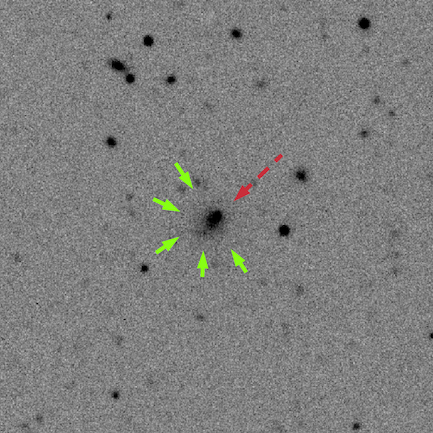} & \includegraphics[width=0.30\linewidth]{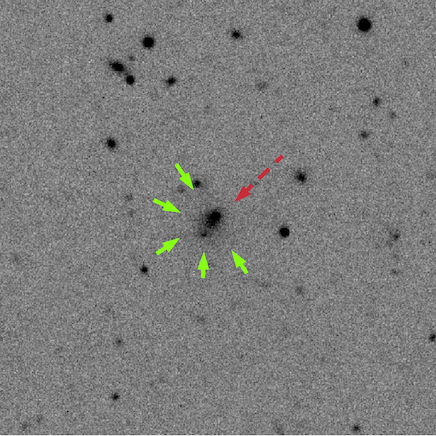} \\
        \includegraphics[width=0.30\linewidth]{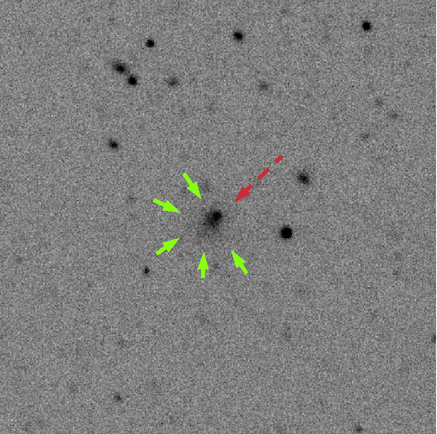} & \includegraphics[width=0.30\linewidth]{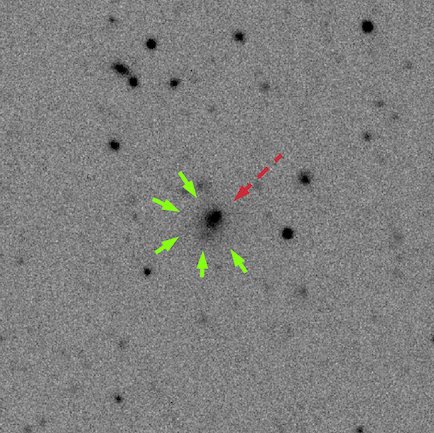} & \includegraphics[width=0.30\linewidth]{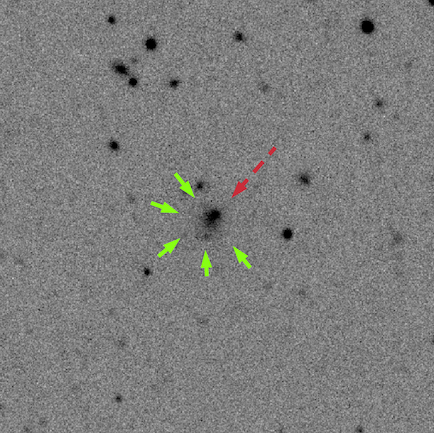} \\
    \end{tabular}
    \caption{\textit{New DCT Observations Gallery.} \og{} imaged December 30, 2019, via the Lowell Observatory 4.3~m Discovery Channel Telescope (Arizona, USA) using the \acf{LMI}. Green arrows trace out a diffuse coma and a dashed red arrow points to the nucleus in each of the six images. Each exposure in the two $g$-$r$-$i$ sequences (top and bottom rows) was 300~s long.}
    \label{og:fig:dctobservations}
\end{figure}

Figure \ref{og:fig:archivalimages} shows six of the archival images in which we originally spotted what appeared to be activity emanating from \og{}. We obtained confirmation first through DECam observations (Figure \ref{og:fig:newobserations}); the coma is more readily apparent in the isophotal contours shown in Figure \ref{og:fig:isophotalcontours}. Figure \ref{og:fig:magellanobservations} shows two additional we took at Magellan provided additional confirmation. Figure \ref{og:fig:dctobservations} shows six images of \og{} we captured a the DCT which enabled us to perform color measurement and radial surface brightness profiling.

\clearpage
\singlespacing
\chapter{Manuscript IV: Recurrent Activity from Active Asteroid (248370) 2005~QN173: A Main-belt Comet} 
\chaptermark{Recurrent Activity from Active Asteroid (248370) 2005~QN173}
\label{chap:2005QN173}
\acresetall

Colin Orion Chandler\footnote{\label{QN:nau}Department of Astronomy and Planetary Science, Northern Arizona University, PO Box 6010, Flagstaff, AZ 86011, USA}, Chadwick A. Trujillo$^\mathrm{\ref{QN:nau}}$, Henry H. Hsieh\footnote{Planetary Science Institute, 1700 East Fort Lowell Rd., Suite 106, Tucson, AZ 85719, USA, Institute of Astronomy and Astrophysics, Academia Sinica, P.O.\ Box 23-141, Taipei 10617, Taiwan}

\textit{This is the Accepted Manuscript version of an article accepted for publication in Astrophysical Journal Letters.  IOP Publishing Ltd is not responsible for any errors or omissions in this version of the manuscript or any version derived from it.  The Version of Record is available online at }\url{https://iopscience.iop.org/article/10.3847/2041-8213/ac365b}\textit{.}

\doublespacing


\section{Abstract}
\label{QN:Abstract}
We present archival observations of main-belt asteroid (248370)~2005~QN$_{173}$ (also designated 433P) that demonstrate this recently discovered active asteroid (a body with a dynamically asteroidal orbit displaying a tail or coma) has had at least one additional apparition of activity near perihelion during a prior orbit. We discovered evidence of this second activity epoch in an image captured 2016 July 22 with the DECam on the 4~m Blanco telescope at the Cerro Tololo Inter-American Observatory in Chile. As of this writing, (248370)~2005~QN$_{173}$ is just the 8th active asteroid demonstrated to undergo recurrent activity near perihelion. Our analyses demonstrate (248370)~2005~QN$_{173}$ is likely a member of the active asteroid subset known as main-belt comets, a group of objects that orbit in the main asteroid belt that exhibit activity that is specifically driven by sublimation. We implement an activity detection technique, \textit{wedge photometry}, that has the potential to detect tails in images of solar system objects and quantify their agreement with computed antisolar and antimotion vectors normally associated with observed tail directions. We present a catalog and an image gallery of archival observations. The object will soon become unobservable as it passes behind the Sun as seen from Earth, and when it again becomes visible (late 2022) it will be farther than 3~au from the Sun. Our findings suggest (248370)~2005~QN$_{173}$ is most active interior to 2.7~au (0.3~au from perihelion), so we encourage the community to observe and study this special object before 2021 December.


\section{Introduction}
\label{QN:introduction}

\begin{figure*}
    \centering
    \includegraphics[width=1.0\linewidth]{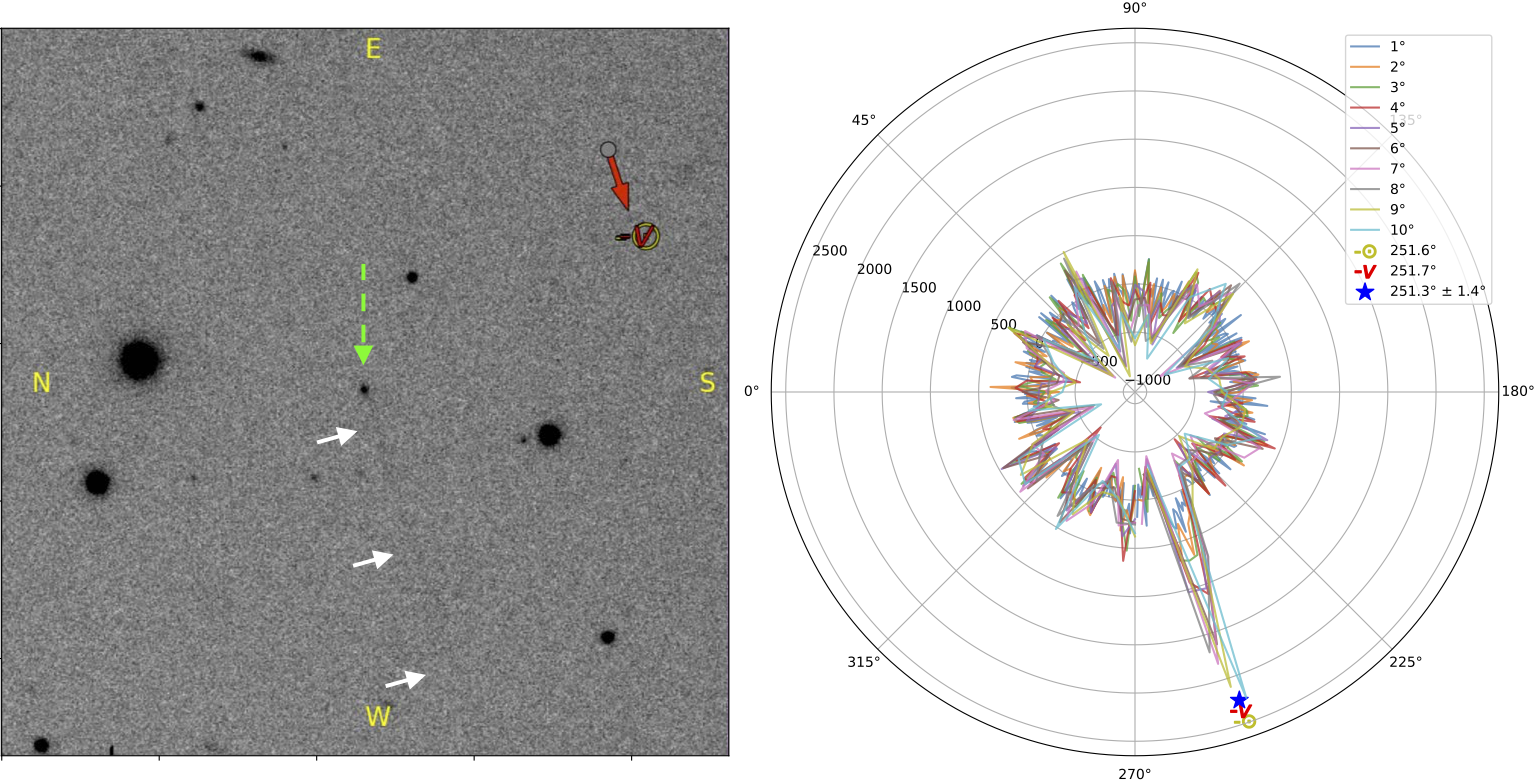}
    \caption{The $126''\times126''$ thumbnail image (left) shows (248370)~2005~QN$_{173}$ (green dashed arrow) at center with a tail (white arrows) oriented towards 5 o'clock. This 89~s $z$-band exposure captured with the DECam is the only image in which we could unambiguously identify activity. We conducted wedge photometry (right) that shows the tail orientation is $251.3^\circ\pm1.4^\circ$ (blue star), in close agreement with the $251.6^\circ$ antisolar angle (yellow $\odot$) and the $251.7^\circ$ antimotion vector (red $v$) as computed by JPL Horizons. The plot shows counts radially outward from the the object center at (0,0).}
    \label{QN:fig:wedgephot}
\end{figure*}

Active asteroids are objects that are dynamically asteroidal but that display comet-like activity such as a tail or coma \citep{hsiehActiveAsteroidsMystery2006}. Activity may be caused by mechanisms unrelated to volatiles (e.g., impact, rotational disruption) or by sublimation as is typically the case with comets. Sublimation driven active objects provide key insights into the present-day volatile distribution in our solar system, as well as clues about the origins of those volatiles and how they arrived on Earth \citep{hsiehPopulationCometsMain2006}. These objects have been persistently difficult to study because of the small numbers detected to date: fewer than 30 active asteroids, of which roughly half are thought to exhibit sublimation driven activity; see \citealt{chandlerSAFARISearchingAsteroids2018} for a summary.

When the aforementioned sublimation driven activity is connected with a main-belt asteroid, the object is classified as a main-belt comet (MBC). MBCs are often characterized by activity near perihelion and the absence of activity elsewhere in the orbit \citep{hsiehMainbeltCometsPanSTARRS12015,agarwalBinaryMainbeltComet2017,hsieh2016ReactivationsMainbelt2018}, suggesting that the primary activity mechanism is sublimation of volatiles such as water ice \citep{snodgrassMainBeltComets2017}.

By contrast, stochastic events like impacts may result in comet-like activity but, in such cases, the appearance of activity is expected to cease once the material dissipates. Roughly 60\% of known active asteroids have been observed to display activity during only a single apparition \citep{chandlerSAFARISearchingAsteroids2018}.

Asteroid (7968), now comet 133P/Elst-Pizarro, was the first active main-belt asteroid to be discovered. While it was unclear at the time whether the activity was sublimation driven \citep{boehnhardtComet1996N21996,boehnhardtImpactInducedActivityAsteroidComet1998} or due to a one-time event \citep{tothImpactgeneratedActivityPeriod2000}, subsequent apparitions showing activity indicated sublimation was the cause \citep{hsiehStrangeCase133P2004,hsiehReturnActivityMainbelt2010}. This example illustrates the importance of detecting additional activity epochs.

Asteroid (248370)~2005~QN$_{173}$ is a 3.2$\pm$0.4~km diameter \citep{hsiehPhysicalCharacterizationMainbelt2021} outer main-belt asteroid ($a$=3.075~au, $e$=0.226, $i$=$0.067^\circ$) that has a 5.37~yr orbit that ranges from a perihelion distance of $q$=2.374~au to an aphelion distance of $Q$=3.761~au. The object first drew particular attention when it was reported as active on 2021 July 9 \citep{fitzsimmons2483702005QN1732021}. Subsequently, Zwicky Transient Facility (ZTF) data were used to help constrain the activity onset to between 2020 July 10 and 2021 June 11 \citep{kelley2483702005QN2021}. 

We set out to locate archival astronomical images of (248370)~2005~QN$_{173}$ in order to characterize prior activity. We made use of solar system object thumbnails (small image cutouts like Figure \ref{QN:fig:wedgephot}) derived from publicly available archival data. We previously demonstrated how our data sources, such as the Dark Energy Camera (DECam), are well suited to discovering and characterizing active objects \citep{chandlerSAFARISearchingAsteroids2018,chandlerSixYearsSustained2019,chandlerCometaryActivityDiscovered2020a}.

Here we report activity of (248370)~2005~QN$_{173}$ on 2016 July 22 \citep{chandler2483702005QN2021a}, an apparition prior to the 2021 outburst. We describe the process by which the activity was identified and examine the implications of this discovery.


\section{Second Activity Epoch}
\label{QN:sec:secondActivityEpoch}

In order to find an additional activity epoch for (248370)~2005~QN$_{173}$, we searched, assessed, and analyzed publicly available archival image data, building upon the methods of \cite{chandlerSAFARISearchingAsteroids2018,chandlerSixYearsSustained2019,chandlerCometaryActivityDiscovered2020a}. 

\subsection{Data Acquisition}
\label{QN:subsec:datamining}

To locate archival images of (248370)~2005~QN$_{173}$, we queried our own database of publicly available observation metadata \citep[see][]{chandlerSAFARISearchingAsteroids2018}. This database, which updates daily, includes observing details such as sky coordinates, exposure time, and filter selection. Additionally, we searched Palomar Transient Factory (PTF) and ZTF data through 2021 August 31 through online search tools (listed in Appendix \ref{QN:sec:equipQuickRef}) as well as a ZTF Alert Stream search and retrieval tool we created for this purpose. All instruments and data sources we made use of are listed in Appendix \ref{QN:sec:equipQuickRef}, and we note that some data were found or retrieved via more than one pathway.

\clearpage
\pagestyle{empty}

\atxy{\dimexpr1in}{.5\paperheight}{\rotatebox[origin=center]{270}{\thepage}}
\begin{sidewaystable}[h]

\centering
\footnotesize
    \caption{(248370)~2005~QN$_{173}$ Observations}
    \begin{tabular}{cclcrcccrrccl}
        Image\footnote{Label in image gallery figures.} & Obs. Date\footnote{UT observing date in year-month-day format.} & Source      & N\footnote{Number of images.} & Exp. [s]\footnote{Exposure time for each image.} & Filter & $V$\footnote{Apparent $V$-band magnitude (Horizons).} & r [au]\footnote{Heliocentric distance.} & STO [$^\circ$]\footnote{Sun--target--observer angle.}  & $\nu$ [$^\circ$]\footnote{True anomaly.} & \%$_{Q\rightarrow q}$\footnote{Percentage to perihelion $q$ from aphelion $Q$.} & Act?\footnote{Activity observed.} & Archive\\
        \hline
a       & 2004-07-08               & MegaPrime   & 3                       & 180                       & \textit{i     } & 20.7                      & 2.74                  & 17.6                    & 287.5                 & 73\%                        & N    & CADC,*      \\
b       & 2005-06-08               & SuprimeCam  & 3                       & 60                        & \textit{W-J-VR} & 21.0                      & 2.42                  & 23.8                    & 18.6                  & 96\%                        & N    & CADC,SMOKA  \\
c       & 2010-06-14               & Pan-STARRS1 & 2                       & 30                        & \textit{z     } & 20.2                      & 2.42                  & 21.1                    & 339.2                 & 96\%                        & N    & CADC        \\
d       & 2010-08-02               & Pan-STARRS1 & 1                       & 45                        & \textit{i     } & 19.0                      & 2.39                  & 3.9                     & 353.5                 & 99\%                        & N    & CADC        \\
e       & 2010-08-05               & Pan-STARRS1 & 1                       & 40                        & \textit{r     } & 18.9                      & 2.39                  & 2.5                     & 355.4                 & 99\%                        & N    & CADC        \\
f       & 2010-08-06               & Pan-STARRS1 & 1                       & 43                        & \textit{g     } & 18.8                      & 2.39                  & 2.0                     & 354.7                 & 99\%                        & N    & CADC        \\
g       & 2010-08-28               & PTF         & 2                       & 60                        & \textit{r     } & 19.2                      & 2.39                  & 8.4                     & 1.2                   & 99\%                        & N    & IRSA/PTF    \\
h       & 2010-08-31               & Pan-STARRS1 & 2                       & 45                        & \textit{i     } & 19.3                      & 2.39                  & 9.7                     & 2.1                   & 99\%                        & N    & CADC        \\
i       & 2010-09-01               & PTF         & 2                       & 60                        & \textit{r     } & 19.3                      & 2.39                  & 10.1                    & 2.3                   & 99\%                        & N    & IRSA/PTF    \\
j       & 2010-09-06               & Pan-STARRS1 & 2                       & 43                        & \textit{g     } & 19.5                      & 2.39                  & 12.2                    & 3.8                   & 99\%                        & N    & CADC        \\
k       & 2010-09-12               & Pan-STARRS1 & 2,2 & 43,40 & \textit{g,r   } & 19.6                      & 2.40                  & 14.5                    & 5.6                   & 98\%                        & N    & CADC        \\
l       & 2010-09-15               & PTF         & 2                       & 60                        & \textit{r     } & 19.7                      & 2.39                  & 15.5                    & 6.5                   & 99\%                        & N    & IRSA/PTF    \\
m       & 2010-10-30               & Pan-STARRS1 & 2                       & 30                        & \textit{z     } & 20.6                      & 2.41                  & 23.8                    & 19.6                  & 97\%                        & N    & CADC        \\
n       & 2011-07-14               & Pan-STARRS1 & 1                       & 40                        & \textit{r     } & 21.8                      & 2.86                  & 15.8                    & 84.2                  & 65\%                        & N    & CADC        \\
o       & 2011-11-24               & Pan-STARRS1 & 2,2 & 43,40 & \textit{g,r   } & 20.6                      & 3.15                  & 4.0                     & 109.3                 & 44\%                        & N    & CADC        \\
p       & 2011-11-30               & Pan-STARRS1 & 2                       & 45                        & \textit{i     } & 20.4                      & 3.16                  & 1.8                     & 110.3                 & 43\%                        & N    & CADC        \\
q       & 2011-12-01               & Pan-STARRS1 & 2                       & 43                        & \textit{g     } & 20.4                      & 3.16                  & 1.4                     & 110.5                 & 43\%                        & N    & CADC        \\
r       & 2014-03-01               & OmegaCam    & 5                       & 360                       & \textit{r     } & 21.4                      & 3.53                  & 7.0                     & 218.2                 & 17\%                        & N    & CADC,ESO    \\
s       & 2016-07-22               & DECam       & 1                       & 89                        & \textit{z     } & 21.2                      & 2.59                  & 22.7                    & 56.5                  & 84\%                        & Y    & CADC,*      \\
t       & 2019-07-03               & DECam       & 9                       & 40                        & \textit{VR    } & 22.6                      & 3.55                  & 14.6                    & 216.9                 & 15\%                        & N    & CADC,*      \\
u       & 2020-02-04               & DECam       & 1                       & 38                        & \textit{r     } & 21.6                      & 3.16                  & 18.1                    & 248.9                 & 43\%                        & N    & CADC,*      \\
v       & 2020-02-10               & DECam       & 1                       & 199                       & \textit{z     } & 21.8                      & 3.15                  & 18.2                    & 249.9                 & 44\%                        & N    & CADC,*      \\
w       & 2020-04-25               & ZTF         & 1,1 & 30,30 & \textit{g,r   } & 20.0                      & 2.99                  & 4.4                     & 263.3                 & 55\%                        & N    & IRSA/ZTF    \\
x       & 2020-05-18               & ZTF         & 1,1 & 30,30 & \textit{g,r   } & 19.9                      & 2.93                  & 4.7                     & 267.7                 & 60\%                        & N    & IRSA/ZTF    \\
x       & 2020-05-27               & ZTF         & 1,1 & 30,30 & \textit{g,r   } & 20.1                      & 2.91                  & 8.4                     & 269.5                 & 61\%                        & N    & IRSA/ZTF    \\
x       & 2020-06-11               & ZTF         & 1,3 & 30,30 & \textit{g,r   } & 20.3                      & 2.88                  & 13.3                    & 272.5                 & 63\%                        & N    & IRSA/ZTF    \\
x       & 2020-06-14               & ZTF         & 1,1 & 30,30 & \textit{g,r   } & 20.4                      & 2.87                  & 14.1                    & 273.1                 & 64\%                        & N    & IRSA/ZTF    \\
x       & 2020-06-17               & ZTF         & 2,2 & 30,30 & \textit{g,r   } & 20.4                      & 2.87                  & 14.9                    & 273.8                 & 64\%                        & N    & IRSA/ZTF    \\
x       & 2020-06-20               & ZTF         & 2,2 & 30,30 & \textit{g,r   } & 20.5                      & 2.86                  & 15.7                    & 274.4                 & 65\%                        & N    & IRSA/ZTF    \\
x       & 2020-06-23               & ZTF         & 2,1 & 30,30 & \textit{g,r   } & 20.5                      & 2.85                  & 16.5                    & 275.0                 & 65\%                        & N    & IRSA/ZTF    \\
x       & 2020-06-26               & ZTF         & 4,1 & 30,30 & \textit{g,r   } & 20.6                      & 2.85                  & 17.2                    & 275.6                 & 65\%                        & N    & IRSA/ZTF\\
\hline
    \end{tabular}
    \footnotesize
    \raggedright
    
    Note \textit{W-J-VR} is a single wide-band filter. See Appendix \ref{QN:sec:equipQuickRef} for image source and archive details.
    
    * indicates the data were obtained from our local repository.
    
    Table entries with multiple comma-separated values contain groups of exposures taken with different filters.
    \label{QN:tab:observations}

\end{sidewaystable}

\clearpage
We identified candidate images where (248370)~2005~QN$_{173}$ was expected to be within the field of view (FOV) based on observation times, pointing centers, and FOV sizes and orientations, downloaded associated data, and extracted image cutouts. We organized data by instrument and observation date and summarize observation details in Table \ref{QN:tab:observations}.

\begin{figure*}
    \centering
    \begin{tabular}{ccccc}
    \begin{overpic}[width=0.17\linewidth]{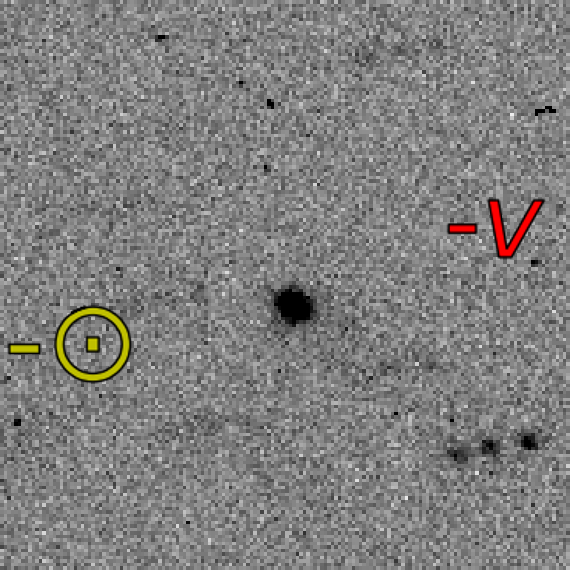}\put (5,7) {\huge\color{green} \textbf{a}}\end{overpic} & 
    \begin{overpic}[width=0.17\linewidth]{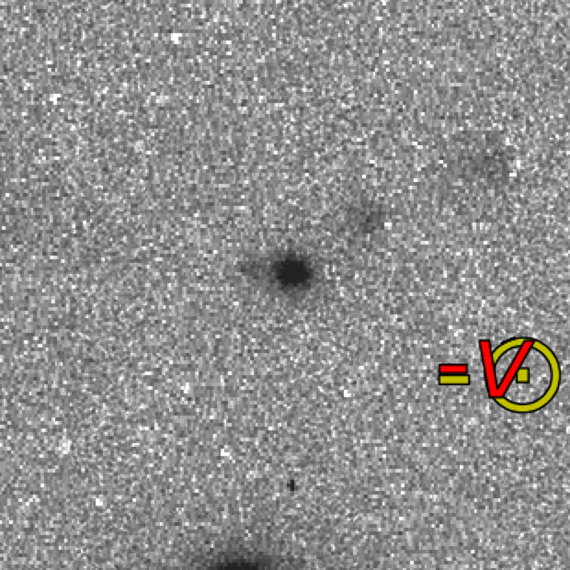}\put (5,7) {\huge\color{green} \textbf{b}}\end{overpic} &
    \begin{overpic}[width=0.17\linewidth]{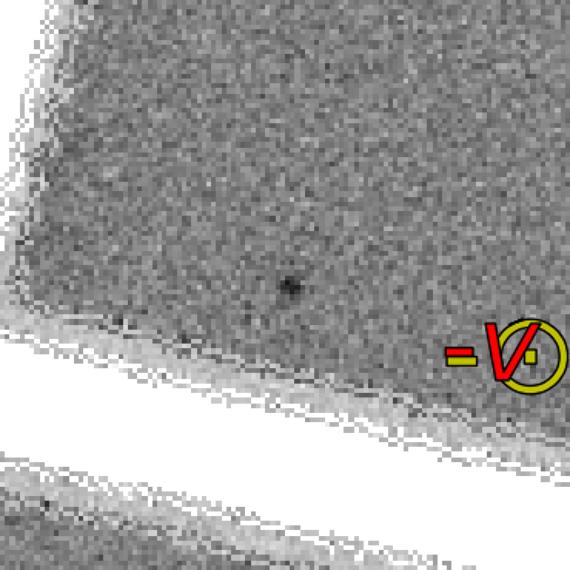}\put (5,7) {\huge\color{green} \textbf{c}}\end{overpic} &
    \begin{overpic}[width=0.17\linewidth]{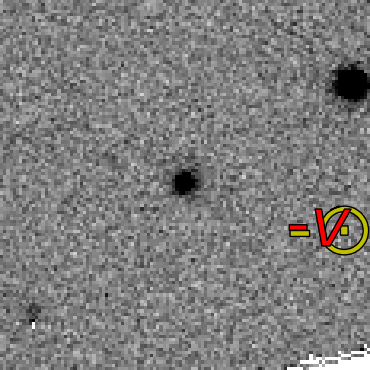}\put (5,7) {\huge\color{green} \textbf{d}}\end{overpic} &
    \begin{overpic}[width=0.17\linewidth]{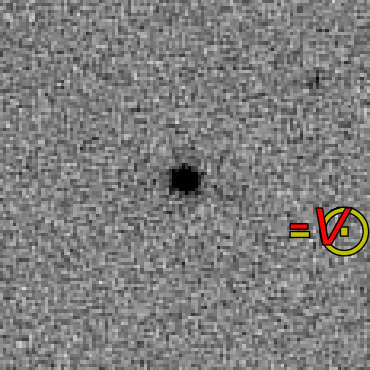}\put (5,7) {\huge\color{green} \textbf{e}}\end{overpic} \\
    \begin{overpic}[width=0.17\linewidth]{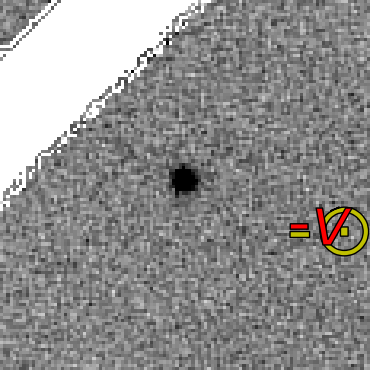}\put (5,7) {\huge\color{green} \textbf{f}}\end{overpic} &
    \begin{overpic}[width=0.17\linewidth]{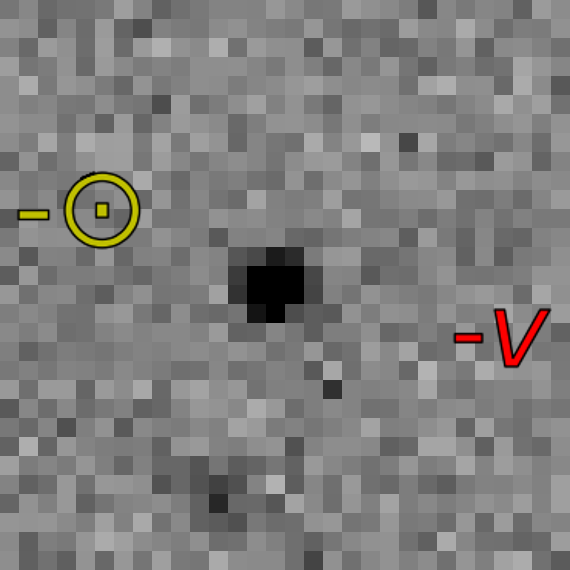} \put (5,7) {\huge\color{green} \textbf{g}}\end{overpic} &
    \begin{overpic}[width=0.17\linewidth]{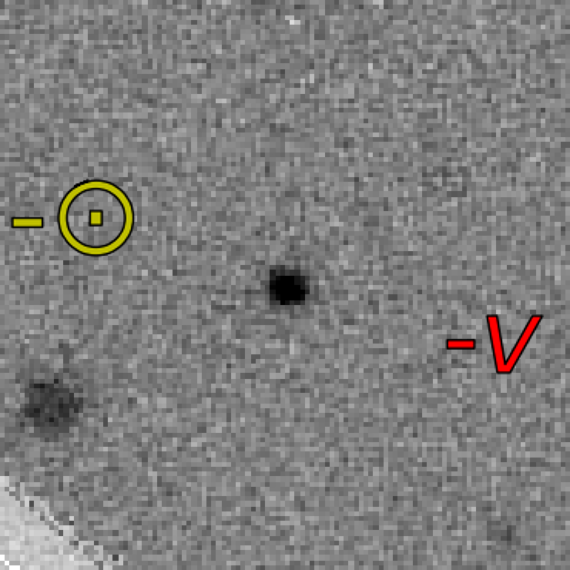}\put (5,7) {\huge\color{green} \textbf{h}}\end{overpic} &
    \begin{overpic}[width=0.17\linewidth]{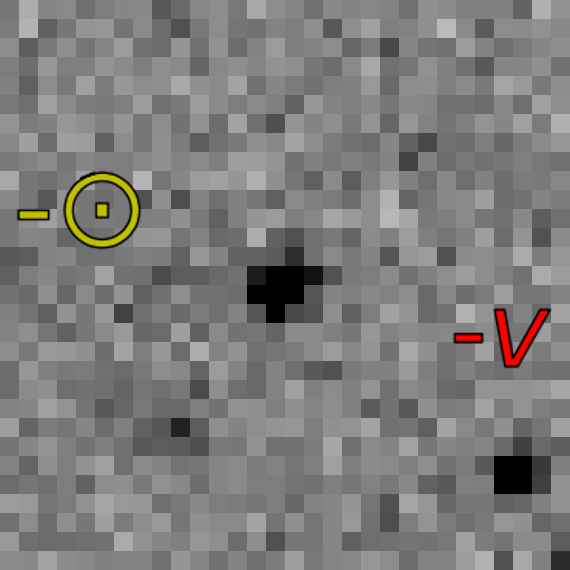} \put (5,7) {\huge\color{green} \textbf{i}}\end{overpic} &
    \begin{overpic}[width=0.17\linewidth]{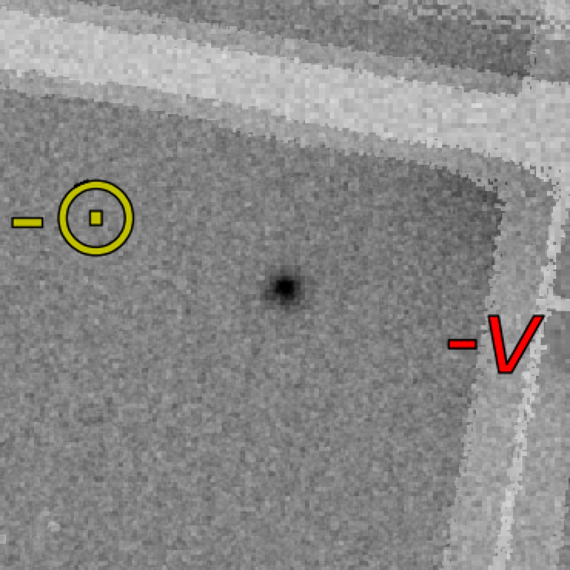} \put (5,7) {\huge\color{green} \textbf{j}}\end{overpic}\\
    \begin{overpic}[width=0.17\linewidth]{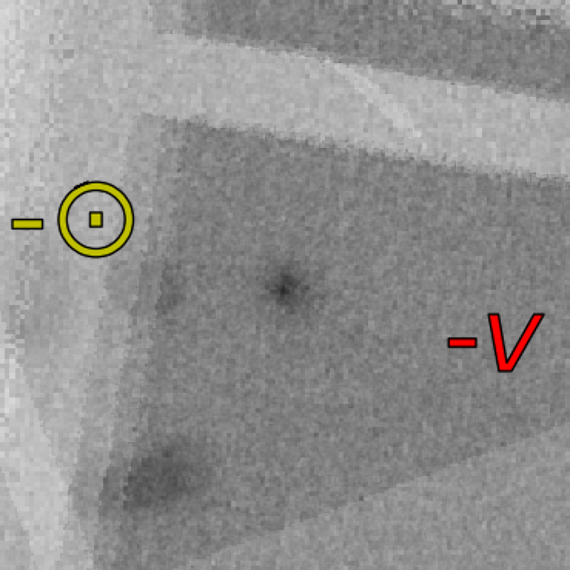} \put (5,7) {\huge\color{green} \textbf{k}}\end{overpic} &
    \begin{overpic}[width=0.17\linewidth]{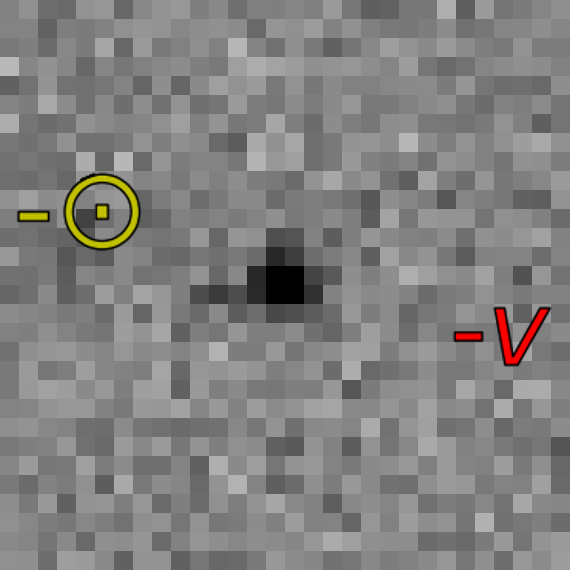} \put (5,7) {\huge\color{green} \textbf{l}}\end{overpic} &
    \begin{overpic}[width=0.17\linewidth]{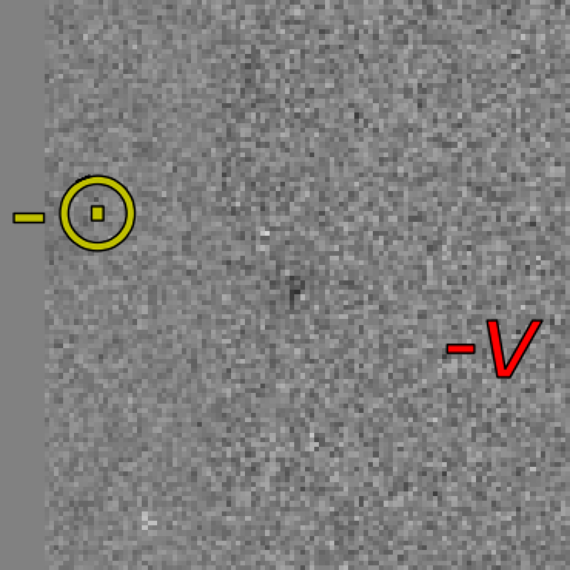} \put (5,7) {\huge\color{green} \textbf{m}}\end{overpic} &
    \begin{overpic}[width=0.17\linewidth]{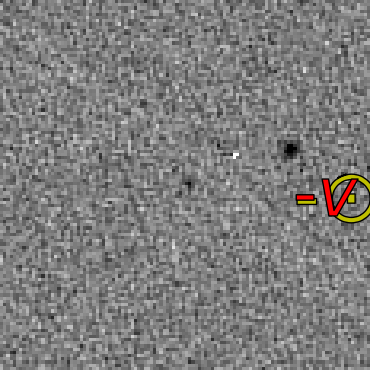} \put (5,7) {\huge\color{green} \textbf{n}}\end{overpic} &
    \begin{overpic}[width=0.17\linewidth]{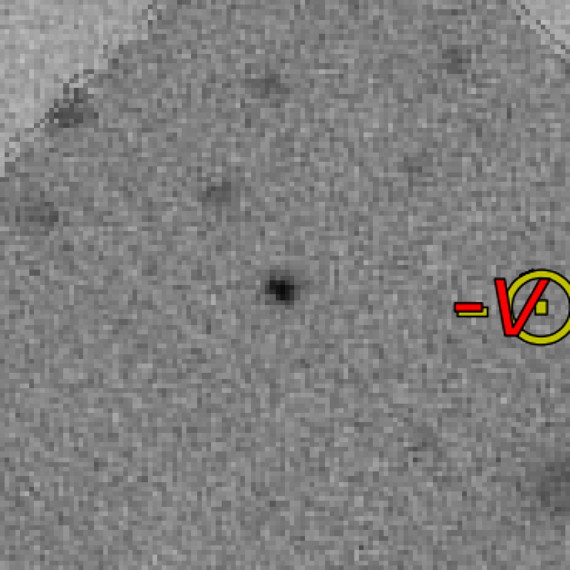}\put (5,7) {\huge\color{green} \textbf{o}}\end{overpic} \\
    \begin{overpic}[width=0.17\linewidth]{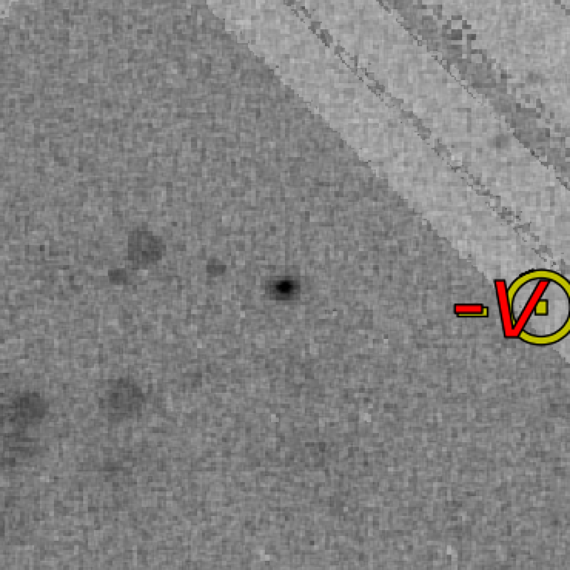}\put (5,7) {\huge\color{green} \textbf{p}}\end{overpic} &
    \begin{overpic}[width=0.17\linewidth]{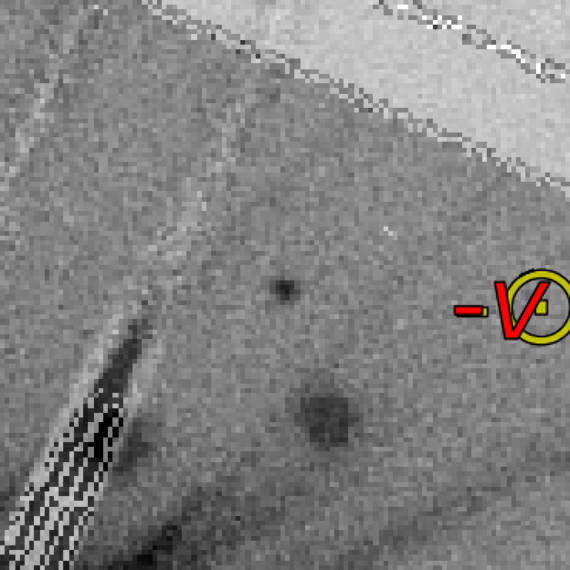}\put (5,7) {\huge\color{green} \textbf{q}}\end{overpic} &
    \begin{overpic}[width=0.17\linewidth]{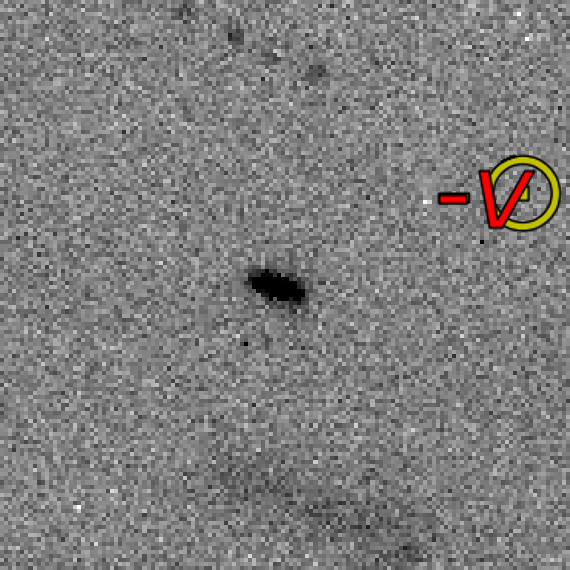}\put (5,7) {\huge\color{green} \textbf{r}}\end{overpic} & 
    \begin{overpic}[width=0.17\linewidth]{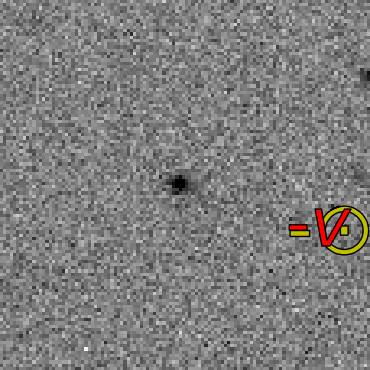}\put (5,7) {\huge\color{green} \textbf{s}}\end{overpic} &
    \begin{overpic}[width=0.17\linewidth]{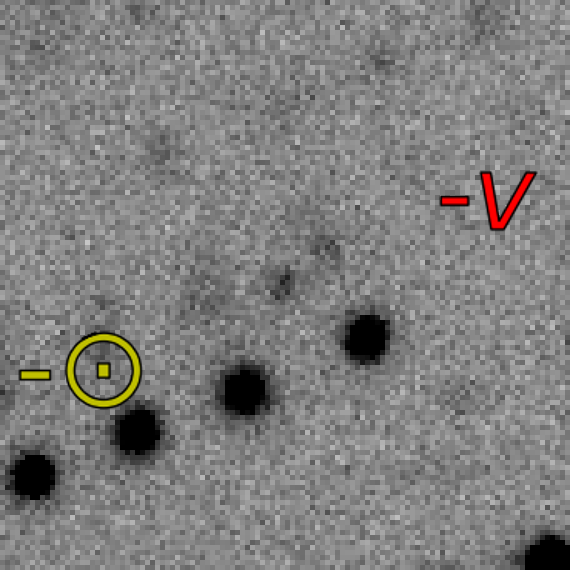}\put (5,7) {\huge\color{green} \textbf{t}}\end{overpic} \\
    \begin{overpic}[width=0.17\linewidth]{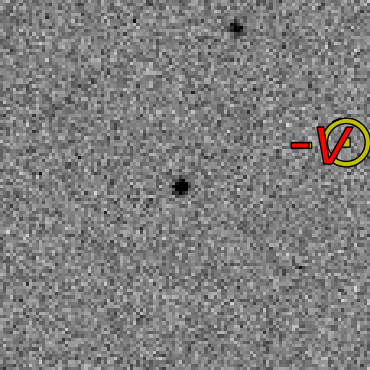}\put (5,7) {\huge\color{green} \textbf{u}}\end{overpic} & 
    \begin{overpic}[width=0.17\linewidth]{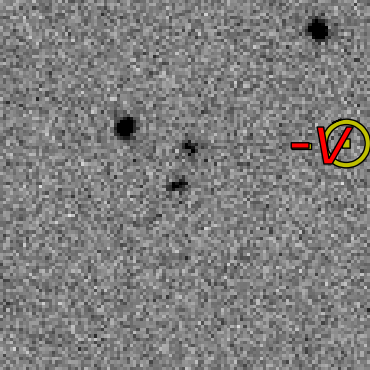}\put (5,7) {\huge\color{green} \textbf{v}}\end{overpic} & 
    \begin{overpic}[width=0.17\linewidth]{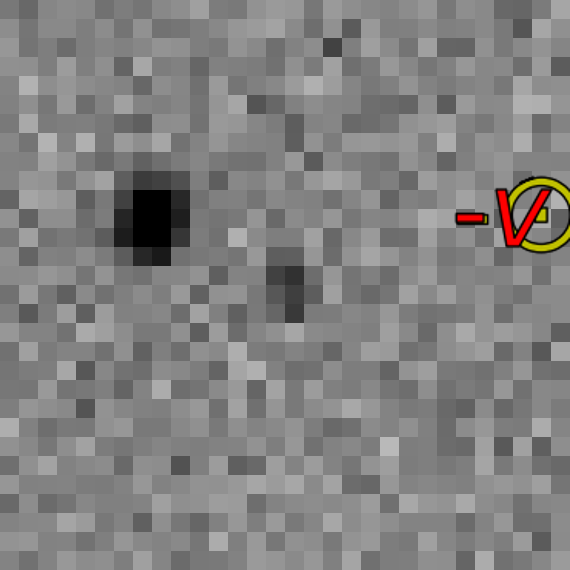}\put (5,7) {\huge\color{green} \textbf{w}}\end{overpic} &
    \begin{overpic}[width=0.17\linewidth]{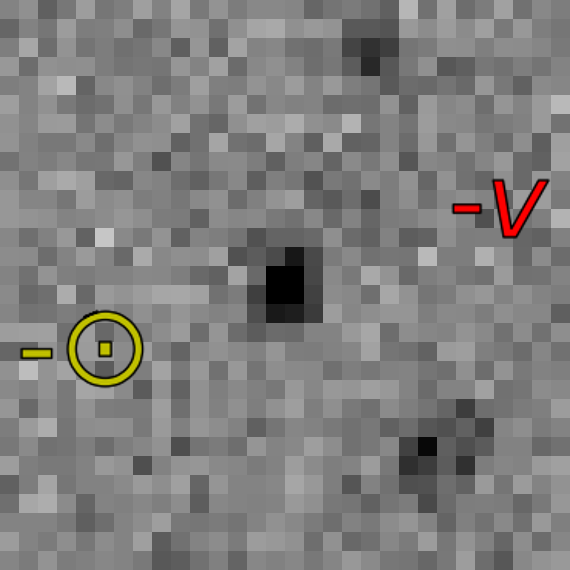}\put (5,7) {\huge\color{green} \textbf{x}}\end{overpic} \\ 
    \end{tabular}
    \caption{Archival images of (248370)~2005~QN$_{173}$ with the best activity detection potential (i.e., sufficient depth and observing conditions) for (248370)~2005~QN$_{173}$. For all images, north is up, east is left, the FOV is $30''\times30''$, the antisolar (yellow -$\odot$) and antimotion (red -$v$) directions are shown with the origin at image center. See Appendix \ref{QN:sec:equipQuickRef} for instrument and archive details. Panel (s) is from the only thumbnail in which we could identify activity unambiguously (Figure \ref{QN:fig:wedgephot}). (a) 2004 July 8 MegaPrime 3$\times$180~s $i$ band. (b) 2005 June 8 SuprimeCam 3$\times$60~s \textit{W-J-VR} band. (c) 2010 June 14 PS1 2$\times$60~s $z$ band. (d) 2010 August 2 PS1 1$\times$45~s $i$ band. (e) 2010 August 5 PS1 1$\times$40~s $r$ band. (f) 2010 August 6 PS1 1$\times$43~s $g$ band. (g) 2010 August 28 PTF 2$\times$60~s $r$ band. (h) 2010 August 31 PS1 $\times$45~s $i$ band. (i) 2010 September 1 PTF 2$\times$60~s $r$ band. (j) 2010 September 6 PS1 2$\times$43~s $g$ band. (k) 2010 September 12 PS1 2$\times$40~s $r$ band + 2$\times$43~s $g$ band. (l) 2010 September 15 PTF 2$\times$60~s $r$ band. (m) 2010 October 30 PS1 2$\times$30~s $z$ band. (n) 2011 July 14 PS1 1$\times$40~s $r$ band. (o) 2011 November 24 PS1 2$\times$40~s $r$ band + 2$\times$43~s $g$ band. (p) 2011 November 30 PS1 2$\times$45~s $i$ band. (q) 2011 December 1 PS1 2$\times$43~s $g$ band. (r) 2014 January 31 OmegaCAM 5$\times$360~s $r$ band. (s) 2016 July 22 1$\times$89~s \textit{z} band. (t) 2019 July 3 DECam 9$\times$40~s \textit{VR} band. (u) 2020 February 4 DECam 1$\times$38~s \textit{r} band. (v) 2020 February 10 DECam 1$\times$199~s $z$ band. (w) 2020 April 25 ZTF 1$\times$30~s \textit{g} band + 1$\times$30~s \textit{r} band. (x) 2020 May 18, 27 + 2020 June 11, 14, 17, 20, 23, \& 26 ZTF 9$\times$30~s \textit{r} band + 12$\times$30~s \textit{g} band.}
    \label{QN:fig:gallery}
\end{figure*}

We extracted eighty-one $480\times480$ pixel thumbnail images (such as the image shown in Figure \ref{QN:fig:wedgephot}) in which we could confidently identify (248370)~2005~QN$_{173}$ (Table \ref{QN:tab:observations}). We coadded images from the same instrument when observations were close enough in time for computed tail orientation to be in close agreement such that coaddition could enhance activity, if present. The thumbnails with the ``best activity detection potential'' -- meaning the images were judged to have observing conditions (e.g., seeing) and depth (i.e., magnitude limit) amenable to activity detection -- are shown in Figure \ref{QN:fig:gallery}. 
To allow for uniform spatial comparisons and to magnify the region of interest around (248370)~2005~QN$_{173}$, all thumbnail images in Figure \ref{QN:fig:gallery} 
are displayed with $30''\times30''$ fields of view.

\subsection{Image Assessment}
\label{QN:subsec:imageassessment}

We vetted each thumbnail by visually confirming (248370)~2005~QN$_{173}$ was visible. In cases where the object could not be readily identified, we employed our pipeline to produce comparison thumbnails derived from DECam data that showed the same region of sky, instrument, broadband filter, and exposure time, but from epochs when the object was not in the FOV. We made use of Gaia DR2 \citep{gaiacollaborationGaiaDataRelease2018} and Sloan Digital Sky Survey Release 9 (SDSS~DR-9) catalogs \citep{ahnNinthDataRelease2012} to visually validate World Coordinate System of images within the SAOImageDS9 Vizier \citep{ochsenbeinVizieRDatabaseAstronomical2000} catalog query system.

We next identified vetted thumbnails that were suitable for coaddition by clustering images based on instrument and date. For compatible image sets that included multiple broadband filters, we carried out coaddition among matching filters as well as combining all images. Finally, we visually examined the results and flagged images with potential activity.

We found a single image with clear evidence of activity (Figure \ref{QN:fig:wedgephot}) in an 89~s $z$-band exposure captured 2016 July 22 by Dustin Lang and Alistair Walker as part of the DECam Legacy Survey \citep[DECaLS;][]{deyOverviewDESILegacy2019}. This discovery makes (248370)~2005~QN$_{173}$ the ninth recurrently active main-belt asteroid to be identified to date. The other objects, 238P/Read, 259P/Garradd, 288P, 311P/PANSTARRS, 313P/Gibbs, 324P/La Sagra, (6478)~Gault, and (7968)~Elst-Pizarro, have all demonstrated recurrent activity near perihelion, with the exception of (6478)~Gault \citep{chandlerSixYearsSustained2019}.

We measured the tail length to be about $2\farcmin14$ ($2.4\times10^5$~km) in this image but a longer tail may well have been revealed with a longer exposure (see \citet{hsiehPhysicalCharacterizationMainbelt2021} for 2021 apparition tail measurement). Applying our wedge photometry technique (Section \ref{QN:subsec:wedgephotometry}), we produced a diagnostic plot (Figure \ref{QN:fig:wedgephot}) and measured a position angle on the sky of $251.3^\circ\pm1.4^\circ$ for the tail, in close agreement with the Horizons computed $251.6^\circ$ antisolar and $251.7^\circ$ antimotion vectors.

\subsection{Wedge Photometry Tail Tool}
\label{QN:subsec:wedgephotometry}

We crafted a new algorithm to (a) identify potentially active objects by detecting likely tails, and (b) quantify alignment between an observed tail and predicted antisolar and antimotion vectors, which are commonly associated with tail direction. Currently, the tool is designed to analyze single tails $<15^\circ$ in angular extent, though we plan to address multiple tails and comae in the future. The technique, which we refer to as \textit{wedge photometry}, sums all pixel values in a variable-width wedge bound between an inner and outer radius and identifies wedges containing excess flux relative to other wedges, if present. A similar approach was used in \cite{2011Icar..215..534S} but we have made improvements in angular resolution and algorithmic efficiency. Excess flux within a particular wedge may indicate the presence of a tail, and testing tail alignment with antisolar and antimotion vectors provides additional weight that a detected tail is real. Here we focused on quantifying tail orientation and position angle agreement.

To optimize the process, we convert Cartesian pixel coordinates ($x$,$y$) to polar coordinates ($r$,$\theta$) where the central thumbnail pixel is defined as (0,0). The resulting three-dimensional array has columns $r$, $\theta$, and $c$ (counts).

For a series of wedge sizes $\theta$ ($1^\circ$--$10^\circ$ in $1^\circ$ increments) we summed pixel values in annular segments spanning an angle $\pm\theta/2$ along a radial component $r$ from an inner bound, $r_0=5$ pixels, to a maximum of $r_\mathrm{max}=50$ pixels ($13''$ for our DECam data), as given by

\begin{equation}
    c(\theta, \Delta\theta) = \sum_{\theta=-\Delta\theta/2}^{\theta=+\Delta\theta/2} \sum_{r=r_0}^{r=r_\mathrm{max}} c(r, \theta).
\end{equation}

\noindent We further optimize the procedure by selecting for the target a starting radius $r_0$ outside the FWHM, and choosing a maximum radius $_\mathrm{max}$ that allows for a wedge length long enough to ensure that all bins have sufficient counts to avoid necessitating resampling of any individual pixels. Thus, pixels are assigned to wedges based solely on their precise pixel center coordinate, and any fractional flux from a pixel that spans a wedge boundary is automatically assigned to the wedge containing the pixel center coordinate. We then compute the mean and standard deviation of the resulting counts for each $\theta$ to compare with the predicted antisolar and antimotion vectors.

We produce a polar plot (Figure \ref{QN:fig:wedgephot}) to aid assessing relative radial flux distribution. Most position angles have a $1\sigma$ of $\sim$200 counts. The tail is clearly identified by our algorithm at $> 7 \sigma$ for several $\Delta \theta$ wedge sizes.


\section{Main-belt Comet Classification}
\label{QN:sec:mainBeltComet}

Once we had identified a previous activity epoch we set out to determine if (248370)~2005~QN$_{173}$ could be an MBC.

\subsection{Prerequisites}
\label{QN:subsec:mbc}
For (248370)~2005~QN$_{173}$ to qualify as an MBC it must (1) be an active asteroid, (2) orbit within the Main Asteroid Belt, and (3) exhibit sublimation driven activity.

(1) To qualify as an active asteroid, a body must typically meet three criteria (see \citealt{jewittActiveAsteroids2012} for discussion): (i) A coma or tail must have been observed visually (as is the case in this work) or, potentially, through alternate means such as spectroscopy \citep[e.g.,][]{kuppersLocalizedSourcesWater2014,busarevNewCandidatesActive2018} or detecting magnetic field enhancements \citep[e.g.,][]{russellInterplanetaryMagneticField1984}. (ii) The semi-major axis $a$ must not be exterior to that of Jupiter ($a_\mathrm{J}\approx$ 5.2~au) as is the case for comets and active Centaurs \citep{jewittActiveCentaurs2009}; (248370)~2005~QN$_{173}$ has $a=$3.1~au. And (iii) the Tisserand parameter with respect to Jupiter, $T_\mathrm{J}$, must be greater than 3; this is because objects with $T_\mathrm{J}<3$ are canonically considered cometary and $T_\mathrm{J}>3$ asteroids \citep{vaghiOriginJupiterFamily1973,vaghiOrbitalEvolutionComets1973}.

$T_\mathrm{J}$ describes how an orbit is related to Jupiter by 

\begin{equation}
    T_\mathrm{J} = \frac{a_\mathrm{J}}{a} + 2 \sqrt{\frac{a\left(1-e^2\right)}{a_\mathrm{J}}}\cos\left(i\right).
\end{equation}

\noindent where $e$ is the eccentricity and $i$ is the orbital inclination. $T_\mathrm{J}$ for (248370)~2005~QN$_{173}$ is 3.192 and thus it qualifies as asteroidal. (248370)~2005~QN$_{173}$ properties are provided in Appendix \ref{QN:sec:ObjectData}, and are established with this criterion.

(2) (248370)~2005~QN$_{173}$ orbits between 2.4~au and 3.76~au and thus does not cross the orbits of either Mars or Jupiter. With a semi-major axis of 3.1~au, (248370)~2005~QN$_{173}$ is an outer main-belt asteroid orbiting between the Kirkwood gaps corresponding to the 7:8 and 2:1 mean motion resonances with Jupiter.

(3) Recurrent activity near perihelion is diagnostic of volatile sublimation as the most likely mechanism responsible for the observed activity  \citep[e.g.,][]{hsiehOpticalDynamicalCharacterization2012}. However, other underlying causes of recurrent activity are known, so this point warrants further investigation.

\subsection{Activity Mechanism}
\label{QN:subsec:mechanism}

We demonstrated in Section \ref{QN:sec:secondActivityEpoch} that (248370)~2005~QN$_{173}$ has been active during at least two epochs. This helps rule out activity mechanisms  such as \textit{impact events} (e.g., (596)~Scheila; \citealt{bodewitsCollisionalExcavationAsteroid2011,ishiguroObservationalEvidenceImpact2011,moreno596ScheilaOutburst2011}) that are only expected to produce one-time outbursts but which can expel dust and produce comet-like activity. Aside from \textit{temperature-correlated volatile sublimation} (which we examine further in Section~\ref{QN:subsec:tempestimation}) other mechanisms for producing recurrent activity have been proposed.

\textit{Rotational destabilization} causes dust to be flung from a body in a potentially multiepisodic manner, as may be the case for (6478)~Gault \citep{chandlerSixYearsSustained2019,kleynaSporadicActivity64782019}. Taxonomic classification can help diagnose rotational destabilization, as with $S$-type (6478)~Gault, because activity from desiccated asteroid classes is unlikely to be sublimation driven. As discussed in Section \ref{QN:subsec:tempestimation}, the taxonomic class of (248370)~2005~QN$_{173}$ is not yet known but it is likely a C-type asteroid. An accurate rotation period for (248370)~2005~QN$_{173}$ is currently unavailable, and as such, we can neither confirm nor rule out destabilization as a contributing factor to the observed activity at this time.

\textit{Rubbing binaries} is a hypothetical scenario whereby two merging asteroids repeatedly collide and eject material. Proposed as a possible mechanism for the activity of 311P/PANSTARRS \citep{hainautContinuedActivity20132014}, the rubbing binary scenario has yet to be confirmed for that object \citep{jewittNucleusActiveAsteroid2018} or any other. As of this writing, there is no evidence that (248370)~2005~QN$_{173}$ is a binary asteroid, and activity spans two epochs separated by 5 yr, so we would expect merging processes to have either finished or that the binary orbit would have stabilized (see \citealt{jewittNucleusActiveAsteroid2018} for additional discussion concerning dissipation timescales). Therefore we find it unlikely that rubbing causes the activity associated with (248370)~2005~QN$_{173}$.

Geminid meteor stream parent (3200)~Phaethon undergoes extreme temperature changes ($\sim$600~K) and peaks at 800~K to 1100~K, well above the 573~K serpentine-phyllosilicate decomposition threshold \citep{ohtsukaSolarRadiationHeatingEffects2009}. These temperatures likely induce \textit{thermal fracture} \citep{licandroNatureCometasteroidTransition2007,kasugaObservations1999YC2008} leading to mass shedding \citep{liRecurrentPerihelionActivity2013,huiResurrection3200Phaethon2017}.

Two mechanisms, thermal fracture and temperature-correlated volatile sublimation warrant further inquiry into the thermophysical properties of (248370)~2005~QN$_{173}$.

\subsection{Temperature Estimation}
\label{QN:subsec:tempestimation}

Estimating temperatures experienced by (248370)~2005~QN$_{173}$ aids us in understanding direct thermal effects (e.g., thermal fracture) as well as assessing long-term volatile survival, especially water. For these reasons, we computed temperatures for an airless body over the course of an orbit similar to that of (248370)~2005~QN$_{173}$.

Following \citet{hsiehMainbeltCometsPanSTARRS12015}, the energy balance equation for a gray body on which water ice sublimation is occurring is 
\begin{equation}
{F_{\odot}\over r_h^2}(1-A) = \chi\left[{\varepsilon\sigma T_{eq}^4 + L f_D\dot m_{w}(T_{eq})}\right]
\label{QN:equation:sublim1}
\end{equation}
where $r_h$ is the object's heliocentric distance, $T_\mathrm{eq}$ is the equilibrium surface temperature, $F_{\odot}=1360$~W~m$^{-2}$ is the solar constant, $r_h$ is in au, $A=0.05$ is the assumed Bond albedo of the body, $\chi$ accounts for the distribution of solar heating over the object's surface, $\sigma$ is the Stefan--Boltzmann constant, and $\varepsilon=0.9$ is the assumed effective infrared emissivity, and $L=2.83$~MJ~kg$^{-1}$ is the latent heat of sublimation of water ice (which we approximate here as being independent of temperature), $f_D$ represents the reduction in sublimation efficiency caused by mantling, where $f_D=1$ in the absence of a mantle, and $\dot m_w$ is the water mass-loss rate due to sublimation of surface ice.

In this equation, $\chi=1$ corresponds to a flat slab facing the Sun, known as the subsolar approximation, and produces the maximum expected temperature for an object, 
while $\chi=4$ applies to objects with fast rotation or low thermal inertia, known as the isothermal approximation, and produces the minimum expected temperature for an object.

Next, the sublimation rate of ice into a vacuum can be computed using
\begin{equation}
\dot m_{w} = P_v(T) \sqrt{\mu\over2\pi k T}
\label{QN:equation:sublim2}
\end{equation}
where $\mu=2.991\cdot 10^{-26}$~kg is the mass of one water molecule, and $k$ is the Boltzmann constant, and the equivalent ice recession rate, $\dot \ell_{i}$, corresponding to $\dot m_{w}$ is given by
$\dot \ell_{i} = \dot m_{w}/ \rho$,
where $\rho$ is the bulk density of the object.

Finally, the Clausius--Clapeyron relation,
\begin{equation}
P_v(T) = 611 \times \exp\left[{{\Delta H_\mathrm{subl}\over R_g}\left({{1\over 273.16} - {1\over T}}\right)}\right]
\label{QN:equation:sublim3}
\end{equation}
gives the vapor pressure of water, $P_v(T)$, in Pa, where $\Delta H_\mathrm{subl}=51.06$~MJ~kmol$^{-1}$ is the heat of sublimation for ice to vapor and $R_g=8314$~J~kmol$^{-1}$~K$^{-1}$ is the ideal gas constant.
Solving these three equations iteratively, one can calculate the equilibrium temperature of a gray body at a given heliocentric distance on which water ice sublimation is occurring.

\begin{figure*}[ht]
	\centering
	\begin{tabular}{c}
	        \hspace{-4mm}\includegraphics[width=0.85\linewidth]{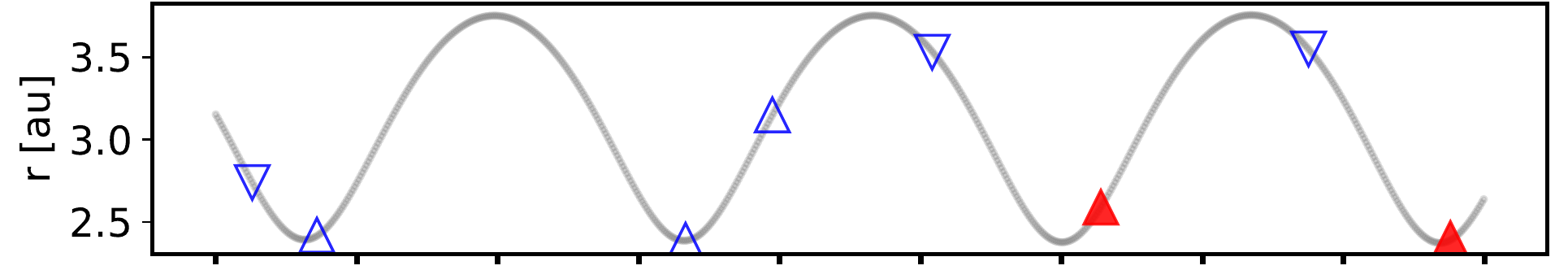}\\
	     	\hspace{9mm}\includegraphics[width=0.92\linewidth]{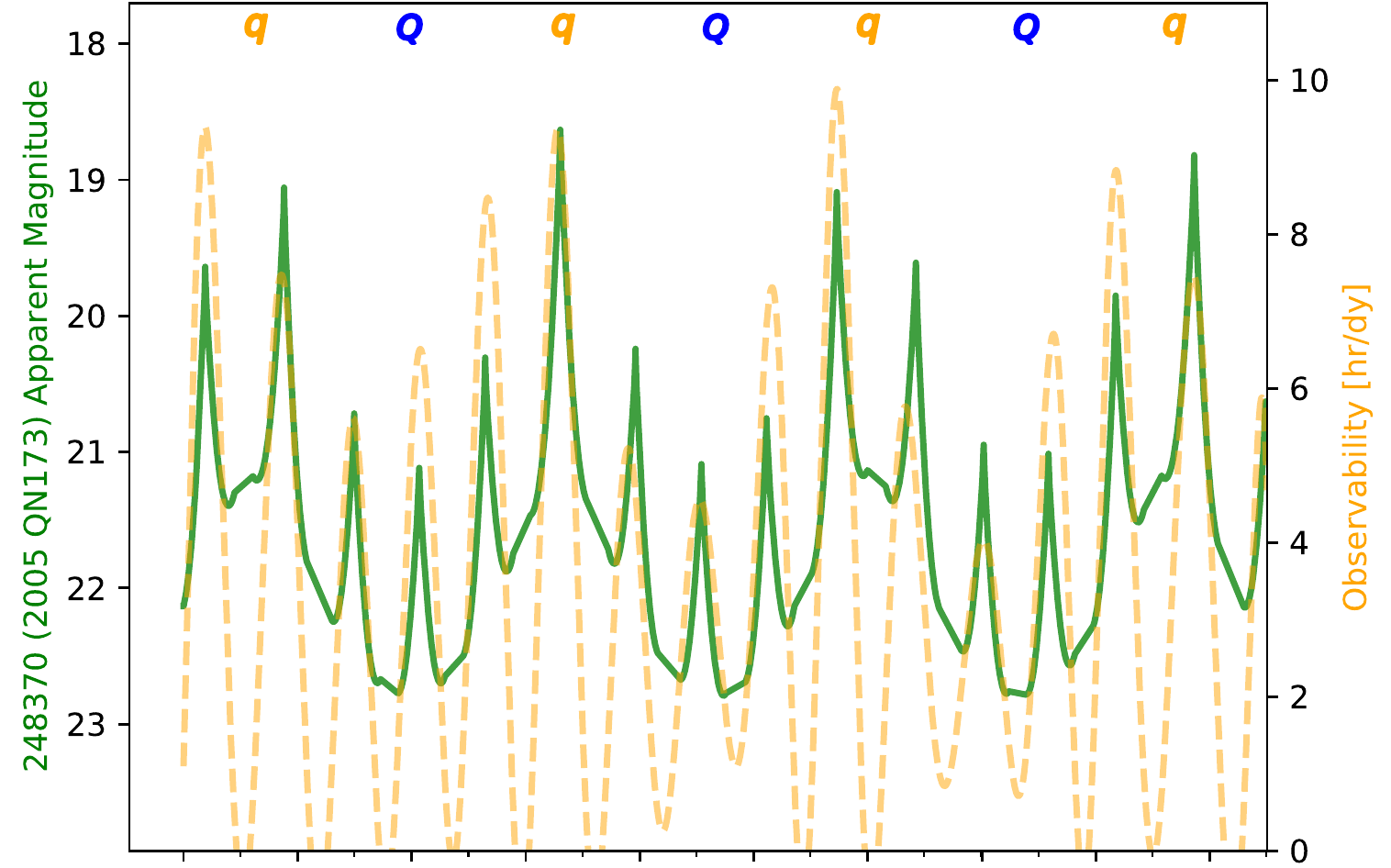}\\
	       \hspace{-4mm}\includegraphics[width=0.845\linewidth]{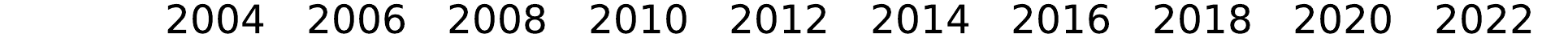}\\
	       \hspace{-1mm}\includegraphics[width=0.845\linewidth]{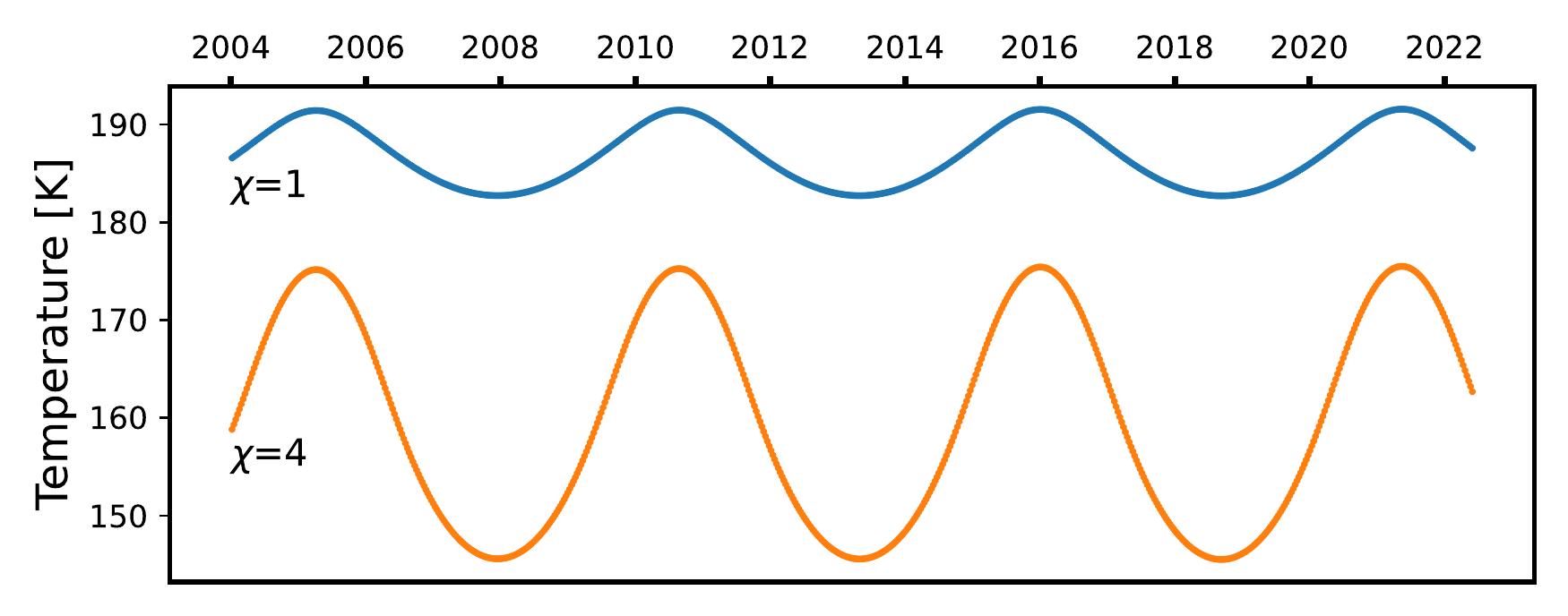}
	\end{tabular}
	\caption{(248370)~2005~QN$_{173}$ heliocentric distance (top plot), observability timeline (middle plot) and temperature (bottom plot), beginning the year of our first archival data (2004) through 2022.
	Top: triangles represent positive (filled red) and negative (unfilled blue) activity detections. Markers indicate when the object was inbound (downward pointing triangles) or outbound (upward pointing triangles). Table \ref{QN:tab:observations} lists observation details. 
	Middle: apparent $V$-band magnitude (solid green line) and our ``observability'' metric (yellow dashed line) that represents hours during a given UT observing date the object is above $> 15^\circ$ elevation while the Sun is below the horizon. Peaks in apparent magnitude coinciding with observability occur during opposition events, and observability troughs indicate solar conjunctions when (248370)~2005~QN$_{173}$ was only above the horizon during daylight. Perihelion (orange $q$) and aphelion (blue $Q$) events are indicated. 
	Bottom: temperature $T$~(K) by date for different $\chi$ values, where $\chi=1$ (top line) represents a ``flat slab'' and $\chi=4$ (bottom line) an isothermal body.
    }
	\label{QN:fig:ActivityTimeline}
\end{figure*}

In Figure~\ref{QN:fig:ActivityTimeline}, we plot the object's expected equilibrium temperature over several orbit cycles, as computed by solving the system of equations above. We plot temperatures computed using both $\chi=1$ and $\chi=4$ to show the full range of possible temperatures.

Figure~\ref{QN:fig:ActivityTimeline} (top) shows two parameters key to observing (248370)~2005~QN$_{173}$ between 2015 and 2023: apparent $V$-band magnitude and ``observability,'' which we define as the number of hours an object remains above $15^\circ$ elevation during nighttime for a given UT observing date. This plot informs us of potential observational biases or geometric effects that may bias activity detection, such as preferential activity discovery during opposition events, as was the case with (6478)~Gault \citep{chandlerSixYearsSustained2019}.

Figure~\ref{QN:fig:ActivityTimeline} (bottom) illustrates (248370)~2005~QN$_{173}$'s heliocentric distance and temperature (as computed using Equations \ref{QN:equation:sublim1}-\ref{QN:equation:sublim3}) over time, plus dates of observed activity and images where no visible activity was conclusively identified. Throughout its entire orbit the surface of (248370)~2005~QN$_{173}$ is consistently warmer than 145~K, the temperature above which water ice is not expected to survive over timescales on the order of the age of the solar system \citep{schorghoferLifetimeIceMain2008,snodgrassMainBeltComets2017}. 

However, it is possible for water ice to remain preserved over long (Gyr) timescales on small asteroids at depths as shallow as a few centimeters to 30~cm below the surface \citep{schorghoferLifetimeIceMain2008,prialnikCanIceSurvive2009}, where present-day activity may be triggered by meter-scale impactors that excavate subsurface ice. We note that water ice has been detected on the surface of main-belt asteroid (24) Themis \citep{campinsWaterIceOrganics2010,rivkinDetectionIceOrganics2010}, but the mechanism by which that water ice is able to persist on its surface -- likely requiring continual replenishment from subsurface volatile reservoirs -- is not well understood, and furthermore may not have the same effectiveness on kilometer-scale objects like (248370)~2005~QN$_{173}$ as it does on the 200~km diameter (24) Themis.

The surface temperature of (248370)~2005~QN$_{173}$ varies at most between 145~K and 190~K over its orbit (Figure \ref{QN:fig:ActivityTimeline}), far less than the 600~K temperature swings peaking at 800--1000~K described in Section \ref{QN:subsec:mechanism}. We consider it is unlikely that thermal fracture is the primary cause of (248370)~2005~QN$_{173}$'s activity.

\subsection{Nondetection of Activity}
\label{QN:subsec:nullresults}

The two known epochs of activity for (248370)~2005~QN$_{173}$ both occurred when the object was interior to a heliocentric distance of 2.7~au. However, (248370)~2005~QN$_{173}$ was observed in 2005 and 2010 when the object was interior to 2.7~au but no activity was detected. We believe the circumstances of these epochs preclude a definitive assessment of activity.

The 2010 Pan-STARRS1 data suffer from image artifacts that are significant enough to obscure activity that may have been present. The 2005 SuprimeCam observations should have been well suited to detecting activity as the 8~m Subaru telescope has a large aperture, exposure times (60~s) were sufficiently long, the \textit{W-J-VR} filter covered a broad wavelength range, and the object was well placed in the sky in terms of airmass/elevation during the observations. However, extinction varied significantly over the observing sequence as the summit log for that night\footnote{\url{https://smoka.nao.ac.jp/calendar/slog/2005/slog_20050608.txt}} indicated that conditions were windy with cirrus clouds and the differential image motion monitor measured significant seeing variation (roughly $0\farcsec8$ to $>3''$)\footnote{\url{https://smoka.nao.ac.jp/calendar/subaruseeing/20050608.gif}}, potentially contributing to a considerable reduction in our ability to detect activity. All sources in the field could be matched to SDSS~DR-9 stars but the faintest stars we were able match to the SDSS~DR-9 catalog were $r\approx21.3$, very similar to the JPL Horizons computed $V$=21.0 for (248370)~2005~QN$_{173}$. The Subaru Exposure Time Calculator estimates an equivalent $r$-band exposure would deliver a signal-to-noise ratio of 66 for a source of equivalent magnitude, but we estimate the images are at best $\sim$0.5 mag deeper than necessary to detect the object and thus we find it unlikely that activity would be detectable unless the object was undergoing a significant outburst at the time.

Although we cannot definitively rule out the presence of activity in 2005 or 2010 from these observations, another possibility is that a triggering event (e.g., impact, rotational destabilization) that started (248370)~2005~QN$_{173}$'s current activity occurred between 2005 June and 2016 July. This would explain the object's apparent inactivity and activity on each of those dates, respectively.

\subsection{Main-belt Comet Membership}
\label{QN:subsec:mbcmembership}
Given the above reasoning, we find it most likely that the activity associated with (248370)~2005~QN$_{173}$ is sublimation driven, in which case the object is an MBC. However, we emphasize that in order to rule out rotational destabilization as the root cause of the observed activity, additional work is needed. Moreover, confirmation of a third activity epoch would lend further evidence favoring sublimation as the primary agency of activity.

\section{Summary and Future Work}
\label{QN:sec:summary}
We harvested eighty-one images of (248370)~2005~QN$_{173}$ (also designated 433P) spanning thirty-one observing dates. We found clear evidence of a previous activity epoch on 2016 July 22. We provide a catalog of archival observations along with an image gallery. Making use of wedge photometry -- a novel tail detection and characterization tool we introduce in this Letter -- we measure tail orientation to be in close agreement with the antisolar and antimotion vectors computed by Horizons. We showed that (248370)~2005~QN$_{173}$ is a likely member of the MBCs, a group of active asteroids orbiting within the Main Asteroid Belt that are active due to volatile sublimation.

The current observing window for this object ends around 2021 December, and when it returns in late 2022 it will be over 3~au from the Sun and less likely to show activity. We did not find any images showing (248370)~2005~QN$_{173}$ active at beyond 3~au, so we call on observers to make use of the present activity apparition while it is still possible to do so. Continued monitoring to study the evolution of the tail's brightness, including surface brightness measurements, can lead to better characterization of ejected dust grain sizes and total mass loss during this apparition. Once activity subsides, time-series observations to measure a rotation period will be especially useful for diagnosing rotational breakup. Preliminary color measurements suggest (248370)~2005~QN$_{173}$ is a C-type asteroid \citep{hsiehPhysicalCharacterizationMainbelt2021}, but a robust taxonomic classification would help further solidify our assessment of the underlying activity mechanism.


\section{Acknowledgements}
\label{QN:sec:acknowledgements}

The authors thank the anonymous referee whose comments greatly improved the quality of this Letter.

We thank Dr.\ Mark Jesus Mendoza Magbanua (University of California San Francisco) for his frequent and timely feedback on the project. 
The authors express their gratitude to 
Prof. Mike Gowanlock (NAU), 
Dr. Annika Gustafsson (NAU, Lowell Observatory, Southwest Research Institute), 
Jay Kueny (Steward Observatory), 
and the Trilling Research Group (NAU), all of whom provided invaluable insights which substantially enhanced this work. The unparalleled support provided by Monsoon cluster administrator Christopher Coffey (NAU) and the High Performance Computing Support team facilitated the scientific process.

This material is based upon work supported by the National Science Foundation Graduate Research Fellowship Program under grant No.\ (2018258765). Any opinions, findings, and conclusions or recommendations expressed in this material are those of the author(s) and do not necessarily reflect the views of the National Science Foundation.  C.O.C., H.H.H. and C.A.T. also acknowledge support from the NASA Solar System Observations program (grant 80NSSC19K0869). 

Computational analyses were run on Northern Arizona University's Monsoon computing cluster, funded by Arizona's Technology and Research Initiative Fund. This work was made possible in part through the State of Arizona Technology and Research Initiative Program. 
World Coordinate System (WCS) corrections facilitated by the \textit{Astrometry.net} software suite \citep{langAstrometryNetBlind2010}.

This research has made use of data and/or services provided by the International Astronomical Union's Minor Planet Center. 
This research has made use of NASA's Astrophysics Data System. 
This research has made use of The Institut de M\'ecanique C\'eleste et de Calcul des \'Eph\'em\'erides (IMCCE) SkyBoT Virtual Observatory tool \citep{berthierSkyBoTNewVO2006}. 
This work made use of the {FTOOLS} software package hosted by the NASA Goddard Flight Center High Energy Astrophysics Science Archive Research Center. 
This research has made use of SAOImageDS9, developed by Smithsonian Astrophysical Observatory \citep{joyeNewFeaturesSAOImage2006}. 
This work made use of the Lowell Observatory Asteroid Orbit Database \textit{astorbDB} \citep{bowellPublicDomainAsteroid1994,moskovitzAstorbDatabaseLowell2021}. 
This work made use of the \textit{astropy} software package \citep{robitailleAstropyCommunityPython2013}.

This project used data obtained with the Dark Energy Camera (DECam), which was constructed by the Dark Energy Survey (DES) collaboration. Funding for the DES Projects has been provided by the U.S. Department of Energy, the U.S. National Science Foundation, the Ministry of Science and Education of Spain, the Science and Technology Facilities Council of the United Kingdom, the Higher Education Funding Council for England, the National Center for Supercomputing Applications at the University of Illinois at Urbana-Champaign, the Kavli Institute of Cosmological Physics at the University of Chicago, Center for Cosmology and Astro-Particle Physics at the Ohio State University, the Mitchell Institute for Fundamental Physics and Astronomy at Texas A\&M University, Financiadora de Estudos e Projetos, Funda\c{c}\~{a}o Carlos Chagas Filho de Amparo, Financiadora de Estudos e Projetos, Funda\c{c}\~ao Carlos Chagas Filho de Amparo \`{a} Pesquisa do Estado do Rio de Janeiro, Conselho Nacional de Desenvolvimento Cient\'{i}fico e Tecnol\'{o}gico and the Minist\'{e}rio da Ci\^{e}ncia, Tecnologia e Inova\c{c}\~{a}o, the Deutsche Forschungsgemeinschaft and the Collaborating Institutions in the Dark Energy Survey. The Collaborating Institutions are Argonne National Laboratory, the University of California at Santa Cruz, the University of Cambridge, Centro de Investigaciones En\'{e}rgeticas, Medioambientales y Tecnol\'{o}gicas–Madrid, the University of Chicago, University College London, the DES-Brazil Consortium, the University of Edinburgh, the Eidgen\"ossische Technische Hochschule (ETH) Z\"urich, Fermi National Accelerator Laboratory, the University of Illinois at Urbana-Champaign, the Institut de Ci\`{e}ncies de l'Espai (IEEC/CSIC), the Institut de Física d'Altes Energies, Lawrence Berkeley National Laboratory, the Ludwig-Maximilians Universit\"{a}t M\"{u}nchen and the associated Excellence Cluster Universe, the University of Michigan, the National Optical Astronomy Observatory, the University of Nottingham, the Ohio State University, the University of Pennsylvania, the University of Portsmouth, SLAC National Accelerator Laboratory, Stanford University, the University of Sussex, and Texas A\&M University.

Based on observations at Cerro Tololo Inter-American Observatory, National Optical Astronomy Observatory (NOAO Prop. ID 2016A-0190, PI: Dey), which is operated by the Association of Universities for Research in Astronomy (AURA) under a cooperative agreement with the National Science Foundation. This research has made use of the NASA/IPAC Infrared Science Archive, which is funded by the National Aeronautics and Space Administration and operated by the California Institute of Technology.

The Legacy Surveys consist of three individual and complementary projects: the Dark Energy Camera Legacy Survey (DECaLS; Proposal ID \#2014B-0404; PIs: David Schlegel and Arjun Dey), the Beijing-Arizona Sky Survey (BASS; NOAO Prop. ID \#2015A-0801; PIs: Zhou Xu and Xiaohui Fan), and the Mayall z-band Legacy Survey (MzLS; Prop. ID \#2016A-0453; PI: Arjun Dey). DECaLS, BASS and MzLS together include data obtained, respectively, at the Blanco telescope, Cerro Tololo Inter-American Observatory, NSF's NOIRLab; the Bok telescope, Steward Observatory, University of Arizona; and the Mayall telescope, Kitt Peak National Observatory, NOIRLab. The Legacy Surveys project is honored to be permitted to conduct astronomical research on Iolkam Du'ag (Kitt Peak), a mountain with particular significance to the Tohono O'odham Nation. BASS is a key project of the Telescope Access Program (TAP), which has been funded by the National Astronomical Observatories of China, the Chinese Academy of Sciences (the Strategic Priority Research Program ``The Emergence of Cosmological Structures'' Grant \# XDB09000000), and the Special Fund for Astronomy from the Ministry of Finance. The BASS is also supported by the External Cooperation Program of Chinese Academy of Sciences (Grant \# 114A11KYSB20160057), and Chinese National Natural Science Foundation (Grant \# 11433005). The Legacy Survey team makes use of data products from the Near-Earth Object Wide-field Infrared Survey Explorer (NEOWISE), which is a project of the Jet Propulsion Laboratory/California Institute of Technology. NEOWISE is funded by the National Aeronautics and Space Administration. The Legacy Surveys imaging of the DESI footprint is supported by the Director, Office of Science, Office of High Energy Physics of the U.S. Department of Energy under Contract No. DE-AC02-05CH1123, by the National Energy Research Scientific Computing Center, a DOE Office of Science User Facility under the same contract; and by the U.S. National Science Foundation, Division of Astronomical Sciences under Contract No. AST-0950945 to NOAO.

Based in part on data collected at Subaru Telescope and obtained from the SMOKA, which is operated by the Astronomy Data Center, National Astronomical Observatory of Japan \citep{2002ASPC..281..298B}.



\section{Appendix}

\subsection{Equipment and Archives}
\label{QN:sec:equipQuickRef}

\begin{sidewaystable}
\caption{Equipment and Archives}
\centering
\footnotesize
    \begin{tabular}{llclcccccc}
Instrument  & Telescope            & Pixel Scale    & Location                & NOIR         & ESO & IRSA         & SMOKA & SSOIS        & STScI  \\
            &                       & [$''$/pix] &                   &   & & &\\
\hline
\hline
DECam       & 4 m Blanco           & 0.263          & Cerro Tololo, Chile     & S,R &     &              &       & S            &        \\
OmegaCAM    & 2.6 m VLT Survey     & 0.214          & Cerro Paranal, Chile    &              & R   &              &       & S            &        \\
GigaPixel1  & 1.8 m Pan-STARRS1    & 0.258          & Haleakalā, Hawaii       &              &     &              &       & S            & R      \\
MegaPrime   & 3.6 m CFHT           & 0.185          & Mauna Kea, Hawaii       &              &     &              &       & S,R &        \\
PTF/CFHT 12K& 48" Samuel Oschin    & 1.010          & Mt. Palomar, California &              &     & S,R &       &              &        \\
SuprimeCam  & 8.2 m Subaru         & 0.200          & Mauna Kea, Hawaii       &              &     &              & R     & S            &        \\
ZTF Camera  & 48" Samuel Oschin    & 1.012          & Mt. Palomar, California &              &     & S,R &       &              &       \\
\end{tabular}
\raggedright
\\
\footnotesize{
R indicates repository for data retrieval. S indicates search capability.\\
NOIR: NSF National Optical Infrared Labs AstroArchive (\url{https://astroarchive.noirlab.edu}).\\
ESO: European Space Organization Archive (\url{https://archive.eso.org}).\\
IRSA: NASA/CalTech Infrared Science Archive (\url{https://irsa.ipac.caltech.edu}).\\
SMOKA: NAOJ Subaru-Mitaka-Okayama-Kiso Archive Science Archive (\url{https://smoka.nao.ac.jp}).\\
SSOIS: CADC Solar System Object Image Search (\url{https://www.cadc-ccda.hia-iha.nrc-cnrc.gc.ca/en/ssois/}).\\
STScI: Space Telescope Science Institute (\url{https://www.stsci.edu/}).}
\label{QN:tab:equipAndArchives}
\end{sidewaystable}

Table \ref{QN:tab:equipAndArchives} lists the instruments and telescopes used in this work, along with their respective pixel scales, locations, and data archives.

\subsection{(248370) 2005 QN173 Data} 
\label{QN:sec:ObjectData}

We provide current information regarding (248370)~2005~QN$_{173}$ below (Table \ref{QN:table:qnproperties}). 

\begin{sidewaystable}[ht]
    \centering
	\caption{(248370)~2005~QN$_{173}$ Properties}
	\begin{tabular}{lll}
		Parameter & Value & Source\\
		\hline\hline
		Discovery Date & 2005 August 29 & Minor Planet Center\\
		Discovery Observers & Near-Earth Asteroid Tracking (NEAT) & Minor Planet Center\\
		Discovery Observatory & Palomar & Minor Planet Center\\
		Activity Discovery Date & 2021 July 7 & CBET 4995\\
		Activity Discoverer(s) & A. Fitzsimmons / ATLAS & CBET 4995 \citep{fitzsimmons2483702005QN1732021}\\
		Orbit Type & Outer Main-belt & IMCCE, AstOrb\\
		Taxonomic Class & C-type (unconfirmed) & \citet{hsiehPhysicalCharacterizationMainbelt2021}\\
		Diameter & $D=3.6$~km; 3.4$\pm$0.4~km & Horizons, {\citet{harrisAsteroidsThermalInfrared2002}} ;  \citet{hsiehPhysicalCharacterizationMainbelt2021}\\ 
		Absolute $V$-band Magnitude & $H=16.02$ & Horizons\\
        Geometric Albedo & 0.054 & Horizons, \citet{mainzerNEOWISEDiametersAlbedos2019}\\
		Rotation Period & unknown & \\
        Orbital Period & $P=5.37$ yr & Horizons \\
		Semimajor Axis & $a=3.075$ au & Horizons\\
		Eccentricity & $e=0.226$ & Horizons\\
		Inclination & $i=0.067^\circ$ & Horizons\\
		Longitude of Ascending Node & $\Omega=174.28$ & Minor Planet Center\\
		Mean Anomaly & $M=8.79^\circ$ & Minor Planet Center\\
		Argument of Perihelion & $\omega=146.09^\circ$ & Horizons\\
		Perihelion Distance & $q=2.374$ au & Horizons\\
		Aphelion Distance & $Q=3.761$ au & Horizons\\
		Tisserand Parameter w.r.t. Jupiter & $T_J=3.192$ & AstOrb\\
	\end{tabular}
	\label{QN:table:qnproperties}
\end{sidewaystable}

\clearpage
\singlespacing
\chapter{Manuscript V: Migratory Outbursting Quasi-Hilda Object 282P/(323137) 2003 BM80}
\chaptermark{Migratory Outbursting Quasi-Hilda Object 282P/(323137) 2003 BM80}
\label{chap:282P}
\acresetall

Colin Orion Chandler\footnote{\label{282P:nau}Department of Astronomy and Planetary Science, Northern Arizona University, PO Box 6010, Flagstaff, AZ 86011, USA}, William J. Oldroyd$^\mathrm{\ref{282P:nau}}$, Chadwick A. Trujillo$^\mathrm{\ref{282P:nau}}$

\textit{This is a preliminary version of an article submitted for publication in Astrophysical Journal Letters.  IOP Publishing Ltd is not responsible for any errors or omissions in this version of the manuscript or any version derived from it.}

\doublespacing

\newcommand{\objnameBM}{282P}
\newcommand{\objnameBMFull}{282P/(323137)~2003~BM$_{80}$} 
\newcommand\blfootnote[1]{%
  \begingroup
  \renewcommand\thefootnote{}\footnote{#1}%
  \addtocounter{footnote}{-1}%
  \endgroup
}

\newcounter{obsnotelabel}
\newcommand{\obsnote}[1]{\refstepcounter{obsnotelabel}\label{#1}}
\setcounter{obsnotelabel}{0}

\section{Abstract}
\label{282P:Abstract}
We report object \objnameBMFull{} is undergoing a sustained activity outburst, lasting over 15 months thus far. These findings stem in part from our \acs{NASA} Partner Citizen Science project \textit{Active Asteroids} (\url{http://activeasteroids.net}), which we introduce here. 
We acquired new observations of \objnameBM{} via our observing campaign (\acf{VATT}, \acf{LDT}, and the Gemini South telescope), confirming \objnameBM{} was active on UT 2022 June 7, some 15 months after 2021 March images showed activity in the 2021--2022 epoch. 
We classify \objnameBM{} as a member of the \acp{QHO}, a group of dynamically unstable objects found in an orbital region similar to, but distinct in their dynamical characteristics to, the Hilda asteroids (objects in 3:2 resonance with Jupiter). Our dynamical simulations show \objnameBM{} has undergone at least five close encounters with Jupiter and one with Saturn over the last 180 years. \objnameBM{} was most likely a Centaur or \ac{JFC} 250 years ago. In 350 years, following some 15 strong Jovian interactions, \objnameBM{} will most likely migrate to become a \ac{JFC} or, less likely, an \acl{OMBA} orbit. These migrations highlight a dynamical pathway connecting Centaurs and \acp{JFC} with Quasi-Hildas and, potentially, active asteroids. Synthesizing these results with our thermodynamical modeling and new activity observations, we find volatile sublimation is the primary activity mechanism. Observations of a quiescent \objnameBM{}, which we anticipate will be possible in 2023, will help confirm our hypothesis by measuring a rotation period and ascertaining spectral type.

\blfootnote{Based on observations obtained at the international Gemini Observatory, a program of NSF’s NOIRLab, which is managed by the Association of Universities for Research in Astronomy (AURA) under a cooperative agreement with the National Science Foundation on behalf of the Gemini Observatory partnership: the National Science Foundation (United States), National Research Council (Canada), Agencia Nacional de Investigaci\'{o}n y Desarrollo (Chile), Ministerio de Ciencia, Tecnolog\'{i}a e Innovaci\'{o}n (Argentina), Minist\'{e}rio da Ci\^{e}ncia, Tecnologia, Inova\c{c}\~{o}es e Comunica\c{c}\~{o}es (Brazil), and Korea Astronomy and Space Science Institute (Republic of Korea).}
\blfootnote{Magellan telescope time was granted by \acs{NSF}’s \acs{NOIRLab}, through the \ac{TSIP}. \ac{TSIP} was funded by \ac{NSF}.}

\section{Introduction}
\label{282P:introduction}

Volatiles are vital to life as we know it and are critically important to future space exploration, yet basic knowledge about where volatiles (e.g., H$_2$O, CO, CH$_4$) are located within our own solar system is still incomplete. Moreover, the origin of solar system volatiles, including terrestrial water, remains inconclusive. Investigating sublimation-driven active solar system bodies can help answer these questions \citep{hsiehPopulationCometsMain2006}.

We define volatile reservoirs as a dynamical class of minor planet that harbors volatile species, such as water ice. Comets have long been known to contain volatiles, but other important reservoirs are coming to light, such as the active asteroids -- objects on orbits normally associated with asteroids, such as those found in the main-belt, that surprisingly display cometary features such as tails and/or comae \citep{jewittActiveAsteroids2015a}. Fewer than 30 active asteroids have been discovered \citep{chandlerSAFARISearchingAsteroids2018} since the first, (4015)~Wilson-Harrington, was discovered in 1949 \citep{cunninghamPeriodicCometWilsonHarrington1950} and, as a result, they remain poorly understood. One scientifically important subset of active asteroids consists of members that display recurrent activity attributed to sublimation: the \acp{MBC} \citep{hsiehMainbeltCometsPanSTARRS12015}. An important diagnostic of indicator sublimating volatiles, like water ice, is recurrent activity near perihelion \citep{hsiehOpticalDynamicalCharacterization2012,snodgrassMainBeltComets2017}, a feature common to the \acp{MBC} \citep{hsiehMainbeltCometsPanSTARRS12015,agarwalBinaryMainbeltComet2017,hsieh2016ReactivationsMainbelt2018}. Fewer than 10 recurrently active \acp{MBC} have been discovered (though others exhibit activity attributed to sublimation), and as a result we know very little about this population.

Another potential volatile reservoir, active Centaurs, came to light after comet 29P/Schwassmann-Wachmann 1 \citep{schwassmannNEWCOMET1927} was identified as a Centaur following the 1977 discovery of (2060)~Chiron \citep{kowalSlowMovingObjectKowal1977}. Centaurs, found between the orbits of Jupiter and Neptune, are cold objects thought to primarily originate in the Kuiper Belt prior to migrating to their current orbits (see review, \citealt{jewittActiveCentaurs2009}). The dynamical properties of these objects are discussed in Section \ref{282P:sec:dynamicalClassification}. Fewer than 20 active Centaurs have been discovered to date, thus they, like the active asteroids, are both rare and poorly understood.

In order to enable the study of active objects in populations not typically associated with activity (e.g., \acp{NEO}, main-belt asteroids), we created a Citizen Science project designed to identify roughly 100 active objects via volunteer identification of activity in images of a known minor planets. The Citizen Science paradigm involves concurrently crowdsourcing tasks yet too complex for computers to perform, while also carrying out an outreach program that engages the public in a scientific endeavor. Launched in Fall 2021, our \ac{NSF} funded, \acs{NASA} partner program \textit{Active Asteroids}\footnote{\url{http://activeasteroids.net}} immediately began yielding results.

\begin{figure*}[ht]
    \centering
    \begin{tabular}{cccc}
    \begin{overpic}[width=0.24\linewidth]{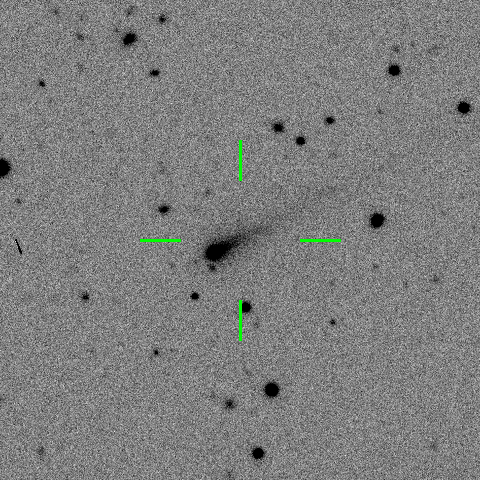}\put (5,7) {\huge\color{green} \textbf{a}}\put (45,8) {\large\color{green} \textbf{2021-03-14}}\end{overpic}  & 
    \begin{overpic}[width=0.24\linewidth]{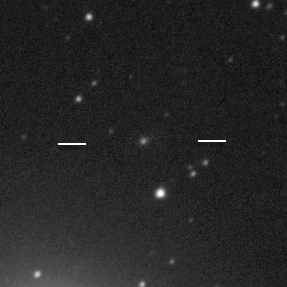}\put (5,7) {\huge\color{green} \textbf{b}}\put (45,8) {\large\color{green} \textbf{2021-03-31}}\end{overpic} &
    \begin{overpic}[width=0.24\linewidth]{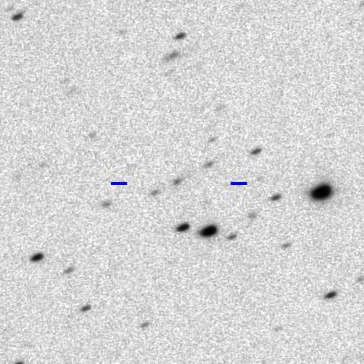}\put (5,7) {\huge\color{green} \textbf{c}}\put (45,8) {\large\color{green} \textbf{2021-04-04}}\end{overpic} &
    \begin{overpic}[width=0.24\linewidth]{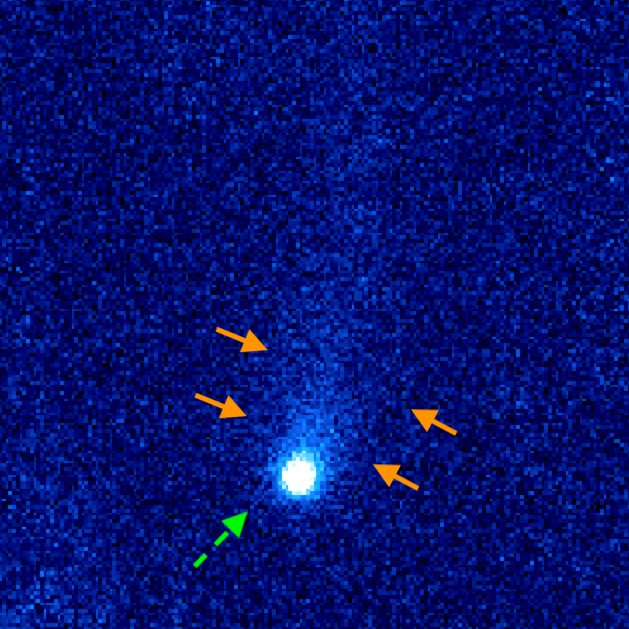}\put (5,7) {\huge\color{green} \textbf{d}}\put (45,8) {\large\color{green} \textbf{2022-06-07}}\end{overpic}\\
    \begin{overpic}[width=0.24\linewidth]{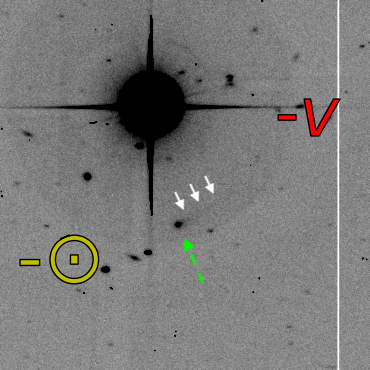}\put (5,7) {\huge\color{green} \textbf{e}}\put (45,8) {\large\color{green} \textbf{2012-03-28}}\end{overpic} & 
    \begin{overpic}[width=0.24\linewidth]{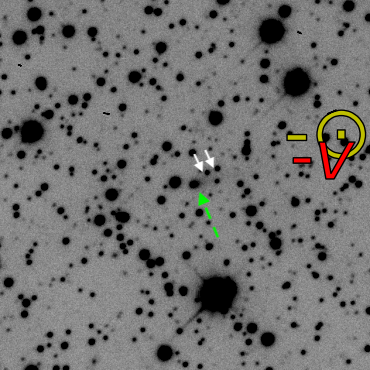}\put (5,7) {\huge\color{green} \textbf{f}}\put (45,8) {\large\color{green} \textbf{2013-05-05}}\end{overpic} &
    \begin{overpic}[width=0.24\linewidth]{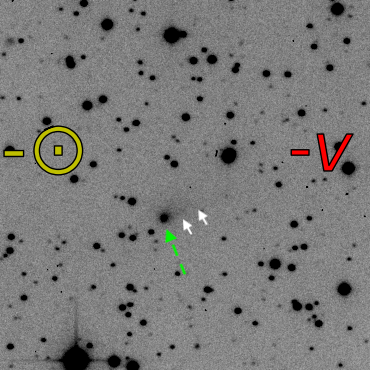}\put (5,7) {\huge\color{green} \textbf{g}}\put (45,8) {\large\color{green} \textbf{2013-06-13}}\end{overpic} &
    \begin{overpic}[width=0.24\linewidth]{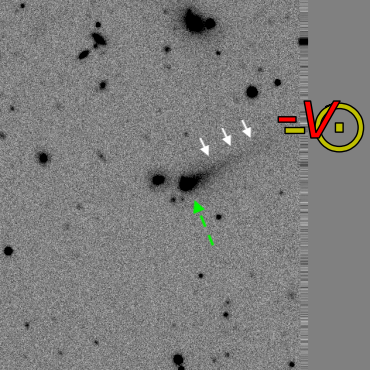}\put (5,7) {\huge\color{green} \textbf{h}}\put (45,8) {\large\color{green} \textbf{2021-03-17}}\end{overpic}
    \end{tabular}
    \caption{
    Top row: four images, spanning 15 months, showing \objnameBMFull{} activity during the recent 2021--2022 activity epoch. 
    \textbf{(a)} Epoch II thumbnail image of \objnameBM{} was classified as ``active'' by 14 of 15 volunteers of our Citizen Science project \textit{Active Asteroids}, a NASA Partner program. This 90~s $i$ band image was taken with the Dark Energy Camera on UT 2021 March 14, Prop. ID 2019A-0305 (\acs{PI} Drlica-Wagner). 
    \textbf{(b)} Epoch II, 12$\times$300~s co-added exposures imaged by Michael Jäger with a QHY600 camera on a 14'' Newtonian telescope in Weißenkirchen, Austria. Image reproduced with permission of Michael Jäger. 
    \textbf{(c)} Epoch II 5$\times$300~s co-added images captured by Roland Fichtl using a CDS cooled Canon 5D Mark III camera on a 16'' Newtonian telescope in Engelhardsberg, Germany. Image reproduced with permission of Roland Fichtl. 
    \textbf{(d)} For this most recent Epoch II image we co-added six 120~s $g'$ band images of \objnameBM{} (green dashed arrow) we acquired on UT 7 June 2022 with the \ac{GMOS} imager on the 8.1~m Gemini South telescope (Prop. ID GS-2022A-DD-103, \acs{PI} Chandler); a tail is clearly visible (orange arrows).
    Bottom row: Archival images of \objnameBM{} that show clear evidence of activity. For each 126\arcsec$\times$126\arcsec thumbnail image, north is up and east is left. With the center of each image as the origin, the antisolar (yellow -$\odot$) and antimotion (red -$v$) directions (often correlated with tail appearance) are indicated. \objnameBM{} is indicated by the green dashed arrow, and visible activity is marked by the white arrows.
    \textbf{(e)} Epoch I image from UT 2012 March 28 MegaPrime 120~s $r$ band, Prop. ID 12AH16 (\acs{PI} Wainscoat). 
    \textbf{(f)} Epoch I image from UT 2013 May 5 DECam 150~s $r$ band, Prop. ID 2013A-0327 (\acs{PI} Rest). 
    \textbf{(g)} Epoch I image from UT 2013 June 13 MegaPrime 120~s $r$ band, Prop. ID 13AH09 (\acs{PI} Wainscoat). 
    \textbf{(h)} Epoch II image from UT 2021 March 17 DECam 90~s $i$ band, Prop. ID 2019A-0305 (\acs{PI} Drlica-Wagner). 
    } 
    \label{282P:fig:282P}
\end{figure*}
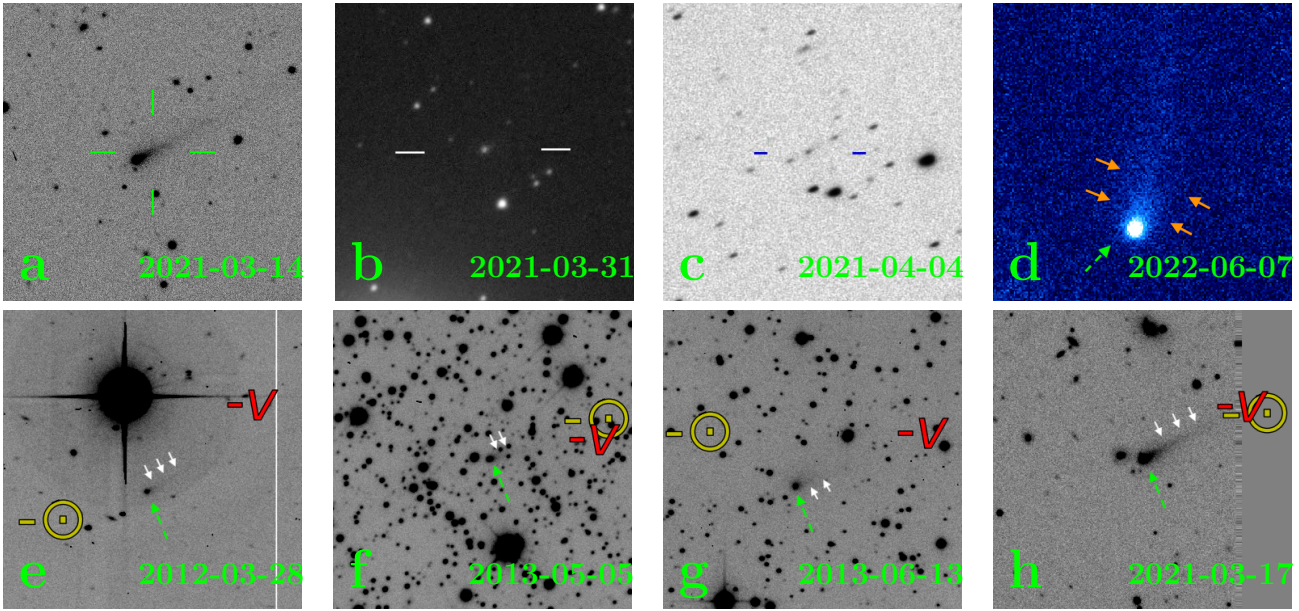

\objnameBMFull{}, hereafter \objnameBM{}, was originally discovered as 2003~BM$_{80}$ on UT 2003 Jan 31 by Brian Skiff of the \ac{LONEOS} survey, and independently as 2003~FV$_{112}$ by \ac{LINEAR} on UT 2003 Apr 18. \objnameBM{} was identified to be active during its 2012--2013 epoch (centered on its perihelion passage) in 2013 \citep{bolinComet2003BM2013}, at which time \objnameBM{} was given the additional identifier 282P. Here, we introduce an additional activity epoch, spanning 2021--2022.

In this work we (1) present our \ac{NASA} Partner Citizen Science project \textit{Active Asteroids}, (2) describe how volunteers identified activity that led to our investigation into \objnameBM{}, (3) present (a) archival images and (b) new observations of \objnameBM{} that show it has undergone periods of activity during at least two epochs (2012--2013 and 2021--2022) spanning consecutive perihelion passages, (4) classify \objnameBM{} as a \ac{QHO}, (5) explore the migratory nature of this object through dynamical modeling, including identification of a dynamical pathway between \acp{QHO} and active asteroids, and (6) determine volatile sublimation as the most probable activity mechanism.

\section{Citizen Science}
\label{282P:subsec:citsci}

We prepared thumbnail images (e.g., Figure \ref{282P:fig:282P}a) for examination by volunteers of our NASA Partner Citizen Science project \textit{Active Asteroids}, hosted on the Zooniverse\footnote{\url{https://www.zooniverse.org}} online Citizen Science platform. First we extract thumbnail images from publicly available \ac{DECam} archival images using a pipeline, \ac{HARVEST}, first described in \cite{chandlerSAFARISearchingAsteroids2018} and expanded upon in \cite{chandlerSixYearsSustained2019,chandlerCometaryActivityDiscovered2020a,chandler2483702005QN2021}. We optimize the Citizen Science process by excluding thumbnail images based on specific criteria, for example when (a) the image depth is insufficient for detecting activity, (b) no source was detected in the thumbnail center, and (c) too many sources were in the thumbnail to allow for reliable target identification.

Our workflow is simple: we show volunteers an image of a known minor planet and ask whether or not they see evidence of activity (like a tail or coma) coming from the object at the center of the image, as marked by a reticle (Figure \ref{282P:fig:282P}a). Each thumbnail is examined by at least 15 volunteers to minimize volunteer bias. To help train volunteers and validate that the project is working as intended, we created a training set of thumbnail images that we positively identified as showing activity, consisting of comets and other active objects, such as active asteroids. Training images are injected at random, though the interval of injection decays over time so that experienced volunteers only see a training image 5\% of the time.

We take the ratio of ``positive for activity'' classifications to the total number of classifications the object received, as a score to estimate the likelihood of the object being active. Members of the science team visually examines all images with a likelihood score of $\ge$80\% and flag candidates that warrant archival image investigation and telescope follow-up (Section \ref{282P:sec:observations}). We also learn of activity candidates through Zooniverse forums where users interact with each other, moderators, and our science team. Volunteers can share images they find interesting which has, in turn, led us directly to discoveries.

As of this writing, over 6,600 volunteers have participated in \textit{Active Asteroids}. They have conducted over 2.8$\times10^6$ classifications, completing assessment of over 171,000 thumbnail images. One image of \objnameBM{} from UT 2021 March 14 (Figure \ref{282P:fig:282P}a) received a score of 93\% after 14 of 15 volunteers classified the thumbnail as showing activity. A second image from UT 2021 March 17 (Figure \ref{282P:fig:282P}h) was classified as active by 15 of 15 volunteers, providing additional strong evidence of activity from 2021 March.

\section{Observations}
\label{282P:sec:observations}

\subsection{Archival Data}
\label{282P:susbec:archivalData}

\begin{figure*}[ht]
	\centering
	\begin{tabular}{c}
	        \hspace{5mm}\includegraphics[width=0.80\linewidth]{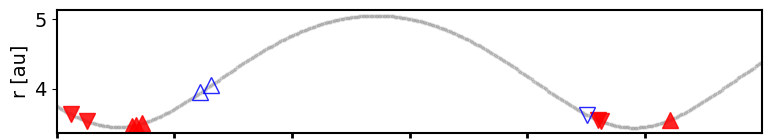}\\
	     	\hspace{9mm}\includegraphics[width=0.92\linewidth]{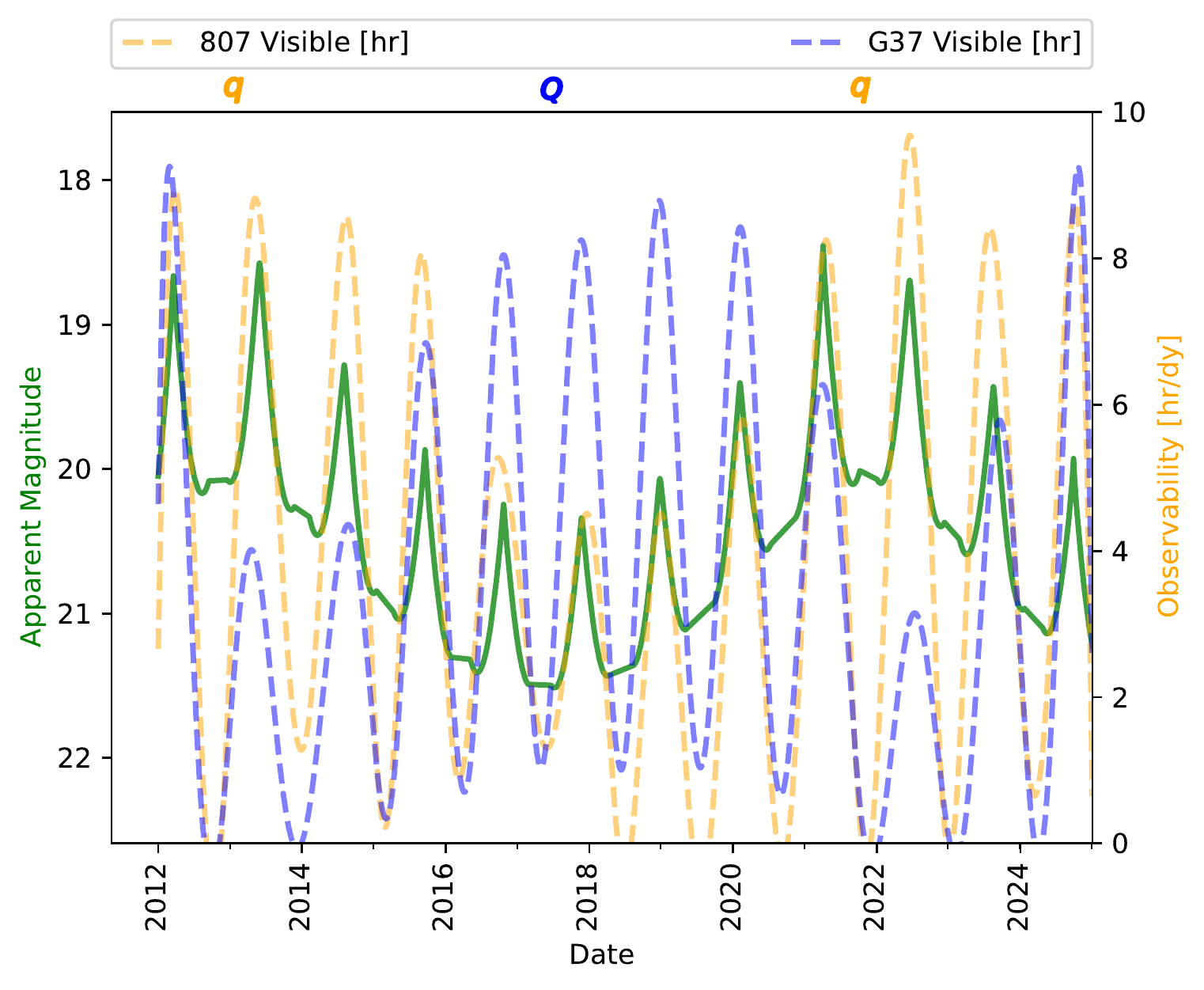}\\
	       \hspace{-5mm}\includegraphics[width=0.85\linewidth]{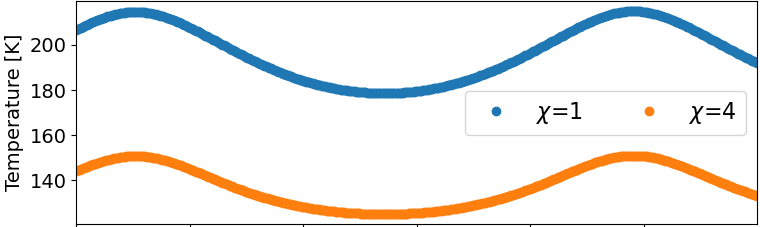} 
	\end{tabular}
	\caption{\objnameBM{} heliocentric distance (top), observability (middle) and temperature (bottom), from 2012 through 2024. 
	\textbf{Top:} activity detections (triangles) are marked as positive (filled red) and negative (unfilled blue) detections and as either
	inbound ($\blacktriangledown$) or outbound ($\blacktriangle$). 
	Observations are cataloged in Table \ref{282P:tab:observationsTable}. 
	\textbf{Middle:} Our observability metric for \acf*{CTIO}, site code 807 (yellow dashed line) and the \acf*{LDT}, site code G37 (blue dashed line), depicting the number of hours \objnameBM{} was observable ($>15\degr$ between sunset and sunrise) during a given \ac{UT} observing date. Opposition events and conjunctions result in concurrent maxima and minima, respectively. Also indicated are perihelion (orange $q$) and aphelion (blue $Q$) passages. 
	\textbf{Bottom:} Modeled temperature by date for the thermophysical extremes: a ``flat slab'' ($\chi=1$, top line), and an isothermal body ($\chi=4$, bottom line).
	}
	\label{282P:fig:ActivityTimeline}
\end{figure*}

For each candidate active object stemming from \textit{Active Asteroids} we conduct an archival data investigation, following the procedure described in \cite{chandler2483702005QN2021}. For this task, we query public astronomical image archives and identify images which may show \objnameBM{} in the \ac{FOV}. We download the data, extract thumbnail images centered on \objnameBM{}, and visually examine all images to search for evidence of activity.

After visually inspecting $>400$ thumbnail images we found 57 images (listed in Table \ref{282P:tab:observationsTable}) in which we could confidently identify \objnameBM{} in the frame. The remaining images either did not probe faintly enough, did not actually capture \objnameBM{} (e.g., \objnameBM{} was not on a detector), or suffered from image artifacts that made the image unsuitable for activity detection. The 57 images span 22 observing dates; nine dates had at least one image we ascertained showed probable activity, five from the 2012--2013 epoch and four dates from the 2021--2022 apparition. Section \ref{282P:sec:observations} provides a complete listing of observations used in this work.

Figure \ref{282P:fig:ActivityTimeline} shows three plots with shared $x$-axes (years). 

Apparent magnitude and observability (the number of hours an object is above the horizon and the Sun is below the horizon) together provide insight into potential observational biases. For example, observations for detecting activity are ideal when \objnameBM{} is brightest, near perihelion, and observable for many hours in an observing night. When contrasting hemispheres, this plot makes it clear that some periods (e.g., 2016 -- 2020) are more favorable for observations in the northern hemisphere, whereas other observation windows (e.g., 2013 -- 2015, 2022) are better suited to southern hemisphere facilities.

\subsection{Follow-up Telescope Observations}
\label{282P:subsec:telescopeobservations}

\paragraph{Magellan} During twilight on UT 2022 March 7 we observed \objnameBM{} with the \ac{IMACS} instrument \citep{dresslerIMACSInamoriMagellanAreal2011} on the Magellan 6.5~m Baade telescope located atop Las Campanas Observatory (Chile). We successfully identified \objnameBM{} in the images, however \objnameBM{} was in front of a dense part of the Milky Way,
preventing us from unambiguously identifying activity. We used these observations to inform our Gemini \ac{SNR} calculations.

\paragraph{VATT} On UT 2022 April 6 we observed \objnameBM{} with the 1.8~m \ac{VATT} at the \ac{MGIO} in Arizona (Proposal ID S165, \ac{PI} Chandler). \objnameBM{} was in an especially dense part of the galaxy so we conducted test observations to assess the viability of activity detection under these conditions. We concluded object detection would be challenging and activity detection essentially impossible in such a dense field.

\paragraph{LDT} On UT 2022 May 21 we observed \objnameBM{} with the \ac{LDT} in Arizona (PI: Chandler). Finding charts indicated \objnameBM{} was in a less dense field compared to our \ac{VATT} observations, however we were hardly able to resolve \objnameBM{} or identify any activity because the field was still too crowded.

\paragraph{Gemini South} On UT 2022 June 7 we observed \objnameBM{} with the \ac{GMOS} South instrument \citep{hookGeminiNorthMultiObjectSpectrograph2004,gimenoOnskyCommissioningHamamatsu2016} on the 8.1~m Gemini South telescope located atop Cerro Pachón in Chile (Proposal ID GS-2022A-DD-103, \acs{PI} Chandler). We timed this observation to take place during a $\sim$10 day window when \objnameBM{} was passing in front of a less dense region of the Milky Way. We acquired eighteen images, six each in $g'$, $r'$, and $i'$. Activity was clearly visible in the reduced data in all filters, with activity appearing strongest in $g'$ (Figure \ref{282P:fig:282P}d). Our observations confirmed \objnameBM{} was still active, 15 months after the 2021 archival data, evidence supporting sublimation as the most likely cause for activity (Section \ref{282P:sec:mechanism}).

\section{Dynamical Modeling}
\label{282P:subsec:dynamicalmodeling}

We analyzed \objnameBM{} orbital characteristics in order to (1) determine its dynamical class (Section \ref{282P:sec:dynamicalClassification}), and (2) inform our activity mechanism assessment (Section \ref{282P:sec:mechanism}). 
We simulated a cloud of 500 \objnameBM{} orbital clones, randomly drawn from Gaussian distributions centered on the current fitted parameters of \objnameBM{}, with widths corresponding to uncertainties of those fits (Table \ref{282P:tab:ObjectData} lists parameters and associated uncertainties), as reported by \acs{JPL} Horizons \citep{giorginiJPLOnLineSolar1996}.

We modeled the gravitational influence of the Sun and the planets (except Mercury) on each orbital clone using the \ac{IAS15} N-body integrator \citep{reinIAS15FastAdaptive2015}, typically accurate to machine precision, with the \texttt{REBOUND} Python package \citep{reinREBOUNDOpensourceMultipurpose2012,reinHybridSymplecticIntegrators2019}. We ran simulations 1,000 years forward and backward through time. Longer integrations were unnecessary because dynamical chaos ensues prior to $\sim$200 years ago and after $\sim$350 years into the future. Beyond these times the orbit of \objnameBM{} is not deterministic due to observational uncertainties.

\begin{figure*}
    \centering
    \begin{tabular}{cc}
        \includegraphics[width=0.36\linewidth]{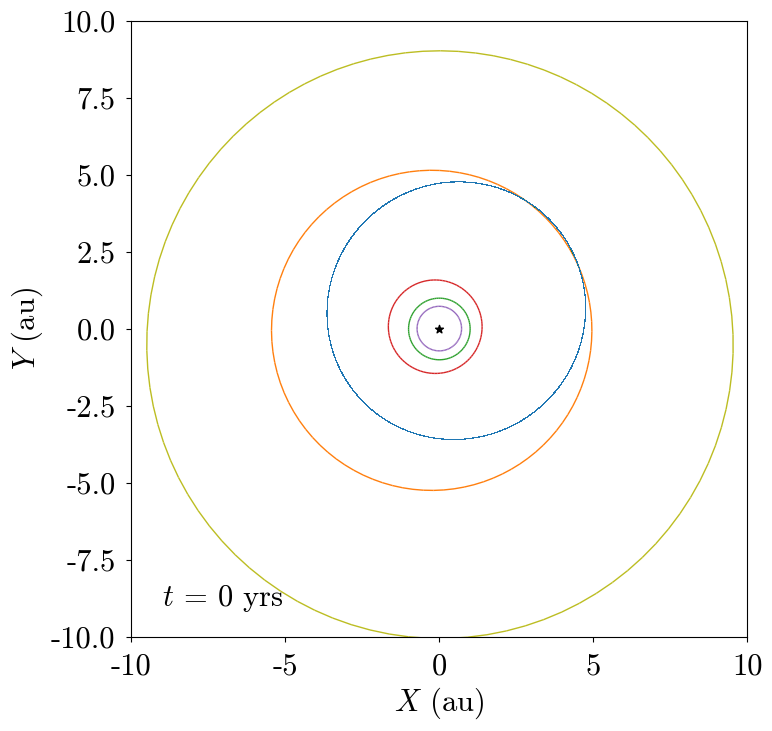} & \includegraphics[width=0.48\linewidth]{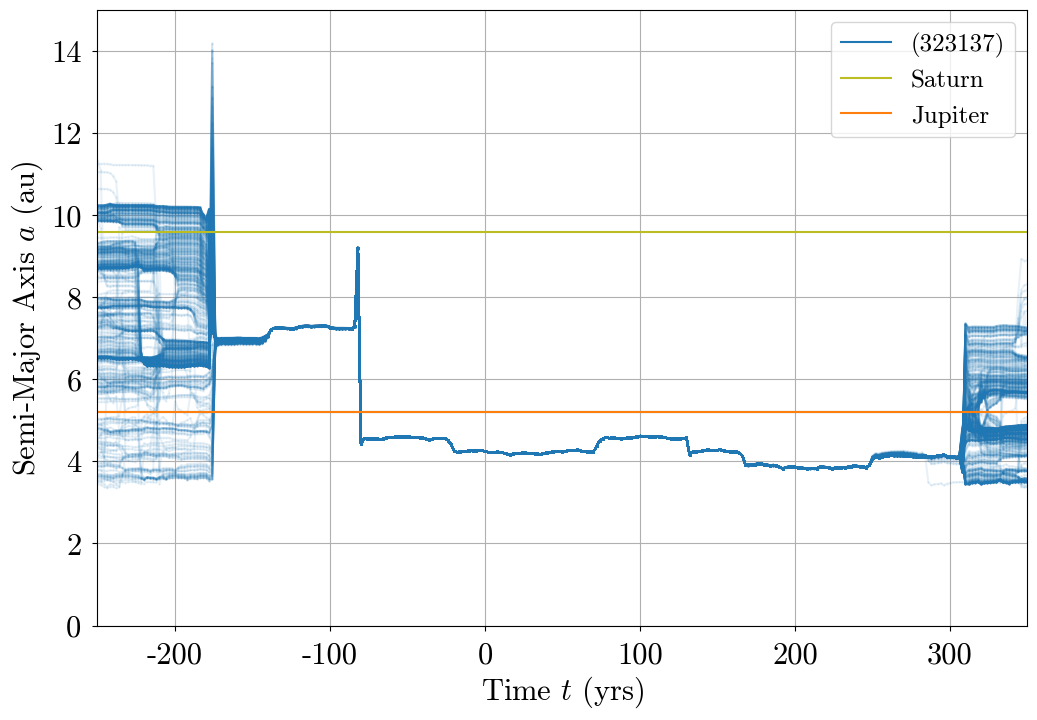}\\
        (a) & (b)\\
        \\
        \includegraphics[width=0.48\linewidth]{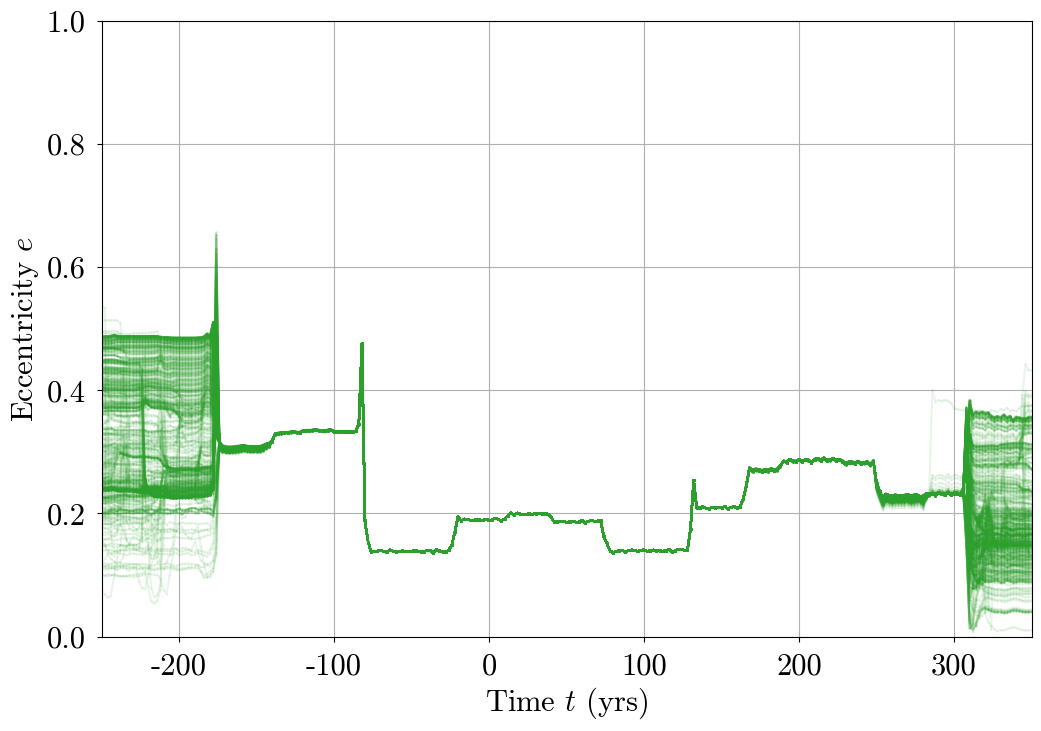} & \includegraphics[width=0.48\linewidth]{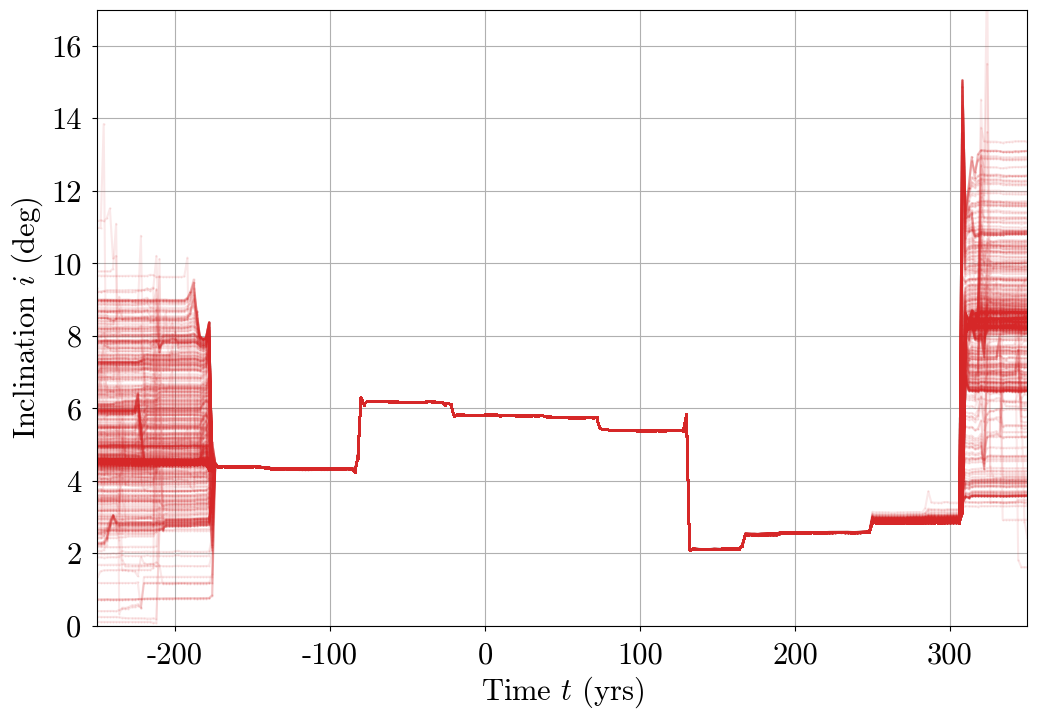}\\
       (c) & (d)\\
    \end{tabular}
    \caption{
        Results from dynamical integration of \objnameBM{} orbital clones. For all plots, time $t=0$ corresponds to UT 2022 January 21. Jovian and Saturnian close encounters prevent accurate orbital parameter determination outside $-180\lesssim t\lesssim300$ yrs, given known orbital uncertainties.  
        \textbf{(a)} Orbital diagram for \objnameBM{} and nearby planets.
        \textbf{(b)} Semi-major axis $a$ evolution.
        \textbf{(c)} Eccentricity $e$ evolution. 
        \textbf{(d)} Inclination $i$ evolution. 
    }
    \label{282P:fig:orbitevolution1}
\end{figure*}

\begin{figure*}
    \centering
    \begin{tabular}{cc}
        \includegraphics[width=0.48\linewidth]{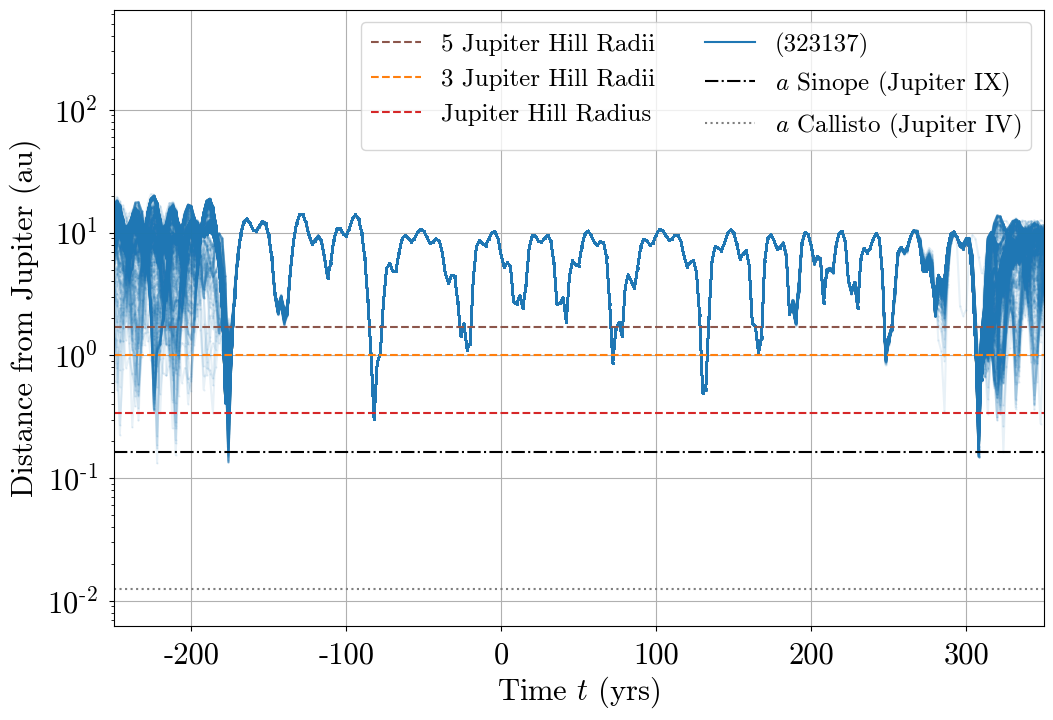} & \includegraphics[width=0.48\linewidth]{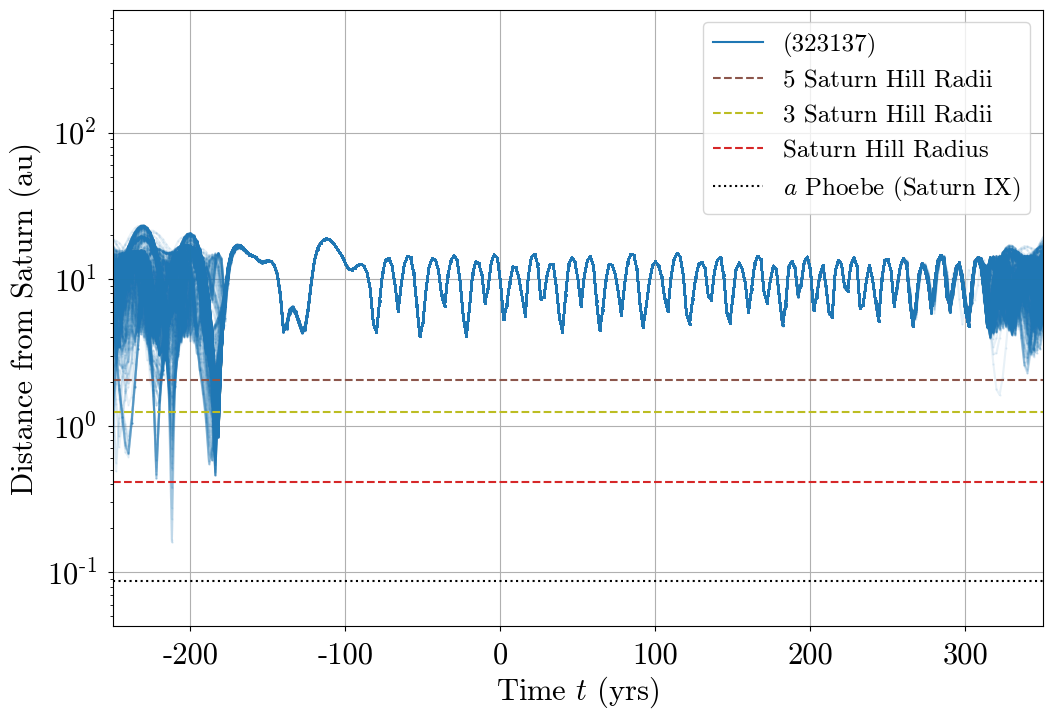}\\
        (a) & (b)\\
        \\
        \includegraphics[width=0.48\linewidth]{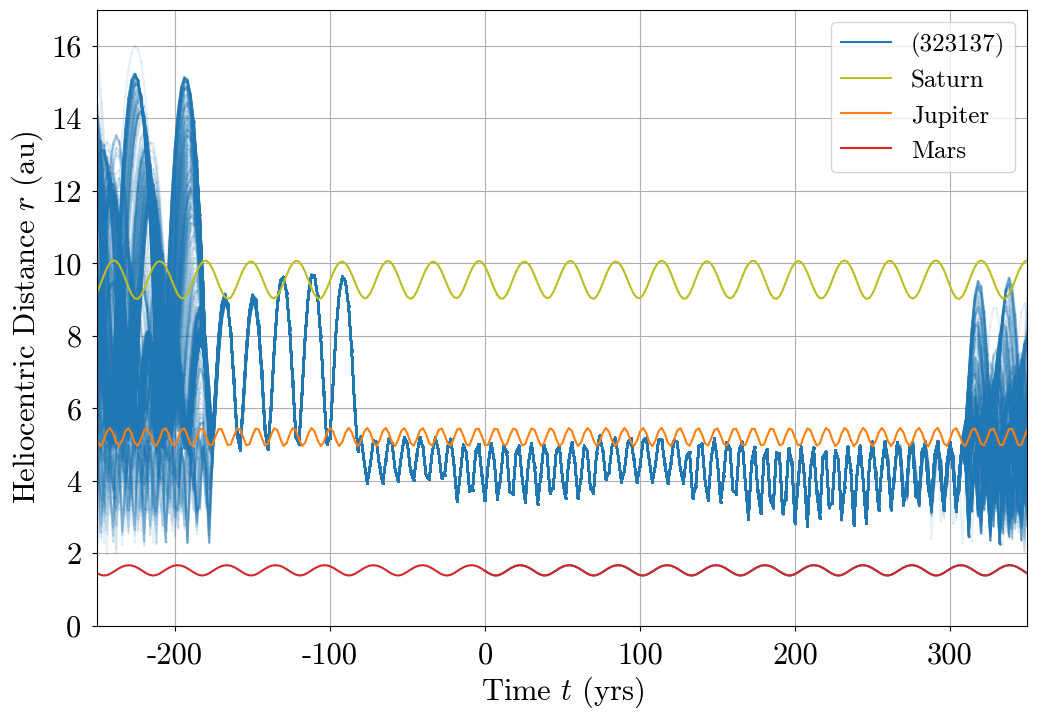} & \includegraphics[width=0.48\linewidth]{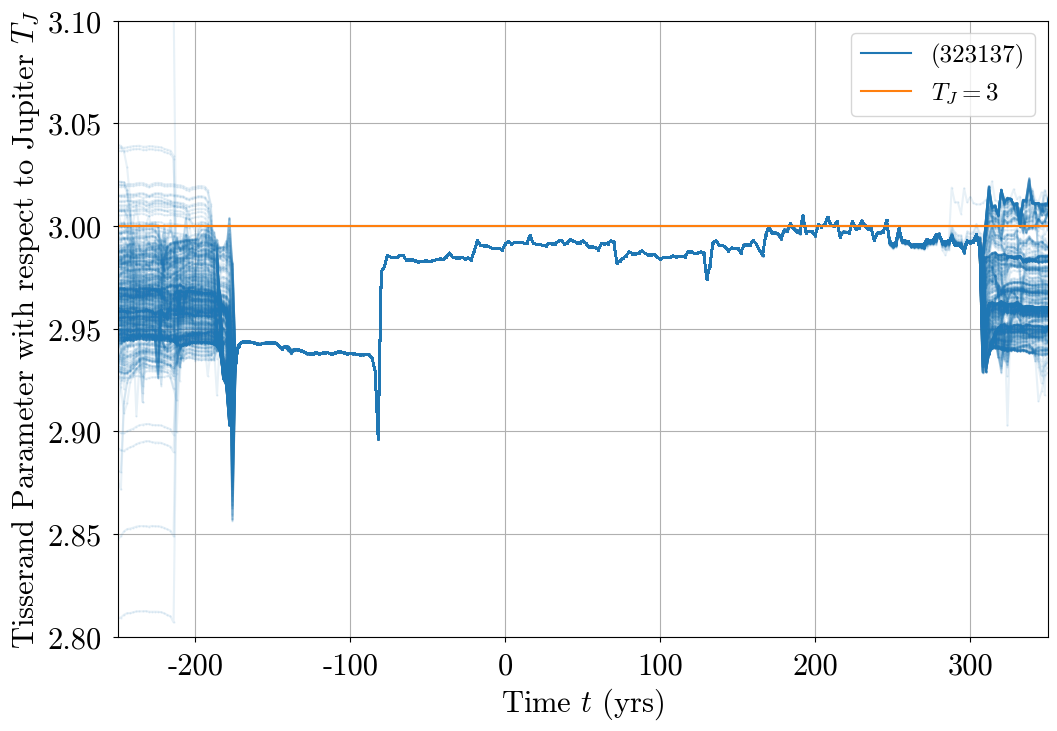}\\
        (c) & (d)\\
    \end{tabular}
    \caption{
    Additional results from dynamical integration of \objnameBM{} orbital clones. For each plot, time $t=0$ is UT 2022 January 21. Close encounters with Jupiter and Saturn are so significant that orbital elements cannot be accurately determined before/after $-180\lesssim t\lesssim300$ yrs, given orbital uncertainties.
    \textbf{(a)} Distance between Jupiter and \objnameBM{} as a function of time. Indicated Hill radii provide references for the degree of orbit alteration imparted by a close encounter. For reference, the semi-major axes of two Jovian moons are shown: Callisto, the outermost Galilean satellite, and Sinope \citep{nicholsonDiscoveryNinthSatellite1914}, a likely captured \citep{gravPhotometricSurveyIrregular2003a} distant irregular and retrograde Jovian moon.
    \textbf{(b)} Distance between Saturn and \objnameBM{} as a function of time. The semi-major axis of the irregular Saturnian moon Phoebe, believed to be captured through close encounter \citep{johnsonSaturnMoonPhoebe2005,jewittIrregularSatellitesPlanets2007}, is given for reference.
    \textbf{(c)} Heliocentric distance $r$ evolution. 
    \textbf{(d)} Tisserand parameter with respect to Jupiter $T_\mathrm{J}$ (Equation \ref{282P:eq:TJ}), where the horizontal orange line representing $T_\mathrm{J}=3$ indicates the widely-adopted boundary between comet-like and asteroid-like orbits.
    }
    \label{282P:fig:orbitevolution2}
\end{figure*}

Results from the dynamical evolution of the \objnameBM{} orbital clones are shown in Figure \ref{282P:fig:orbitevolution1} and Figure \ref{282P:fig:orbitevolution2}. For all plots, time $t=0$ corresponds to \ac{JD} 2459600.5 (UT 2022 Jan 21) and time ranges from $t=-250$ to $t=+350$ (1772--2372 AD). Horizontal lines at distances of one, three, and five Hill radii (Equation \ref{282P:eq:rH}) from Jupiter and Saturn are shown in Figure \ref{282P:fig:orbitevolution2} panels a and b. The Hill Radius \citep{hillResearchesLunarTheory1878} $r_H$ is a metric of orbital stability and indicates the region where a secondary body (e.g., a planet) has dominant gravitational influence over a tertiary body (e.g., a moon), with both related to a primary body, such as the Sun. At pericenter, the Hill radius of the secondary body can be approximated as

\begin{equation}
    r_\mathrm{H} \approx a(1-e)(m/3M)^{1/3},
    \label{282P:eq:rH}
\end{equation}

\noindent where $a$, $e$, and $m$ are the semi-major axis, eccentricity and mass of the secondary (Jupiter or Saturn in our case), respectively, and $M$ is the mass of the primary (here, the Sun). Close passages of a small body within a few Hill radii of a planet are generally considered to be significant perturbations and may drastically alter the orbit of the small body (see \citealt{hamiltonOrbitalStabilityZones1992} Section 2.1.2 for discussion).

From $\sim$180 years ago until $\sim$300 years in the future, the orbit of \objnameBM{} is well-constrained in our simulations. Figure \ref{282P:fig:orbitevolution2}a illustrates that \objnameBM{} has roughly 10 close encounters (within $\sim$2 au) with Jupiter, and one with Saturn, over the range $-250<t<350$ yr. These encounters have a strong effect on the semi-major axis $a$ of \objnameBM{} (Figure \ref{282P:fig:orbitevolution1}b), and, as illustrated by Figure \ref{282P:fig:orbitevolution2}d, a noticeable influence on its Tisserand parameter with respect to Jupiter $T_\mathrm{J}$, 

\begin{equation}
	T_\mathrm{J} = \frac{a_\mathrm{J}}{a} + 2\cos(i)\sqrt{\frac{a}{a_\mathrm{J}}\left(1-e^2\right)},
	\label{282P:eq:TJ}
\end{equation}

\noindent where $a_\mathrm{J}$ is the semi-major axis of Jupiter and $a$, $e$ and $i$ are the semi-major axis, eccentricity and inclination of the body, respectively. $T_\mathrm{J}$ essentially describes an object's close approach speed to Jupiter or, in effect, the degree of dynamical perturbation an object will experience as a consequence of Jovian influence. $T_\mathrm{J}$ is often described as invariant \citep{kresakJacobianIntegralClassificational1972} or conserved, meaning that changes in orbital parameters still result in the same $T_\mathrm{J}$, although, in practice, its value does change slightly as a result of close encounters (see Figure \ref{282P:fig:orbitevolution2}d).

Due to the small Jupiter-centric distances of \objnameBM{} during these encounters, compounded by its orbital uncertainties, the past orbit of \objnameBM{}, prior to $t\approx-180$ yrs, is not deterministic. Dynamical chaos is plainly evident in all panels as orbital clones take a multitude of paths within the parameter space, resulting in a broad range of possible orbital outcomes due only to slight variations in initial \objnameBM{} orbital parameters.

A consequential encounter with Saturn occurred around 1838 ($t\approx-184$~yr; Figure \ref{282P:fig:orbitevolution2}b), followed by another interaction with Jupiter in 1846 ($t=-176$ yr; Figure \ref{282P:fig:orbitevolution2}a). After these encounters \objnameBM{} was a \ac{JFC} (100\% of orbital clones) with a semi-major axis between Jupiter's and Saturn's semi-major axes (Figure \ref{282P:fig:orbitevolution1}b), and crossing the orbits of both planets (Figure \ref{282P:fig:orbitevolution2}c). These highly perturbative passages placed \objnameBM{} on the path that would lead to its current Quasi-Hilda orbit.

In 1940 ($t=-82$~yr), \objnameBM{} had a very close encounter with Jupiter, at a distance of 0.3~au -- interior to one Hill radius. As seen in Figure \ref{282P:fig:orbitevolution1}a, this encounter dramatically altered \objnameBM{}'s orbit, shifting \objnameBM{} from an orbit primarily exterior to Jupiter to an orbit largely interior to Jupiter (Figure \ref{282P:fig:orbitevolution1}b). This same interaction also caused \objnameBM{}'s orbit to migrate from Jupiter- and Saturn-crossing to only a Jupiter-crossing orbit (Figure \ref{282P:fig:orbitevolution2}c). This step in the orbital evolution of \objnameBM{} also changed its $T_\mathrm{J}$ (Figure \ref{282P:fig:orbitevolution2}d) to be close to the traditional $T_\mathrm{J}=3$ comet--asteroid dynamical boundary. At this point in time, \objnameBM{} remained a \ac{JFC} (100\% of orbital clones) despite its dramatic change in orbit.

Around $t\approx200$ yr, \objnameBM{} crosses the $T_\mathrm{J}=3$ boundary dividing the \ac{JFC}s and the asteroids on the order of 10 times. Although no major changes in the orbit \objnameBM{} occur during this time, because of the stringency of this boundary, relatively minor perturbations result in oscillation between dynamical classes.

After a major encounter with Jupiter around 2330 AD ($t\approx308$ yrs), dynamical chaos again becomes dominant and remains so for the rest of the simulation. Following this encounter, the orbit of \objnameBM{} does not converge around any single solution. Slight diffusion following the previous several Jupiter passages are also visible in Figure \ref{282P:fig:orbitevolution1}b-d and Figure \ref{282P:fig:orbitevolution2}a-d, and these also add uncertainty concerning encounters around 2301 to 2306 ($t\approx280$ to $285$ yrs). Although we are unable to precisely determine past and future orbits of \objnameBM{} outside of $-180\lesssim t\lesssim300$ because of dynamical chaos, we are able to examine the fraction of orbital clones that finish the simulation (forwards and backwards) on orbits associated with different orbital classes.

\section{Dynamical Classifications: Past, Present and Future}
\label{282P:sec:dynamicalClassification}

Minor planets are often classified dynamically, based on orbital characteristics such as semi-major axis. \objnameBM{} was labeled a \ac{JFC} by \cite{hsiehMainbeltCometsPanSTARRS12015}, in agreement with a widely adopted system that classifies objects dynamically based on their Tisserand parameter with respect to Jupiter, $T_\mathrm{J}$ (Equation \ref{282P:eq:TJ}).

Via Equation \ref{282P:eq:TJ}, Jupiter's $T_\mathrm{J}$ is 2.998 given $a_\mathrm{J}=5.20$, $e_\mathrm{J}=0.049$, and $i_\mathrm{J}=0.013$. Notably, objects with $T_\mathrm{J}>3$ cannot cross the Jovian orbit, thus their orbits are entirely interior or exterior to Jupiter's orbit \citep{levisonCometTaxonomy1996}. 
Objects with $T_\mathrm{J}<3$ are considered cometary \citep{levisonCometTaxonomy1996}, while those with $T_\mathrm{J}>3$ are not \citep{vaghiOriginJupiterFamily1973,vaghiOrbitalEvolutionComets1973}, a classification approach first suggested by \cite{carusiHighOrderLibrationsHalleyType1987,carusiCometTaxonomy1996}. \acp{JFC} have $2<T_\mathrm{J}<3$ (see e.g., \citealt{jewittActiveCentaurs2009}), 
and Damocloids and have $T_\mathrm{J}<2$ \citep{jewittFirstLookDamocloids2005}. 
We note, however, that the traditional $T_\mathrm{J}$ asteroid -- \ac{JFC} -- Damocloid continuum does not include (or exclude) \acp{QHO}.

As discussed in Section \ref{282P:introduction}, we adopt the \cite{jewittActiveCentaurs2009} definition of Centaur, which stipulates that a Centaur has an orbit entirely exterior to Jupiter, with both $q$ and $a$ interior to Neptune, and the body is not in 1:1 resonance with a planet. \objnameBM{} has a semi-major axis $a=4.240$~au, well interior to Jupiter's $a_\mathrm{J}=5.2$~au. This disqualifies \objnameBM{} as presently on a Centaurian orbit.

Active objects other than comets orbiting interior to Jupiter are primarily the active asteroids, defined as (1)  $T_\mathrm{J}>3$, (2) displaying comet-like activity, and (3) orbiting outside of mean-motion resonance with any of the planets. This last stipulation rules out the Jupiter Trojans (1:1 resonance) and the Hildas (3:2 resonance with Jupiter), even though both classes have members above and below the  $T_\mathrm{J}=3.0$ asteroid--comet transition line. We compute $T_\mathrm{J}=2.99136891\pm(3.73\times10^{-8})$ for \objnameBM{} (see Table \ref{282P:tab:ObjectData} for a list of orbital parameters). These values do not exceed the traditional $T_\mathrm{J}=3$ cutoff; thus \objnameBM{} cannot be considered an active asteroid in its current orbit. \acp{MBC} are an active asteroid subset defined as orbiting entirely within the main asteroid belt  \citep{hsiehMainbeltCometsPanSTARRS12015}. Figure \ref{282P:fig:orbitevolution2}c shows that \objnameBM{}'s heliocentric distance does not stay within the boundaries of the Asteroid Belt (i.e., between the orbits of Mars and Jupiter), and so \objnameBM{} does not qualify as a \ac{MBC}.

\begin{figure*}
    \centering
    \begin{tabular}{ccc}
         \includegraphics[width=0.32\linewidth]{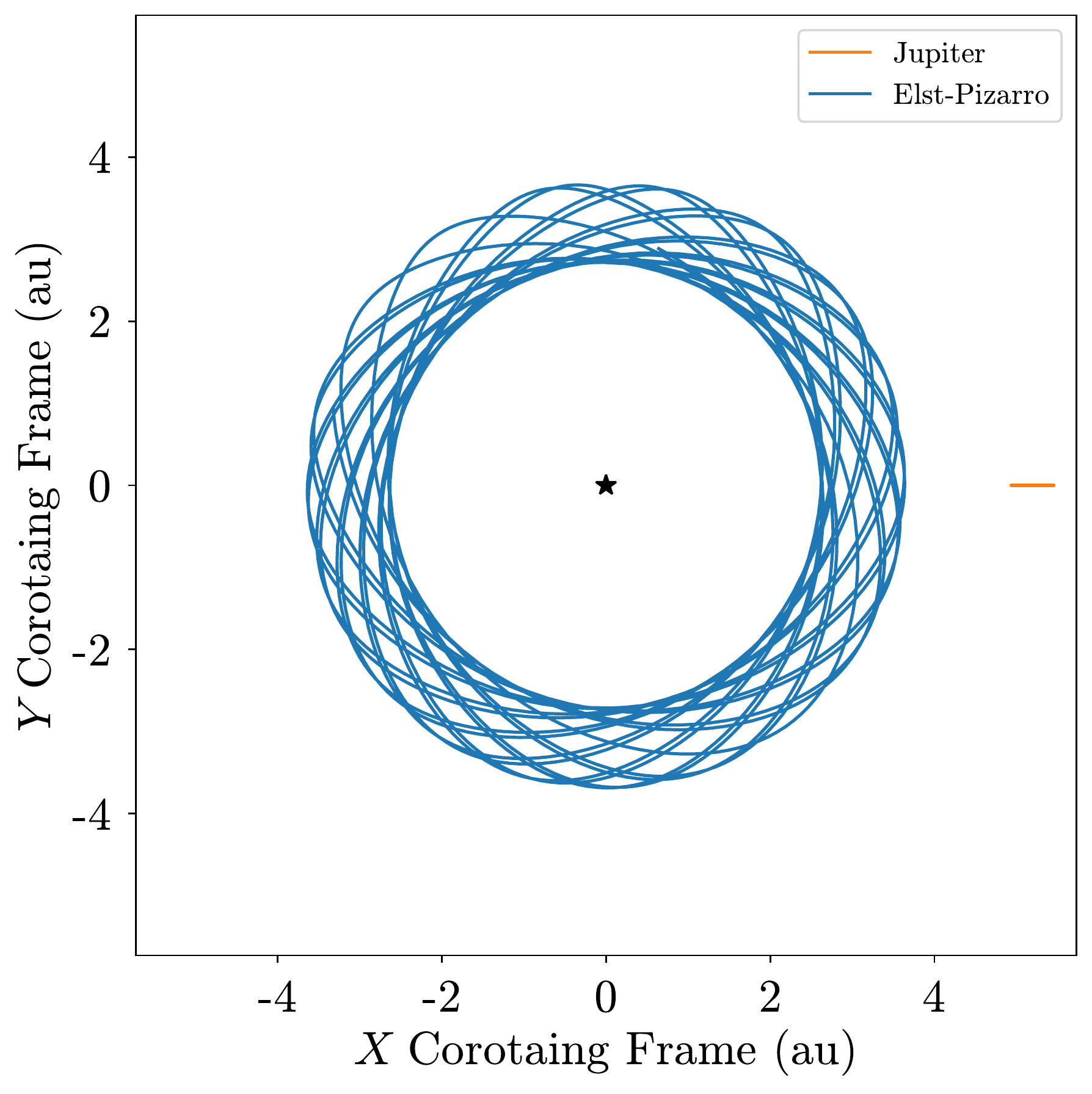} &
         \includegraphics[width=0.32\linewidth]{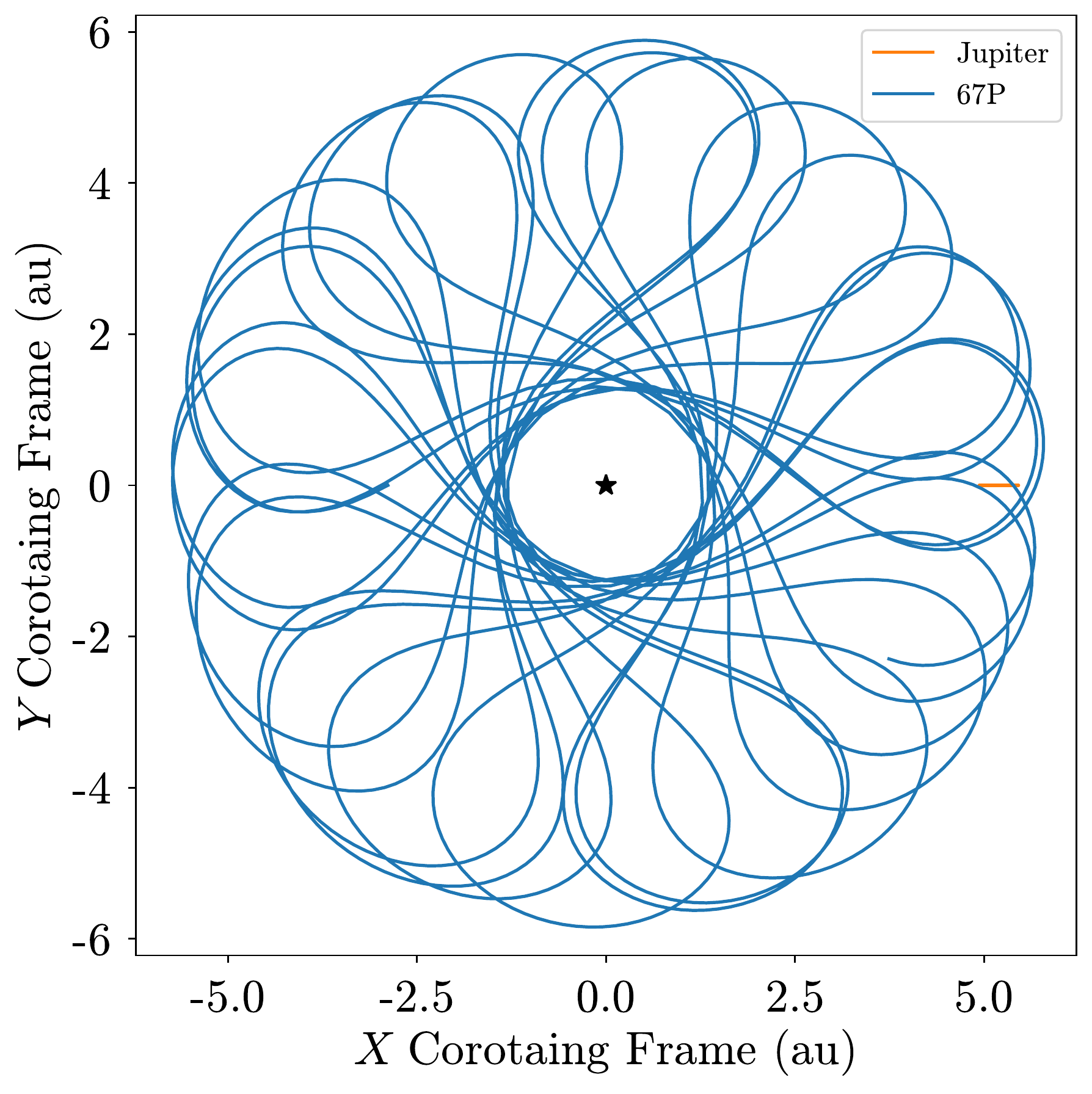} &
         \includegraphics[width=0.32\linewidth]{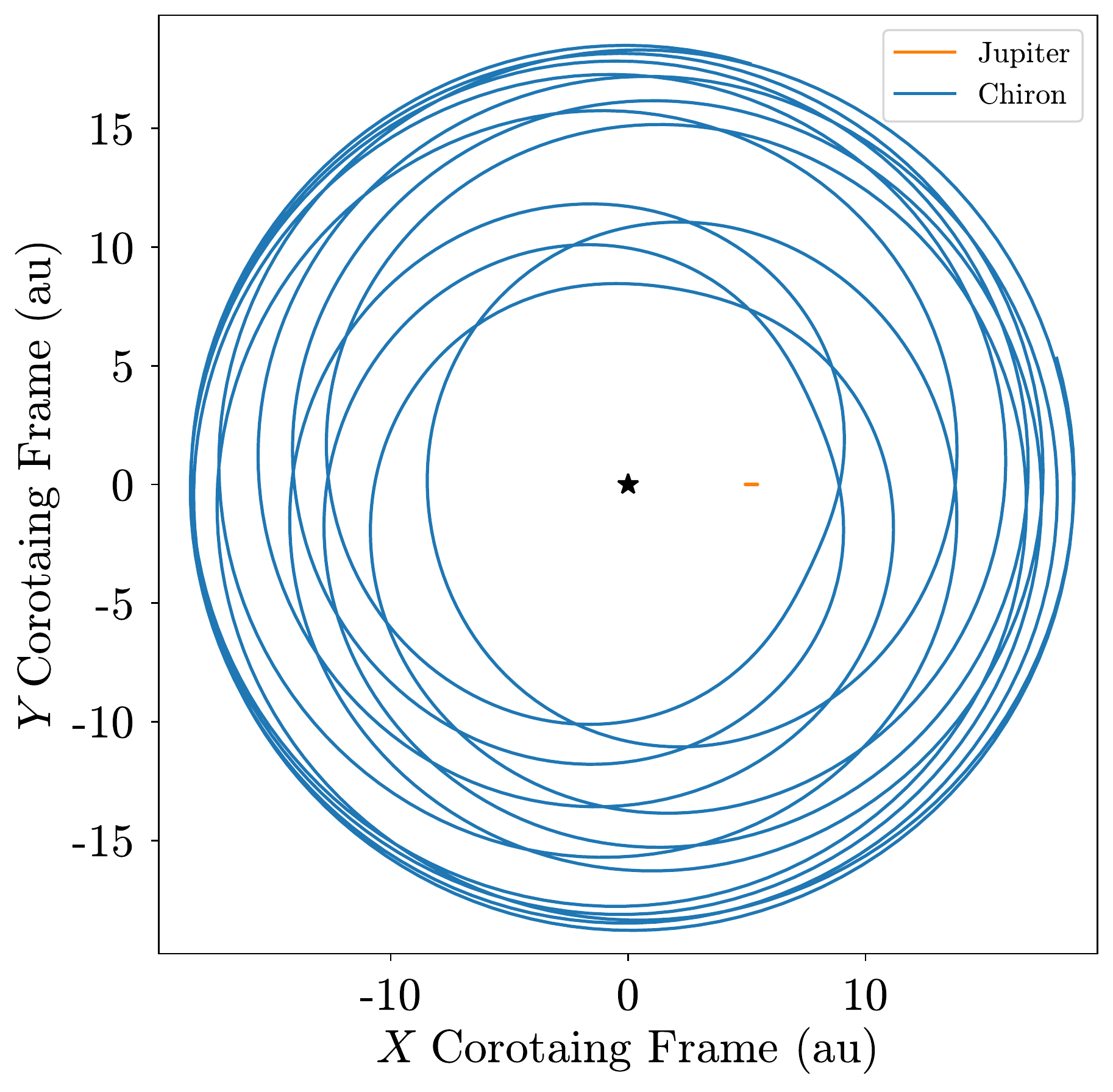}\\
         (a) & (b) & (c)\\
          \\
          \includegraphics[width=0.32\linewidth]{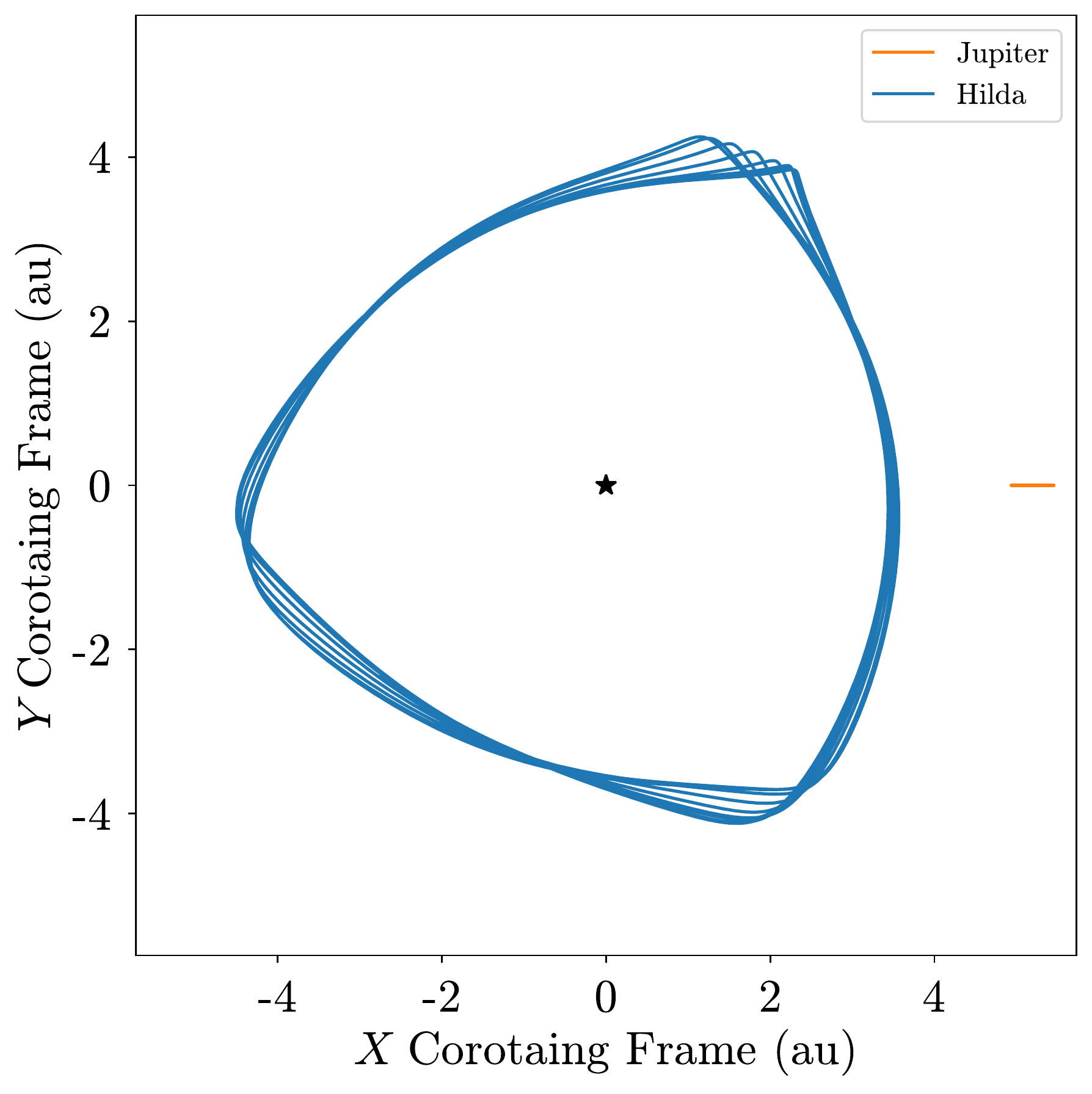} &
          \includegraphics[width=0.32\linewidth]{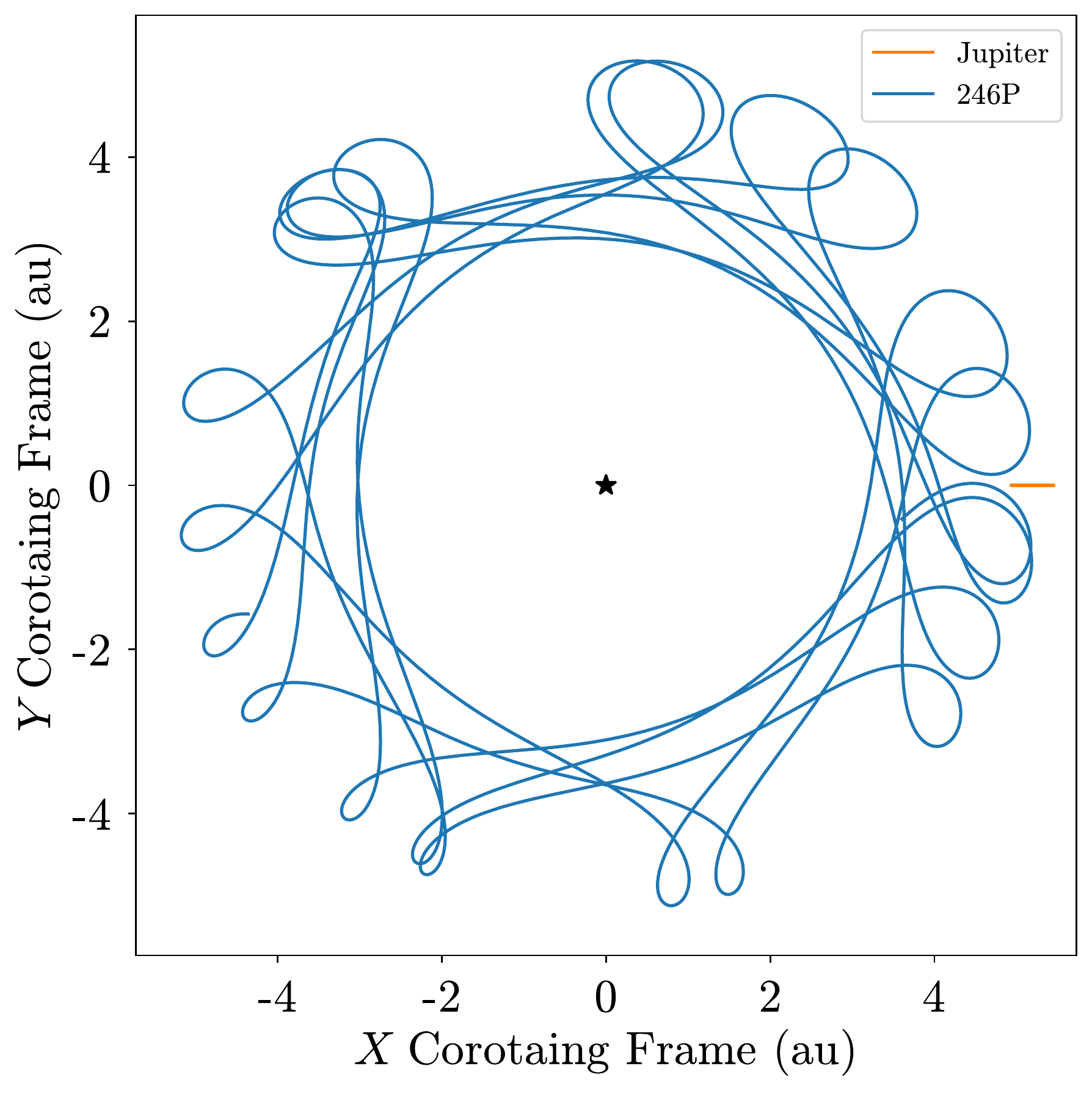} &
          \includegraphics[width=0.32\linewidth]{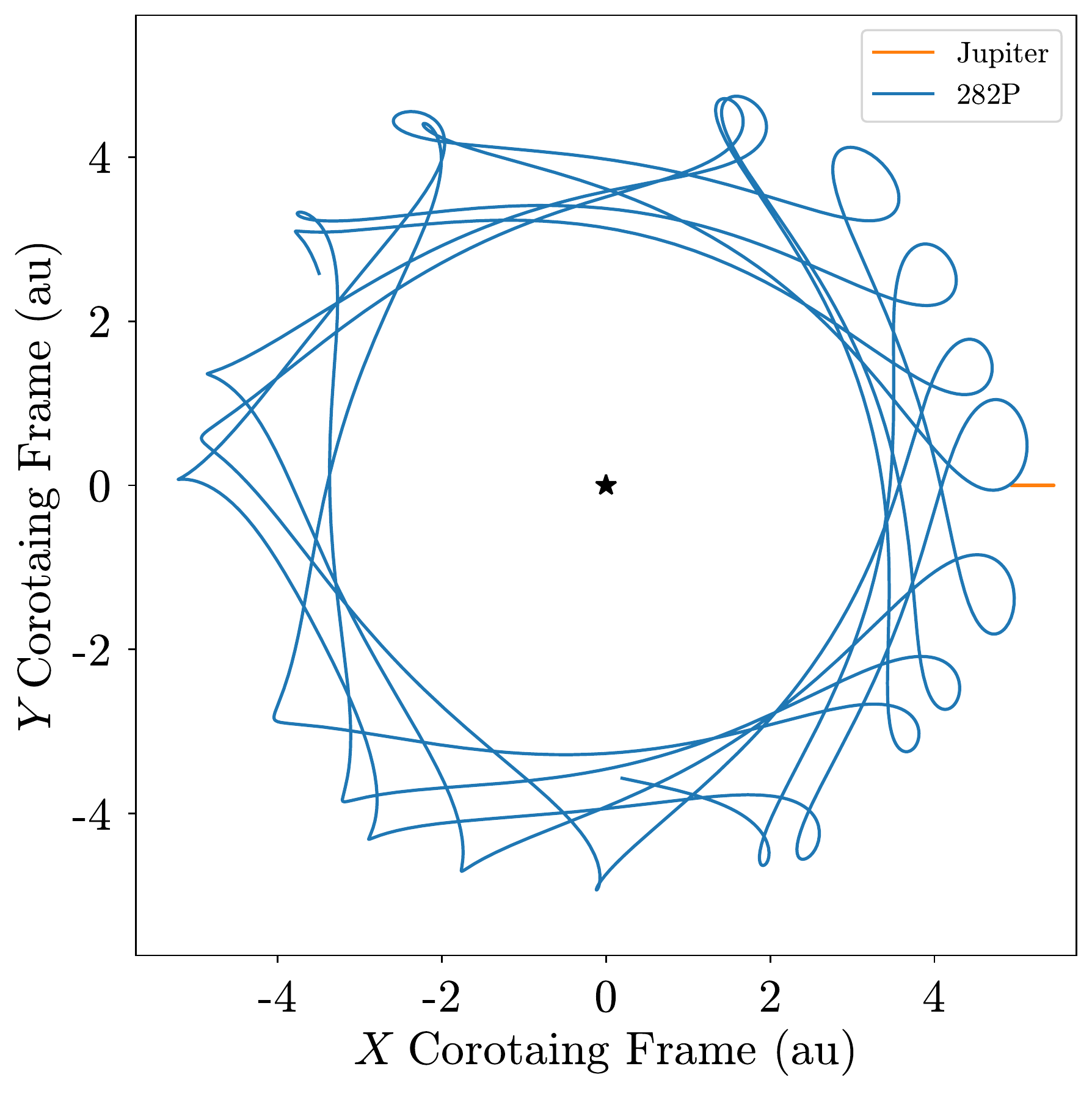} \\
          (d) & (e) & (f)\\
    \end{tabular}
    \caption{The orbital motion of minor planets (blue lines) as seen in the reference frame corotating with Jupiter (orange lines at right edge of plots). 
    (a) \ac{MBC} (7968)~Elst-Pizarro (133P).
    (b)  \ac{JFC} 67P/Churyumov-Gerasimenko (previously visited by the \ac{ESA} Rosetta Spacecraft).
    (c) Centaur (2060)~Chiron (95P).
    (d) (153)~Hilda, the namesake of the Hilda dynamical class, in the 3:2 interior mean-motion resonance with Jupiter. 
    (e) Quasi-Hilda 246P/\acs{NEAT}, also designated 2010~V$_2$ and 2004~F$_3$. 
    (f) Our object of study, \objnameBM{}, in its Quasi-Hilda orbit.
    }
    \label{282P:fig:corotatingFrame}
\end{figure*}

Blurring the lines between \ac{JFC} and Hilda is the Quasi-Hilda regime. A Quasi-Hilda, also referred to as a \ac{QHO}, \ac{QHA} \citep{jewittOutburstingQuasiHildaAsteroid2020}, or \ac{QHC}, is a minor planet on an orbit similar to a Hilda \citep{tothQuasiHildaSubgroupEcliptic2006,gil-huttonCometCandidatesQuasiHilda2016}. Hildas are defined by their 3:2 interior mean-motion resonance with Jupiter, however Quasi-Hildas are not in this resonance, though they do orbit near it. Quasi-Hildas likely migrated from the \ac{JFC} region (see discussion, \citealt{jewittOutburstingQuasiHildaAsteroid2020}). We favor the term \ac{QHO} or \ac{QHA} over \ac{QHC}, given that fewer than 15 Quasi-Hildas have been found to be active, while the remainder of the $>270$ identified Quasi-Hildas \citep{gil-huttonCometCandidatesQuasiHilda2016} have not been confirmed to be active. Notable objects from the Quasi-Hilda class are 39P/Oterma \citep{otermaNEWCOMETOTERMA1942}, an object that was a Quasi-Hilda prior to 1963, when a very close (0.095~au) encounter with Jupiter redirected the object into a Centuarian orbit. Another notable Quasi-Hilda was D/Shoemaker-Levy~9, which famously broke apart and impacted Jupiter in 1994 (e.g., \citealt{weaverHubbleSpaceTelescope1995}).

Quasi-Hildas have orbital parameters similar to that of the Hildas, approximately $3.7 \lesssim a \lesssim 4.2$~au, $e\le0.3$, and $i\le20\degr$. In rough agreement, \objnameBM{} has $a=4.24$~au, $e=0.188$, and $i=5.8\degr$ (Table \ref{282P:tab:ObjectData}). Hildas are also known for their trilobal orbits as viewed in the Jupiter corotating frame (caused by their residence in the 3:2 interior mean motion resonance with Jupiter), especially the namesake asteroid (153)~Hilda (Figure \ref{282P:fig:corotatingFrame}d). Because (153)~Hilda is in a stable 3:2 resonant orbit with Jupiter, its orbit remains roughly constant, with a small amount of libration over time. By contrast, Quasi-Hildas like 246P/\acs{NEAT} (Figure \ref{282P:fig:corotatingFrame}e) are near the same resonance and show signs of this characteristic trilobal pattern, however their orbits drift considerably on timescales of hundreds of years. \objnameBM{} (Figure \ref{282P:fig:corotatingFrame}f) also displays a typical Quasi-Hilda orbit as viewed in the Jupiter corotating reference frame.

In the past, prior to 250~yr ago, 52\% (260) of the 500 orbital clones were \acp{JFC}, 48\% (239) were Cenaturs, 5\% (26) were already \acp{QHO}, and one (0.2\%) was an \ac{OMBA}. The most probable scenario prior to 250 years ago was that was either a \ac{JFC} or Centaur, both classes that trace their origins to the Kuiper Belt (see reviews, \citealt{morbidelliKuiperBeltFormation2020} and \citealt{jewittActiveCentaurs2009}, respectively).

In the future, after 350 years time, 81\% (403) of clones become \acp{JFC}, 18\% (90) remain \acp{QHO}, 14\% (69) become \acp{OMBA}, and 5.6\% (28) return to Centaurian orbits. Clearly the most likely scenario is that \objnameBM{} will become a \ac{JFC}, however there are still significant possibilities that \objnameBM{} remains a \ac{QHO} or becomes an active \ac{OMBA}.

\section{Thermodynamical Modeling}
\label{282P:sec:thermo}

In order to understand the approximate temperature ranges that \objnameBM{} experiences over the course of its present orbit in order to (1) understand what role, if any, thermal fracture may play in the activity we observe, and (2) evaluate the likelihood of ices surviving on the surface, albeit with limited effect because of the narrow window ($\sim$500 years) of dynamically well-determined orbital parameters available (Section \ref{282P:subsec:dynamicalmodeling}).

Following the procedure of \cite{chandler2483702005QN2021} (originally adapted from \citealt{hsiehMainbeltCometsPanSTARRS12015}), we compute the surface equilibrium temperature $T_\mathrm{eq}$ for \objnameBM{} as a gray airless body. $\chi$ describes the distribution of heat over the surface of a body, with $\chi=1$ the isothermal approximation (i.e., a fast-rotating body) and $\chi=4$ the ``slab'' case, where an object has one side that always faces the Sun; these cases result in the minimum and maximum expected temperatures for the body, respectively.

Solving Equations 3 -- 5 of \cite{chandler2483702005QN2021} (energy balance for a gray airless body, sublimation mass loss rate, and the Clausius--Clapeyron relationship) for the body's heliocentric distance $r_\mathrm{h}$ (in au) as a function of equilibrium temperature $T_\mathrm{eq}$ and $\chi$, where 
$A=0.05$ is the assumed typical bond albedo, 
$\Delta H_\mathrm{subl}=51.06\ \mathrm{MJ}/\mathrm{kmol}$ is the ice-to-gas heat of sublimation, 
$\epsilon = 0.9$ is the assumed typical effective infrared emissivity, 
$f_\mathrm{D}=1$ accounts for sublimation efficiency dampening due to mantling (unity in the absence of mantle), 
$F_\odot$ is the solar constant 1360~$\mathrm{W}/\mathrm{m}^2$, 
$L=2.83\ \mathrm{MJ}/\mathrm{kg}$ is the latent heat of H$_2$O ice (approximated here as temperature independent), 
$R_\mathrm{G}=8314\ \mathrm{J}\ \mathrm{kmol}^{-1}\ \mathrm{K}^{-1}$ is the ideal gas constant, 
and 
$\sigma=5.67037\times10^{-8}\ \mathrm{W}\ \mathrm{m}^{-2}\ \mathrm{K}^{-4}$ is the Stefan--Boltzmann constant: 

\begin{equation} 
    r_\mathrm{h}(T_\mathrm{eq},\chi) = \frac{F_\odot\left(1-A\right)\chi^{-1}}{\epsilon\sigma T_\mathrm{eq}^4 + L f_\mathrm{D} \cdot 611\ e^{\frac{\Delta H_\mathrm{subl}}{R_\mathrm{G}}\left(\frac{1}{273.16\mathrm{K}} - \frac{1}{T_\mathrm{eq}}\right)}}
    \label{282P:eq:teq}
\end{equation}

We translate Equation \ref{282P:eq:teq} to a function of equilibrium temperature $T_\mathrm{eq}$ by computing $r_\mathrm{h}$ for an array of values (100~K to 300~K in this case), then fit a model to these data with a \texttt{SciPy} \citep{virtanenSciPyFundamentalAlgorithms2020} (Python package) univariate spline. Using this model we compute \objnameBM{} temperatures for \objnameBM{} heliocentric distances from perihelion and aphelion.
with this function to arrive at temperatures for \objnameBM{} over the course of its orbit.

Figure \ref{282P:fig:ActivityTimeline} (bottom panel) shows the temperature evolution for the maximum and minimum solar heating distribution scenarios ($\chi=1$ and $\chi=4$, respectively) for \objnameBM{} from 2012 through 2024. Temperatures range between roughly 175~K and 220~K for $\chi=1$, or 130~K and 160~K for $\chi=4$, with a $\sim45$~K maximum temperature variation in any one orbit. \objnameBM{} spends some ($\chi=4$) or all ($\chi=1$) of its time with surface temperatures above 145~K. Water ice is not expected to survive above this temperature on Gyr timescales \citep{schorghoferLifetimeIceMain2008,snodgrassMainBeltComets2017}, however we showed in Section \ref{282P:subsec:dynamicalmodeling} that, prior to $\sim80$ years ago, \objnameBM{} had a semi-major axis of $a>6$~au, a region much colder than 145~K. Even if \objnameBM{} had spent most of its life with temperatures at the high end of our computed temperatures ($>220$~K), water ice can survive on Gyr timescales at shallow (a few cm) depths \citep{schorghoferLifetimeIceMain2008,prialnikCanIceSurvive2009}. Some bodies, such as (24)~Themis, have been found to have surface ices \citep{campinsWaterIceOrganics2010,rivkinDetectionIceOrganics2010} that suggest that an unknown mechanism may replenish surface ice with subsurface volatiles. In this case the ice lifetimes could be greatly extended.

\section{Activity Mechanism}
\label{282P:sec:mechanism}

Infrequent stochastic events, such as impacts (e.g., (596)~Scheila, \citealt{bodewitsCollisionalExcavationAsteroid2011,ishiguroObservationalEvidenceImpact2011,moreno596ScheilaOutburst2011}), are highly unlikely to be the activity mechanism given the multi-epoch nature of the activity we identified in this work. Moreover, it is unlikely that activity ceased during the 15 month interval between the UT 2021 March 14 archival activity and our UT 7 June 2022 Gemini South activity observations (Section \ref{282P:sec:observations}), when \objnameBM{} was at a heliocentric distance $r_\mathrm{H}=3.548$~au and $r_\mathrm{H}$=3.556~au, respectively, and \objnameBM{} was only closer to the Sun in the interim. Similarly, our archival data shows activity lasted $\sim15$ months during the 2012 -- 2013 apparition.

Recurrent activity is most commonly caused by volatile sublimation (e.g., 133P, \citealt{boehnhardtComet1996N21996,hsiehStrangeCase133P2004}) or rotational instability (e.g., (6478)~Gault, \citealt{kleynaSporadicActivity64782019,chandlerSixYearsSustained2019}). Rotational instability is impossible to rule out entirely for \objnameBM{} because its rotation period is unknown. However, (1) no activity attributed to rotational stability for any object has been observed to be continuous for as long as the 15 month episodes we report, and (2) rotational instability is not correlated with perihelion passage. It is worth noting that there are not yet many known objects with activity attributed to rotational disruption, so it is still difficult to draw firm conclusions about the behavior of those objects. In any case it would be useful to measure a rotation period for \objnameBM{} to help assess potential influence of rotational instability in the observed activity of \objnameBM{}. The taxonomic class of \objnameBM{} is unknown, but should \objnameBM{} be classified as a member of a desiccated spectral class (e.g., S-type), then sublimation would not likely be the underlying activity mechanism. Color measurements or spectroscopy when \objnameBM{} is quiescent would help determine its spectral class.

A caveat, however, is that many of our archival images were taken when \objnameBM{} was significantly fainter than the images showing activity 
(Figure \ref{282P:fig:ActivityTimeline}), 
thereby making activity detection more difficult than if \objnameBM{} was brighter. Consequently, archival images showing \objnameBM{} were predominitely taken near its perihelion passage. The farthest evidently quiescent image of \objnameBM{} was taken when it was at $\sim$4~au (Figure \ref{282P:fig:ActivityTimeline}). Thus we cannot state with total certainty that \objnameBM{} was inactive elsewhere in its orbit.

Thermal fracture can cause repeated activity outbursts. For example, (3200)~Phaethon undergoes 600~K temperature swings, peaking at 800~K -- 1100~K, exceeding the serpentine-phyllosilicate decomposition threshold of 574~K \citep{ohtsukaSolarRadiationHeatingEffects2009}, and potentially causing thermal fracture \citep{licandroNatureCometasteroidTransition2007,kasugaObservations1999YC2008} including mass loss \citep{liRecurrentPerihelionActivity2013,huiResurrection3200Phaethon2017}. Temperatures on \objnameBM{} reach at most $\sim220$~K (Figure \ref{282P:fig:ActivityTimeline}), with $\sim45$~K the maximum variation. Considering the relatively low temperatures and mild temperature changes we (1) consider it unlikely that \objnameBM{} activity is due to thermal fracture, and (2) reaffirm that thermal fracture is generally considered a nonviable mechanism for any objects other than \acp{NEO}.

Overall, we find volatile sublimation on \objnameBM{} the most likely activity mechanism, because (1) it is unlikely that an object originating from the Kuiper Belt such as \objnameBM{} would be desiccated
, (2) archival and new activity observations are from when \objnameBM{} was near perihelion (Figure \ref{282P:fig:ActivityTimeline}), a characteristic diagnostic of sublimation-driven activity \citep[e.g.,][]{hsiehOpticalDynamicalCharacterization2012}, and (3) 15 months of continuous activity has not been reported for any other activity mechanism (e.g., rotational instability, impact events) to date, let alone two such epochs.

\section{Summary and Future Work}
\label{282P:sec:summary}

This study was prompted by Citizen Scientists from the NASA Partner program \textit{Active Asteroids} classifying two images of \objnameBM{} from 2021 March as showing activity. Two additional images by astronomers Roland Fichtl and Michael Jäger brought the total number of images (from UT 2021 March 31 and UT 2021 April 4) to four. We conducted follow-up observations with the Gemini South 8.1~m telescope on UT 2022 June 7 and found \objnameBM{} still active, indicating it has been active for $>15$ months during the current 2021 -- 2022 activity epoch. Our archival investigation revealed the only other known apparition, from 2012--2013, also spanned $\sim15$ months. Together, our new and archival data demonstrate \objnameBM{} has been active during two consecutive perihelion passages, consistent with sublimation-driven activity.

We conducted extensive dynamical modeling and found \objnameBM{} has experienced a series of $\sim5$ strong interactions with Jupiter and Saturn in the past, and that \objnameBM{} will again have close encounters with Jupiter in the near future. These interactions are so strong that dynamical chaos dominates our simulations prior to 180 years ago and beyond 350 years in the future, but we are still able to statistically quantify a probable orbital class for \objnameBM{} prior to $-180$ yr (52\% \acp{JFC}, 48\% Centaur) and after $+350$ yr (81\% \acp{JFC}, 18\% \ac{QHO}, 14\% \ac{OMBA}). We classify present-day \objnameBM{} as a \acf{QHO}.

We carried out thermodynamical modeling that showed \objnameBM{} undergoes temperatures ranging at most between 135~K and 220~K, too mild for thermal fracture but warm enough that surface water ice would not normally survive on timescales of the solar system lifetime. However, \objnameBM{} arrived at its present orbit recently; prior to 1941 \objnameBM{} was primarily exterior to Jupiter's orbit and, consequently, sufficiently cold for water ice to survive on its surface. Given that both activity apparitions (Epoch I: 2012 -- 2013 and Epoch II: 2021 -- 2022) each lasted over 15 months, and both outbursts spanned perihelia passage, we determine the activity mechanism to most likely be volatile sublimation.

Coma likely accounts for the majority of the reflected light we observe emanating from \objnameBM{}, so it is infeasible to determine the color of the nucleus and, consequently, \objnameBM{}'s spectral class (e.g., C-type, S-type). Measuring its rotational period would also help assess what (if any) role rotational instability plays in the observed activity. Specifically, a rotation period faster than the spin-barrier limit of two hours would indicate breakup.

Most images of \objnameBM{} were taken when it was near perihelion passage (3.441~au), though there were observations from Epoch I that showed \objnameBM{} clearly, without activity, when it was beyond $\sim$4~au. \objnameBM{} is currently outbound and will again be beyond 4~au in mid-2023 and, thus, likely inactive; determining if/when \objnameBM{} returns to a quiescent state would help bolster the case for sublimation-driven activity because activity occurring preferentially near perihelion, and a lack of activity elsewhere, is characteristic of sublimation-driven activity.

\objnameBM{} is currently observable, especially from the southern hemisphere, however the object is passing in front of dense regions of the Milky Way until the end of 2022 November (see Lowell \texttt{AstFinder}\footnote{\url{https://asteroid.lowell.edu/astfinder/}} finding charts). \objnameBM{} will be in a less dense region of the Milky Way and be observable, in a similar fashion to our Gemini South observations (Section \ref{282P:sec:observations}) on UT 2022 September 26 for $\sim$12 days, carefully timed for sky regions with fewer stars. As Earth's orbit progresses around the Sun, \objnameBM{} becomes observable for less time each night through 2022 November, until UT 2022 December 26, when it becomes observable only during twilight. Observations during this window would help constrain the timeframe for periods of quiescence.


\section{Acknowledgements}
\label{282P:sec:acknowledgements}

We thank Dr.\ Mark Jesus Mendoza Magbanua of \ac{UCSF} for his frequent and timely feedback on the project. Many thanks for the helpful input from Henry Hsieh of the \ac{PSI} and David Jewitt of \ac{UCLA}. 

The authors express their gratitude to Prof. Mike Gowanlock (\acs{NAU}), Jay Kueny of \ac{UA} and Lowell Observatory, and the Trilling Research Group (\acs{NAU}), all of whom provided invaluable insights which substantially enhanced this work. The unparalleled support provided by Monsoon cluster administrator Christopher Coffey (\acs{NAU}) and the High Performance Computing Support team facilitated the scientific process.

We thank Gemini Observatory Director Jennifer Lotz for granting our \ac{DDT} request for observations, German Gimeno for providing science support, and Pablo Prado for observing. Proposal ID GS-2022A-DD-103, \acs{PI} Chandler.

The VATT referenced herein refers to the Vatican Observatory’s Alice P. Lennon Telescope and Thomas J. Bannan Astrophysics Facility. We are grateful to the Vatican Observatory for the generous time allocations (Proposal ID S165, \acs{PI} Chandler). We especially thank Vatican Observatory Director Br. Guy Consolmagno, S.J. for his guidance, Vice Director for Tucson Vatican Observatory Research Group Rev.~Pavel Gabor, S.J. for his frequent assistance, Astronomer and Telescope Scientist Rev. Richard P. Boyle, S.J. for patiently training us to use the \ac{VATT} and for including us in minor planet discovery observations, Chris Johnson (\ac{VATT} Facilities Management and Maintenance) for many consultations that enabled us to resume observations, Michael Franz (\acs{VATT} Instrumentation) and Summer Franks (\ac{VATT} Software Engineer) for on-site troubleshooting assistance, and Gary Gray (\ac{VATT} Facilities Management and Maintenance) for everything from telescope balance to building water support, without whom we would have been lost.

This material is based upon work supported by the \acs{NSF} \ac{GRFP} under grant No.\ 2018258765. Any opinions, findings, and conclusions or recommendations expressed in this material are those of the author(s) and do not necessarily reflect the views of the \acl{NSF}. The authors acknowledge support from the \acs{NASA} Solar System Observations program (grant 80NSSC19K0869, PI Hsieh) and grant 80NSSC18K1006 (PI: Trujillo).

Computational analyses were run on Northern Arizona University's Monsoon computing cluster, funded by Arizona's \ac{TRIF}. This work was made possible in part through the State of Arizona Technology and Research Initiative Program. 
\acf{WCS} corrections facilitated by the \textit{Astrometry.net} software suite \citep{langAstrometryNetBlind2010}.

This research has made use of data and/or services provided by the \ac{IAU}'s \ac{MPC}. 
This research has made use of \acs{NASA}'s Astrophysics Data System. 
This research has made use of The \acf{IMCCE} SkyBoT Virtual Observatory tool \citep{berthierSkyBoTNewVO2006}. 
This work made use of the \texttt{FTOOLS} software package hosted by the \acs{NASA} Goddard Flight Center High Energy Astrophysics Science Archive Research Center. 
\ac{SAO} \ac{DS9}: This research has made use of \texttt{\acs{SAO}Image\acs{DS9}}, developed by \acl{SAO} \citep{joyeNewFeaturesSAOImage2006}. \acf{WCS} validation was facilitated with Vizier catalog queries \citep{ochsenbeinVizieRDatabaseAstronomical2000} of the Gaia \ac{DR} 2 \citep{gaiacollaborationGaiaDataRelease2018} and the \acf{SDSS DR-9} \citep{ahnNinthDataRelease2012} catalogs. 
This work made use of AstOrb, the Lowell Observatory Asteroid Orbit Database \textit{astorbDB} \citep{bowellPublicDomainAsteroid1994,moskovitzAstorbDatabaseLowell2021}. 
This work made use of the \texttt{astropy} software package \citep{robitailleAstropyCommunityPython2013}.

Based on observations at \ac{CTIO}, \acs{NSF}’s \acs{NOIRLab} (\acs{NOIRLab} Prop. ID 2019A-0305; \acs{PI}: A. Drlica-Wagner, \acs{NOIRLab} Prop. ID 2013A-0327; \acs{PI}: A. Rest), which is managed by the \acf{AURA} under a cooperative agreement with the \acl{NSF}. 
This project used data obtained with the \acf{DECam}, which was constructed by the \acf{DES} collaboration. Funding for the \acs{DES} Projects has been provided by the US Department of Energy, the US \acl{NSF}, the Ministry of Science and Education of Spain, the Science and Technology Facilities Council of the United Kingdom, the Higher Education Funding Council for England, the National Center for Supercomputing Applications at the University of Illinois at Urbana-Champaign, the Kavli Institute for Cosmological Physics at the University of Chicago, Center for Cosmology and Astro-Particle Physics at the Ohio State University, the Mitchell Institute for Fundamental Physics and Astronomy at Texas A\&M University, Financiadora de Estudos e Projetos, Fundação Carlos Chagas Filho de Amparo à Pesquisa do Estado do Rio de Janeiro, Conselho Nacional de Desenvolvimento Científico e Tecnológico and the Ministério da Ciência, Tecnologia e Inovação, the Deutsche Forschungsgemeinschaft and the Collaborating Institutions in the Dark Energy Survey. The Collaborating Institutions are Argonne National Laboratory, the University of California at Santa Cruz, the University of Cambridge, Centro de Investigaciones Enérgeticas, Medioambientales y Tecnológicas–Madrid, the University of Chicago, University College London, the \acs{DES}-Brazil Consortium, the University of Edinburgh, the Eidgenössische Technische Hochschule (ETH) Zürich, Fermi National Accelerator Laboratory, the University of Illinois at Urbana-Champaign, the Institut de Ciències de l’Espai (IEEC/CSIC), the Institut de Física d’Altes Energies, Lawrence Berkeley National Laboratory, the Ludwig-Maximilians Universität München and the associated Excellence Cluster Universe, the University of Michigan, \acs{NSF}’s \acs{NOIRLab}, the University of Nottingham, the Ohio State University, the OzDES Membership Consortium, the University of Pennsylvania, the University of Portsmouth, \ac{SLAC} National Accelerator Laboratory, Stanford University, the University of Sussex, and Texas A\&M University.

These results made use of the \acf{LDT} at Lowell Observatory. Lowell is a private, non-profit institution dedicated to astrophysical research and public appreciation of astronomy and operates the \acs{LDT} in partnership with Boston University, the University of Maryland, the University of Toledo, \acf{NAU} and Yale University. The \acf{LMI} was built by Lowell Observatory using funds provided by the \acf{NSF} (AST-1005313).

\ac{VST} OMEGACam \citep{arnaboldiVSTVLTSurvey1998,kuijkenOmegaCAM16k16k2002,kuijkenOmegaCAMESONewest2011} data were originally acquired as part of the \ac{KIDS} \citep{dejongFirstSecondData2015}.

The \acs{Pan-STARRS}1 Surveys (PS1) and the PS1 public science archive have been made possible through contributions by the Institute for Astronomy, the University of Hawaii, the \acs{Pan-STARRS} Project Office, the Max-Planck Society and its participating institutes, the Max Planck Institute for Astronomy, Heidelberg and the Max Planck Institute for Extraterrestrial Physics, Garching, The Johns Hopkins University, Durham University, the University of Edinburgh, the Queen's University Belfast, the Harvard-Smithsonian Center for Astrophysics, the \ac{LCOGT} Network Incorporated, the National Central University of Taiwan, the \acl{STScI}, the \acl{NASA} under Grant No. NNX08AR22G issued through the Planetary Science Division of the \acs{NASA} Science Mission Directorate, the \acf{NSF} Grant No. AST-1238877, the University of Maryland, \ac{ELTE}, the Los Alamos National Laboratory, and the Gordon and Betty Moore Foundation.

Based on observations obtained with MegaPrime/MegaCam, a joint project of \ac{CFHT} and \ac{CEA}/\ac{DAPNIA}, at the \ac{CFHT} which is operated by the \acf{NRC} of Canada, the Institut National des Science de l'Univers of the \acf{CNRS} of France, and the University of Hawaii. The observations at the \acf{CFHT} were performed with care and respect from the summit of Maunakea which is a significant cultural and historic site.

Magellan observations made use of the \ac{IMACS} instrument \citep{dresslerIMACSInamoriMagellanAreal2011}.

This research has made use of the \acs{NASA}/\ac{IPAC} \ac{IRSA}, which is funded by the \acl{NASA} and operated by the California Institute of Technology.

\subsection{Facilities}
    Astro Data Archive, 
    Blanco (DECam),
    CFHT (MegaCam), 
    Gaia,
    Gemini-South (GMOS-S),
    IRSA, 
    LDT (LMI),
    Magellan: Baade (TSIP),
    PO:1.2m (PTF, ZTF), 
    PS1, 
    Sloan,
    VATT (VATT4K),
    VST (OmegaCAM)

\subsection{Software}
{\tt astropy} \citep{robitailleAstropyCommunityPython2013},
        {\tt astrometry.net} \citep{langAstrometryNetBlind2010},
        {\tt FTOOLS}\footnote{\url{https://heasarc.gsfc.nasa.gov/ftools/}},
        {\tt IAS15} integrator \citep{reinIAS15FastAdaptive2015},
        {\tt JPL Horizons} \citep{giorginiJPLOnLineSolar1996},
        {\tt Matplotlib} \citep{hunterMatplotlib2DGraphics2007},
        {\tt NumPy} \citep{harrisArrayProgrammingNumPy2020},
        {\tt pandas} \citep{mckinneyDataStructuresStatistical2010,rebackPandasdevPandasPandas2022},
        {\tt REBOUND} \citep{reinREBOUNDOpensourceMultipurpose2012,reinHybridSymplecticIntegrators2019},
        {\tt SAOImageDS9} \citep{joyeNewFeaturesSAOImage2006},
        {\tt SciPy} \citep{virtanenSciPyFundamentalAlgorithms2020},
        {\tt Siril}\footnote{\url{https://siril.org}},
        {\tt SkyBot} \citep{berthierSkyBoTNewVO2006},
        {\tt termcolor}\footnote{\url{https://pypi.org/project/termcolor}},
        {\tt tqdm} \citep{costa-luisTqdmFastExtensible2022},
        {\tt Vizier} \citep{ochsenbeinVizieRDatabaseAstronomical2000}

\clearpage
\thispagestyle{empty}

\section{Appendix}
\label{282P:sec:appendix}
\clearpage
\footnotesize

\atxy{\dimexpr1in}{.5\paperheight}{\rotatebox[origin=center]{270}{\thepage}}

\singlespacing
\begin{sidewaystable}
\caption{Table of Observations}
\label{282P:tab:observationsTable}
\centering

\begin{tabular}{cccccrccccrrc}
Fig.$^\mathrm{a}$ & Act.$^\mathrm{b}$ & Obs. Date$^\mathrm{c}$   & Source              & $N^\mathrm{d}$  & Exp. [s]$^\mathrm{e}$ & Filter(s) & V$^\mathrm{f}$    & $r$ [au]$^\mathrm{g}$    & STO [$\degr$]$^\mathrm{h}$  & $\nu$ [$\degr$]$^\mathrm{i}$    & \%$_{Q\rightarrow q}^\mathrm{j}$ & Note$^\mathrm{k}$\\
\hline\hline
                                &   & 04 Feb 2011 & PS1             & 2     &   40  & $r$       & 19.7 & 4.26 &  3.3 & 258.1 & 84\% & \ref{20120224}  \\
                                &   & 16 Feb 2012 & PS1             & 1     &   45  & $i$       & 19.3 & 3.69 &  8.9 & 306.4 & 95\% & \ref{20120224}  \\
                                &   & 24 Feb 2012 & PS1             & 1,1   &43, 40 & $g$, $r$  & 19.2 & 3.68 &  6.9 & 307.6 & 95\% & \obsnote{20120224}\ref{20120224}  \\
                                &   & 26 Feb 2012 & PS1             & 2     &   40  & $r$       & 19.2 & 3.67 &  6.3 & 307.9 & 96\% & \ref{20120224}\\
\ref{282P:fig:282P}e            & Y & 28 Mar 2012 & MegaPrime       & 2     &  120  & $r$       & 18.9 & 3.64 &  3.2 & 312.6 & 96\% & \obsnote{20120328}\ref{20120328}\\
                                & Y & 05 Jul 2012 & OmegaCAM        & 4     &  240  & $i$       & 20.1 & 3.54 & 16.1 & 328.0 & 98\% & \obsnote{20120705}\ref{20120705} \\
                                &   & 14 Apr 2013 & PS1             & 2     &   45  & $i$       & 19.3 & 3.47 & 12.8 &  14.9 & 99\% & \ref{20120224}\\ 
                                &   & 22 Apr 2013 & PS1             & 2     &   30  & $z$       & 19.2 & 3.47 & 11.1 &  16.3 & 99\% & \ref{20120224}\\
\ref{282P:fig:282P}f            & Y & 05 May 2013 & \acs{DECam}     & 2     &  150  & $r$       & 19.5 & 3.48 & 12.9 & 318.4 & 97\% & \obsnote{20130505}\ref{20130505}   \\
                                & Y & 15 May 2013 & PS1             & 1     &   43  & $g$       & 18.8 & 3.48 &  5.2 &  20.1 & 99\% & \ref{20120224}\\
\ref{282P:fig:282P}g            & Y & 13 Jun 2013 & MegaPrime       & 10    &  120  & $r$       & 18.8 & 3.50 &  4.9 &  24.8 & 99\% & \obsnote{20130613}\ref{20130613}  \\
                                &   & 03 Aug 2013 & PS1             & 2     &  80,60  & $y$, $z$  & 19.6 & 3.54 & 15.3 &  33.0 & 98\% & \ref{20120224}\\
                                &   & 11 Jun 2014 & PS1             & 2     &   45  & $i$       & 20.0 & 3.95 & 12.5 &  78.1 & 90\% & \ref{20120224}\\
                                &   & 14 Aug 2014 & PS1             & 3     &   45  & $i$       & 19.4 & 4.05 &  3.3 &  86.1 & 88\% & \ref{20120224}\\
                                &   & 15 Aug 2014 & PS1             & 4     &   45  & $i$       & 19.4 & 4.04 &  3.5 &  86.2 & 88\% & \ref{20120224}\\
                                &   & 04 Jan 2021 & \acs{ZTF}       & 1     &   30  & $r$       & 20.0 & 3.63 & 15.7 & 312.5 & 96\% & \obsnote{20210104}\ref{20210104}\\
                                &   & 07 Jan 2021 & \acs{ZTF}       & 1     &   30  & $g$       & 20.0 & 3.63 & 15.7 & 312.9 & 96\% & \ref{20210104}\\
                                &   & 09 Jan 2021 & \acs{ZTF}       & 1     &   30  & $r$       & 20.0 & 3.62 & 15.7 & 313.2 & 96\% & \ref{20210104}\\
\ref{282P:fig:282P}a            & Y & 14 Mar 2021 & \acs{DECam}     & 1     &   90  & $i$       & 18.9 & 3.55 &  6.1 & 323.1 & 98\% & \obsnote{20210314}\ref{20210314}    \\
\ref{282P:fig:282P}h            & Y & 17 Mar 2021 & \acs{DECam}     & 1     &   90  & $i$       & 18.9 & 3.55 &  5.2 & 323.6 & 98\% & \ref{20210314} \\
\ref{282P:fig:282P}b            & Y & 31 Mar 2021 & QHY600          & 1     & 2160  & UV/IR     & 18.5 & 3.54 &  0.7 & 325.9 & 98\% & \obsnote{20210331}\ref{20210331}\\
\ref{282P:fig:282P}c            & Y & 04 Apr 2021 & CDS-5D          & 1     & 1500  & (none)    & 18.5 & 3.54 &  0.5 & 326.4 & 98\% & \obsnote{20210404}\ref{20210404} \\
                                &   & 07 Mar 2022 & \acs{IMACS}     & 5     &   10  & WB4800-7800&20.0 & 3.48 & 16.3 &  22.3 & 99\% & \obsnote{20220307}\ref{20220307}\\
                                &   & 21 May 2022 & \acs{LDT}       & 3     &   90  & VR, $i$   & 19.1 & 3.54 &  8.3 &  34.4 & 98\% & \obsnote{20220521}\ref{20220521}\\
\ref{282P:fig:282P}d            & Y & 07 Jun 2022 & \ac{GMOS}-S     & 6,6,6 & 120   & $g$, $r$, $i$&18.8&3.56 &  3.8 &  37.2 & 98\% & \obsnote{20220607}\ref{20220607}\\
\end{tabular}
\footnotesize
\noindent
\raggedright\\
$^\mathrm{a}$Figure showing the image. $^\mathrm{b}$Activity identified in image(s). $^\mathrm{c}$UT date of observation. $^\mathrm{d}$Number of images. $^\mathrm{e}$Exposure time. $^\mathrm{f}$Apparent $V$-band magnitude (Horizons). $^\mathrm{g}$Heliocentric distance. $^\mathrm{h}$Sun--target--observer angle. 
$^\mathrm{i}$True anomaly. 
$^\mathrm{j}$Percentage to perihelion $q$ from aphelion $Q$, defined by $\%_{T\rightarrow q} = \left(\frac{Q - r}{Q-q}\right)\cdot 100\mathrm{\%}$. 
$^\mathrm{k}$Note number. \\

Notes: 
\ref{20120224}: PS1 is the \acf{Pan-STARRS} One. 
\ref{20120328}: Prop. ID 12AH16, \acs{PI} Wainscoat. 
\ref{20120705}: Prop. ID 177.A-3016(D), \acs{PI} Kuijken. 
\ref{20130505}: \acf{DECam}; Prop. ID 2013A-0327, \acs{PI} Rest. 
\ref{20130613}: Prop. ID 13AH09, \acs{PI} Wainscoat. 
\ref{20210104}: \acf{ZTF}; Prop. ID 1467501130115, \acs{PI} Kulkarni; data acquired through \acs{ZTF} Alert Stream service \citep{pattersonZwickyTransientFacility2019}. 
\ref{20210314}: Prop. ID 2019A-0305, \acs{PI} Drlica-Wagner. 
\ref{20210331}: Michael Jäger (Weißenkirchen, Austria), QHY600 \ac{CCD} on a 14'' Newtonian, . 
\ref{20210404}: Roland Fichtl (Engelhardsberg, Germany), Central DS brand modified cooled Canon 5D Mark III on a 0.4~m f/2.5 Newtonian;  \url{http://www.dieholzhaeusler.de/Astro/comets/0282P.htm}. 
\ref{20220307}: \acf{IMACS}; PI Trujillo. 
\ref{20220521}: \acf{IMACS}; PI Trujillo. 
\ref{20220607}: \acf{GMOS}; Prop. ID GS-2022A-DD-103, \acs{PI} Chandler. 
\end{sidewaystable}
\normalsize

\doublespacing

\clearpage
\thispagestyle{empty}

\atxy{\dimexpr1in}{.5\paperheight}{\rotatebox[origin=center]{270}{\thepage}}

\begin{sidewaystable}
\caption{Equipment and Archives}
\label{282P:tab:equipQuickRef}
\centering

\footnotesize
    \begin{tabular}{llclccccc}
Instrument  & Telescope            & Pixel Scale    & Location                & \texttt{AstroArchive}         & \acs{ESO} & \acs{SSOIS}        & \acs{STScI}  & \acs{IRSA}\\
            &                       & [\arcsec/pix] &                   &   & & &\\
\hline
\hline
\acs{DECam}       & 4.0~m Blanco           & 0.263          & Cerro Tololo, Chile     & S,R &     &               S            &        \\
\acs{GMOS}-S        & 8.1~m Gemini South & 0.080        & Cerro Pachón, Chile &  &\\
\acs{IMACS}         & 6.5~m Baade       & 0.110          & Las Campanas, Chile & & &  \\ 
OmegaCAM    & 2.6~m \acs{VLT} Survey     & 0.214          & Cerro Paranal, Chile    &              & R   &           S            &        \\
GigaPixel1  & 1.8 m \acs{Pan-STARRS}1    & 0.258          & Haleakalā, Hawaii       &              &     &              S            & R      \\
\acs{LMI} & 4.3~m \acs{LDT}             & 0.120         & Happy Jack, Arizona       & & & \\
MegaPrime   & 3.6~m \acs{CFHT}           & 0.185          & Mauna Kea, Hawaii       &              &     &             S,R &        \\
\acs{PTF}/\acs{CFHT}12K & 48" Samuel Oschin    & 1.010          & Mt. Palomar, California &              &     &  &       &                 S,R   \\
\acs{ZTF} Camera  & 48" Samuel Oschin    & 1.012          & Mt. Palomar, California &              &     &  &       &                 S,R  \\
\acs{VATT}4K \acs{CCD}        & 1.8~m \acs{VATT}            & 0.188         & Mt. Graham, Arizona       & & & &\\
\end{tabular}
\raggedright
\footnotesize{\\
R indicates repository for data retrieval. S indicates search capability.\\
\texttt{AstroArchive}: \ac{NSF} \ac{NOIRLab} \texttt{AstroArchive} (\url{https://astroarchive.noirlab.edu}).\\
\ac{ESO}: \acl{ESO} (\url{https://archive.eso.org}).\\
\ac{IRSA}: \acs{NASA}/CalTech \ac{IRSA} (\url{https://irsa.ipac.caltech.edu}).\\
\acs{PTF}: The \ac{PTF}. 
\acs{SSOIS}: The \ac{SSOIS} (\citealt{gwynSSOSMovingObjectImage2012}, \url{https://www.cadc-ccda.hia-iha.nrc-cnrc.gc.ca/en/ssois/}).\\
\ac{STScI}: \url{https://www.stsci.edu/}.
}
\label{282P:tab:equipAndArchives}
\end{sidewaystable}
\doublespacing

\clearpage

\atxy{\dimexpr1in}{.5\paperheight}{\rotatebox[origin=center]{270}{\thepage}}

\begin{sidewaystable}
    \caption{282P/(323137) 2003 BM80 Data}
    \label{282P:tab:ObjectData}
    \singlespacing
    \centering
    \footnotesize
	\begin{tabular}{lll}
		Parameter & Value & Source\\
		\hline\hline
		Designations & (323137), 2003~BM$_{80}$, 2003~FV$_{112}$, 282P & \acs{JPL} \acs{SBDB}, \acs{MPC}\\
		Discovery Date & 2003 January 31 & \acs{JPL} \acs{SBDB}, \acs{MPC}\\
		Discovery Observer(s) & \ac{LONEOS} & \acs{JPL} \acs{SBDB}, \acs{MPC}\\
		Discovery Observatory & Lowell Observatory & \acs{JPL} \acs{SBDB}, \acs{MPC}\\
		Discovery Site & Anderson Mesa Station, Arizona & \acs{JPL} \acs{SBDB}, \acs{MPC}\\
		Discovery Site Code & 688 & \acs{MPC} \\
		Activity Discovery Date & 2013 June 12 & \acs{CBET} 3559 \citep{bolinComet2003BM2013}\\
		Activity Discoverer(s) & Bryce Bolin, Larry Denneau, Peter Veres & \acs{CBET} 3559 \citep{bolinComet2003BM2013}\\
        Orbit Type & \acf{QHO} & this work\\ 
		Diameter & $D=$3.4$\pm$0.4~km & {\citet{harrisAsteroidsThermalInfrared2002}}\\ 
		Absolute $V$-band Magnitude & $H=13.63$ & \acs{MPC} (MPO648742)\\
        Geometric Albedo & Unknown & \\
        Assumed Geometric Albedo & 4\% & \cite{snodgrassSizeDistributionJupiter2011}\\
        Rotation Period & Unknown & \\
        Orbital Period & $P=8.732\pm(2.174\times10^{-7})$~yr & \acs{JPL} \acs{SBDB} \\ 
		Semi-major Axis & $a=4.240\pm(7.039\times10^{-8})$~au & \acs{JPL} \acs{SBDB}\\ 
		Perihelion Distance & $q=3.441\pm(3.468\times10^{-7})$~au & \acs{JPL} \acs{SBDB}\\ 
		Aphelion Distance & $Q=5.039\pm(8.366\times10^{-8})$~au & \acs{JPL} \acs{SBDB}\\ 
		Eccentricity & $e=0.188\pm(7.790\times10^{-8})$ & \acs{JPL} \acs{SBDB}\\ 
		Inclination & $i=5.812\degr\pm(1.166\degr\times10^{-5})$ & \acs{JPL} \acs{SBDB}\\ 
		Argument of Perihelion & $\omega=217.626\degr\pm(7.816\degr\times10^{-5})$ & \acs{JPL} \acs{SBDB}\\ 
		Longitude of Ascending Node & $\Omega=9.297\degr\pm(5.974\degr\times10^{-5})$ & \acs{JPL} \acs{SBDB}\\ 
		Mean Anomaly & $M=9.979\degr\pm(3.815\degr\times10^{-5})$ & \acs{JPL} \acs{SBDB}\\ 
		Tisserand Parameter w.r.t. Jupiter & $T_\mathrm{J}=2.99136891\pm\left(3.73\times10^{-8}\right)$ & this work\\ 
		Orbital Solution Date & 2021 October 8 & \acs{JPL} \acs{SBDB}\\
	\end{tabular}
	\raggedright\footnotesize

    Notes: 
    \ac{CBET} \footnote{\url{http://www.cbat.eps.harvard.edu}}. 
    \ac{JPL} \ac{SBDB} is the \acs{NASA} \acs{JPL} \acl{SBDB}\footnote{\url{https://ssd.jpl.nasa.gov/tools/sbdb_lookup.html}}. 
    \acs{MPC} is the \acl{MPC}\footnote{\url{https://minorplanetcenter.net}}. 
\end{sidewaystable}
\doublespacing

\clearpage
\chapter{Overall Discussion}
\label{chap:discussion}
\acresetall

In order to help fill key knowledge gaps about solar system volatiles, such as where volatiles are found throughout the solar system, we set out to find objects like active asteroids and active Centaurs, bodies that display cometary properties such as tails and comae even though they are not classified as comets. In furtherance of this goal we designed and launched a Citizen Science project, \textit{Active Asteroids}, that carries out an outreach program of public engagement while concurrently identifying new members of known and unknown active minor planet classes. Before and after launch, \textit{Active Asteroids} has led to discoveries ranging from identifying new active asteroids to uncovering new epochs of activity associated with known active objects. Some of these discoveries are in this dissertation, while others will be the subject of future investigations by our team.

Prior to constructing the Citizen Science project, we carried out a proof-of-concept (Chapter \ref{chap:SAFARI}) that demonstrated the suitability of \ac{DECam} data for identifying activity emanating from minor planets. First we created a pipeline, \ac{HARVEST}, that extracts small thumbnail images centered on a known solar system object. Without \textit{a priori} knowledge of which object was shown in each thumbnail image, we visually examined all \allthumbsSAFARI{} thumbnails we had extracted from \fitscountSAFARI{} \ac{FITS} files. Of the \uniquethumbsSAFARI{} unique minor planets visible in the thumbnails, three were previously identified as active asteroids. However, (1)~Ceres and (779)~Nina have not been observed to be visibly active from Earth. The third, (62412)~2000~SY$_{178}$ \citep{sheppardDiscoveryCharacteristicsRapidly2015}, we successfully identified as active in a thumbnail image (Figure \ref{safari:fig:62412}). This one in \uniquethumbsSAFARI{} occurrence rate agreed with other works that estimated activity takes place in approximately one in ten thousand asteroids \citep{jewittActiveAsteroids2015a,hsiehMainbeltCometsPanSTARRS12015}.

\begin{figure*}
    \centering
    \includegraphics[width=1.0\linewidth]{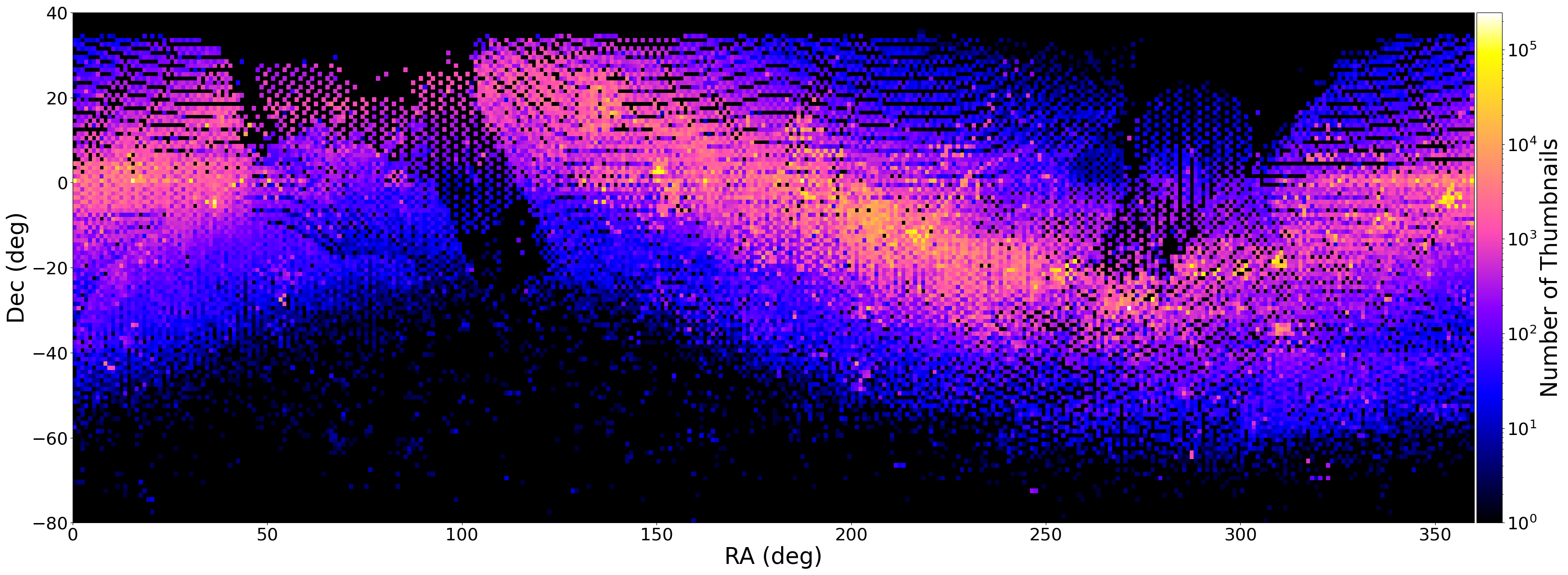}
    \caption{The distribution in each square degree on sky of the roughly 18 million automatically vetted thumbnail images \ac{HARVEST} produced.} 
    \label{discussion:fig:asteroidsOnSky}
\end{figure*}

My \ac{NSF} \ac{GRFP} proposal was selected for funding and we expanded the \ac{HARVEST} pipeline to work with all publicly available \ac{DECam} archival image data. We implemented a vetting scheme to exclude images that, for example, did not probe faintly enough to detect activity. Figure \ref{discussion:fig:asteroidsOnSky} shows the distribution on sky of the $\sim$18 million automatically vetted thumbnail images in the \ac{HARVEST} database.

In 2019 January asteroid (6478)~Gault was observed to be active \citep{smith6478Gault2019}. We made use of \ac{HARVEST} to find additional images of Gault in \ac{DECam} archival data. We discovered Gault had been active during two prior orbits: 2013 and 2016. We introduced a new metric, observability, to help identify potential biases by describing the number of hours an object is observable at a specific location for a given UT observing date. Notably, Gault's activity appeared unrelated to heliocentric distance (Chapter \ref{Gault:subsec:imageanalysis}), indicating the activity was probably not caused by volatile sublimation.

Gault is a core member of the Phocaea family \citep{knezevicProperElementCatalogs2003}, a group of predominantly (75\%) S-type asteroids \citep{carvanoSpectroscopicSurveyHungaria2001}. S-type asteroids are composed primarily of non-primitive, desiccated silica \citep{demeoExtensionBusAsteroid2009}, material highly unlikely to harbor volatile substances. The majority of active asteroids found to date (see Table \ref{safari:Table:TheAAs}) have been composed of primitive materials (e.g., C-type, which are carbonaceous and volatile-rich), amenable to sublimation-driven activity associated with perihelion passage \citep{hsieh2016ReactivationsMainbelt2018}. Very few (around four, or $\sim$20\%) known active asteroids are S-type, and their activity is thought to be caused by impact, such as (596)~Scheila \citep{bodewitsCollisionalExcavationAsteroid2011,jewittHubbleSpaceTelescope2011}, or rotational instability, as with 311P/PanSTARRS, \citep{jewittExtraordinaryMultitailedMainbelt2013}, rather than sublimation \citep{hsiehAsteroidFamilyAssociations2018}.

Given the combination of (1) Gault's probable desiccated S-type spectral class, (2) rotational breakup as the likely activity mechanism \citep{morenoDustPropertiesDoubletailed2019, yeMultipleOutburstsAsteroid2019,yeContinuedActivity64782019,kleynaSporadicActivity64782019}, and (3) our discovery of recurrent activity, we proposed that Gault is a new type of active asteroid, one that is persistently active because of spin-up induced rotational instability. We also identified Gault as the first active asteroid with recurrent and sustained activity throughout its orbit. We anticipated that Gault would remain active for some time, and although it did not catastrophically disintegrate (cf.\ \citet{morenoDustPropertiesDoubletailed2019}), Gault became inactive in 2020 \citep{2021MNRAS.505..245D} and has remained so through at least 7 January 2022. 

During ongoing Citizen Science project preparations we found indicators suggestive of activity in thumbnails of Centaur 2014~OG$_{392}$ (Figure \ref{og:fig:archivalimages}). We carried out an observing campaign that confirmed the presence of activity based on a result stemming from the \ac{HARVEST} pipeline. We uncovered archival images of 2014~OG$_{392}$ that indicated it had been active for more than two years, effectively ruling out stochastic events (e.g., impact).

We introduced a new technique that synthesizes sublimation modeling with dynamical modeling to help estimate which molecules are most likely responsible for activity. My simple sublimation model is not intended to account for the complex behaviors of ice mixtures \citep{grundySolarGardeningSeasonal2000} or amorphous--crystalline ice transitions \citep{jewittActiveCentaurs2009}, rather it treats bodies as airless and uniformly covered in a single species of ice. We found carbon dioxide and/or ammonia most likely to be sublimating, while other surface species would have sublimated completely over its 13 kyr to 1.8 Myr dynamical lifetime (carbon monoxide, molecular nitrogen, methane), or were unable to appreciably sublimate at all (water, methanol).

Colors we measured from our observations revealed that 2014~OG$_{392}$ was about one magnitude redder than the Sun in the visible spectrum and, with a $B$-$R$ color of \ogColor{}, the object is considered a red Centaur \citep{peixinhoCentaursComets402020}. However, both species we identified as likely to be responsible for sublimation (carbon dioxide, ammonia) are spectrally neutral in visible wavelengths, as many ices are, indicating the involvement of an as-yet unidentified reddening agent. The \acp{TNO} also have an unknown reddening agent \citep{boehnhardtVisibleNearIRObservations2001}. We computed a new absolute magnitude of $H\approx11.3$, fainter than previously reported by about 0.5 magnitudes. This $H$ value indicates a diameter of $\sim$20~km, assuming a slope $G=0.15$ as is typical for a dark surface \citep{bowellApplicationPhotometricModels1989}. We measured a coma mass of roughly $\sim2.4\times10^{15}$~g, assuming a grain density of 1~g/cm$^3$, 1~mm grain radii, and an albedo of $A=0.1$.

In 2021 July, prior to launching the \textit{Active Asteroids} Citizen Science project, activity was discovered coming from (248370) 2005~QN$_{173}$ \citep{fitzsimmons2483702005QN1732021}. We conducted an archival investigation (described in Chapters \ref{methods:subsec:archivalInvestigation} and \ref{QN:sec:secondActivityEpoch}) and located 81 images (spanning 31 observations) in which the object was positively identifiable. From these we found a single image (Figure \ref{QN:fig:wedgephot}) from UT 2016 July 22 that clearly showed the object had been active during this orbit. Our analysis indicated that (248370)~2005~QN$_{173}$ is most likely a member of the \acp{MBC} because of (1) the recurrent nature of its activity, (2) its orbit with the Main Asteroid Belt, and (3) its likely C-type spectral class \citep{hsiehPhysicalCharacterizationMainbelt2021}.

Along with my archival investigation we introduced a tail angle measurement tool, wedge photometry, that may also be used for activity detection. We showed the tail orientation of (248370)~2005~QN$_{173}$ was in close agreement with the anti-Solar and anti-Motion vectors (coincident at the time) as computed by \ac{JPL} Horizons.

We launched the Citizen Science project \textit{Active Asteroids}\footnote{\url{http://activeasteroids.net}}, a NASA Partner\footnote{\url{https://science.nasa.gov/citizenscience}}, on 31 August 2022. The initial subject set of $\sim$18,000 thumbnail images was completely classified by volunteers within days of launch. Although we did produce statistical evidence that volunteers could identify activity based on the training set
, discoveries of activity from objects not yet known to be active came to light within 24 hours of launch.

One example, 2015~TC$_1$ (Figure \ref{methods:fig:comparison}), is a \ac{JFC}. We had coincidentally observed this object previously with the twin 8.5~m \ac{LBT} on 2021 April 4 (Prop. ID AZ-2021A-506, PI Oldroyd). Observations were carried out in poor weather conditions and the object was not near perihelion, and we did not find any conclusive evidence of activity. The next perihelion passage for 2015~TC$_1$ takes place in 2022 December and, per the \ac{JPL} \ac{SBDB}, the object will have an especially close (0.1~au) approach to Jupiter in November 2145.

\begin{figure*}[h]
    \centering
    \begin{tabular}{cc}
         \includegraphics[width=0.45\linewidth]{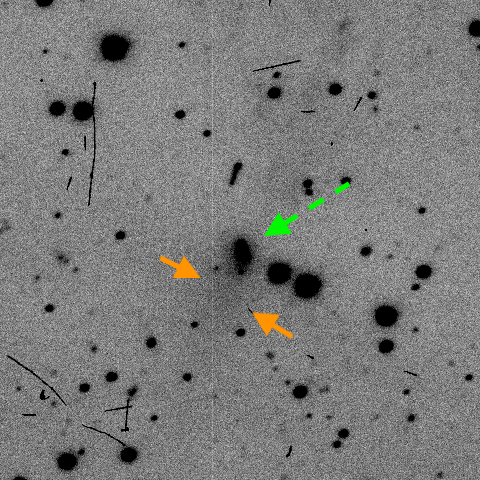} & \includegraphics[width=0.45\linewidth]{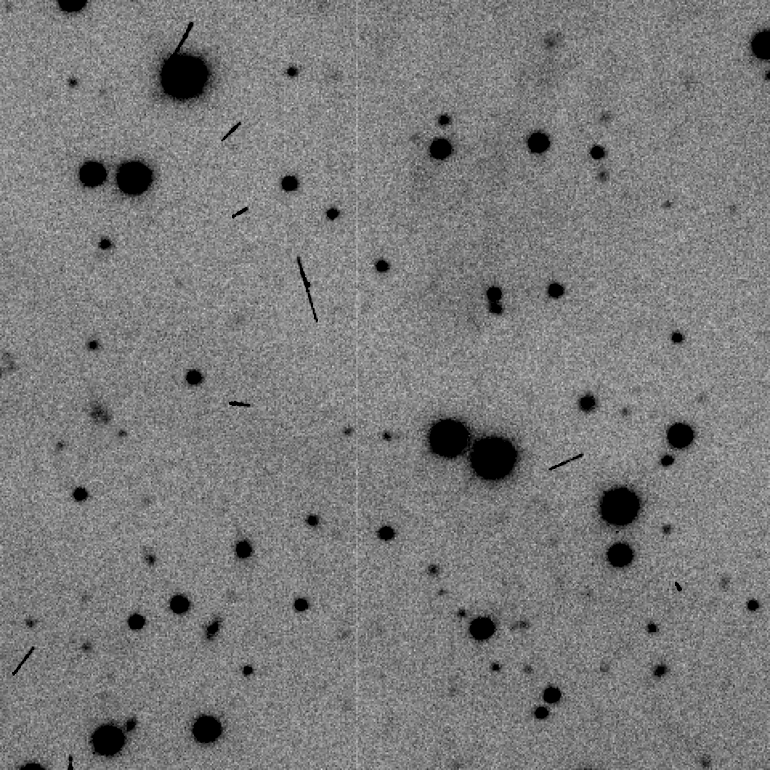}  \\
    \end{tabular}
    \caption{
    \textbf{Left:} This image of \ac{JFC} 2017~QN$_{84}$ (green dashed arrow), originally taken 2017 December 23 (Prop. ID 2017B-0307, PI Sheppard), was flagged as active by 14 of 15 \textit{Active Asteroids} (\url{http://activeasteroids.net}) volunteers.
    \textbf{Right:} this comparison image, from an image captured on 2017 December 24 (Prop. ID 2017B-0307, PI Sheppard), shows that there is no comparable object or extended feature (e.g., a galaxy) that could be mistaken for the activity visible in the image on the left (orange arrows).}
    \label{discussion:fig:2017QN84} 
\end{figure*}

Another example of a Citizen Science--informed discovery was activity emanating from 2017~QN$_{84}$ (Figure \ref{discussion:fig:2017QN84}). 14 of 15 volunteers classified a thumbnail of this \ac{JFC} as showing activity. Unfortunately, this was the only image available of the object, which makes identifying activity more challenging. However, as seen in Figure \ref{discussion:fig:2017QN84}, a comparison image (technique described in Chapter \ref{methods:subsec:comparisonImages}) indicated that a chance diffuse background object was not responsible for the observed activity. The object had just completed its 2017 perihelion passage, and it was relatively bright at $V\approx20$.

We are currently investigating other potentially active minor planets. Here we mention both anticipated and unexpected impediments to confirming the presence of activity are worth mentioning here. As described above, (1) not all images have unambiguous activity, and (2) there may only be a single image with activity indicators. Moreover, (3) some objects may not be observable (i.e., below the horizon at night)
. (4) Objects may be very faint, and thus difficult or impossible to observe even with the largest telescopes; this is especially problematic for detecting comae and tails which may appear several magnitudes fainter than the nucleus. (5) An object may now be far from perihelion and, consequently, less likely to be active; depending on the object's orbit and time elapsed from the thumbnail image, it may take years or decades for an object to again approach perihelion. (6) Objects with poorly constrained orbits may be lost by the time we identify activity and set out to observe the object. (7) A candidate may be transiting a rich area on sky (e.g., the Milky Way), potentially for months, effectively preventing activity detection and hampering study of the object. (8) Objects with strong activity indicators may still not meet the ``new comet'' \ac{MPC} requirements, as described in Chapter \ref{intro:subsec:visibleActivity}.

\begin{figure}
    \centering
    \includegraphics[width=4in]{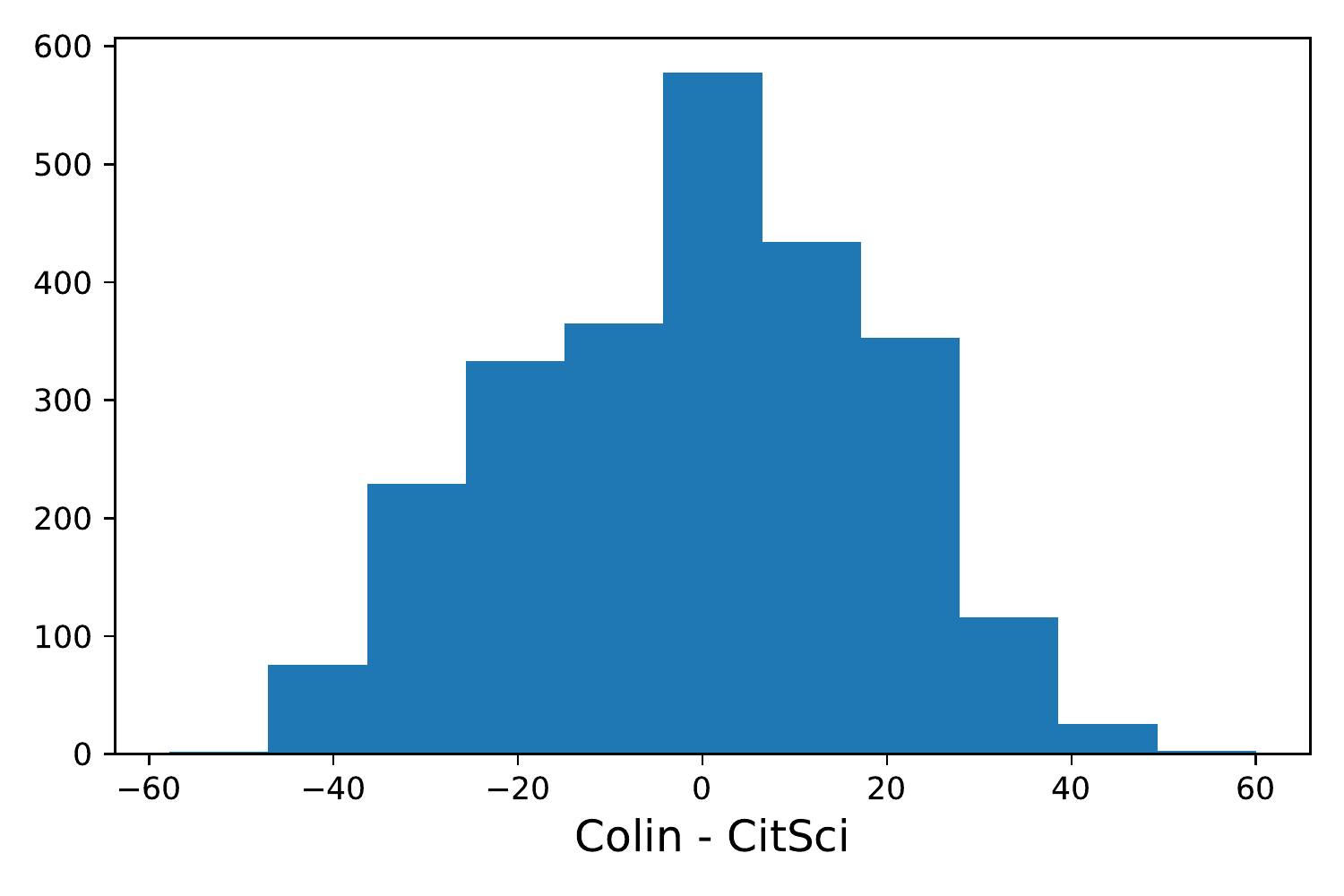}
    \caption{Agreement between the percentage of Citizen Scientists that identified a thumbnail as showing activity, and my own measure of how active an object appears in an image. Zero represents complete agreement, negative numbers indicates a high fraction of volunteers identified a thumbnail as active as compared to my score, and positive numbers indicate that my activity score was higher than the fraction of volunteers who classified the image as showing activity.}
    \label{discussion:fig:colinVsCitSci}
\end{figure}

We plan to derive many useful data products from the \textit{Active Asteroids} project. For example, a training dataset stems from the images we classified for the project training set (Chapter \ref{methods:subset:trainingSet}). This dataset will have two dimensions: (1) my own scores (described in Chapter \ref{methods:subset:trainingSet}), and (2) high numbers ($>1000$) of volunteer classifications (because training data are classified many times). Another data product will contain classification statistics accompanying all images submitted for examination. These types of products will be of use for for surveys such as the \ac{LSST} \citep{verac.rubinobservatorylsstsolarsystemsciencecollaborationScientificImpactVera2021}. For example, an outstanding need is for tools capable of automatically detecting activity, especially for petabyte-scale datasets \citep{kelleyCommunityChallengesEra2021}. \Ac{ML}--assisted activity detection tools can be trained using products resulting from our own classifications as well as those of Citizen Science volunteers \citep{breivikDataSoftwareScience2022}, though it is worth mentioning that the quality of the data are important to the effectiveness of \ac{ML}-based algorithms \citep{kuminskiCombiningHumanMachine2014}. Figure \ref{discussion:fig:colinVsCitSci} shows relative agreement between the two systems (expert inspection, Section \ref{methods:subset:trainingSet}) and the fraction of volunteers that classified a training thumbnail as active.

\begin{figure}
    \centering
    \includegraphics[width=0.45\linewidth]{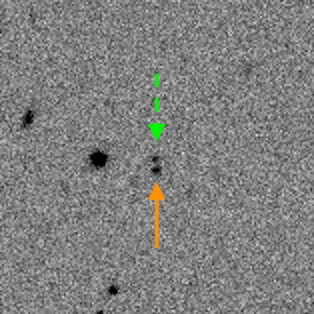}
    \caption{
    The two components (green dashed and solid orange arrows) of \acf{TNO} 2012~KU$_{50}$ are fully resolved in this 600~s \textit{VR} filter \ac{DECam} image from UT 2014 March 27 (Prop. ID 2014A-0479, PI Sheppard). 
    }
    \label{discussion:fig:additionalAvenues}
\end{figure}

An avenue of inquiry stemming from \textit{Active Asteroids} preparations unexpectedly supplementing of our original science goals is the discovery of companions to minor planets, including \acp{TNO} and Jupiter Trojans. \ac{TNO} binaries are of special interest because they hold potential clues to the formation of the trans-Neptunian region and the solar system as a whole. At present we are preparing a paper about one such discovery \citep{chandler5540992012KU502021} made during project preparations, that of \ac{TNO} 2012~KU$_{50}$ (Figure \ref{discussion:fig:additionalAvenues}). Companions are readily distinguishable in \ac{DECam} data, especially for widely separated objects in good seeing conditions. Binaries are especially prominent in image sequences of a given object, thus one potential avenue to facilitate companion discovery would be a Citizen Science workflow that presents animated \ac{GIF} files to volunteers and asks if they see a co-moving companion to the object at the center of the images.

\section{Project Status and Assessment}
\label{discussion:sec:projStatus}

\subsection{Volatile Distribution}
\label{discussion:subsec:volatiles}

Here we consider the present-day volatile distribution and delivery of water to Earth. This is by no means meant to be an exhaustive review or in-depth investigation; rather, this section serves as a rough guide. A word of caution: as stated in Chapter \ref{chap:intro}, the purpose of this dissertation and project is to discover more active objects to enable meaningful study of active objects as populations; at present, the numbers remain low (see Section \ref{discussion:subsubsec:activeAsteroidCompleteness}). As an example, the 1 in 10,000 occurrence rate of active objects in the Asteroid Belt has been extrapolated from studies as small as $\sim$10,000 minor planets \citep{chandlerSAFARISearchingAsteroids2018}, less than 1\% of the Asteroid Belt population.

\subsubsection{Terrestrial Water Origins}
\label{discussion:subsubsec:volatilesOnEarth}

We start by considering the total mass of water on Earth today, $M_{\mathrm{W},\oplus}$. We can roughly estimate this quantity via

\begin{equation}
    M_{\mathrm{W},\oplus} \approx M_{\mathrm{W},\mathrm{accretion}} + M_{\mathrm{W},\mathrm{delivered}} - M_{\mathrm{W},\mathrm{lost}},
\end{equation}

\noindent where $M_{\mathrm{W},\mathrm{accretion}}$ is the mass of water on Earth that formed \textit{in situ} during accretion, $M_{\mathrm{W},\mathrm{lost}}$ is the mass of water lost from Earth (due to, for example, a large-scale impact event such as the one that formed the Moon), and the delivered water mass $M_{\mathrm{W},\mathrm{delivered}}$, described by

\begin{equation}
    M_{\mathrm{W},\mathrm{delivered}} \approx \sum_{s=1}^{s=n} N_s \cdot \frac{4}{3}\pi \left(\bar{r}_s\right)^3 \cdot \bar{\rho}_s \cdot \bar{f}_{s,\mathrm{W}} \cdot f_{s,\mathrm{impact}},
\end{equation}

\noindent where, for a given source class $s$ (e.g., comets, asteroids
), $N_s$ is the number of objects, $\bar{r}_s$ is the average radius, $\bar{\rho}_s$ is the average density, $\bar{f}_{s,\mathrm{W}}$ is the average fraction of material that is water, and $f_{s,\mathrm{impact}}$ the fraction of those objects that impacted Earth through present-day. A more robust treatment would take into account the size, mass, and compositional distributions through time for each of these elements, but these are very poorly known and they are outside the scope of this dissertation.

Our efforts focus primarily on active body populations as volatile sources ($N_s$), so for the purpose of a broad examination of an active body class let us only consider the modern Asteroid Belt. Modern Earth holds around $5\times10^{-4} M_\oplus$ of water (e.g., \citealt{morbidelliSourceRegionsTime2000} and references therein), where $1M_\oplus\approx 6 \times10^{24}$~kg. The most massive active asteroid, (1) Ceres, is also the most massive body in the Asteroid Belt, with a mass of $1.6\times10^{-4} M_\oplus$. If active asteroids are composed of 50\% water (probably a very generous estimate), then the mass of water on Ceres would be $M_\mathrm{W,Ceres}\approx 0.8\times10^{-4} M_\oplus$. The mass of water on Earth could be measured as $M_{\mathrm{W},\oplus}\approx6M_\mathrm{W,Ceres}$. Considering the hypothesized population of $\sim$100 active asteroids ($\sim$1/10,000 of the $>1$~km population), for them to contain enough water to account for all of modern Earth's water they would each need to be roughly 6\% the mass of Ceres. These bodies would be comparable in size to (10)~Hygeia ($\sim$300~km diameter), the fourth largest asteroid in the solar system. To date the vast majority of recurrently active sublimation-driven active asteroids have diameters $< 10$~km (e.g., 133P/Elst-Pizarro at 3.8~km; \citealt{hsiehAlbedosMainBeltComets2009}), with the notable exception of (1) Ceres. This calculation, while rudimentary, nevertheless leaves little doubt that the active asteroids alone cannot at present hold enough water to supply all of the terrestrial water found on Earth today. However, active asteroids may have played a more significant role in the past given its much larger (150 to 250 times) population in the primordial solar system \citep{bottkeFossilizedSizeDistribution2005} .

\subsection{Citizen Science Project}
\label{discussion:subsec:citSciStatus}

The NASA Partner project \textit{Active Asteroids} was launched on the Citizen Science platform Zooniverse on 2022 August 31. Here we discuss the overall status of the project, make some rough estimates about what the project can deliver based on performance thus far, and consider potential enhancements to the project.

\subsubsection{Thumbnail Classification Rate and Completeness}
\label{discussion:subsubsec:classificationRate}

As of this writing, classifications have taken place at approximate rates varying between 12,000 classifications/day and 120,000 classifications/day. The exact cause of the variation is as-yet undetermined, but factors may include media coverage, social media activity (e.g., tweets from \@ActiveAsteroids), time of year (e.g., summer break), and timing of electronic newsletters to project volunteers. Our goal is to increase participation over time, and we hope the upcoming publications stemming from the project, including peer-reviewed journal articles and press releases, will help. Meanwhile we assess completeness making use of the aforementioned rates.

\paragraph{Time to Complete Existing Thumbnails} At present \ac{HARVEST} has produced about 18 million algorithmically vetted thumbnail images. Given each thumbnail is examined by 15 volunteers, the total number of classifications needed to examine the entire dataset is $15\times1.8\times10^7=2.7\times10^8$ classifications. At the maximum rate (120,000 classifications/day) this works out to 2,250 days, or about six years.

\paragraph{Staying Current} \ac{DECam} started operations in 2012 September, and data are released regularly through the \ac{NOIRLab} AstroArchive. The average number of vetted thumbnails produced per day is $\sim$5000, averaged over all dates, not just dates the telescope acquired data. It is worth noting that, in addition to normal weather and engineering telescope time losses, \ac{COVID-19} restrictions led to telescope shutdown for an extended period in 2020, thus reducing the overall average data output. At the maximum classification rate (120,000 classifications/day), it would require 15 hours of volunteer classifications per day stay current.

\paragraph{Time to Completion and Staying Current} Considering \textbf{Staying Current} would require 15 hours/day at the maximum classification rate, only nine hours (0.375 days) remain daily for classifying the existing data pool, or 120,000 classifications/day $\times$ 0.375 = 45,000 classifications/day. Each image is classified 15 times, so the rate in images/day is 3000. To classify. all 18 million thumbnails at this rate would require $>$16 years. Afterwards, the project would be able to stay current with \ac{DECam} output at 75,000 classifications/day, lower than the maximum classification rate seen to date.

\paragraph{Triage Options} Parameters used to select thumbnails for Citizen Science classification (Section \ref{methods:subsubsec:thumbnailSelection}) begin with constraining ranges of $\Delta_\mathrm{mag}$, the number of magnitudes fainter an exposure probes than the apparent magnitude of the minor planet (recall $\Delta_\mathrm{mag} = V_\mathrm{JPL} - V_\mathrm{ITC}$, with $V_\mathrm{JPL}$ the Horizons-provided apparent $V$-band magnitude for the object and $V_\mathrm{ITC}$ the $V$-band magnitude depth we computed for the exposure; Section \ref{methods:subsubsec:deltaMagLim}). The number of existing and new thumbnails needing classification can be reduced significantly by altering this parameter alone. For example, requiring $\Delta_\mathrm{mag}<-2$ (instead of the default $\Delta_\mathrm{mag}<-1$) reduces the number of thumbnails to be classified from 18 million to 16 million. Requiring $\Delta_\mathrm{mag}<-3$ lowers the number of thumbnails for classification to 9 million, a 50\% reduction in the classification workload as compared to the original 18 million thumbnails. This action would exacerbate the existing bias towards observing bright objects which, in turn, favors images of objects close to perihelion and disfavors distant objects such as Centaurs. For example, assuming the typical magnitude limit of \ac{DECam} is $\sim V=23$ (Chapter \ref{chap:SAFARI}), requiring $\Delta_\mathrm{mag}<-3$ results in images showing objects brighter than $V=20$. 

\begin{table}
    \centering
    \caption{Classification Rates and Potential Optimizations}
    \label{discussion:tab:ratesAndCompleteness}
    \begin{tabular}{c}
    \textbf{This Work}\\
    \begin{tabular}{ccrrrllrr}
    Description    & $N_\mathrm{current}$ & $t_\mathrm{current,min}$ & $t_\mathrm{current,ave}$ & $N_\mathrm{daily}$ & $t_\mathrm{daily,min}$ & $t_\mathrm{daily,ave}$ & $t_\mathrm{all,min}$ & $t_\mathrm{all,ave}$ \\
                    & & [days] & [days] & & [days] & [days] & [days] & [days]\\
    \hline
    Vetted & 1.57E+7     & 1960        & 23500        & 4300 & 0.54        & 6.51$^b$        & 4280                 & $\infty$\\
    $\Delta_\mathrm{mag}<-2$ & 1.11E+7     & 1380        & 16600        & 3073 & 0.384        & 4.61$^b$        & 2250                 & $\infty$\\
    $\Delta_\mathrm{mag}<-3$ & 6.10E+6     & 763        & 9150        & 1693  & 0.212        & 2.54$^b$        & 967                 & $\infty$ \\
    $\%_{q\rightarrow Q} \ge$ 80\% & 2.16E+6     & 270        & 3240        & 600 & 0.075        & 0.9        & 292                 & 32500 \\
    Wedge Phot.$^a$    & 2.16E+5     & 27        & 324        & 60 & 7.50E-03        & 0.09        & 27                 & 356                    \\
    Mach. Learn.$^a$             & 2.16E+3     & 0.27        & 3        & 0.60 & 7.50E-05        & 0.0009        & 0.27                 & 3.25  
    \end{tabular}\\
    \\
    \textbf{LSST}\\
        \begin{tabular}{ccrr}
        Description & $N_\mathrm{daily}$ & $t_\mathrm{daily,min}$ & $t_\mathrm{daily,ave}$\\
         & & [days] & [days]\\
        \hline
        \acs{LSST} baseline                    & 1.36E+7 & 1700$^b$ & 20400$^b$ \\
        \acs{HARVEST} Vetting                     & 4.08E+6 & 510$^b$ & 6120$^b$ \\
        $\Delta_\mathrm{mag}<-2$             & 2.89E+6 & 361$^b$ & 4330$^b$ \\
        $\Delta_\mathrm{mag}<-3$             & 1.59E+6 & 199$^b$ & 2386$^b$ \\
        $\%_{q\rightarrow Q} \ge$ 80\%          & 5.65E+5 & 71$^b$  & 846 $^b$ \\
        Wedge Phot.$^a$                             & 5.65E+5 & 7$^b$   & 85$^b$   \\
        Mach. Learn.$^a$                                & 5.65E+3 & 0.1    & 0.8
        \end{tabular}
\end{tabular}
\\
    \raggedright $N_\mathrm{current}$ indicates the number of \ac{DECam} images in the \ac{HARVEST}-produced database. Daily indicates the average number of thumbnail images produced on a daily basis. Minimum (min) and average (ave) values estimated based on peak (120,000) and mean (10,000) classifications per day, respectively. Vetted: the number of \ac{DECam} thumbnail images after vetting by the \acf{HARVEST} pipeline (Section \ref{methods:sec:methods:pipeline}). $\Delta_\mathrm{mag}$ is how faint an exposure probes versus the apparent magnitude of the object (Section \ref{methods:subsubsec:deltaMagLim}); relative brightness of the object in contrast with the overall exposure depth increases in the negative direction. $^a$Not easily implemented. $^b$Will never catch up. Hypothetical application of the Wedge Photometry is a technique that searches for tails (Section \ref{QN:fig:wedgephot}). Hypothetical application of \acf{ML}-based activity detection technique. The baseline number of \acf{LSST} thumbnails per night, estimated from an average of 8,000 minor planets per field, with 1,700 fields per night, depending on final cadence selection.
\end{table}

Table \ref{discussion:tab:ratesAndCompleteness} shows the impact of applying different optimization parameters to the \ac{HARVEST}-produced thumbnail images as well as rough estimates for implications for the \ac{LSST}. The number of vetted images in the \ac{HARVEST} dataset is roughly 16 million as of these calculations. Applying a $\Delta_\mathrm{mag}<-3$ constraint reduces the number of images by a factor of two, and likewise if we only allow thumbnail images showing objects at or above 80\% of the distance from aphelion to perihelion $\%_{q\rightarrow Q}$ (Section \ref{og:eq:percentperi}). To illustrate potential benefits of additional screening tools, a hypothetical factor of ten improvement is shown for Wedge Photometry (Section \ref{QN:fig:wedgephot}) application, and a factor of 100$\times$ improvement for an as yet undeveloped \ac{ML}-based activity detection tool. Even at our average classification rate of 10,000 classifications/day, the entire \ac{DECam} dataset would be processed and current in $\sim$3.25 days.

A hypothetical scenario for \ac{LSST} is also shown in Table \ref{discussion:tab:ratesAndCompleteness}. Although the final cadence has yet to be confirmed, we compute a baseline number of thumbnails from $\sim$1,700 fields imaged per night with an average of $\sim$8,000 minor planets per field \citep{ivezicLSSTScienceDrivers2019} as $1400\ \mathrm{fields/night}\ \cdot 8000\ \mathrm{objects/field}\ = 1.36\times10^7$ thumbnail images per night. \ac{HARVEST} has an approximate 70\% reduction in thumbnail images following our vetting routines, which still leaves over 45 million minor planets imaged per night. The only scenario in which the Citizen Scientists can keep up with the \ac{LSST} is if the classifications are kept at the rate of 120,000 classifications/night and all of the hypothetical vetting tools provide a combined 1,000$\times$ reduction in the number of thumbnail images needing examination over this work. It is important to note that classification should commence immediately because it will be highly impractical to ``catch up'' after \ac{LSST} has started acquiring data. It is also worth remembering that all times are dependent on classification rates, thus improvements would result from increased participation in the Citizen Science project.

\paragraph{Human Lifetimes} Classification requires time contributed by human volunteers, so we take care to optimize the time spent on our project (e.g., excluding thumbnails with an object we compute to be too faint to identify activity; Section \ref{methods:subsubsec:deltaMagLim}). One measure of how much time is required for a given task is \textit{human lifetimes}, which is 73.4 years as of 2019 \citep{worldhealthorganizationWHOMethodsData2019}, or $\sim4\times10^7$~s. Classifications take, on average, 4.7~s per thumbnail. Assuming very generously that each person is awake and classifying for 2/3 of their day, and classifies for 70 years of their life, then

\begin{equation}
    \frac{70\ \mathrm{yr}}{\mathrm{life}} \cdot\frac{5\ \mathrm{classifications}}{\mathrm{s}} \cdot\frac{365 \mathrm{d}}{1~\mathrm{yr}} \cdot{\frac{2}{3}}\cdot \frac{24\ \mathrm{hr}}{\mathrm{d}} \cdot\frac{60\ \mathrm{min}}{\mathrm{hr}}\cdot\frac{60\ \mathrm{s}}{\mathrm{min}} \approx \frac{10^8\ \mathrm{classifications}}{\mathrm{life}},
\end{equation}

\noindent or roughly 100 million classifications per lifetime. Via Table \ref{discussion:tab:ratesAndCompleteness}, \textit{Active Asteroids} \ac{DECam} data could be completed in $\sim0.1$ lifetimes (7 yr). However, \ac{LSST} in the same \ac{HARVEST} vetted state would require, on average, $\sim0.03$ lifetimes per day to classify all of the thumbnail images. Given the nominal survey duration for \ac{LSST} is 10 years (3,650 days), over 120  human lifetimes would be required to classify all of the \ac{HARVEST} vetted thumbnail images. Even with the $\Delta_\mathrm{mag}<-3$ and $\%_{q\rightarrow Q} \ge$ 80\% filters applied, the \ac{LSST} data would still require 16 lifetimes (at 565,000 thumbnails/day for 10 years) without the as-yet unimplemented Wedge Photometry and/or \ac{ML} filters.

\subsubsection{Active Asteroid Detection Rate and Completeness}
\label{discussion:subsubsec:activeAsteroidCompleteness}

As discussed in Chapter \ref{chap:intro}, the occurrence rate of activity among asteroids (due to any cause) is estimated to be very roughly around 1 in 10,000. Considering only the Asteroid Belt, there are roughly 1 million asteroids $>1$~km in diameter (e.g., \citealt{tedescoInfraredSpaceObservatory2002}). This suggests there should be roughly 100 active asteroids in the Asteroid Belt alone. As of this writing about 20 have been observed with visual evidence of activity, so there should be about 80 undiscovered active asteroids in the Asteroid Belt.

Thus far \textit{Active Asteroids} volunteers have discovered one previously unknown active asteroid (paper in preparation) out of the $\sim$78,000 Asteroid Belt objects examined. While this seems to be a rate lower than 1 in 10,000, we have (so far) provided images of objects without selecting for proximity to perihelion passage, so it is likely one or more active objects have been examined but during their presumably quiescent period near aphelion. If we continue to provide images unbiased with respect to orbital position, then volunteers would need to examine 6.2 million thumbnails before identifying 80 active asteroids. Given current classification rates (Section \ref{discussion:subsubsec:classificationRate}) we intend to start prioritizing objects near perihelion passage, so we expect the detection frequency to increase.

\subsubsection{Project Plans}
\label{discussion:projectPlans}

A large number of discoveries stem from the \textit{Active Asteroids} project and \ac{HARVEST} pipeline, such as four published peer-reviewed publications thus far, one article currently in review, and three additional manuscripts are in preparation. These results make it clear the project is successful and should continue. Moreover, the public engagement aspect of the project should not be overlooked, with over 6,000 volunteers participating to date. As discussed in Section \ref{discussion:subsubsec:classificationRate}, the project could use additional participation to achieve completeness sooner, but adjusting thumbnail selection parameters can help in the interim. \ac{HARVEST} has already been upgraded to incorporate other instrument archives, such as the \ac{CFHT} MegaPrime. It is evident that additional selection criteria are necessary to further filter the thumbnail images selected for \textit{Active Asteroids} in order to keep the quantity viable with current classification rates.

One possible way to enhance thumbnail selection for Citizen Scientist examination would be to introduce additional automated detection techniques that identify possible activity and, in combination, result in an activity likelihood score. This will require combining existing detection techniques, such as \ac{PSF} analysis (Section \ref{intro:sec:activityDiscovery}), Wedge Photometry (Section \ref{QN:subsec:wedgephotometry}), and \ac{ML}-based image recognition informed by classification data from \textit{Active Asteroids}. Without this additional activity likelihood vetting, projects that produce massive amounts of data, such as \ac{LSST}, will not be able to benefit fully from Citizen Science projects like \textit{Active Asteroids}.

It is worth noting that the project did take several years to develop, and that involvement by this author and others were critical to project success. The vast majority of work on the \ac{HARVEST} pipeline, the Citizen Science project, the follow-up archival and telescope observations, are analyses performed by one individual (this author), with assistance provided by at least one additional member of the science team (Section \ref{methods:sec:citsci}) at any given time. For ongoing operations, project moderators (Section \ref{methods:subsec:talk}) play a key role in interfacing with volunteers through online forums. Citizen Science project development required additional efforts and involved one additional person (Jay Kueny), and development of an analysis pipeline to optimize classification data evaluation was carried out by another individual (Will Burris). The high volume of \textit{Active Asteroids} discoveries warrants additional full-time commitment by at least one more person, bringing the number of full-time dedicated individuals needed to run the project efficiently to at least two. More would be necessary if, for example, development of new methods are needed.

\subsubsection{Next Steps}
\label{discussion:nextSteps}
The multi-pronged approach to finding and characterizing active solar system bodies presented in this work has proven effective and continues to yield results. Discoveries stemming from the \textit{Active Asteroids} project arise each time we upload a new subject set, with seven batches completed as of this writing. At an activity occurrence rate of roughly one in 10,000 asteroids, there should be around 100 active asteroids in the Asteroid Belt. We hope to double the number of known active asteroids within the next two years by finding an additional 30 active asteroids. This will allow for a more comprehensive study of active bodies, thus enabling the small body community to draw meaningful conclusions about the solar system volatile distribution, the potential volume of volatiles held within each reservoir, and the implications for the origin of terrestrial water. Anyone who can see images on an internet-connected device can participate in \textit{Active Asteroids} by visiting \url{http://activeasteroids.net}.

%
\clearpage
\singlespacing
\chapter{Acronyms}
\label{chap:acronyms}
\begin{acronym}
\acro{API}{Application Programming Interface}
\acro{APT}{Aperture Photometry Tool}
\acro{ARO}{Atmospheric Research Observatory}
\acro{ASU}{Arizona State University}
\acro{ASU}{Arizona Statue University}
\acro{AURA}{Association of Universities for Research in Astronomy}
\acro{AstOrb}{Asteroid Orbital Elements Database}
\acro{BLT}{Barry Lutz Telescope}
\acro{CADC}{Canadian Astronomy Data Centre}
\acro{CASU}{Cambridge Astronomy Survey Unit}
\acro{CATCH}{Comet Asteroid Telescopic Catalog Hub}
\acro{CBAT}{Central Bureau for Astronomical Telegrams}
\acro{CBET}{Central Bureau for Electronic Telegrams}
\acro{CCD}{charge-coupled device}
\acro{CEA}{Commissariat a l'Energes Atomique}
\acro{CFHT}{Canada France Hawaii Telescope}
\acro{CFHT}{Canada-France-Hawaii Telescope}
\acro{CNRS}{Centre National de la Recherche Scientifique}
\acro{COVID-19}{corona virus disease 2019} 
\acro{CSBN}{Committee for Small Bodies Nomenclature}
\acro{CTIO}{Cerro Tololo Inter-American Observatory}
\acro{DAPNIA}{Département d'Astrophysique, de physique des Particules, de physique Nucléaire et de l'Instrumentation Associée}
\acro{DART}{Double Asteroid Redirection Test}
\acro{DCT}{Discovery Channel Telescope}
\acro{DDT}{Director's Discretionary Time}
\acro{DECam}{Dark Energy Camera}
\acro{DESTINY+}{Demonstration and Experiment of Space Technology for INterplanetary voYage with Phaethon fLyby and dUst Science}
\acro{DES}{Dark Energy Survey}
\acro{DiRAC}{Data Intensive Research in Astrophysics \& Cosmology}
\acro{DPRC}{Disability Program Resource Center}
\acro{DR}{Data Release}
\acro{DS9}{Deep Space Nine}
\acro{Dec}{Declination}
\acro{ELTE}{Eotvos Lorand University}
\acro{ESA}{European Space Agency}
\acro{ESO}{European Space Organization}
\acro{ETC}{exposure time calculator}
\acro{FAQ}{Frequently Asked Questions}
\acro{FITS}{Flexible Image Transport System}
\acro{FOV}{field of view}
\acro{GEODSS}{Ground-Based Electro-Optical Deep Space Surveillance}
\acro{GIF}{Graphic Interchange Format}
\acro{GMOS}{Gemini Multi-Object Spectrograph}
\acro{GRFP}{Graduate Research Fellowship Program}
\acro{HARVEST}{Hunting for Activity in Repositories with Vetting-Enhanced Search Techniques}
\acro{HPC}{High Performance Computing}
\acro{HSC}{Hyper Suprime-Cam}
\acro{IAS15}{Integrator with Adaptive Step-size control, 15th order}
\acro{IAU}{International Astronomical Union}
\acro{IMACS}{Inamori-Magellan Areal Camera and Spectrograph}
\acro{IMB}{inner Main-belt}
\acro{IMCCE}{Institut de Mécanique Céleste et de Calcul des Éphémérides}
\acro{INT}{Isaac Newton Telescopes}
\acro{IPAC}{Infrared Processing and Analysis Center}
\acro{IP}{Internet Protocol}
\acro{IRSA}{Infrared Science Archive}
\acro{ITC}{integration time calculator}
\acro{JAXA}{Japan Aerospace Exploration Agency}
\acro{JD}{Julian Date}
\acro{JFC}{Jupiter Family Comet}
\acro{JPL}{Jet Propulsion Laboratory}
\acro{KBO}{Kuiper Belt object}
\acro{KIDS}{Kilo-Degree Survey}
\acro{KOA}{Keck Observatory Archive}
\acro{KPNO}{Kitt Peak National Observatory}
\acro{LBCB}{Large Binocular Camera Blue}
\acro{LBCR}{Large Binocular Camera Red}
\acro{LBC}{Large Binocular Camera}
\acro{LBT}{Large Binocular Telescope}
\acro{LCO}{Los Cumbres Observatory}
\acro{LCOGT}{Las Cumbres Observatory Global Telescope}
\acro{LDT}{Lowell Discovery Telescope}
\acro{LINEAR}{Lincoln Near-Earth Asteroid Research}
\acro{LMI}{Large Monolithic Imager}
\acro{LONEOS}{Lowell Observatory Near-Earth-Object Search}
\acro{LSST}{Legacy Survey of Space and Time}
\acro{MBC}{Main-belt Comet}
\acro{MGIO}{Mount Graham International Observatory}
\acro{ML}{machine learning}
\acro{MMB}{middle Main-belt}
\acro{MOST}{Moving Object Search Tool}
\acro{MPC}{Minor Planet Center}
\acro{NASA}{National Aeronautics and Space Administration}
\acro{NAU}{Northern Arizona University}
\acro{NEAT}{Near-Earth Asteroid Tracking}
\acro{NEA}{near-Earth asteroid}
\acro{NEO}{near-Earth object}
\acro{NIHTS}{Near-Infrared High-Throughput Spectrograph}
\acro{NOAO}{National Optical Astronomy Observatory}
\acro{NOIRLab}{National Optical and Infrared Laboratory}
\acro{NONCOM}{Not Orbitally a Nominal Comet but Overtly a Minor planet}
\acro{NRC}{National Research Council}
\acro{NSF}{National Science Foundation}
\acro{OMBA}{outer main-belt asteroid}
\acro{OMB}{outer Main-belt}
\acro{OSIRIS-REx}{Origins, Spectral Interpretation, Resource Identification, Security, Regolith Explorer}
\acro{PAC}{Physics and Astronomy Club}
\acro{PANSTARRS}{Panoramic Survey Telescope and Rapid Response System}
\acro{PI}{Principal Investigator}
\acro{PNG}{Portable Network Graphics}
\acro{PS1}{Panoramic Survey Telescope and Rapid Response System 1}
\acro{PSF}{point spread function}
\acro{PSI}{Planetary Science Institute}
\acro{PTF}{Palomar Transient Factory}
\acro{Pan-STARRS}{Panoramic Survey Telescope and Rapid Response System}
\acro{QHA}{Quasi-Hilda Asteroid}
\acro{QHC}{Quasi-Hilda Comet}
\acro{QHO}{Quasi-Hilda Object}
\acro{RA}{Right Ascension}
\acro{SAFARI}{Searching Asteroids For Activity Revealing Indicators}
\acro{SAO}{Smithsonian Astrophysical Observatory}
\acro{SBDB}{Small Body Database}
\acro{SDSS DR-9}{Sloan Digital Sky Survey Data Release Nine}
\acro{SDSS}{Sloan Digital Sky Survey}
\acro{SDSU}{San Diego State University}
\acro{SFSU}{San Francisco State University}
\acro{SLAC}{Stanford Linear Accelerator Center}
\acro{SMOKA}{Subaru Mitaka Okayama Kiso Archive}
\acro{SNR}{signal to noise ratio}
\acro{SOAR}{Southern Astrophysical Research Telescope}
\acro{SOFIA}{Stratospheric Observatory for Infrared Astronomy}
\acro{SQL}{Structured Query Language}
\acro{SSOIS}{Solar System Object Information Search}
\acro{STScI}{Space Telescope Science Institute}
\acro{SUP}{Suprime Cam}
\acro{SwRI}{Southwestern Research Institute}
\acro{TNO}{Trans-Neptunian object}
\acro{TRIF}{Technology and Research Initiative Fund} 
\acro{TSIP}{Telescope System Instrumentation Program}
\acro{TTU}{Texas Tech University}
\acro{UA}{University of Arizona}
\acro{UCB}{University of Colorado Boulder}
\acro{UCLA}{University of California Los Angeles}
\acro{UCSC}{University of California Santa Cruz}
\acro{UCSD}{University of California San Diego}
\acro{UCSF}{University of California San Francisco}
\acro{UT}{Universal Time}
\acro{UW}{University of Washington}
\acro{VATT}{Vatican Advanced Technology Telescope}
\acro{VIRCam}{VISTA InfraRed Camera}
\acro{VISTA}{Visible and Infrared Survey Telescope for Astronomy}
\acro{VLT}{Very Large Telescope}
\acro{VST}{Very Large Telescope (VLT) Survey Telescope}
\acro{WCS}{World Coordinate System}
\acro{WFC}{Wide Field Camera}
\acro{WGSBN}{Working Group for Small Bodies Nomenclature}
\acro{WIRCam}{Wide-field Infrared Camera}
\acro{WISE}{Wide-field Infrared Survey Explorer}
\acro{YORP}{Yarkovsky--O'Keefe--Radzievskii--Paddack}
\acro{ZTF}{Zwicky Transient Facility}
\end{acronym}
%
\singlespacing
\bibliography{ColinPhD.bib}

\end{document}